\documentclass[prd,showpacs,showkeys,floatfix,onecolumn,amsmath,amssymb,floatfix]{revtex4}
\usepackage{graphicx,color,dcolumn,booktabs,bm}
\usepackage{longtable,lscape}
\usepackage{amssymb}
\usepackage{indentfirst}
\usepackage{feynmf}   
\usepackage{epstopdf}   
\usepackage{slashed}  
\usepackage{cases}
\usepackage{color}
\usepackage{multirow}
\usepackage[colorlinks,citecolor=blue,anchorcolor=red,menucolor=red,linkcolor=red,filecolor=red,runcolor=red,urlcolor=blue,frenchlinks=red]{hyperref}
\usepackage{dcolumn}
\usepackage{bm}
\usepackage{enumerate}
\allowdisplaybreaks[4]

\begin{document}

\title{Decay properties of $P$-wave charmed baryons from light-cone QCD sum rules}

\author{Hua-Xing Chen$^1$}
\author{Qiang Mao$^{1,2}$}
\author{Wei Chen$^3$}
\email{wec053@mail.usask.ca}
\author{Atsushi Hosaka$^{4,5}$}
\email{hosaka@rcnp.osaka-u.ac.jp}
\author{Xiang Liu$^{6,7}$}
\email{xiangliu@lzu.edu.cn}
\author{Shi-Lin Zhu$^{8,9,10}$}
\email{zhusl@pku.edu.cn}
\affiliation{
$^1$School of Physics and Beijing Key Laboratory of Advanced Nuclear Materials and Physics, Beihang University, Beijing 100191, China \\
$^2$Department of Electrical and Electronic Engineering, Suzhou University, Suzhou 234000, China\\
$^3$Department of Physics and Engineering Physics, University of Saskatchewan, Saskatoon, Saskatchewan, S7N 5E2, Canada \\
$^4$Research Center for Nuclear Physics, Osaka University, Ibaraki 567--0047, Japan \\
$^5$ Advanced Science Research Center, Japan Atomic Energy Agency, Tokai, Ibaraki, 319-1195 Japan \\
$^6$School of Physical Science and Technology, Lanzhou University, Lanzhou 730000, China\\
$^7$Research Center for Hadron and CSR Physics, Lanzhou University and Institute of Modern Physics of CAS, Lanzhou 730000, China\\
$^8$School of Physics and State Key Laboratory of Nuclear Physics and Technology, Peking University, Beijing 100871, China \\
$^9$Collaborative Innovation Center of Quantum Matter, Beijing 100871, China \\
$^{10}$Center of High Energy Physics, Peking University, Beijing 100871, China}

\begin{abstract}
We study decay properties of the $P$-wave charmed baryons using the method of light-cone QCD sum rules, including the $S$-wave decays of the flavor $\mathbf{\bar 3}_F$ $P$-wave charmed baryons into ground-state charmed baryons accompanied by a pseudoscalar meson ($\pi$ or $K$) or a vector meson ($\rho$ or $K^*$), and the $S$-wave decays of the flavor $\mathbf{6}_F$ $P$-wave charmed baryons into ground-state charmed baryons accompanied by a pseudoscalar meson ($\pi$ or $K$). We study both two-body and three-body decays which are kinematically allowed. We find two mixing solutions from internal $\rho$- and $\lambda$-mode excitations, which can well describe both masses and decay properties of the $\Lambda_c(2595)$, $\Lambda_c(2625)$, $\Xi_c(2790)$ and $\Xi_c(2815)$. We also discuss the possible interpretations of $P$-wave charmed baryons for the $\Sigma_c(2800)$, $\Xi_c(2930)$, $\Xi_c(2980)$, and the recently observed $\Omega_c(3000)$, $\Omega_c(3050)$, $\Omega_c(3066)$, $\Omega_c(3090)$, and $\Omega_c(3119)$.
\end{abstract}
\pagenumbering{arabic}
\maketitle

\section{Introduction}
\label{sec:intro}

Recently, the LHCb Collaboration observed five excited $\Omega_c$ states in the $\Xi_c^+ K^-$ mass spectrum~\cite{Aaij:2017nav}, i.e., the $\Omega_c(3000)^0$, $\Omega_c(3050)^0$, $\Omega_c(3066)^0$, $\Omega_c(3090)^0$, and $\Omega_c(3119)^0$.
Their masses and widths were measured to be
\begin{eqnarray*}
\Omega_c(3000)^0&:& M = 3000.4 \pm 0.2 \pm 0.1 ^{+0.3}_{-0.5}~{\rm MeV} \, , \, \Gamma = 4.5 \pm 0.6 \pm 0.3~{\rm MeV} \, ,
\\ \Omega_c(3050)^0&:& M = 3050.2 \pm 0.1 \pm 0.1 ^{+0.3}_{-0.5}~{\rm MeV} \, , \, \Gamma = 0.8 \pm 0.2 \pm 0.1~{\rm MeV} \, ,
\\ \Omega_c(3066)^0&:& M = 3065.6 \pm 0.1 \pm 0.3 ^{+0.3}_{-0.5}~{\rm MeV} \, , \, \Gamma = 3.5 \pm 0.4 \pm 0.2~{\rm MeV} \, ,
\\ \Omega_c(3090)^0&:& M = 3090.2 \pm 0.3 \pm 0.5 ^{+0.3}_{-0.5}~{\rm MeV} \, , \, \Gamma = 8.7 \pm 1.0 \pm 0.8~{\rm MeV} \, ,
\\ \Omega_c(3119)^0&:& M = 3119.1 \pm 0.3 \pm 0.9 ^{+0.3}_{-0.5}~{\rm MeV} \, , \, \Gamma = 1.1 \pm 0.8 \pm 0.4~{\rm MeV} \, .
\end{eqnarray*}
These excited $\Omega_c$ states are good $P$-wave charmed baryon candidates. Besides them, the $\Lambda_c(2595)$ ($J^P=1/2^-$), $\Lambda_c(2625)$ ($J^P=3/2^-$), $\Xi_c(2790)$ ($J^P=1/2^-$) and $\Xi_c(2815)$ ($J^P=3/2^-$)
can be well interpreted as the $P$-wave charmed baryons completing two flavor $\mathbf{\bar 3}_F$ multiplets of $J^P=1/2^-$ and $3/2^-$~\cite{pdg,Frabetti:1993hg,Albrecht:1993pt,Edwards:1994ar,Alexander:1999ud};
the $\Sigma_c(2800)$ ($J^P=?^?$), $\Xi_c(2930)$ ($J^P=?^?$) and $\Xi_c(2980)$ ($J^P=?^?$) are also $P$-wave charmed baryon candidates of the flavor $\mathbf{6}_F$~\cite{Mizuk:2004yu,Chistov:2006zj,Aubert:2007dt,Aubert:2008ax,Yelton:2016fqw,Kato:2016hca,Aaij:2017vbw}.

These charmed baryons are interesting in a theoretical point of view, and many phenomenological methods/models were proposed to study them~\cite{Chen:2016spr}, such as
various quark models~\cite{Capstick:1986bm,Ebert:2007nw,Garcilazo:2007eh,Ebert:2011kk,Ortega:2012cx,Shah:2016nxi},
various dynamical models~\cite{GarciaRecio:2012db,Liang:2014eba,Liang:2014kra,Liang:2016ydj,Liang:2016exm,Lu:2014ina},
the hyperfine interaction~\cite{Copley:1979wj,Karliner:2008sv},
and the Lattice QCD~\cite{Bowler:1996ws,Burch:2008qx,Brown:2014ena}, etc~\cite{Roncaglia:1995az,Jenkins:1996de,Roberts:2007ni,Chen:2014nyo,Lu:2016ctt}.
Their productions and decay properties were studied in Refs.~\cite{Cheng:2006dk,Chen:2007xf,Zhong:2007gp,Kim:2014qha,Nagahiro:2016nsx,Xie:2015zga,Huang:2016ygf}.
See reviews in Refs.~\cite{Korner:1994nh,Bianco:2003vb,Klempt:2009pi,Crede:2013sze,Cheng:2015rra,Chen:2016qju,Chen:2016spr} for their recent progress.

We have also systematically studied the mass spectra of these excited heavy baryons~\cite{Liu:2007fg,Chen:2015kpa,Mao:2015gya,Chen:2016phw},
using the method of QCD sum rules~\cite{Shifman:1978bx,Reinders:1984sr} in the framework of heavy quark effective theory (HQET)~\cite{Eichten:1989zv,Grinstein:1990mj,Falk:1990yz}.
The HQET works well in the bottom sector but not so good in the charm sector. Hence, in Refs.~\cite{Liu:2007fg,Chen:2015kpa,Mao:2015gya,Chen:2016phw} we have taken into account the ${\mathcal O}(1/m_Q)$ corrections ($m_Q$ is the heavy quark mass), and did find them to be non-negligible. In Ref.~\cite{Ivanov:1999bu} it was also found that the finite quark mass corrections to the form factors and the rates of semileptonic transitions are important for heavy-to-light (charm-to-strange) transitions and not negligible for heavy-to-heavy (bottom-to-charm) transitions. In a different framework based on the Dyson-Schwinger equation~\cite{Ivanov:1998ms} it was similarly concluded that the heavy-quark expansion is accurate for the bottom quark while it provides a poor approximation for the charm quark.
More studies on heavy mesons and baryons using this scheme can be found in Refs.~\cite{Shuryak:1981fza,Bagan:1991sg,Neubert:1991sp,Broadhurst:1991fc,Grozin:1992td,Neubert:1993mb,Bagan:1993ii,Dai:1993kt,Huang:1994zj,Dai:1995bc,Dai:1996yw,Dai:1996qx,Groote:1996em,Colangelo:1998ga,Lee:2000tb,Huang:2000tn,Wang:2003zp,Dai:2003yg,Zhou:2014ytp,Zhou:2015ywa},
and others using the method of QCD sum rules but not in HQET can be found in Refs.~\cite{Bagan:1991sc,Bagan:1992tp,Duraes:2007te,Wang:2007sqa,Chen:2015moa}.

Based on the heavy quark effective theory, we can classify the $P$-wave charmed baryons into eight charmed baryon multiplets, including four of the flavor $\mathbf{\bar 3}_F$ \big($[\mathbf{\bar 3}_F, 0, 1, \rho]$, $[\mathbf{\bar 3}_F, 1, 1, \rho]$, $[\mathbf{\bar 3}_F, 2, 1, \rho]$ and $[\mathbf{\bar 3}_F, 1, 0, \lambda]$\big) and four of the flavor $\mathbf{6}_F$ \big($[\mathbf{6}_F, 1, 0, \rho]$, $[\mathbf{6}_F, 0, 1, \lambda]$, $[\mathbf{6}_F, 1, 1, \lambda]$ and $[\mathbf{6}_F, 2, 1, \lambda]$\big). See Sec.~\ref{sec:input} for the explanations of these symbols.
For each set of multiples, the first one of $j_l = 0$ forms a heavy quark singlet, while the other three of $j_l = 1$ form heavy-quark doublets.
These multiplets provide lots of $P$-wave charmed baryons. For example, there can be as many as seven $P$-wave $\Omega_c$ states theoretically, including three $J^P=1/2^-$, three $J^P=3/2^-$, and one $J^P=5/2^-$ states. Previously, it seems impossible to observe all these $P$-wave $\Omega_c$ states experimentally. However, the recent LHCb experiment observed as many as five excited $\Omega_c$ states at the same time~\cite{Aaij:2017nav} (actually this number is six if the $\Omega_c(3188)^0$ is included), suggesting that
``{\it An ideal platform to study these structures (the gross, fine and hyperfine structures of the strong interaction) is the heavy hadrons containing one charm or bottom quark~\cite{Chen:2016spr}.}''

In this paper we shall further use the method of light-cone QCD sum rules~\cite{Balitsky:1989ry,Braun:1988qv,Chernyak:1990ag,Ball:1998je,Ball:2006wn,Ball:2004rg,Ball:1998kk,Ball:1998sk,Ball:1998ff,Ball:2007rt,Ball:2007zt,Wang:2007mc,Wang:2009hra,Aliev:2010uy,Sun:2010nv,Khodjamirian:2011jp,Han:2013zg,Offen:2013nma,Meissner:2013hya,Aliev:2016xvq} to study the decay properties of these $P$-wave charmed baryons, the method of which is also based on the heavy quark effective theory (HQET).
We shall only study their $S$-wave decay properties, and note that their $D$-wave decays can also happen but these contributions may not be significant.
Our sum rule calculations will be done separately for the above eight charmed baryon multiplets. However, because the heavy quark symmetry is not perfect, the physical states are probably mixed states containing various components with different inner quantum numbers. Hence, we shall also discuss the mixing of these charmed baryon multiplets in this paper, where we shall find that the decay properties of the $P$-wave charmed baryons are quite sensitive to this.
We refer to Refs.~\cite{Dai:1996xv,Zhu:2000py,Wei:2005ag,Huang:2009zy,Liu:2009sn,Huang:2009is,Huang:2010qa,Huang:2010dc} for earlier studies using the method of light-cone QCD sum rules in the framework of HQET.
The decay properties of heavy baryons have also been studied using many other methods, such as in Ref.~\cite{Hussain:1999sp} where the one-pion transitions between heavy baryons were investigated in the constituent quark model based on the heavy quark symmetry.

This paper is organized as follows.
First in Sec.~\ref{sec:input} we reevaluated and listed the input parameters for the present study, including the masses, decay constants, and interpolating fields of the ground-state and $P$-wave charmed baryons.
Then in Sec.~\ref{sec:sumrule1} we investigate the decay properties of the flavor $\mathbf{\bar 3}_F$ $P$-wave charmed baryons within the method of light-cone QCD sum rules, and we shall study their $S$-wave decays into ground-state charmed baryons accompanied by a pseudoscalar meson ($\pi$ or $K$) or a vector meson ($\rho$ or $K^*$).
Using the same procedures, in Sec.~\ref{sec:sumrule2} we investigate the decay properties of the flavor $\mathbf{6}_F$ $P$-wave charmed baryons, and we shall study their $S$-wave decays into ground-state charmed baryons accompanied by a pseudoscalar meson ($\pi$ or $K$).
The results are summarized and discussed in Sec.~\ref{sec:summary}.

\section{Ground-state and $P$-wave charmed baryons}
\label{sec:input}

To study the decay properties of charmed baryons in the method of light-cone QCD sum rules,
we need some parameters of these states, such as their masses ($m_{\mathcal{B}}$), interpolating fields ($J^{\alpha_1\cdots\alpha_{j-1/2}}$), decay constants ($f_{\mathcal{B}}$), and threshold values ($\omega_c$), etc.
These parameters are defined below, while their values are listed separately in the following three subsections for both ground-state charmed baryons, $P$-wave charmed baryons of flavor $\mathbf{\bar 3}_F$ and $P$-wave charmed baryons of flavor $\mathbf{6}_F$.

The coupling of the interpolating field $J^{\alpha_1\cdots\alpha_{j-1/2}}(x)$ to the charmed baryon $\mathcal{B}$ of spin $j$ is defined to be
\begin{eqnarray}
\langle 0| J^{\alpha_1\cdots\alpha_{j-1/2}}(x) | \mathcal{B} \rangle = f_{\mathcal{B}} u^{\alpha_1\cdots\alpha_{j-1/2}}(x) \, ,
\end{eqnarray}
where $f_{\mathcal{B}}$ is the decay constant, and $u^{\alpha_1\cdots\alpha_j}$ is the relevant spinor.

Then the two-point correlation function at the hadron level can be written as
\begin{eqnarray}
\Pi^{\alpha_1\cdots\alpha_{j-1/2},\beta_1\cdots\beta_{j-1/2}} (\omega) &=& i \int d^4 x e^{i k x} \langle 0 | T[J^{\alpha_1\cdots\alpha_{j-1/2}}(x) \bar J^{\beta_1\cdots\beta_{j-1/2}}(0)] | 0 \rangle
\label{eq:pole}
\\ \nonumber &=& \mathbb{S} [ g_t^{\alpha_1 \beta_1} \cdots g_t^{\alpha_{j-1/2} \beta_{j-1/2}} ] \times {1 + v\!\!\!\slash \over 2} \times \Pi_{F,j_l,s_l,\rho\rho/\lambda\lambda/\rho\lambda} (\omega) \, ,
\\ \nonumber &=& \mathbb{S} [ g_t^{\alpha_1 \beta_1} \cdots g_t^{\alpha_{j-1/2} \beta_{j-1/2}} ] \times {1 + v\!\!\!\slash \over 2} \times \Big( {f_{\mathcal{B}}^{2} \over \overline{\Lambda}_{\mathcal{B}} - \omega} + \mbox{higher states} \Big) \, .
\end{eqnarray}
Here $\mathbb{S} [\cdots]$ denotes symmetrization and subtracting the trace
terms in the sets $(\alpha_1 \cdots \alpha_{j-1/2})$ and $(\beta_1 \cdots
\beta_{j-1/2})$; $\omega$ is the external off-shell energy $\omega = v \cdot k$; $\overline{\Lambda}_{\mathcal{B}}$ is defined to be
\begin{eqnarray}
\overline{\Lambda}_{\mathcal{B}} \equiv \lim_{m_Q \rightarrow \infty} (m_{\mathcal{B}} - m_Q) \, ,
\end{eqnarray}
where $m_Q$ is the heavy quark mass, and $m_{\mathcal{B}}$ is the mass of the lowest-lying charmed baryon state coupling  with $J^{\alpha_1\cdots\alpha_{j-1/2}}(x)$.

At the quark-gluon level, one can calculate the two-point correlation function, Eq.~(\ref{eq:pole}), using the method of QCD operator product expansion. By assuming the contribution from the continuum states (higher states) can be approximated well by the OPE spectral density above a threshold value $\omega_c$, one can arrive at the mass sum rule relation which can be used to calculate masses and decay constants of charmed baryons.
See Refs.~\cite{Liu:2007fg,Chen:2015kpa,Mao:2015gya} for detailed discussions, and their results are reevaluated and listed below.

\subsection{Ground-state charmed baryons}

The masses and decay constants of the $S$-wave bottom baryons have been systematically investigated in Ref.~\cite{Liu:2007fg} using the method of QCD sum rules in HQET. We replace the bottom quark by the charm quark, reevaluate their parameters, and shortly summarize the results here. For completeness, we first list masses of the ground-state charmed baryons from PDG~\cite{pdg}:
\begin{eqnarray}
\nonumber \Lambda_c(1/2^+) &:& m = 2286.46 {\rm~MeV} \, ,
\\ \nonumber \Xi_c(1/2^+) &:& {m} = 2469.34 {\rm~MeV}  \, ,
\\ \nonumber \Sigma_c(1/2^+) &:& {m} = 2453.54 {\rm~MeV}  \, , {\Gamma} = 1.86 {\rm~MeV}  \, , {g}_{\Sigma_c \Lambda_c \pi} = 3.94~{\rm GeV}^{-1} \, ,
\\ \Xi_c^\prime(1/2^+) &:& {m} = 2576.8 {\rm~MeV}  \, ,
\\ \nonumber \Omega_c(1/2^+) &:& m = 2695.2 {\rm~MeV}  \, ,
\\ \nonumber \Sigma_c^*(3/2^+) &:& {m} = 2518.1 {\rm~MeV} \, , {\Gamma} = 14.7 {\rm~MeV} \, , {g}_{\Sigma_c^* \Lambda_c \pi} = 7.39~{\rm GeV}^{-1}  \, ,
\\ \nonumber \Xi_c^*(3/2^+) &:& {m} = 2645.9 {\rm~MeV} \, , {\Gamma} \leq 4.3 {\rm~MeV} \, , {g}_{\Xi_c^* \Xi_c \pi} < 4.90~{\rm GeV}^{-1} \, ,
\\ \nonumber \Omega_c^{*}(3/2^+) &:& m = 2765.9 {\rm~MeV}  \, ,
\end{eqnarray}
whose values have been averaged over isospin. We also list masses of the ground-state pseudoscalar and vector mesons~\cite{pdg}:
\begin{eqnarray}
\nonumber \pi(0^-) &:& {m} = 138.04 {\rm~MeV} \, ,
\\ K(0^-) &:& {m} = 495.65 {\rm~MeV} \, ,
\\ \nonumber \rho(1^-) &:& {m} = 775.21 {\rm~MeV} \, , {\Gamma} = 148.2 {\rm~MeV} \, , {g}_{\rho \pi \pi} = 5.94 \, ,
\\ \nonumber K^*(1^-) &:& {m} = 893.57 {\rm~MeV} \, , {\Gamma} = 49.1 {\rm~MeV} \, , {g}_{K^* K \pi} = 6.40 \, .
\end{eqnarray}
In these equations there are five coupling constants, ${g}_{\Sigma_c \Lambda_c \pi}$, ${g}_{\Sigma_c^* \Lambda_c \pi}$, ${g}_{\Xi_c^* \Xi_c \pi}$, ${g}_{\rho \pi \pi}$, and ${g}_{K^* K \pi}$, which are evaluated using the experimental decay widths of the $\Sigma_c(1/2^+)$, $\Sigma_c^*(3/2^+)$, $\Xi_c^*(3/2^+)$, $\rho(1^-)$ and $K^*(1^-)$~\cite{pdg} through the following Lagrangians
\begin{eqnarray}
\nonumber \mathcal{L}_{\Sigma_c \Lambda_c \pi} &=& {g}_{\Sigma_c \Lambda_c \pi} \overline{\Sigma}^{+}_c \gamma_\mu \gamma_5 \Lambda_c^+ \partial^\mu \pi^0 + \cdots \, ,
\\ \nonumber \mathcal{L}_{\Sigma_c^* \Lambda_c \pi} &=& {g}_{\Sigma_c^* \Lambda_c \pi} \overline{\Sigma}_{c\mu}^{*+} \Lambda_c^+ \partial^\mu \pi^0 + \cdots \, ,
\\ \mathcal{L}_{\Xi_c^* \Xi_c \pi} &=& {g}_{\Xi_c^* \Xi_c \pi} \overline{\Xi}_{c\mu}^{*+} \Xi_c^+ \partial^\mu \pi^0 + \cdots \, ,
\label{lag:swave}
\\ \nonumber \mathcal{L}_{\rho \pi \pi} &=& {g}_{\rho \pi \pi} \times \left( \rho_\mu^0 \pi^+ \partial^\mu \pi^- + \rho_\mu^0 \pi^- \partial^\mu \pi^+ \right) + \cdots \, ,
\\ \nonumber \mathcal{L}_{K^* K \pi} &=& {g}_{K^* K \pi} K^{*+}_\mu K^- \partial^\mu \pi^0 + \cdots \, ,
\end{eqnarray}
where $\cdots$ contain their isospin partners as well as their hermitian conjugate.

The ground-state charmed baryons have been systematically investigated in Ref.~\cite{Liu:2007fg}, which compose one flavor $\mathbf{\bar 3}_F$ multiplet of $J^P = 1/2^+$, one flavor $\mathbf{6}_F$ multiplet of $J^P = 1/2^+$, and one flavor $\mathbf{6}_F$ multiplet of $J^P = 3/2^+$. The two flavor $\mathbf{6}_F$ multiplets of $J^P = 1/2^+$ and $3/2^+$ compose one charmed baryon multiplet where the spin of the two light quarks is $s_l=1$, and the flavor $\mathbf{\bar 3}_F$ multiplet of $J^P = 1/2^+$ composes another charmed baryon multiplet where the spin of the two light quarks is $s_l=0$. The results of their mass sum rules are~\cite{Liu:2007fg}:
\begin{enumerate}

\item The flavor $\mathbf{\bar 3}_F$ multiplet of $J^P = 1/2^+$ contains $\Lambda_c^+(1/2^+)$, $\Xi_c^+(1/2^+)$ and $\Xi_c^0(1/2^+)$, which can be well coupled by the following interpolating fields:
\begin{eqnarray}
J_{\Lambda_c^+}(x)&=&\epsilon_{abc}[u^{aT}(x)C\gamma_{5}d^{b}(x)]h_{v}^{c}(x) \, ,
\\
J_{\Xi_c^+}(x)&=&\epsilon_{abc}[u^{aT}(x)C\gamma_{5}s^{b}(x)]h_{v}^{c}(x) \, ,
\\
J_{\Xi_c^0}(x)&=&\epsilon_{abc}[d^{aT}(x)C\gamma_{5}s^{b}(x)]h_{v}^{c}(x) \, ,
\end{eqnarray}
where $a$, $b$ and $c$ are color indices; $\epsilon_{abc}$ is the totally antisymmetric tensor; $C$ is the charge-conjugation operator; the superscript $T$ represents the transpose of the Dirac indices only; $u(x)$, $d(x)$ and $s(x)$ are the light quark fields at location $x$; $h_v(x)$ is the heavy quark field at location $x$.
Based on the results of Ref.~\cite{Liu:2007fg}, we evaluate their parameters to be $\bar\Lambda_{\Lambda_c^+} = 0.773$ GeV, $\bar\Lambda_{\Xi_c^+} = \bar\Lambda_{\Xi_c^0} = 0.908$ GeV, $f_{\Lambda_c^+} = 0.0255$ GeV$^3$ and $f_{\Xi_c^+} = f_{\Xi_c^0} = 0.0258$ GeV$^3$, with the threshold values $\omega_{\Lambda_c^+} = 1.1$ GeV and $\omega_{\Xi_c^+} = \omega_{\Xi_c^0} = 1.25$ GeV.

\item The flavor $\mathbf{6}_F$ multiplet of $J^P = 1/2^+$ contains $\Sigma_c^{++}(1/2^+)$, $\Sigma_c^{+}(1/2^+)$, $\Sigma_c^{0}(1/2^+)$, $\Xi_c^{\prime+}(1/2^+)$, $\Xi_c^{\prime0}(1/2^+)$ and $\Omega_c^{0}(1/2^+)$, which can be well coupled by
\begin{eqnarray}
J_{\Sigma_c^{++}}(x)&=&\epsilon_{abc}[u^{aT}(x)C\gamma_{\mu}u^{b}(x)]\gamma_{t}^{\mu}\gamma_{5}h_{v}^{c}(x) \, ,
\label{eq:sigmapp}
\\
J_{\Sigma_c^{+}}(x)&=&\epsilon_{abc}[u^{aT}(x)C\gamma_{\mu}d^{b}(x)]\gamma_{t}^{\mu}\gamma_{5}h_{v}^{c}(x) \, ,
\\
J_{\Sigma_c^{0}}(x)&=&\epsilon_{abc}[d^{aT}(x)C\gamma_{\mu}d^{b}(x)]\gamma_{t}^{\mu}\gamma_{5}h_{v}^{c}(x) \, ,
\\
J_{\Xi_c^{\prime+}}(x)&=&\epsilon_{abc}[u^{aT}(x)C\gamma_{\mu}s^{b}(x)]\gamma_{t}^{\mu}\gamma_{5}h_{v}^{c}(x)\, ,
\\
J_{\Xi_c^{\prime0}}(x)&=&\epsilon_{abc}[d^{aT}(x)C\gamma_{\mu}s^{b}(x)]\gamma_{t}^{\mu}\gamma_{5}h_{v}^{c}(x) \, ,
\\
J_{\Omega_c^{0}}(x)&=&\epsilon_{abc}[s^{aT}(x)C\gamma_{\mu}s^{b}(x)]\gamma_{t}^{\mu}\gamma_{5}h_{v}^{c}(x) \, .
\end{eqnarray}
Based on the results of Ref.~\cite{Liu:2007fg}, we evaluate their parameters to be $\bar\Lambda_{\Sigma_c^{++}} = \bar\Lambda_{\Sigma_c^{+}} = \bar\Lambda_{\Sigma_c^{0}} = 0.950$ GeV, $\bar\Lambda_{\Xi_c^{\prime+}} = \bar\Lambda_{\Xi_c^{\prime0}} = 1.042$ GeV, $\bar\Lambda_{\Omega_c^{0}} = 1.169$ GeV, ${1\over\sqrt2}f_{\Sigma_c^{++}} = f_{\Sigma_c^{+}} = {1\over\sqrt2}f_{\Sigma_c^{0}} = 0.0432$ GeV$^3$, $f_{\Xi_c^{\prime+}} = f_{\Xi_c^{\prime0}} = 0.0435$ GeV$^3$ and ${1\over\sqrt2}f_{\Omega_c^{0}} = 0.0438$ GeV$^3$,
with the threshold values $\omega_{\Sigma_c^{++}} = \omega_{\Sigma_c^{+}} = \omega_{\Sigma_c^{0}} = 1.3$ GeV, $\omega_{\Xi_c^{\prime+}} = \omega_{\Xi_c^{\prime0}} = 1.4$ GeV and $\omega_{\Omega_c^{0}} = 1.55$ GeV.

\item The flavor $\mathbf{6}_F$ multiplet of $J^P = 3/2^+$ contains $\Sigma_c^{*++}(3/2^+)$, $\Sigma_c^{*+}(3/2^+)$, $\Sigma_c^{*0}(3/2^+)$, $\Xi_c^{*+}(3/2^+)$, $\Xi_c^{*0}(3/2^+)$ and $\Omega_c^{*0}(3/2^+)$, which can be well coupled by
\begin{eqnarray}
J^{\mu}_{\Sigma_c^{*++}}(x)&=&\epsilon_{abc}[u^{aT}(x)C\gamma_{\nu}u^{b}(x)]\Big(-g_{t}^{\mu\nu}+\frac{1}{3} \gamma^{\mu}_{t}\gamma_{t}^{\nu}\Big)h_{v}^{c}(x) \, ,
\\
J^{\mu}_{\Sigma_c^{*+}}(x)&=&\epsilon_{abc}[u^{aT}(x)C\gamma_{\nu}d^{b}(x)]\Big(-g_{t}^{\mu\nu}+\frac{1}{3} \gamma^{\mu}_{t}\gamma_{t}^{\nu}\Big)h_{v}^{c}(x) \, ,
\\
J^{\mu}_{\Sigma_c^{*0}}(x)&=&\epsilon_{abc}[d^{aT}(x)C\gamma_{\nu}d^{b}(x)]\Big(-g_{t}^{\mu\nu}+\frac{1}{3} \gamma^{\mu}_{t}\gamma_{t}^{\nu}\Big)h_{v}^{c}(x) \, ,
\\
J^{\mu}_{\Xi_c^{*+}}(x)&=&\epsilon_{abc}[u^{aT}(x)C\gamma_{\nu}s^{b}(x)]\Big(-g_{t}^{\mu\nu}+\frac{1}{3} \gamma^{\mu}_{t}\gamma_{t}^{\nu}\Big)h_{v}^{c}(x) \, ,
\\
J^{\mu}_{\Xi_c^{*0}}(x)&=&\epsilon_{abc}[d^{aT}(x)C\gamma_{\nu}s^{b}(x)]\Big(-g_{t}^{\mu\nu}+\frac{1}{3} \gamma^{\mu}_{t}\gamma_{t}^{\nu}\Big)h_{v}^{c}(x) \, ,
\\
J^{\mu}_{\Omega_c^{*0}}(x)&=&\epsilon_{abc}[s^{aT}(x)C\gamma_{\nu}s^{b}(x)]\Big(-g_{t}^{\mu\nu}+\frac{1}{3} \gamma^{\mu}_{t}\gamma_{t}^{\nu}\Big)h_{v}^{c}(x) \, .
\end{eqnarray}
Based on the results of Ref.~\cite{Liu:2007fg}, we evaluate their parameters to be $\bar\Lambda_{\Sigma_c^{*++}} = \bar\Lambda_{\Sigma_c^{*+}} = \bar\Lambda_{\Sigma_c^{*0}} = 0.950$ GeV, $\bar\Lambda_{\Xi_c^{*+}} = \bar\Lambda_{\Xi_c^{*0}} = 1.042$ GeV, $\bar\Lambda_{\Omega_c^{*0}} = 1.169$ GeV, ${1\over\sqrt2}f_{\Sigma_c^{*++}} = f_{\Sigma_c^{*+}} = {1\over\sqrt2}f_{\Sigma_c^{*0}} = {1\over\sqrt3}0.0432$ GeV$^3$, $f_{\Xi_c^{*+}} = f_{\Xi_c^{*0}} = {1\over\sqrt3}0.0435$ GeV$^3$ and ${1\over\sqrt2}f_{\Omega_c^{*0}} = {1\over\sqrt3}0.0438$ GeV$^3$,
with the threshold values $\omega_{\Sigma_c^{*++}} = \omega_{\Sigma_c^{*+}} = \omega_{\Sigma_c^{*0}} = 1.3$ GeV, $\omega_{\Xi_c^{*+}} = \omega_{\Xi_c^{*0}} = 1.4$ GeV and $\omega_{\Omega_c^{*0}} = 1.55$ GeV.

\end{enumerate}
We list all these values in Table~\ref{tab:swave}.

\begin{table}[hbt]
\begin{center}
\renewcommand{\arraystretch}{1.5}
\caption{The parameters of the ground-state charmed baryons. The two flavor $\mathbf{6}_F$ multiplets of $J^P = 1/2^+$ and $3/2^+$ compose one charmed baryon multiplet where the spin of the two light quarks is $s_l=1$, while the flavor $\mathbf{\bar 3}_F$ multiplet of $J^P = 1/2^+$ composes another charmed baryon multiplet where the spin of the two light quarks is $s_l=0$.
}
\begin{tabular}{c | c c c c c}
\hline\hline
Multiplets & ~~~Baryon~~~ & ~~~Mass (MeV)~~~ & ~~~$\omega_c$ (GeV)~~~ & ~~~$\overline{\Lambda}$ (GeV)~~~ & ~~~$f$ (GeV$^{3}$)~~~
\\ \hline\hline
\multirow{3}{*}{$[\mathbf{\bar 3}_F,{1\over2}^+]$} & $\Lambda_c^+({1/2}^+)$ & 2286.46 & 1.10 & $0.773$ & $0.0255$
\\
                                                   & $\Xi_c^+({1/2}^+)$     & 2467.8  & 1.25 & $0.908$ & $0.0258$
\\
                                                   & $\Xi_c^0({1/2}^+)$     & 2470.88 & 1.25 & $0.908$ & $0.0258$
\\ \hline
\multirow{6}{*}{$[\mathbf{6}_F,{1\over2}^+]$}    & $\Sigma_c^{++}({1/2}^+)$    & 2453.98 & 1.30 & $0.950$ & $\sqrt2 \times 0.0432$
\\
                                                 & $\Sigma_c^{+}({1/2}^+)$     & 2452.9  & 1.30 & $0.950$ & $0.0432$
\\
                                                 & $\Sigma_c^{0}({1/2}^+)$     & 2453.74 & 1.30 & $0.950$ & $\sqrt2 \times 0.0432$
\\
                                                 & $\Xi_c^{\prime+}({1/2}^+)$  & 2575.6  & 1.40 & $1.042$ & $0.0435$
\\
                                                 & $\Xi_c^{\prime0}({1/2}^+)$  & 2577.9  & 1.40 & $1.042$ & $0.0435$
\\
                                                 & $\Omega_c^{+}({1/2}^+)$     & 2695.2  & 1.55 & $1.169$ & $\sqrt2 \times 0.0438$
\\ \hline
\multirow{6}{*}{$[\mathbf{6}_F,{3\over2}^+]$}    & $\Sigma_c^{*++}({1/2}^+)$ & 2517.9  & 1.30 & $0.950$ & $\sqrt{2\over3} \times 0.0432$
\\
                                                 & $\Sigma_c^{*+}({1/2}^+)$  & 2517.5  & 1.30 & $0.950$ & $\sqrt{1\over3} \times 0.0432$
\\
                                                 & $\Sigma_c^{*0}({1/2}^+)$  & 2518.8  & 1.30 & $0.950$ & $\sqrt{2\over3} \times 0.0432$
\\
                                                 & $\Xi_c^{*+}({1/2}^+)$     & 2645.9  & 1.40 & $1.042$ & $\sqrt{1\over3} \times 0.0435$
\\
                                                 & $\Xi_c^{*0}({1/2}^+)$     & 2645.9  & 1.40 & $1.042$ & $\sqrt{1\over3} \times 0.0435$
\\
                                                 & $\Omega_c^{*+}({1/2}^+)$  & 2765.9  & 1.55 & $1.169$ & $\sqrt{2\over3} \times 0.0438$
\\ \hline \hline
\end{tabular}
\label{tab:swave}
\end{center}
\end{table}

\subsection{$P$-wave charmed baryons of flavor $\mathbf{\bar 3}_F$}

The four observed states $\Lambda_c(2595)$ ($J^P=1/2^-$), $\Lambda_c(2625)$ ($J^P=3/2^-$), $\Xi_c(2790)$ ($J^P=1/2^-$) and $\Xi_c(2815)$ ($J^P=3/2^-$) probably complete two flavor $\mathbf{\bar 3}_F$ multiplets of $J^P=1/2^-$ and $3/2^-$~\cite{pdg}. Accordingly, in the present study we assume masses of the $P$-wave charmed baryon states of flavor $\mathbf{\bar 3}_F$ to be
\begin{eqnarray}
\nonumber \Lambda_c({1/2}^-) &:& m = 2592.25 {\rm~MeV} \, ,
\\ \Lambda_c({3/2}^-) &:& m = 2628.11 {\rm~MeV} \, ,
\\ \nonumber \Xi_c({1/2}^-) &:& m = 2790.5 {\rm~MeV} \, ,
\\ \nonumber \Lambda_c({3/2}^-) &:& m = 2818.1 {\rm~MeV} \, ,
\end{eqnarray}
which values will be used in Sec.~\ref{sec:sumrule1} to evaluate their decay widths.

The masses and decay constants of the flavor $\mathbf{\bar 3}_F$ $P$-wave charmed baryons have been systematically investigated in Refs.~\cite{Chen:2015kpa,Mao:2015gya} using the method of QCD sum rules in HQET, and our results suggested that the  $\Lambda_c(2595)$, $\Lambda_c(2625)$, $\Xi_c(2790)$ and $\Xi_c(2815)$ can be well described by the baryon doublet $[\mathbf{\bar 3}_F, 1, 1, \rho]$ and they complete two $\mathbf{\bar 3}_F$ multiplets of $J^P=1/2^-$ and $3/2^-$; while the results obtained using the baryon doublet $[\mathbf{\bar 3}_F, 1, 0, \lambda]$ seem also consistent with the data.
We note that the definition of the external off-shell energy $\omega$ in the present study is different from that used in Refs.~\cite{Chen:2015kpa,Mao:2015gya}. Hence, in the present study we also reevaluate their parameters, and shortly summarize their results in the following.

Based on HQET, we use $J^{\alpha_1\cdots\alpha_{j-1/2}}_{j,P,F,j_l,s_l,\rho/\lambda}$ to denote the $P$-wave charmed baryon field coupling to $|j,P,F,j_l,s_l,\rho/\lambda\rangle$,
where $j$, $P$, and $F$ are the total angular momentum, parity and flavor representation (either $\mathbf{\bar 3}_F$ or $\mathbf{6}_F$) of the charmed baryons,
and $j_l$ and $s_l$ are the total angular momentum and spin angular momentum of the light components.
We use $l_\rho$ to denote the orbital angular momentum between the two light quarks, $l_\lambda$ to denote the orbital angular momentum between the charm quark and the two-light-quark system,
and then $\rho$ to denote $l_\rho = 1$ and $l_\lambda = 0$, while $\lambda$ to denote $l_\rho = 0$ and $l_\lambda = 1$.
We have the relations $L = l_\lambda \otimes l_\rho$, $j_l = L \otimes s_l$ and $j = j_l \otimes s_Q$, where $s_Q = 1/2$ is the spin of the heavy quark.
This field $J^{\alpha_1\cdots\alpha_{j-1/2}}_{j,P,F,j_l,s_l,\rho/\lambda}$ belongs to the baryon multiplet $[F,j_l,s_l,\rho/\lambda]$.

There are altogether four $P$-wave charmed baryon multiplets of the flavor $\mathbf{\bar 3}_F$, $[\mathbf{\bar 3}_F, 0, 1, \rho]$, $[\mathbf{\bar 3}_F, 1, 1, \rho]$, $[\mathbf{\bar 3}_F, 2, 1, \rho]$, and $[\mathbf{\bar 3}_F, 1, 0, \lambda]$. The results of their mass sum rules are~\cite{Chen:2015kpa,Mao:2015gya}:
\begin{enumerate}

\item The $[\mathbf{\bar 3}_F, 0, 1, \rho]$ multiplet contains $\Lambda_c^+({1\over2}^-)$ and $\Xi_c^{+,0}({1\over2}^-)$, which are coupled by
\begin{eqnarray}
J_{1/2,-,\mathbf{\bar3}_F,0,1,\rho} &=& i \epsilon_{abc} \Big ( [\mathcal{D}_t^{\mu} q^{aT}] C \gamma_t^\mu q^b -  q^{aT} C \gamma_t^\mu [\mathcal{D}_t^{\mu} q^b] \Big ) h_v^c \, ,
\label{eq:current3}
\end{eqnarray}
where $D^\mu = \partial^\mu - i g A^\mu$ is the gauge-covariant derivative. We can further explicitly denote the quark contents by simply replacing $\mathbf{\bar3}_F$ by $\Lambda_c$ and $\Xi_c$. For example, We use $J_{1/2,-,\Xi_c^{+},0,1,\rho}$ to denote $J_{1/2,-,\mathbf{\bar3}_F,0,1,\rho}$ with the quark contents $usc$:
\begin{eqnarray}
J_{1/2,-,\Xi_c^{+},0,1,\rho} &=& i \epsilon_{abc} \Big ( [\mathcal{D}_t^{\mu} u^{aT}] C \gamma_t^\mu s^b -  u^{aT} C \gamma_t^\mu [\mathcal{D}_t^{\mu} s^b] \Big ) h_v^c \, .
\end{eqnarray}
Based on the results of Refs.~\cite{Chen:2015kpa,Mao:2015gya}, we evaluate their parameters to be $\bar\Lambda_{\Lambda_c^+({1\over2}^-)} = 0.987$ GeV, $\bar\Lambda_{\Xi_c^{+,0}({1\over2}^-)} = 1.181$ GeV, $f_{\Lambda_c^+({1\over2}^-)} = 0.0198$ GeV$^4$ and $f_{\Xi_c^{+,0}({1\over2}^-)} = 0.0296$ GeV$^4$,
with the threshold values $\omega_{\Lambda_c^+({1\over2}^-)} = 1.75$ GeV and $\omega_{\Xi_c^{+,0}({1\over2}^-)} = 1.55$ GeV.

\item The $[\mathbf{\bar 3}_F, 1, 1, \rho]$ multiplet contains $\Lambda_c^+({1\over2}^-/{3\over2}^-)$ and $\Xi_c^{+,0}({1\over2}^-/{3\over2}^-)$, which are coupled by
\begin{eqnarray}
J_{1/2,-,\mathbf{\bar3}_F,1,1,\rho} &=& i \epsilon_{abc} \Big ( [\mathcal{D}_t^{\mu} q^{aT}] C \gamma_t^\nu q^b -  q^{aT} C \gamma_t^\nu [\mathcal{D}_t^{\mu} q^b] \Big ) \sigma_t^{\mu\nu} h_v^c \, ,
\label{eq:current4}
\\ J^{\alpha}_{3/2,-,\mathbf{\bar3}_F,1,1,\rho} &=&
\label{eq:current5}
i \epsilon_{abc} \Big ( [\mathcal{D}_t^{\mu} q^{aT}] C \gamma_t^\nu q^b - q^{aT} C \gamma_t^\nu [\mathcal{D}_t^{\mu} q^b] \Big )
\\ \nonumber && \times \Big ( g_t^{\alpha\mu} \gamma_t^{\nu} \gamma_5 - g_t^{\alpha\nu} \gamma_t^{\mu} \gamma_5 - {1 \over 3} \gamma_t^{\alpha} \gamma_t^{\mu} \gamma_t^{\nu} \gamma_5 + {1 \over 3} \gamma_t^{\alpha} \gamma_t^{\nu} \gamma_t^{\mu} \gamma_5 \Big ) h_v^c \, .
\end{eqnarray}
Based on the results of Refs.~\cite{Chen:2015kpa,Mao:2015gya}, we evaluate their parameters to be $\bar\Lambda_{\Lambda_c^+({1\over2}^-/{3\over2}^-)} = 1.164$ GeV, $\bar\Lambda_{\Xi_c^{+,0}({1\over2}^-/{3\over2}^-)} = 1.349$ GeV, $f_{\Lambda_c^+({1\over2}^-)} = 0.0523$ GeV$^4$, $f_{\Xi_c^{+,0}({1\over2}^-)} = 0.0788$ GeV$^4$, $f_{\Lambda_c^+({3\over2}^-)} = 0.0523$ GeV$^4$ and $f_{\Xi_c^{+,0}({3\over2}^-)} = 0.0788$ GeV$^4$,
with the threshold values $\omega_{\Lambda_c^+({1\over2}^-/{3\over2}^-)} = 1.55$ GeV and $\omega_{\Xi_c^{+,0}({1\over2}^-/{3\over2}^-)} = 1.8$ GeV.

\item The $[\mathbf{\bar 3}_F, 2, 1, \rho]$ multiplet contains $\Lambda_c^+({3\over2}^-/{5\over2}^-)$ and $\Xi_c^{+,0}({3\over2}^-/{5\over2}^-)$, which are coupled by
\begin{eqnarray}
J^{\alpha}_{3/2,-,\mathbf{\bar3}_F,2,1,\rho} &=&
\label{eq:current6}
i \epsilon_{abc} \Big ( [\mathcal{D}_t^{\mu} q^{aT}] C \gamma_t^\nu q^b - q^{aT} C \gamma_t^\nu [\mathcal{D}_t^{\mu} q^b] \Big )
\times \Big ( g_t^{\alpha\mu} \gamma_t^{\nu} \gamma_5 + g_t^{\alpha\nu} \gamma_t^{\mu} \gamma_5 - {2 \over 3} g_t^{\mu\nu} \gamma_t^{\alpha} \gamma_5 \Big ) h_v^c \, ,
\\ J^{\alpha_1\alpha_2}_{5/2,-,\mathbf{\bar3}_F,2,1,\rho} &=&
\label{eq:current7}
i \epsilon_{abc} \Big ( [\mathcal{D}_t^{\mu} q^{aT}] C \gamma_t^{\nu} q^b - q^{aT} C \gamma_t^{\nu} [\mathcal{D}_t^{\mu} q^b] \Big ) \times \Gamma^{\alpha_1\alpha_2,\mu\nu} h_v^c \, ,
\end{eqnarray}
where $\Gamma^{\alpha_1\alpha_2,\mu\nu}$ is the projection operator:
\begin{eqnarray}
\Gamma^{\alpha\beta,\mu\nu} &=& g_t^{\alpha\mu} g_t^{\beta\nu} + g_t^{\alpha\nu} g_t^{\beta\mu} - {2 \over 15} g_t^{\alpha\beta} g_t^{\mu\nu}
 - {1 \over 3} g_t^{\alpha\mu} \gamma_t^{\beta}\gamma_t^{\nu} - {1 \over 3} g_t^{\alpha\nu} \gamma_t^{\beta}\gamma_t^{\mu}
 - {1 \over 3} g_t^{\beta\mu} \gamma_t^{\alpha}\gamma_t^{\nu} - {1 \over 3} g_t^{\beta\nu} \gamma_t^{\alpha}\gamma_t^{\mu}
\\ \nonumber && + {1 \over 15} \gamma_t^{\alpha} \gamma_t^{\mu} \gamma_t^{\beta} \gamma_t^{\nu} + {1 \over 15} \gamma_t^{\alpha} \gamma_t^{\nu} \gamma_t^{\beta} \gamma_t^{\mu}
 + {1 \over 15} \gamma_t^{\beta} \gamma_t^{\mu} \gamma_t^{\alpha} \gamma_t^{\nu} + {1 \over 15} \gamma_t^{\beta} \gamma_t^{\nu} \gamma_t^{\alpha} \gamma_t^{\mu} \, .
\end{eqnarray}
Based on the results of Refs.~\cite{Chen:2015kpa,Mao:2015gya}, we evaluate their parameters to be $\bar\Lambda_{\Lambda_c^+({3\over2}^-/{5\over2}^-)} = 1.339$ GeV, $\bar\Lambda_{\Xi_c^{+,0}({3\over2}^-/{5\over2}^-)} = 1.510$ GeV, $f_{\Lambda_c^+({3\over2}^-)} = 0.0578$ GeV$^4$, $f_{\Xi_c^{+,0}({3\over2}^-)} = 0.0901$ GeV$^4$, $f_{\Lambda_c^+({5\over2}^-)} = {1\over\sqrt5}0.0578$ GeV$^4$ and $f_{\Xi_c^{+,0}({5\over2}^-)} = {1\over\sqrt5}0.0901$ GeV$^4$,
with the threshold values
$\omega_{\Lambda_c^+({3\over2}^-/{5\over2}^-)} = 1.8$ GeV and $\omega_{\Xi_c^{+,0}({3\over2}^-/{5\over2}^-)} = 2.0$ GeV.

\item The $[\mathbf{\bar 3}_F, 1, 0, \lambda]$ multiplet contains $\Lambda_c^+({1\over2}^-/{3\over2}^-)$ and $\Xi_c^{+,0}({1\over2}^-/{3\over2}^-)$, which are coupled by
\begin{eqnarray}
J_{1/2,-,\mathbf{\bar3}_F,1,0,\lambda} &=& i \epsilon_{abc} \Big ( [\mathcal{D}_t^{\mu} q^{aT}] C \gamma_5 q^b +  q^{aT} C \gamma_5 [\mathcal{D}_t^{\mu} q^b] \Big ) \gamma_t^{\mu} \gamma_5 h_v^c \, ,
\label{eq:current8}
\\ J^{\alpha}_{3/2,-,\mathbf{\bar3}_F,1,0,\lambda} &=& i \epsilon_{abc} \Big ( [\mathcal{D}_t^{\mu} q^{aT}] C \gamma_5 q^b +  q^{aT} C \gamma_5 [\mathcal{D}_t^{\mu} q^b] \Big ) \Big ( g_t^{\alpha\mu} - {1\over3} \gamma_t^{\alpha} \gamma_t^{\mu} \Big ) h_v^c \, .
\label{eq:current9}
\end{eqnarray}
Based on the results of Refs.~\cite{Chen:2015kpa,Mao:2015gya}, we evaluate their parameters to be $\bar\Lambda_{\Lambda_c^+({1\over2}^-/{3\over2}^-)} = 0.961$ GeV, $\bar\Lambda_{\Xi_c^{+,0}({1\over2}^-/{3\over2}^-)} = 1.057$ GeV, $f_{\Lambda_c^+({1\over2}^-)} = 0.0201$ GeV$^4$, $f_{\Xi_c^{+,0}({1\over2}^-)} = 0.0255$ GeV$^4$, $f_{\Lambda_c^+({3\over2}^-)} = {1\over\sqrt3}0.0201$ GeV$^4$ and $f_{\Xi_c^{+,0}({3\over2}^-)} = {1\over\sqrt3}0.0255$ GeV$^4$,
with the threshold values $\omega_{\Lambda_c^+({1\over2}^-/{3\over2}^-)} = 1.45$ GeV and $\omega_{\Xi_c^{+,0}({1\over2}^-/{3\over2}^-)} = 1.55$ GeV.

\end{enumerate}
We also list these values in Table~\ref{tab:pwave1}.


\begin{table}[hbt]
\begin{center}
\renewcommand{\arraystretch}{1.5}
\caption{The parameters of the $P$-wave charmed baryons of flavor $\mathbf{\bar 3}_F$. In Ref.~\cite{Chen:2015kpa} we have systematically evaluated masses of the $P$-wave charmed baryons, and our results suggested the four observed states $\Lambda_c(2595)$ ($J^P=1/2^-$), $\Lambda_c(2625)$ ($J^P=3/2^-$), $\Xi_c(2790)$ ($J^P=1/2^-$) and $\Xi_c(2815)$ ($J^P=3/2^-$) can be well described by the baryon doublet $[\mathbf{\bar 3}_F, 1, 1, \rho]$ and they complete two $\mathbf{\bar 3}_F$ multiplets of $J^P=1/2^-$ and $3/2^-$, while the currents belonging to the baryon doublet $[\mathbf{\bar 3}_F, 1, 0, \lambda]$ seem also consistent with the data.
}
\begin{tabular}{c | c c c c c}
\hline\hline
Multiplets & ~~~Baryon~~~ & ~~~$\omega_c$ (GeV)~~~ & ~~~$T$ (GeV)~~~ & ~~~$\overline{\Lambda}$ (GeV)~~~ & ~~~$f$ (GeV$^{4}$)~~~
\\ \hline\hline
\multirow{3}{*}{$[\mathbf{\bar 3}_F, 0, 1, \rho]$} & $\Lambda_c^+({1\over2}^-)$ & 1.75 & $\sim0.37$ & $0.987$ & $0.0198$
\\
                                                   & $\Xi_c^+({1\over2}^-)$     & 1.55 & $\sim0.38$ & $1.181$ & $0.0296$
\\
                                                   & $\Xi_c^0({1\over2}^-)$     & 1.55 & $\sim0.38$ & $1.181$ & $0.0296$
\\ \hline
\multirow{3}{*}{$[\mathbf{\bar 3}_F, 1, 1, \rho]$} & $\Lambda_c^+({1\over2}^-/{3\over2}^-)$ & 1.55 & $0.27<T<0.30$ & $1.164$ & $0.0523$/$0.0523$
\\
                                                   & $\Xi_c^+({1\over2}^-/{3\over2}^-)$     & 1.80 & $0.27<T<0.32$ & $1.349$ & $0.0788$/$0.0788$
\\
                                                   & $\Xi_c^0({1\over2}^-/{3\over2}^-)$     & 1.80 & $0.27<T<0.32$ & $1.349$ & $0.0788$/$0.0788$
\\ \hline
\multirow{3}{*}{$[\mathbf{\bar 3}_F, 2, 1, \rho]$} & $\Lambda_c^+({3\over2}^-/{5\over2}^-)$ & 1.80 & $\sim0.30$ & $1.339$ & $0.0578$/$\sqrt{1\over5} \times 0.0578$
\\
                                                   & $\Xi_c^+({3\over2}^-/{5\over2}^-)$     & 2.00 & $\sim0.33$ & $1.510$ & $0.0901$/$\sqrt{1\over5} \times 0.0901$
\\
                                                   & $\Xi_c^0({3\over2}^-/{5\over2}^-)$     & 2.00 & $\sim0.33$ & $1.510$ & $0.0901$/$\sqrt{1\over5} \times 0.0901$
\\ \hline
\multirow{3}{*}{$[\mathbf{\bar 3}_F, 1, 0, \lambda]$} & $\Lambda_c^+({1\over2}^-/{3\over2}^-)$ & 1.45 & $\sim0.30$ & $0.961$ & $0.0201$/$\sqrt{1\over3} \times 0.0201$
\\
                                                      & $\Xi_c^+({1\over2}^-/{3\over2}^-)$     & 1.55 & $\sim0.32$ & $1.057$ & $0.0255$/$\sqrt{1\over3} \times 0.0255$
\\
                                                      & $\Xi_c^0({1\over2}^-/{3\over2}^-)$     & 1.55 & $\sim0.32$ & $1.057$ & $0.0255$/$\sqrt{1\over3} \times 0.0255$
\\ \hline \hline
\end{tabular}
\label{tab:pwave1}
\end{center}
\end{table}

\subsection{$P$-wave charmed baryons of flavor $\mathbf{6}_F$}

The three observed states $\Sigma_c(2800)$ ($J^P=?^?$), $\Xi_c(2930)$ ($J^P=?^?$) and $\Xi_c(2980)$ ($J^P=?^?$) may be $P$-wave charmed baryon states of the flavor $\mathbf{6}_F$. Besides them, there are five excited $\Omega_c$ states recently observed by LHCb~\cite{Aaij:2017nav}: the $\Omega_c(3000)$, $\Omega_c(3050)$, $\Omega_c(3066)$, $\Omega_c(3090)$, and $\Omega_c(3119)$, which may also be interpreted as $P$-wave charmed baryon states of the flavor $\mathbf{6}_F$. Accordingly, in the present study we assume the masses of the $P$-wave charmed baryon fields to be
\begin{eqnarray}
\nonumber \Sigma_c\Big({1\over2}^-/{3\over2}^-\Big) &:& m \approx 2800 {\rm~MeV} \, ,
\\ \nonumber \Xi^\prime_c\Big({1\over2}^-/{3\over2}^-\Big) &:& m \approx 2950 {\rm~MeV} \, ,
\\ \nonumber \Omega_c\Big({1\over2}^-\Big) &:& m \approx 3100 {\rm~MeV} \, ,
\\ \nonumber \Omega_c\Big({3\over2}^-\Big) &:& m \approx 3120 {\rm~MeV} \, ,
\end{eqnarray}
which values will be used in Sec.~\ref{sec:sumrule2} to evaluate their decay widths.

The masses and decay constants of the flavor $\mathbf{6}_F$ $P$-wave charmed baryons have been systematically investigated in Refs.~\cite{Chen:2015kpa,Mao:2015gya} using the method of QCD sum rules in HQET, and our results suggested that the baryon doublet $[\mathbf{6}_F, 1, 0, \rho]$ contains $\Sigma_c(1/2^-,3/2^-)$, $\Xi^\prime_c(1/2^-,3/2^-)$, and $\Omega_c(1/2^-,3/2^-)$, and its obtained results are consistent with the observed states $\Sigma_c(2800)$ ($J^P=?^?$) and $\Xi_c(2980)$ ($J^P=?^?$), while the results obtained by using the baryon doublet $[\mathbf{6}_F, 2, 1, \lambda]$ are also consistent with them. We shortly summarize their results in the following.

There are altogether four $P$-wave charmed baryon multiplets of the flavor $\mathbf{6}_F$, $[\mathbf{6}_F, 1, 0, \rho]$, $[\mathbf{6}_F, 0, 1, \lambda]$, $[\mathbf{6}_F, 1, 1, \lambda]$ and $[\mathbf{6}_F, 2, 1, \lambda]$, and the results of their mass sum rules are~\cite{Chen:2015kpa,Mao:2015gya}:
\begin{enumerate}

\item The $[\mathbf{6}_F, 1, 0, \rho]$ multiplet contains $\Sigma_c^{++,+,0}({1\over2}^-/{3\over2}^-)$, $\Xi_c^{\prime+,0}({1\over2}^-/{3\over2}^-)$ and $\Omega_c^{0}({1\over2}^-/{3\over2}^-)$, which are coupled by
\begin{eqnarray}
J_{1/2,-,\mathbf{6}_F,1,0,\rho} &=& i \epsilon_{abc} \Big ( [\mathcal{D}_t^{\mu} q^{aT}] C \gamma_5 q^b -  q^{aT} C \gamma_5 [\mathcal{D}_t^{\mu} q^b] \Big ) \gamma_t^{\mu} \gamma_5 h_v^c \, ,
\label{eq:current1}
\\ J^{\alpha}_{3/2,-,\mathbf{6}_F,1,0,\rho} &=& i \epsilon_{abc} \Big ( [\mathcal{D}_t^{\mu} q^{aT}] C \gamma_5 q^b -  q^{aT} C \gamma_5 [\mathcal{D}_t^{\mu} q^b] \Big ) \Big ( g_t^{\alpha\mu} - {1\over3} \gamma_t^{\alpha} \gamma_t^{\mu} \Big ) h_v^c \, .
\label{eq:current2}
\end{eqnarray}
Based on the results of Refs.~\cite{Chen:2015kpa,Mao:2015gya}, we evaluate their parameters to be $\bar\Lambda_{\Sigma_c^{++,+,0}({1\over2}^-/{3\over2}^-)} = 1.224$ GeV, $\bar\Lambda_{\Xi_c^{\prime+,0}({1\over2}^-/{3\over2}^-)} = 1.422$ GeV, $\bar\Lambda_{\Omega_c^{0}({1\over2}^-/{3\over2}^-)} = 1.641$ GeV, ${1\over\sqrt2}f_{\Sigma_c^{++,0}({1\over2}^-)} = f_{\Sigma_c^{+}({1\over2}^-)} = 0.0437$ GeV$^4$, $f_{\Xi_c^{\prime+,0}({1\over2}^-)} = 0.0680$ GeV$^4$, ${1\over\sqrt2}f_{\Omega_c^{0}({1\over2}^-)} = 0.108$ GeV$^4$, ${1\over\sqrt2}f_{\Sigma_c^{++,0}({3\over2}^-)} = f_{\Sigma_c^{+}({3\over2}^-)} = {1\over\sqrt3}0.0437$ GeV$^4$, $f_{\Xi_c^{\prime+,0}({3\over2}^-)} = {1\over\sqrt3}0.0680$ GeV$^4$ and ${1\over\sqrt2}f_{\Omega_c^{0}({3\over2}^-)} = {1\over\sqrt3}0.108$ GeV$^4$,
with the threshold values $\omega_{\Sigma_c^{++,+,0}({1\over2}^-/{3\over2}^-)} = 1.7$ GeV, $\omega_{\Xi_c^{\prime+,0}({1\over2}^-/{3\over2}^-)} = 1.95$ GeV and $\omega_{\Omega_c^{0}({1\over2}^-/{3\over2}^-)} = 2.2$ GeV.

\item The $[\mathbf{6}_F, 0, 1, \lambda]$ multiplet contains $\Sigma_c^{++,+,0}({1\over2}^-)$, $\Xi_c^{\prime+,0}({1\over2}^-)$, and $\Omega_c^{0}({1\over2}^-)$, which are coupled by
\begin{eqnarray}
J_{1/2,-,\mathbf{6}_F,0,1,\lambda} &=& i \epsilon_{abc} \Big ( [\mathcal{D}_t^{\mu} q^{aT}] C \gamma_t^\mu q^b +  q^{aT} C \gamma_t^\mu [\mathcal{D}_t^{\mu} q^b] \Big ) h_v^c \, .
\label{eq:current10}
\end{eqnarray}
Based on the results of Refs.~\cite{Chen:2015kpa,Mao:2015gya}, we evaluate their parameters to be $\bar\Lambda_{\Sigma_c^{++,+,0}({1\over2}^-)} = 1.100$ GeV, $\bar\Lambda_{\Xi_c^{\prime+,0}({1\over2}^-)} = 1.295$ GeV, $\bar\Lambda_{\Omega_c^{0}({1\over2}^-)} = 1.504$ GeV, ${1\over\sqrt2}f_{\Sigma_c^{++,0}({1\over2}^-)} = f_{\Sigma_c^{+}({1\over2}^-)} = 0.0344$ GeV$^4$, $f_{\Xi_c^{\prime+,0}({1\over2}^-)} = 0.0512$ GeV$^4$ and ${1\over\sqrt2}f_{\Omega_c^{0}({1\over2}^-)} = 0.0804$ GeV$^4$,
with the threshold values $\omega_{\Sigma_c^{++,+,0}({1\over2}^-)} = 1.45$ GeV, $\omega_{\Xi_c^{\prime+,0}({1\over2}^-)} = 1.7$ GeV and $\omega_{\Omega_c^{0}({1\over2}^-)} = 1.95$ GeV.

\item The $[\mathbf{6}_F, 1, 1, \lambda]$ multiplet $\Sigma_c^{++,+,0}({1\over2}^-/{3\over2}^-)$, $\Xi_c^{\prime+,0}({1\over2}^-/{3\over2}^-)$, and $\Omega_c^{0}({1\over2}^-/{3\over2}^-)$, which are coupled by
\begin{eqnarray}
J_{1/2,-,\mathbf{6}_F,1,1,\lambda} &=& i \epsilon_{abc} \Big ( [\mathcal{D}_t^{\mu} q^{aT}] C \gamma_t^\nu q^b +  q^{aT} C \gamma_t^\nu [\mathcal{D}_t^{\mu} q^b] \Big ) \sigma_t^{\mu\nu} h_v^c \, ,
\label{eq:current11}
\\ J^{\alpha}_{3/2,-,\mathbf{6}_F,1,1,\lambda} &=&
\label{eq:current12}
i \epsilon_{abc} \Big ( [\mathcal{D}_t^{\mu} q^{aT}] C \gamma_t^\nu q^b + q^{aT} C \gamma_t^\nu [\mathcal{D}_t^{\mu} q^b] \Big )
\\ \nonumber && \times \Big ( g_t^{\alpha\mu} \gamma_t^{\nu} \gamma_5 - g_t^{\alpha\nu} \gamma_t^{\mu} \gamma_5 - {1 \over 3} \gamma_t^{\alpha} \gamma_t^{\mu} \gamma_t^{\nu} \gamma_5 + {1 \over 3} \gamma_t^{\alpha} \gamma_t^{\nu} \gamma_t^{\mu} \gamma_5 \Big ) h_v^c \, .
\end{eqnarray}
Based on the results of Refs.~\cite{Chen:2015kpa,Mao:2015gya}, we evaluate their parameters to be $\bar\Lambda_{\Sigma_c^{++,+,0}({1\over2}^-/{3\over2}^-)} = 1.066$ GeV, $\bar\Lambda_{\Xi_c^{\prime+,0}({1\over2}^-/{3\over2}^-)} = 1.181$ GeV, $\bar\Lambda_{\Omega_c^{0}({1\over2}^-/{3\over2}^-)} = 1.270$ GeV, ${1\over\sqrt2}f_{\Sigma_c^{++,0}({1\over2}^-)} = f_{\Sigma_c^{+}({1\over2}^-)} = 0.0349$ GeV$^4$, $f_{\Xi_c^{\prime+,0}({1\over2}^-)} = 0.0451$ GeV$^4$, ${1\over\sqrt2}f_{\Omega_c^{0}({1\over2}^-)} = 0.0546$ GeV$^4$, ${1\over\sqrt2}f_{\Sigma_c^{++,0}({3\over2}^-)} = f_{\Sigma_c^{+}({3\over2}^-)} = 0.0349$ GeV$^4$, $f_{\Xi_c^{\prime+,0}({3\over2}^-)} = 0.0451$ GeV$^4$ and ${1\over\sqrt2}f_{\Omega_c^{0}({3\over2}^-)} = 0.0546$ GeV$^4$,
with the threshold values $\omega_{\Sigma_c^{++,+,0}({1\over2}^-/{3\over2}^-)} = 1.75$ GeV, $\omega_{\Xi_c^{\prime+,0}({1\over2}^-/{3\over2}^-)} = 1.75$ GeV and $\omega_{\Omega_c^{0}({1\over2}^-/{3\over2}^-)} = 1.75$ GeV.

\item The $[\mathbf{6}_F, 2, 1, \lambda]$ multiplet $\Sigma_c^{++,+,0}({3\over2}^-/{5\over2}^-)$, $\Xi_c^{\prime+,0}({3\over2}^-/{5\over2}^-)$, and $\Omega_c^{0}({3\over2}^-/{5\over2}^-)$, which are coupled by
\begin{eqnarray}
J^{\alpha}_{3/2,-,\mathbf{6}_F,2,1,\lambda} &=&
\label{eq:current13}
i \epsilon_{abc} \Big ( [\mathcal{D}_t^{\mu} q^{aT}] C \gamma_t^\nu q^b + q^{aT} C \gamma_t^\nu [\mathcal{D}_t^{\mu} q^b] \Big )
\times \Big ( g_t^{\alpha\mu} \gamma_t^{\nu} \gamma_5 + g_t^{\alpha\nu} \gamma_t^{\mu} \gamma_5 - {2 \over 3} g_t^{\mu\nu} \gamma_t^{\alpha} \gamma_5 \Big ) h_v^c \, ,
\\ J^{\alpha_1\alpha_2}_{5/2,-,\mathbf{6}_F,2,1,\lambda} &=&
\label{eq:current14}
i \epsilon_{abc} \Big ( [\mathcal{D}_t^{\mu} q^{aT}] C \gamma_t^{\nu} q^b + q^{aT} C \gamma_t^{\nu} [\mathcal{D}_t^{\mu} q^b] \Big ) \times \Gamma^{\alpha_1\alpha_2,\mu\nu} h_v^c \, .
\end{eqnarray}
Based on the results of Refs.~\cite{Chen:2015kpa,Mao:2015gya}, we evaluate their parameters to be $\bar\Lambda_{\Sigma_c^{++,+,0}({3\over2}^-/{5\over2}^-)} = 1.099$ GeV, $\bar\Lambda_{\Xi_c^{\prime+,0}({3\over2}^-/{5\over2}^-)} = 1.254$ GeV, $\bar\Lambda_{\Omega_c^{0}({3\over2}^-/{5\over2}^-)} = 1.461$ GeV, ${1\over\sqrt2}f_{\Sigma_c^{++,0}({3\over2}^-)} = f_{\Sigma_c^{+}({3\over2}^-)} = 0.0395$ GeV$^4$, $f_{\Xi_c^{\prime+,0}({3\over2}^-)} = 0.0599$ GeV$^4$, ${1\over\sqrt2}f_{\Omega_c^{0}({3\over2}^-)} = 0.0976$ GeV$^4$, ${1\over\sqrt2}f_{\Sigma_c^{++,0}({5\over2}^-)} = f_{\Sigma_c^{+}({5\over2}^-)} = {1\over\sqrt5}0.0395$ GeV$^4$, $f_{\Xi_c^{\prime+,0}({5\over2}^-)} = {1\over\sqrt5}0.0599$ GeV$^4$ and ${1\over\sqrt2}f_{\Omega_c^{0}({5\over2}^-)} = {1\over\sqrt5}0.0976$ GeV$^4$,
with the threshold values $\omega_{\Sigma_c^{++,+,0}({3\over2}^-/{5\over2}^-)} = 1.5$ GeV, $\omega_{\Xi_c^{\prime+,0}({3\over2}^-/{5\over2}^-)} = 1.75$ GeV and $\omega_{\Omega_c^{0}({3\over2}^-/{5\over2}^-)} = 2.0$ GeV.

\end{enumerate}
We also list all these values in Table \ref{tab:pwave2}.

\begin{table}[hbt]
\begin{center}
\renewcommand{\arraystretch}{1.5}
\caption{The parameters of the $P$-wave charmed baryons of flavor $\mathbf{6}$. In Ref.~\cite{Chen:2015kpa} we have systematically evaluated the masses of the $P$-wave charmed baryons, and our results suggested that the baryon doublet $[\mathbf{6}_F, 1, 0, \rho]$ contains $\Sigma_c(1/2^-,3/2^-)$, $\Xi^\prime_c(1/2^-,3/2^-)$, and $\Omega_c(1/2^-,3/2^-)$, and its obtained results are consistent with the observed states $\Sigma_c(2800)$ ($J^P=?^?$) and $\Xi_c(2980)$ ($J^P=?^?$), while the results obtained by using the baryon doublet $[\mathbf{6}_F, 2, 1, \lambda]$ are also consistent with them.
}
\begin{tabular}{c | c c c c c}
\hline\hline
Multiplets & ~~~Baryon~~~ & ~~~$\omega_c$ (GeV)~~~ & ~~~$T$ (GeV)~~~ & ~~~$\overline{\Lambda}$ (GeV)~~~ & ~~~$f$ (GeV$^{4}$)~~~
\\ \hline\hline
\multirow{6}{*}{$[\mathbf{6}_F, 1, 0, \rho]$}    & $\Sigma_c^{++}({1\over2}^-/{3\over2}^-)$    & 1.70 & $0.26<T<0.32$ & $1.224$ & $\sqrt2 \times 0.0437$/$\sqrt{2\over3} \times 0.0437$
\\
                                                 & $\Sigma_c^{+}({1\over2}^-/{3\over2}^-)$     & 1.70 & $0.26<T<0.32$ & $1.224$ & $0.0437$/$\sqrt{1\over3} \times 0.0437$
\\
                                                 & $\Sigma_c^{0}({1\over2}^-/{3\over2}^-)$     & 1.70 & $0.26<T<0.32$ & $1.224$ & $\sqrt2 \times 0.0437$/$\sqrt{2\over3} \times 0.0437$
\\
                                                 & $\Xi_c^{\prime+}({1\over2}^-/{3\over2}^-)$  & 1.95 & $0.26<T<0.35$ & $1.422$ & $0.0680$/$\sqrt{1\over3} \times 0.0680$
\\
                                                 & $\Xi_c^{\prime0}({1\over2}^-/{3\over2}^-)$  & 1.95 & $0.26<T<0.35$ & $1.422$ & $0.0680$/$\sqrt{1\over3} \times 0.0680$
\\
                                                 & $\Omega_c^{+}({1\over2}^-/{3\over2}^-)$     & 2.20 & $0.25<T<0.39$ & $1.641$ & $\sqrt2 \times 0.108$/$\sqrt{2\over3} \times 0.108$
\\ \hline
\multirow{6}{*}{$[\mathbf{6}_F, 0, 1, \lambda]$}    & $\Sigma_c^{++}({1\over2}^-)$    & 1.45 & $\sim0.29$ & $1.100$ & $\sqrt2 \times 0.0344$
\\
                                                    & $\Sigma_c^{+}({1\over2}^-)$     & 1.45 & $\sim0.29$ & $1.100$ & $0.0344$
\\
                                                    & $\Sigma_c^{0}({1\over2}^-)$     & 1.45 & $\sim0.29$ & $1.100$ & $\sqrt2 \times 0.0344$
\\
                                                    & $\Xi_c^{\prime+}({1\over2}^-)$  & 1.70 & $0.27<T<0.32$ & $1.295$ & $0.0512$
\\
                                                    & $\Xi_c^{\prime0}({1\over2}^-)$  & 1.70 & $0.27<T<0.32$ & $1.295$ & $0.0512$
\\
                                                    & $\Omega_c^{+}({1\over2}^-)$     & 1.95 & $0.27<T<0.33$ & $1.504$ & $\sqrt2 \times 0.0804$
\\ \hline
\multirow{6}{*}{$[\mathbf{6}_F, 1, 1, \lambda]$}    & $\Sigma_c^{++}({1\over2}^-/{3\over2}^-)$    & 1.75 & $0.32<T<0.34$ & $1.066$ & $\sqrt2 \times 0.0349$/$\sqrt2 \times 0.0349$
\\
                                                    & $\Sigma_c^{+}({1\over2}^-/{3\over2}^-)$     & 1.75 & $0.32<T<0.34$ & $1.066$ & $0.0349$/$0.0349$
\\
                                                    & $\Sigma_c^{0}({1\over2}^-/{3\over2}^-)$     & 1.75 & $0.32<T<0.34$ & $1.066$ & $\sqrt2 \times 0.0349$/$\sqrt2 \times 0.0349$
\\
                                                    & $\Xi_c^{\prime+}({1\over2}^-/{3\over2}^-)$  & 1.75 & $\sim0.35$ & $1.181$ & $0.0451$/$0.0451$
\\
                                                    & $\Xi_c^{\prime0}({1\over2}^-/{3\over2}^-)$  & 1.75 & $\sim0.35$ & $1.181$ & $0.0451$/$0.0451$
\\
                                                    & $\Omega_c^{+}({1\over2}^-/{3\over2}^-)$     & 1.75 & $\sim0.36$ & $1.270$ & $\sqrt2 \times 0.0546$/$\sqrt2 \times 0.0546$
\\ \hline
\multirow{6}{*}{$[\mathbf{6}_F, 2, 1, \lambda]$}    & $\Sigma_c^{++}({3\over2}^-/{5\over2}^-)$    & 1.50 & $0.27<T<0.29$ & $1.099$ & $\sqrt2 \times 0.0395$/$\sqrt{2\over5} \times 0.0395$
\\
                                                    & $\Sigma_c^{+}({3\over2}^-/{5\over2}^-)$     & 1.50 & $0.27<T<0.29$ & $1.099$ & $0.0395$/$\sqrt{1\over5} \times 0.0395$
\\
                                                    & $\Sigma_c^{0}({3\over2}^-/{5\over2}^-)$     & 1.50 & $0.27<T<0.29$ & $1.099$ & $\sqrt2 \times 0.0395$/$\sqrt{2\over5} \times 0.0395$
\\
                                                    & $\Xi_c^{\prime+}({3\over2}^-/{5\over2}^-)$  & 1.75 & $0.26<T<0.32$ & $1.254$ & $0.0599$/$\sqrt{1\over5} \times 0.0599$
\\
                                                    & $\Xi_c^{\prime0}({3\over2}^-/{5\over2}^-)$  & 1.75 & $0.26<T<0.32$ & $1.254$ & $0.0599$/$\sqrt{1\over5} \times 0.0599$
\\
                                                    & $\Omega_c^{+}({3\over2}^-/{5\over2}^-)$     & 2.00 & $0.26<T<0.36$ & $1.461$ & $\sqrt2 \times 0.0976$/$\sqrt{2\over5} \times 0.0976$
\\ \hline \hline
\end{tabular}
\label{tab:pwave2}
\end{center}
\end{table}

\section{Decay Properties of flavor $\mathbf{\bar 3}_F$ $P$-wave charmed baryons}
\label{sec:sumrule1}

In this section we use the method of light-cone QCD sum rules to study decay properties of the flavor $\mathbf{\bar 3}_F$ $P$-wave charmed baryons.
We only study their $S$-wave decays into ground-state charmed baryons accompanied by a pseudoscalar meson ($\pi$ or $K$) or a vector meson ($\rho$ or $K^*$).
Because the masses of the flavor $\mathbf{\bar 3}_F$ $P$-wave charmed baryons are sometimes below the two-body decay thresholds (such as $\Lambda_c(3/2^-) \rightarrow \Sigma_c^{*}(3/2^+) + \pi$), we also study their three-body decays, which are kinematically allowed (such as $\Lambda_c(3/2^-) \rightarrow \Sigma_c^{*}(3/2^+) + \pi \rightarrow \Lambda_c(1/2^+) + \pi + \pi$).

The possible decay channels are:
\begin{eqnarray}
&(a)& {\bf \Gamma\Big[}\Lambda_c(1/2^-) \rightarrow \Sigma_c(1/2^+) + \pi{\Big ]}
\\ \nonumber &=& {\bf \Gamma\Big[}\Lambda_c^+(1/2^-) \rightarrow \Sigma_c^{+}(1/2^+) + \pi^0{\Big ]}
+ 2 \times {\bf \Gamma\Big[}\Lambda_c^+(1/2^-) \rightarrow \Sigma_c^{++}(1/2^+) + \pi^- \rightarrow \Lambda_c^{+}(1/2^+) + \pi^+ + \pi^-{\Big ]} \, ,
\\ &(b)& {\bf \Gamma\Big[}\Lambda_c(3/2^-) \rightarrow \Sigma_c^{*}(3/2^+) + \pi \rightarrow \Lambda_c(1/2^+) + \pi + \pi{\Big ]}
\\ \nonumber &=& 3 \times {\bf \Gamma\Big[}\Lambda_c^+(3/2^-) \rightarrow \Sigma_c^{*++}(3/2^+) + \pi^- \rightarrow \Lambda_c^{+}(1/2^+) + \pi^+ + \pi^-{\Big ]} \, ,
\\ &(c)& {\bf \Gamma\Big[}\Xi_c(1/2^-) \rightarrow \Xi_c(1/2^+) + \pi{\Big ]}
= {3\over2} \times {\bf \Gamma\Big[}\Xi_c^0(1/2^-) \rightarrow \Xi_c^{+}(1/2^+) + \pi^-{\Big ]} \, ,
\\ &(d)& {\bf \Gamma\Big[}\Xi_c(1/2^-) \rightarrow \Lambda_c(1/2^+) + K{\Big ]}
= {\bf \Gamma\Big[}\Xi_c^0(1/2^-) \rightarrow \Lambda_c^{+}(1/2^+) + K^-{\Big ]} \, ,
\\ &(e)& {\bf \Gamma\Big[}\Xi_c(1/2^-) \rightarrow \Xi_c(1/2^+) + \rho \rightarrow \Xi_c(1/2^+) + \pi + \pi{\Big ]}
\\ \nonumber &=& {3\over2} \times {\bf \Gamma\Big[}\Xi_c^{0}(1/2^-) \rightarrow \Xi_c^{+}(1/2^+) + \rho^- \rightarrow \Xi_c^{+}(1/2^+) + \pi^0 + \pi^-{\Big ]} \, ,
\\ &(f)& {\bf \Gamma\Big[}\Xi_c(1/2^-) \rightarrow \Xi_c^{\prime}(1/2^+) + \pi{\Big ]}
= {3\over2} \times {\bf \Gamma\Big[}\Xi_c^0(1/2^-) \rightarrow \Xi_c^{\prime+}(1/2^+) + \pi^-{\Big ]} \, ,
\\ &(g)& {\bf \Gamma\Big[}\Xi_c(3/2^-) \rightarrow \Xi_c(1/2^+) + \rho^- \rightarrow \Xi_c(1/2^+) + \pi + \pi{\Big ]}
\\ \nonumber &=& {3\over2} \times {\bf \Gamma\Big[}\Xi_c^{0}(3/2^-) \rightarrow \Xi_c^{+}(1/2^+) + \rho^- \rightarrow \Xi_c^{+}(1/2^+) + \pi^0 + \pi^-{\Big ]} \, ,
\\ &(h)& {\bf \Gamma\Big[}\Xi_c(3/2^-) \rightarrow \Xi_c^{*}(3/2^+) + \pi{\Big ]}
= {3\over2} \times {\bf \Gamma\Big[}\Xi_c^0(3/2^-) \rightarrow \Xi_c^{*+}(3/2^+) + \pi^-{\Big ]} \, ,
\end{eqnarray}
which can be calculated through the following Lagrangians
\begin{eqnarray}
\nonumber &(a)& \mathcal{L}_{\Lambda_c[{1\over2}^-] \rightarrow \Sigma_c \pi} = g_{\Lambda_c^+[{1\over2}^-] \rightarrow \Sigma_c^{++} \pi^-} {\bar \Lambda_c^+({1/2}^-)} \Sigma_c^{++} \pi^- + \cdots \, ,
\\
\nonumber &(b)& \mathcal{L}_{\Lambda_c[{3\over2}^-] \rightarrow \Sigma_c^* \pi} = g_{\Lambda_c^+[{3\over2}^-] \rightarrow \Sigma_c^{*++} \pi^-} {\bar \Lambda_{c\mu}^+({3/2}^-)} \Sigma_{c\mu}^{*++} \pi^- + \cdots \, ,
\\
\nonumber &(c)& \mathcal{L}_{\Xi_c[{1\over2}^-] \rightarrow \Xi_c \pi} = g_{\Xi_c^0[{1\over2}^-] \rightarrow \Xi_c^{+} \pi^-} {\bar \Xi_c^0({1/2}^-)} \Xi_c^{+} \pi^- + \cdots \, ,
\\
&(d)& \mathcal{L}_{\Xi_c[{1\over2}^-] \rightarrow \Lambda_c K} = g_{\Xi_c^0[{1\over2}^-] \rightarrow \Lambda_c^{+} K^-} {\bar \Xi_c^0({1/2}^-)} \Lambda_c^{+} K^- + \cdots \, ,
\label{lag:ah}
\\
\nonumber &(e)& \mathcal{L}_{\Xi_c[{1\over2}^-] \rightarrow \Xi_c \rho} = g_{\Xi_c^0[{1\over2}^-] \rightarrow \Xi_c^{+} \rho^-} {\bar \Xi_c^0({1/2}^-)} \gamma_\mu \gamma_5 \Xi_c^{+} \rho_\mu^- + \cdots \, ,
\\
\nonumber &(f)& \mathcal{L}_{\Xi_c[{1\over2}^-] \rightarrow \Xi_c^\prime \pi} = g_{\Xi_c^0[{1\over2}^-] \rightarrow \Xi_c^{\prime+}\pi^-} {\bar \Xi_c^0({1/2}^-)} \Xi_c^{\prime+} \pi^- + \cdots \, ,
\\
\nonumber &(g)& \mathcal{L}_{\Xi_c[{3\over2}^-] \rightarrow \Xi_c \rho} = g_{\Xi_c^0[{3\over2}^-] \rightarrow \Xi_c^{+} \rho^-} {\bar \Xi_{c\mu}^0({3/2}^-)} \Xi_c^{+} \rho_\mu^- + \cdots \, ,
\\
\nonumber &(h)& \mathcal{L}_{\Xi_c[{3\over2}^-] \rightarrow \Xi_c^* \pi} = g_{\Xi_c^0[{3\over2}^-] \rightarrow \Xi_c^{*+} \pi^-} {\bar \Xi_{c\mu}^0({3/2}^-)} \Xi_{c\mu}^{*+} \pi^- + \cdots \, .
\end{eqnarray}
We note that the mass of the $\Lambda_c(2595)$ is above the threshold of $\Sigma_c^{+} \pi^0$ but below the thresholds of $\Sigma_c^{++} \pi^-$ and $\Sigma_c^{0} \pi^+$, so we evaluate both its two-body decay $\Lambda_c(2595) \rightarrow \Sigma_c^{+} \pi^0$ and three-body decays $\Lambda_c(2595) \rightarrow \Sigma_c^{++} \pi^- \rightarrow \Lambda_c^{+} \pi^+ \pi^-$ and $\Lambda_c(2595) \rightarrow \Sigma_c^{0} \pi^+ \rightarrow \Lambda_c^{+} \pi^+ \pi^-$, using~\cite{pdg}:
\begin{eqnarray}
\nonumber && \Sigma_c^{++,0} : m \approx 2453.86~{\rm MeV}  \, , \, \Sigma_c^{+} : m = 2452.9~{\rm MeV}  \, ,
\\ \nonumber && \pi^\pm : m = 139.57~{\rm MeV} \, , \,  \pi^0 : m = 134.98~{\rm MeV} \, .
\end{eqnarray}
Besides these channels, we also assume masses of the $\Lambda_c(5/2^-)$ and $\Xi_c(5/2^-)$ to be around
\begin{eqnarray}
\Lambda_c({5/2}^-) &:& m \sim 2850 {\rm~MeV} \, ,
\label{mass:spin52}
\\ \nonumber \Xi_c({5/2}^-) &:& m \sim 3000 {\rm~MeV} \, ,
\end{eqnarray}
so that the following decay channels are kinematically allowed
\begin{eqnarray}
&(i)& {\bf \Gamma\Big[}\Lambda_c(5/2^-) \rightarrow \Sigma_c^{*}(3/2^+) + \rho  \rightarrow \Sigma_c^{*}(3/2^+) + \pi + \pi{\Big ]}
\\ \nonumber &=& 3 \times {\bf \Gamma\Big[}\Lambda_c^+(5/2^-) \rightarrow \Sigma_c^{*++}(3/2^+) + \rho^- \rightarrow \Sigma_c^{*++}(3/2^+) + \pi^0 + \pi^-{\Big ]} \, ,
\\ &(j)& {\bf \Gamma\Big[}\Xi_c(5/2^-) \rightarrow \Xi_c^{*}(3/2^+) + \rho \rightarrow \Xi_c^{*}(3/2^+) + \pi + \pi{\Big ]}
\\ \nonumber &=& {3\over2} \times {\bf \Gamma\Big[}\Xi_c^0(5/2^-) \rightarrow \Xi_c^{*+}(3/2^+) + \rho^- \rightarrow \Xi_c^{*+}(3/2^+) + \pi^0 + \pi^-{\Big ]} \, ,
\end{eqnarray}
and can be calculated through the following Lagrangians
\begin{eqnarray}
&(i)& \mathcal{L}_{\Lambda_c[{5\over2}^-] \rightarrow \Sigma_c^* \rho} = g_{\Lambda_c^+[{5\over2}^-] \rightarrow \Sigma_c^{*++} \rho^-} {\bar \Lambda_{c\mu\nu}^+({5/2}^-)} \Sigma_{c\mu}^{*++} \rho_\nu^- + \cdots \, ,
\label{lag:ij}
\\
\nonumber &(j)& \mathcal{L}_{\Xi_c[{5\over2}^-] \rightarrow \Xi_c^* \rho} = g_{\Xi_c^0[{5\over2}^-] \rightarrow \Xi_c^{*+} \rho^-} {\bar \Xi_{c\mu\nu}^0({5/2}^-)} \Xi_{c\mu}^{*+} \rho_\nu^- + \cdots \, .
\end{eqnarray}

As an example, we shall first study the $S$-wave decay of the $\Xi_c^0({1/2}^-)$ belonging to $[\mathbf{\bar 3}_F, 1, 1, \rho]$ into $\Xi_c^{\prime+}(1/2^+)$ and $\pi^-(0^-)$ in the next subsection, and then separately investigate the four charmed baryon multiplets of flavor $\mathbf{\bar 3}_F$, $[\mathbf{\bar 3}_F, 0, 1, \rho]$, $[\mathbf{\bar 3}_F, 1, 1, \rho]$, $[\mathbf{\bar 3}_F, 2, 1, \rho]$ and $[\mathbf{\bar 3}_F, 1, 0, \lambda]$, in the following subsections.

\subsection{$\Xi_c^0({1/2}^-)$ of $[\mathbf{\bar 3}_F, 1, 1, \rho]$ decaying into $\Xi_c^{\prime+}(1/2^+)$ and $\pi^-(0^-)$}

As an example, we evaluate the following three-point correlation function to study the $S$-wave decay of the $\Xi_c^0({1/2}^-)$ belonging to $[\mathbf{\bar 3}_F, 1, 1, \rho]$ into $\Xi_c^{\prime+}(1/2^+)$ and $\pi^-(0^-)$:
\begin{eqnarray}
\Pi(\omega, \, \omega^\prime) &=& \int d^4 x e^{-i k \cdot x} \langle 0 | J_{1/2,-,\Xi_c^0,1,1,\rho}(0) \bar J_{\Xi_c^{\prime+}}(x) | \pi^- \rangle
\\ \nonumber &=& {1+v\!\!\!\slash\over2} G_{\Xi_c^0[{1\over2}^-] \rightarrow \Xi_c^{\prime+}\pi^-} (\omega, \omega^\prime) \, ,
\end{eqnarray}
where
\begin{eqnarray}
k^\prime = k + q \, ,  \, \omega^\prime = v \cdot k^\prime \, ,  \, \omega = v \cdot k \, .
\end{eqnarray}
The currents $J_{1/2,-,\Xi_c^0,1,1,\rho}$ and $J_{\Xi_c^{\prime+}}$ have been defined in Eqs.~(\ref{eq:current4}) and (\ref{eq:sigmapp}),
and couple to $\Xi_c^0({1/2}^-)$ belonging to $[\mathbf{\bar 3}_F, 1, 1, \rho]$ and $\Xi_c^{\prime+}(1/2^+)$, respectively.
The function $G_{\Xi_c^0[{1\over2}^-] \rightarrow \Xi_c^{\prime+}\pi^-}$ has the following pole terms at the hadronic level from double dispersion relation:
\begin{eqnarray}
G_{\Xi_c^0[{1\over2}^-] \rightarrow \Xi_c^{\prime+}\pi^-} (\omega, \omega^\prime) &=& g_{\Xi_c^0[{1\over2}^-] \rightarrow \Xi_c^{\prime+}\pi^-} \times { f_{\Xi_c^0[{1\over2}^-]} f_{\Xi_c^{\prime+}} \over (\bar \Lambda_{\Xi_c^0[{1\over2}^-]} - \omega^\prime) (\bar \Lambda_{\Xi_c^{\prime+}} - \omega)} + { c^\prime \over \bar \Lambda_{\Xi_c^0[{1\over2}^-]} - \omega^\prime} + { c \over \bar \Lambda_{\Xi_c^{\prime+}} - \omega} \, , \label{G0C}
\end{eqnarray}
where the $S$-wave coupling constants $g_{\Xi_c^0[{1\over2}^-] \rightarrow \Xi_c^{\prime+}\pi^-}$ is defined through the following Lagrangian
\begin{eqnarray}
\label{eq:Lagexample}
\mathcal{L}_{\Xi_c[{1\over2}^-] \rightarrow \Xi_c^\prime \pi} &=& g_{\Xi_c^0[{1\over2}^-] \rightarrow \Xi_c^{\prime+}\pi^-} {\bar \Xi_c^0({1/2}^-)} \Xi_c^{\prime+} \pi^- + \cdots \, .
\end{eqnarray}
$c$ and $c^\prime$ in Eq. \eqref{G0C} are free parameters which can be suppressed by the Borel transformation. The three-point correlation function $\Pi(\omega, \omega^\prime)$ can also be calculated at the quark-gluon level using the QCD operator product expansion (in our calculations we have used the software {\it Mathematica} with the $FeynCalc$ package~\cite{Mertig:1990an}):
\begin{eqnarray}
\label{eq:sumrule}
&& G_{\Xi_c^0[{1\over2}^-] \rightarrow \Xi_c^{\prime+}\pi^-} (\omega, \omega^\prime)
= g_{\Xi_c^0[{1\over2}^-] \rightarrow \Xi_c^{\prime+}\pi^-} \times { f_{\Xi_c^0[{1\over2}^-]} f_{\Xi_c^{\prime+}} \over (\bar \Lambda_{\Xi_c^0[{1\over2}^-]} - \omega^\prime) (\bar \Lambda_{\Xi_c^{\prime+}} - \omega)}
\\ \nonumber &=&
\int_0^\infty dt \int_0^1 du e^{i (1-u) \omega^\prime t} e^{i u \omega t} \times 4 \times \Big (
\frac{3 i f_\pi v \cdot q }{2 \pi^2 t^4} \phi_{2;\pi}(u)
+ \frac{3 i f_\pi v \cdot q}{32 \pi^2 t^2} \phi_{4;\pi}(u)
+ \frac{3 i f_\pi}{2 \pi^2 t^4 v \cdot q} \psi_{4;\pi}(u)
\\ \nonumber &&
+ \frac{i f_\pi m_\pi^2 v \cdot q}{24 (m_u + m_d)} {\langle \bar s s \rangle} \phi^\sigma_{3;\pi}(u)
+ \frac{i f_\pi m_\pi^2 t^2 v \cdot q}{384 (m_u + m_d)} \langle g_s \bar s \sigma G s\rangle \phi_{3;\pi}^\sigma(u)
+ \frac{i f_\pi m_\pi^2 v \cdot q}{8 \pi^2 t^2 (m_u + m_d)} m_s \phi_{3;\pi}^\sigma(u)
\\ \nonumber &&
+ \frac{i f_\pi v \cdot q}{16} m_s {\langle \bar s s \rangle} \phi_{2;\pi}(u)
+ \frac{i f_\pi t^2 v\cdot q}{256} m_s {\langle \bar s s \rangle} \phi_{4;\pi}(u)
+ \frac{i f_\pi }{16 v\cdot q} m_s {\langle \bar s s \rangle} \psi_{4;\pi}(u) \Big )
\\ \nonumber &+&
\int_0^\infty dt \int_0^1 du \int \mathcal{D}\underline{\alpha} e^{i \omega^\prime t (\alpha_2 + u\alpha_3)} e^{i \omega t(1-\alpha_2-u\alpha_3)} \times 4 \times \Big (
\\ \nonumber &&
\frac{3 i f_\pi v\cdot q}{8 \pi^2 t^2} \Phi_{4;\pi}(\underline{\alpha})
- \frac{i f_\pi v\cdot q}{4 \pi^2 t^2} \Psi_{4;\pi}(\underline{\alpha})
+ \frac{i f_\pi v\cdot q}{8 \pi^2 t^2} \widetilde \Phi_{4;\pi}(\underline{\alpha})
+ \frac{i f_\pi v\cdot q}{4 \pi^2 t^2} \widetilde \Psi_{4;\pi}(\underline{\alpha})
+ \frac{i f_\pi u v \cdot q }{4 \pi^2 t^2} \Phi_{4;\pi}(\underline{\alpha})
+ \frac{i f_\pi u v\cdot q}{2 \pi^2 t^2} \Psi_{4;\pi}(\underline{\alpha}) \Big ) \, ,
\end{eqnarray}
which contains many light-cone distribution amplitudes, whose definitions and explicit forms can be found in Refs.~\cite{Ball:1998je,Ball:2006wn,Ball:2004rg,Ball:1998kk,Ball:1998sk,Ball:1998ff,Ball:2007rt,Ball:2007zt}.
As examples, we list the light-cone distribution amplitudes of the $K$ meson in Appendix~\ref{sec:wavefunction}.
Their values can also be found in these references, and in the present study we work at the renormalization scale 1 GeV.
The condensates contained in this sum rules take the following
values~\cite{pdg,Yang:1993bp,Hwang:1994vp,Ovchinnikov:1988gk,Jamin:2002ev,Ioffe:2002be,Narison:2002pw,Gimenez:2005nt,colangelo}:
%
\begin{eqnarray}
\nonumber && \langle \bar qq \rangle = - (0.24 \mbox{ GeV})^3 \, ,
\\ \nonumber && \langle \bar ss \rangle = (0.8\pm 0.1)\times \langle\bar qq \rangle \, ,
\\ &&\langle g_s^2GG\rangle =(0.48\pm 0.14) \mbox{ GeV}^4\, ,
\label{eq:condensates}
\\ \nonumber && \langle g_s \bar q \sigma G q \rangle = M_0^2 \times \langle \bar qq \rangle\, ,
\\ \nonumber && \langle g_s \bar s \sigma G s \rangle = M_0^2 \times \langle \bar ss \rangle\, ,
\\ \nonumber && M_0^2= 0.8 \mbox{ GeV}^2\, .
\end{eqnarray}
After Wick rotations and making double Borel transformation with the variables $\omega$ and $\omega^\prime$ to be $T_1$ and $T_2$, we obtain
\begin{eqnarray}
&& g_{\Xi_c^0[{1\over2}^-] \rightarrow \Xi_c^{\prime+}\pi^-} f_{\Xi_c^0[{1\over2}^-]} f_{\Xi_c^{\prime+}} e^{- {\bar \Lambda_{\Xi_c^0[{1\over2}^-]} \over T_1}} e^{ - {\bar \Lambda_{\Xi_c^{\prime+}} \over T_2}}
\\ \nonumber &=& 4 \times \Big ( - \frac{3 f_\pi}{2 \pi^2} T^6 f_5({\omega_c \over T}) {d\phi_{2;\pi}(u_0)\over du} + \frac{3 f_\pi}{32 \pi^2} T^4 f_3({\omega_c \over T}) {d \phi_{4;\pi}(u_0) \over du} - \frac{3 f_\pi}{2 \pi^2} T^4 f_3({\omega_c \over T}) \int_0^{u_0} \psi_{4;\pi}(u)du
\\ \nonumber && - \frac{f_\pi m_\pi^2}{24 (m_u + m_d)} {\langle \bar s s \rangle} T^2 f_1({\omega_c \over T}) {d \phi_{3;\pi}^\sigma(u_0) \over du} + \frac{f_\pi m_\pi^2}{384 (m_u + m_d)} \langle g_s \bar s \sigma G s\rangle {d \phi_{3;\pi}^\sigma(u_0) \over du}
\\ \nonumber && + \frac{f_\pi m_\pi^2 m_s}{8 \pi^2 (m_u + m_d)} T^4 f_3({\omega_c \over T}) {d \phi_{3;\pi}^\sigma(u_0) \over du} - \frac{f_\pi m_s}{16} {\langle \bar s s \rangle} T^2 f_1({\omega_c \over T}) {d \phi_{2;\pi}(u_0) \over du}
\\ \nonumber && + \frac{f_\pi m_s}{256} {\langle \bar s s \rangle} {d \phi_{4;\pi}(u_0) \over du}
- \frac{f_\pi m_s}{16} {\langle \bar s s \rangle} {d \psi_{4;\pi}(u_0) \over du} \Big )
\\ \nonumber &+& \int_0^{u_0} d\alpha_2 \times {f_\pi\over 2\pi^2 \alpha_3} T^4 f_3({\omega_c \over T}) \times \Big ( 3 \Phi_{4;\pi}(\underline{\alpha}) - 2 \Psi_{4;\pi}(\underline{\alpha}) + \widetilde \Phi_{4;\pi}(\underline{\alpha}) + 2 \widetilde \Psi_{4;\pi}(\underline{\alpha}) \Big ) \Big |_{\alpha_1 = u_0, \alpha_3 = 1 - u_0 - \alpha_2}
\\ \nonumber &-& \int_0^{u_0} d\alpha_1 \times {f_\pi\over 2\pi^2 \alpha_3} T^4 f_3({\omega_c \over T}) \times \Big ( 3 \Phi_{4;\pi}(\underline{\alpha}) - 2 \Psi_{4;\pi}(\underline{\alpha}) + \widetilde \Phi_{4;\pi}(\underline{\alpha}) + 2 \widetilde \Psi_{4;\pi}(\underline{\alpha}) \Big ) \Big |_{\alpha_2 = u_0, \alpha_3 = 1 - u_0 - \alpha_1}
\\ \nonumber &+& \int_0^{u_0} d\alpha_2 \times {f_\pi\over \pi^2 \alpha_3} T^4 f_3({\omega_c \over T}) \times \Big ( \Phi_{4;\pi}(\underline{\alpha}) + 2\Psi_{4;\pi}(\underline{\alpha}) \Big ) \Big |_{\alpha_1 = u_0, \alpha_3 = 1 - u_0 - \alpha_2}
\\ \nonumber &-& \int_0^{u_0} d\alpha_2 \int_{u_0-\alpha_2}^{1-\alpha_2} d\alpha_3 \times {f_\pi\over \pi^2\alpha_3^2} T^4 f_3({\omega_c \over T}) \times \Big ( \Phi_{4;\pi}(\underline{\alpha}) + 2 \Psi_{4;\pi}(\underline{\alpha}) \Big ) \Big |_{\alpha_1 = u_0}
\, ,
\end{eqnarray}
where $u_0 = {T_1 \over T_1 + T_2}$, $T = {T_1 T_2 \over T_1 + T_2}$ and $f_n(x) = 1 - e^{-x} \sum_{k=0}^n {x^k \over k!}$.

\begin{figure}[htb]
\begin{center}
\scalebox{0.6}{\includegraphics{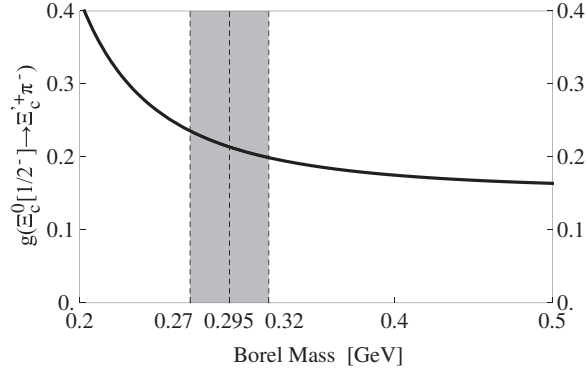}}
\end{center}
\caption{The coupling constant $g_{\Xi_c^0[{1\over2}^-] \rightarrow \Xi_c^{\prime+}\pi^-}$ as a function of the Borel mass $T$.
The current $J_{1/2,-,\Xi_c^0,1,1,\rho}$ belonging to the baryon doublet $[\mathbf{\bar 3}_F, 1, 1, \rho]$ is used here.
\label{fig:311rhox1}}
\end{figure}

We work at the symmetric point $T_1 = T_2 = 2T$, so $u_0 =1/2$. Now the coupling constant $g_{\Xi_c^0[{1\over2}^-] \rightarrow \Xi_c^{\prime+}\pi^-}$ only depends on two free parameters, the threshold value $\omega_c$ and the Borel mass $T$. After choosing $\omega_c = 1.60$ GeV (the average of the thresholds of the $\Xi_c(1/2^-)$ and $\Xi_c^{\prime+}$ mass sum rules), we show $g_{\Xi_c^0[{1\over2}^-] \rightarrow \Xi_c^{\prime+}\pi^-}$ as a function of $T$ in Fig.~\ref{fig:311rhox1}.
The working region for $T$ has been reevaluated and listed in Table~\ref{tab:pwave1} to be $0.27$ GeV $<T<0.32$ GeV, where we obtain
\begin{eqnarray}
&(f)& g_{\Xi_c^0[{1\over2}^-] \rightarrow \Xi_c^{\prime+}\pi^-} = 0.21~{^{+0.15}_{-0.07}} = 0.21~{^{+0.03}_{-0.01}}~{^{+0.06}_{-0.04}}~{^{+0.13}_{-0.06}}~{^{+0.00}_{-0.00}} \, .
\end{eqnarray}
Using this value and the parameters listed in Sec.~\ref{sec:input}, we further obtain
\begin{eqnarray}
&(f)& \Gamma_{\Xi_c^0[{1\over2}^-] \rightarrow \Xi_c^{\prime+}\pi^-} = 1.6~{^{+2.7}_{-0.9}} {\rm~MeV} = 1.6~{^{+0.5}_{-0.1}}~{^{+1.0}_{-0.5}}~{^{+2.5}_{-0.8}}~{^{+0.00}_{-0.00}} {\rm~MeV} \, ,
\end{eqnarray}
where the uncertainties mainly come from the Borel mass ($0.27$ GeV $<T<0.32$ GeV), the parameters of the $\Xi_c^{\prime+}$ \big($\omega_{\Xi_c^{\prime+}} = 1.4 \pm 0.1$ GeV, $\overline{\Lambda}_{\Xi_c^{\prime+}} = 1.042 \pm 0.080$ GeV, and $f_{\Xi_c^{\prime+}} = 0.0435\pm0.0080$ GeV$^3$\big), the parameters of the $\Xi_c^0[{1\over2}^-]$ \big($\omega_{\Xi_c^0[{1\over2}^-]} = 1.8 \pm 0.1$ GeV, $\overline{\Lambda}_{\Xi_c^0[{1\over2}^-]} = 1.349 \pm 0.130$ GeV, and $f_{\Xi_c^0[{1\over2}^-]} = 0.0788\pm0.0280$ GeV$^4$\big), and various quark masses and condensates listed in Eq.~(\ref{eq:condensates}), respectively.
We note that the ${\mathcal O}(1/m_Q)$ corrections ($m_Q$ is the heavy quark mass) have not not considered in the present study, which can cause some theoretical uncertainties (but the ${\mathcal O}(1/m_Q)$ corrections to the masses of the heavy baryons have been taken into account in Refs.~\cite{Liu:2007fg,Chen:2015kpa,Mao:2015gya,Chen:2016phw}).
Totally, the results can be three times larger or smaller than those we have obtained, i.e., $\Gamma_{\Xi_c^0[{1\over2}^-] \rightarrow \Xi_c^{\prime+}\pi^-} = 1.6^{+200\%}_{-67\%}$ MeV.
We shall not estimate the uncertainties of other coupling constants, but just note that their uncertainties are at the same level.

Following these procedures, we separately investigate the four multiplets of flavor $\mathbf{\bar 3}_F$, $[\mathbf{\bar 3}_F, 0, 1, \rho]$, $[\mathbf{\bar 3}_F, 1, 1, \rho]$, $[\mathbf{\bar 3}_F, 2, 1, \rho]$ and $[\mathbf{\bar 3}_F, 1, 0, \lambda]$, in the following subsections.

\subsection{The baryon singlet $[\mathbf{\bar 3}_F, 0, 1, \rho]$}

\begin{figure}[htb]
\begin{center}
\scalebox{0.6}{\includegraphics{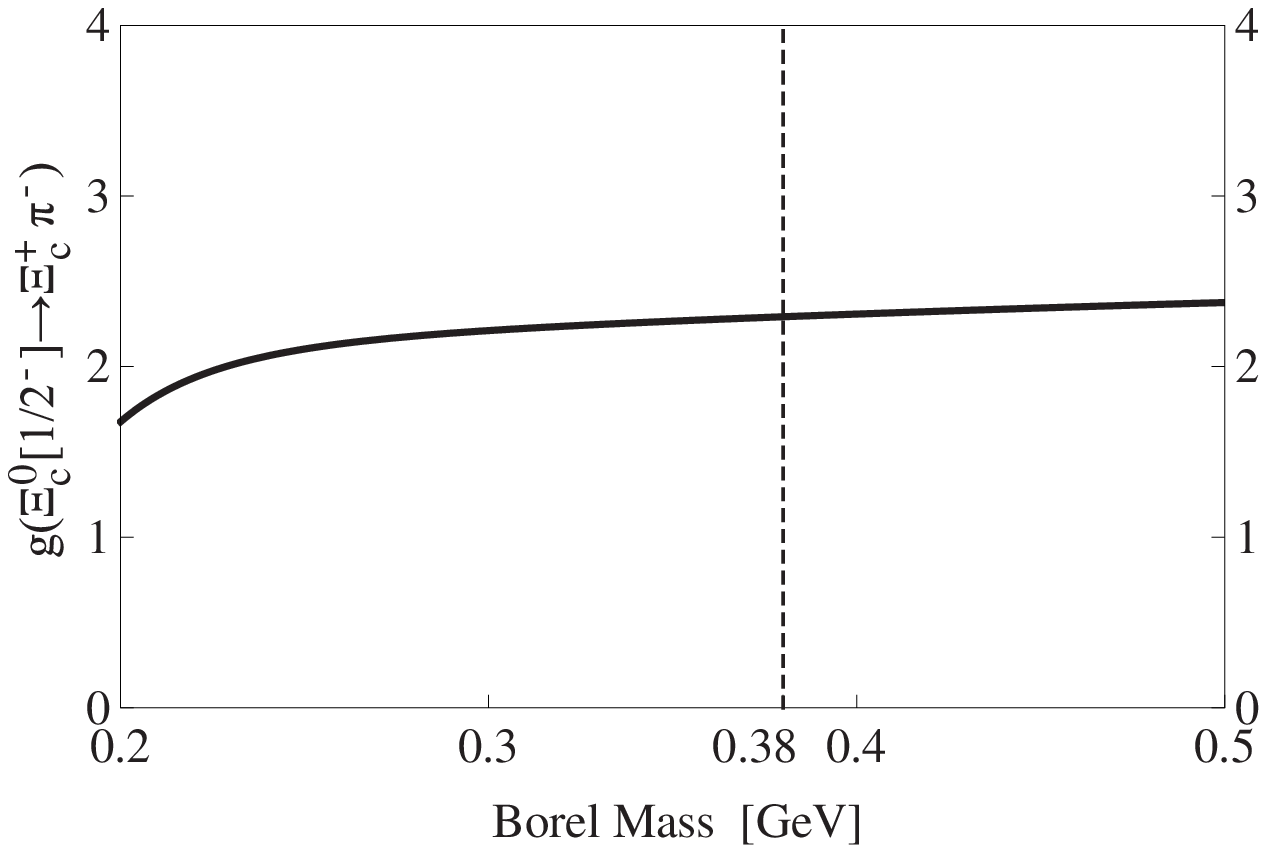}}
\scalebox{0.6}{\includegraphics{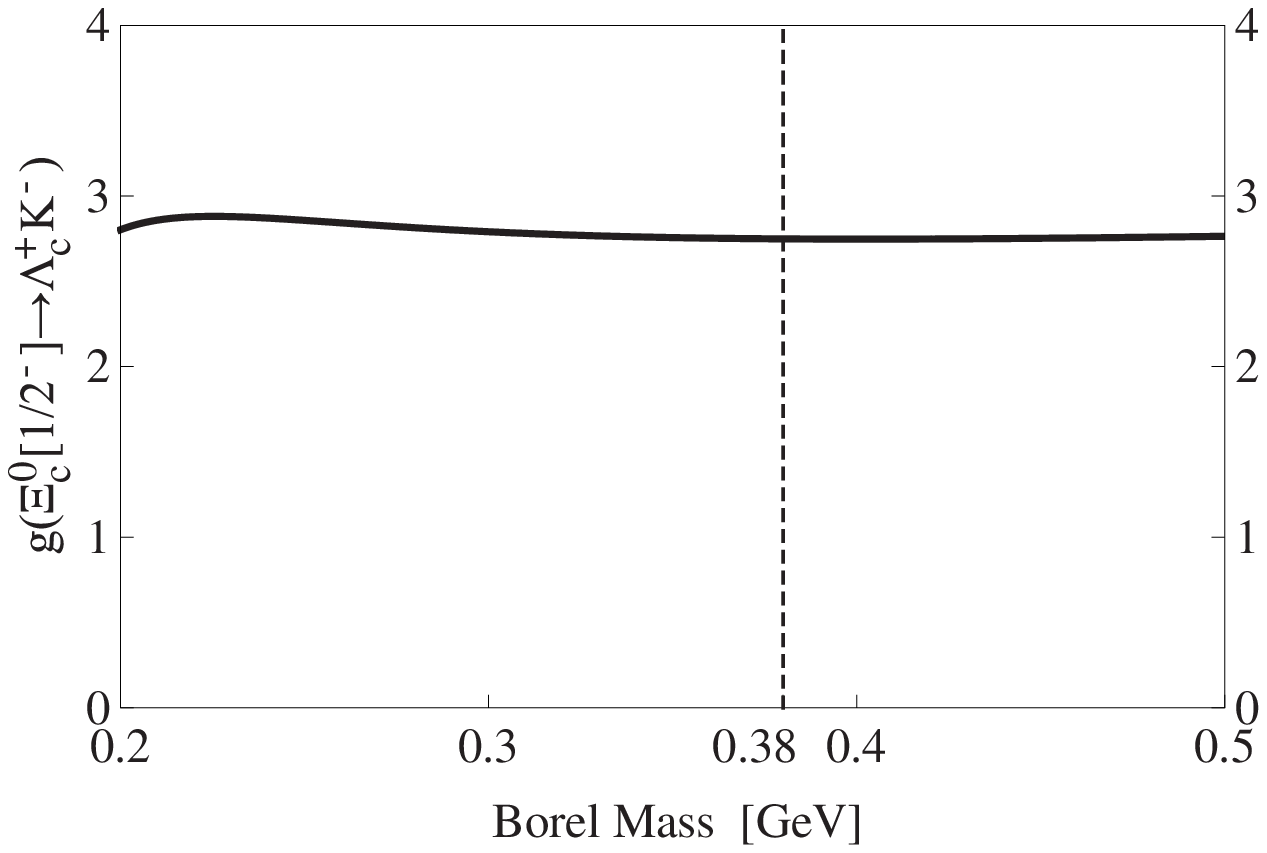}}
\end{center}
\caption{The coupling constants $g_{\Xi_c^0[{1\over2}^-] \rightarrow \Xi_c^{+}\pi^-}$ and $g_{\Xi_c^0[{1\over2}^-] \rightarrow \Lambda_c^{+}K^-}$ as functions of the Borel mass $T$.
The currents belonging to the baryon singlet $[\mathbf{\bar 3}_F, 0, 1, \rho]$ are used here.
\label{fig:301rho}}
\end{figure}

The $[\mathbf{\bar 3}_F, 0, 1, \rho]$ multiplet contains $\Lambda_c({1\over2}^-)$ and $\Xi_c({1\over2}^-)$. Their sum rules are listed in Appendix~\ref{sec:301rho}, suggesting that their possible decay channels are $(c)$ and $(d)$,
while the other three channels $(a)$, $(e)$ and $(f)$ vanish.
We show the two coupling constants, $g_{\Xi_c^0[{1\over2}^-] \rightarrow \Xi_c^{+} \pi^-}$ and $g_{\Xi_c^0[{1\over2}^-] \rightarrow \Lambda_c^{+} K^-}$, as functions of the Borel mass $T$ in Fig.~\ref{fig:301rho}. Using the values of $T$ listed in Table~\ref{tab:pwave1}, we obtain
\begin{eqnarray}
&(c)& g_{\Xi_c^0[{1\over2}^-] \rightarrow \Xi_c^{+} \pi^-} = 2.3 \, ,
\\
\nonumber &(d)& g_{\Xi_c^0[{1\over2}^-] \rightarrow \Lambda_c^{+} K^-} = 2.7 \, .
\end{eqnarray}
Using these values and the parameters listed in Sec.~\ref{sec:input}, we further obtain
\begin{eqnarray}
&(c)& \Gamma_{\Xi_c[{1\over2}^-] \rightarrow \Xi_c \pi} = 300 {\rm~MeV} \, ,
\\
\nonumber &(d)& \Gamma_{\Xi_c[{1\over2}^-] \rightarrow \Lambda_c K} = 82 {\rm~MeV} \, .
\end{eqnarray}

\subsection{The baryon doublet $[\mathbf{\bar 3}_F, 1, 1, \rho]$}

\begin{figure}[htb]
\begin{center}
\scalebox{0.6}{\includegraphics{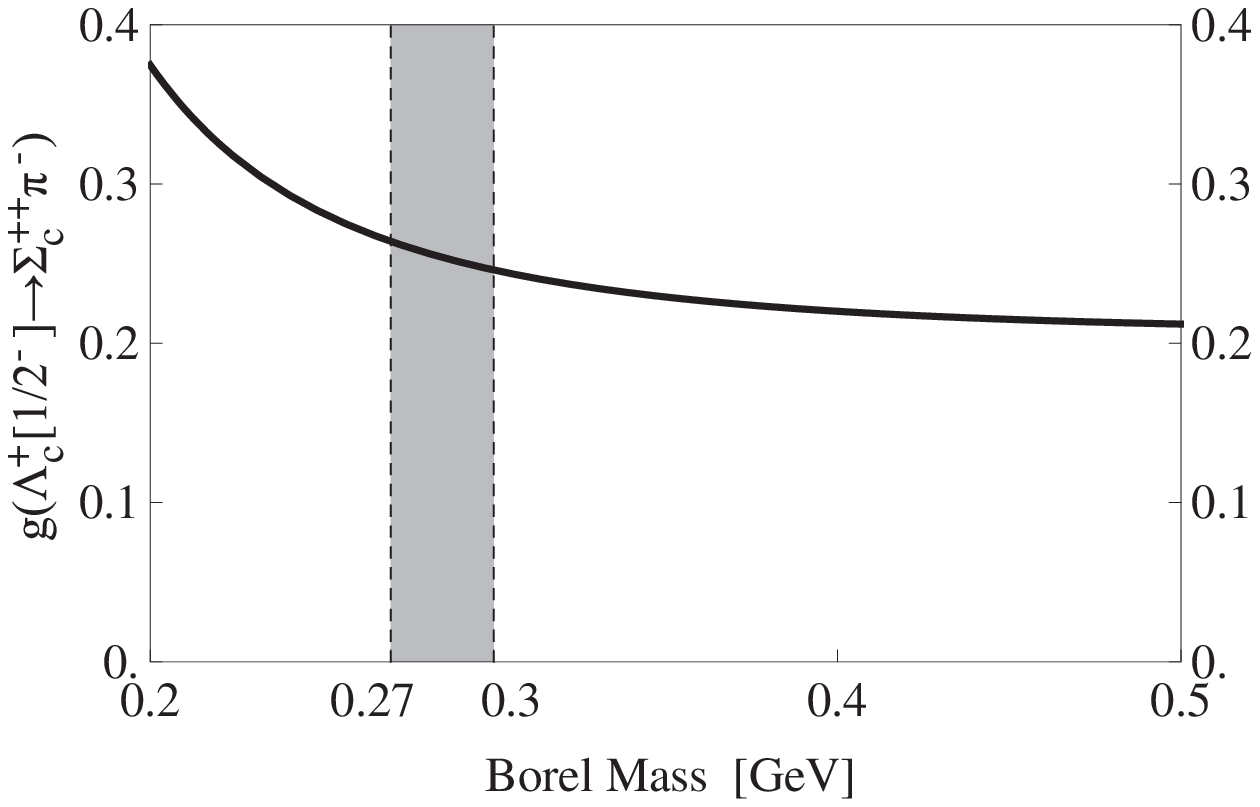}}
\scalebox{0.6}{\includegraphics{R311PX1X1p.eps}}
\\
\scalebox{0.6}{\includegraphics{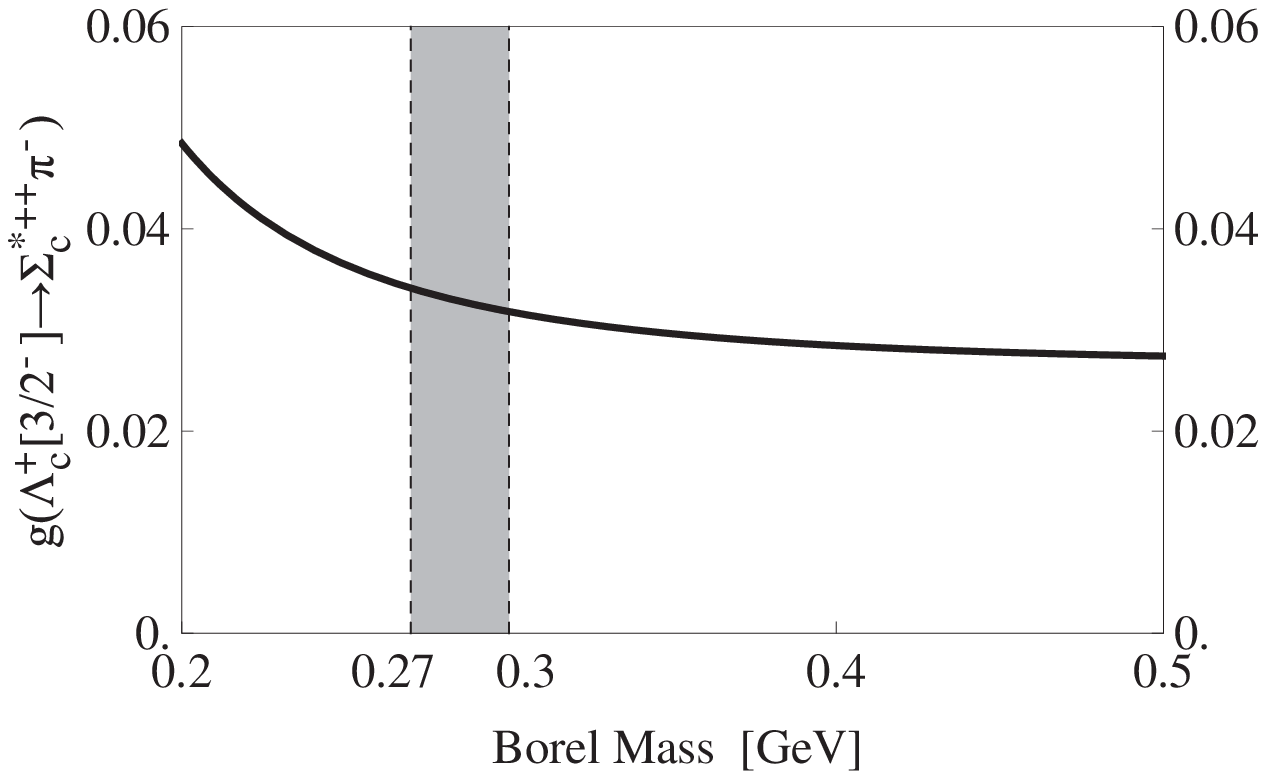}}
\scalebox{0.6}{\includegraphics{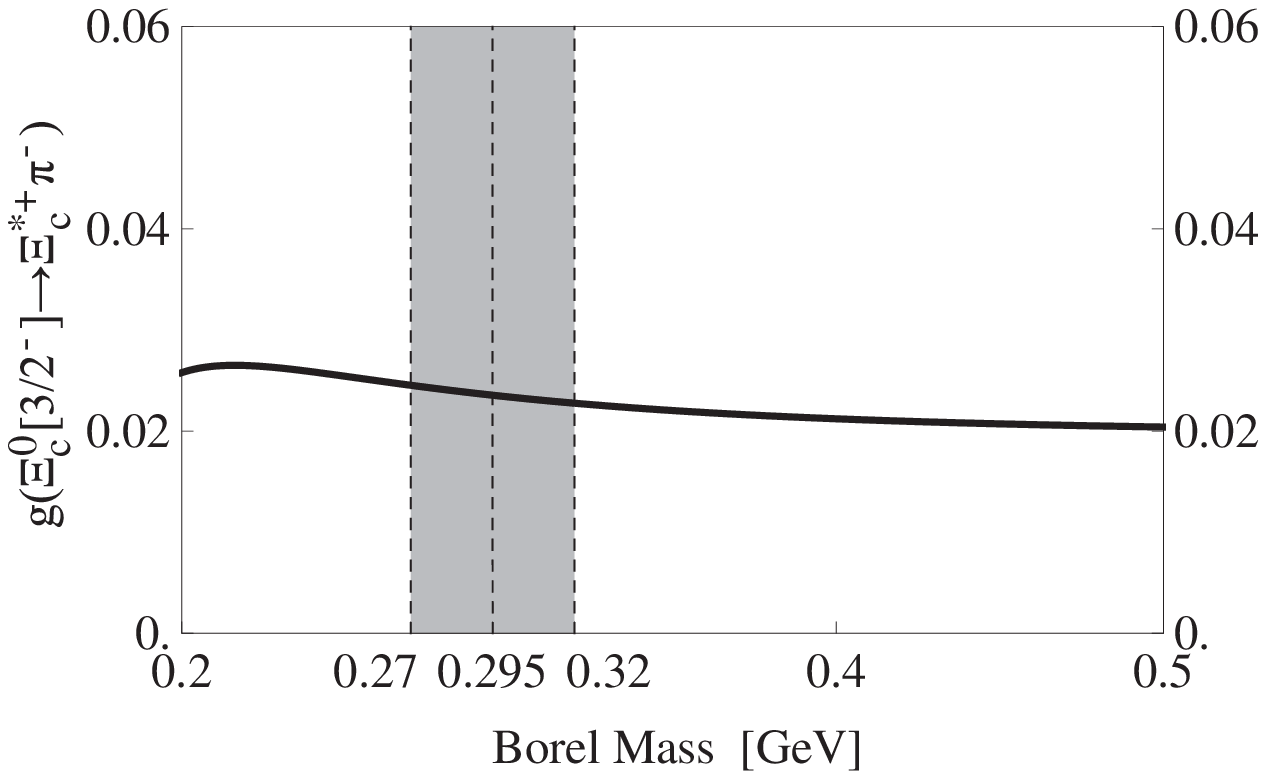}}
\\
\scalebox{0.6}{\includegraphics{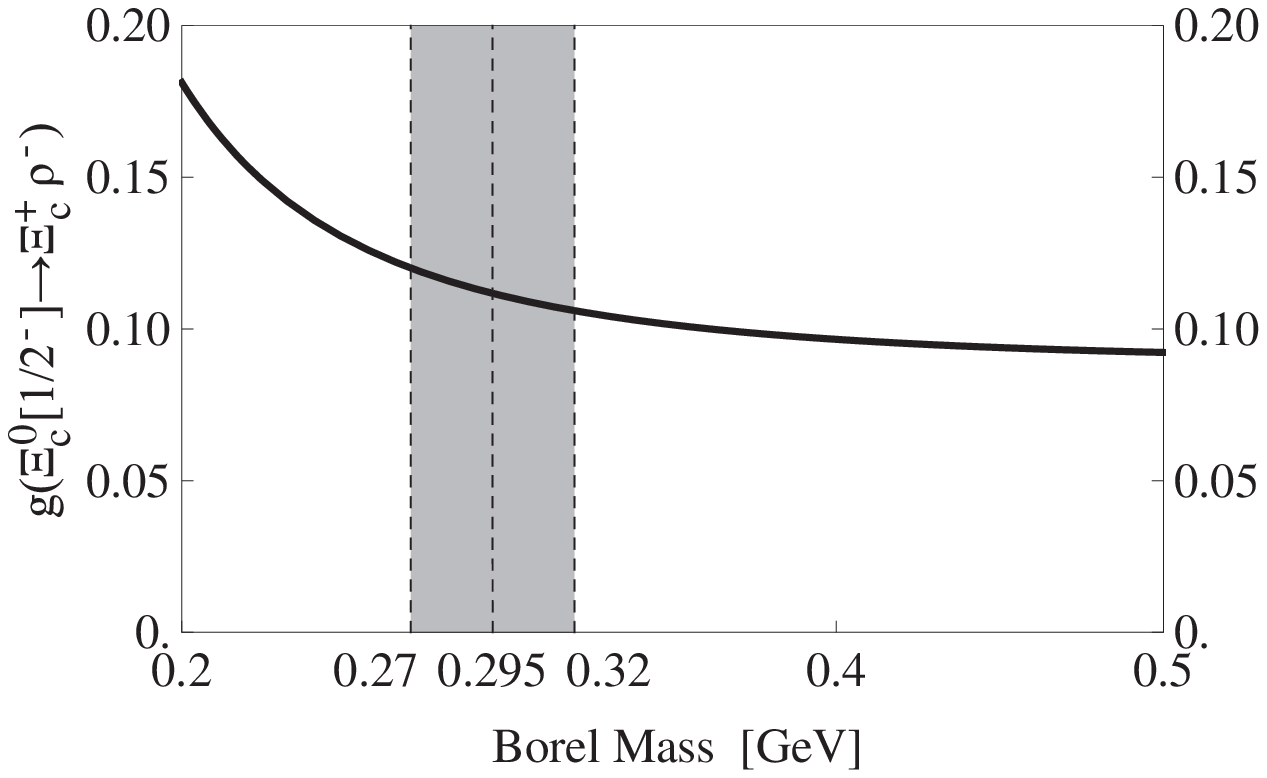}}
\scalebox{0.6}{\includegraphics{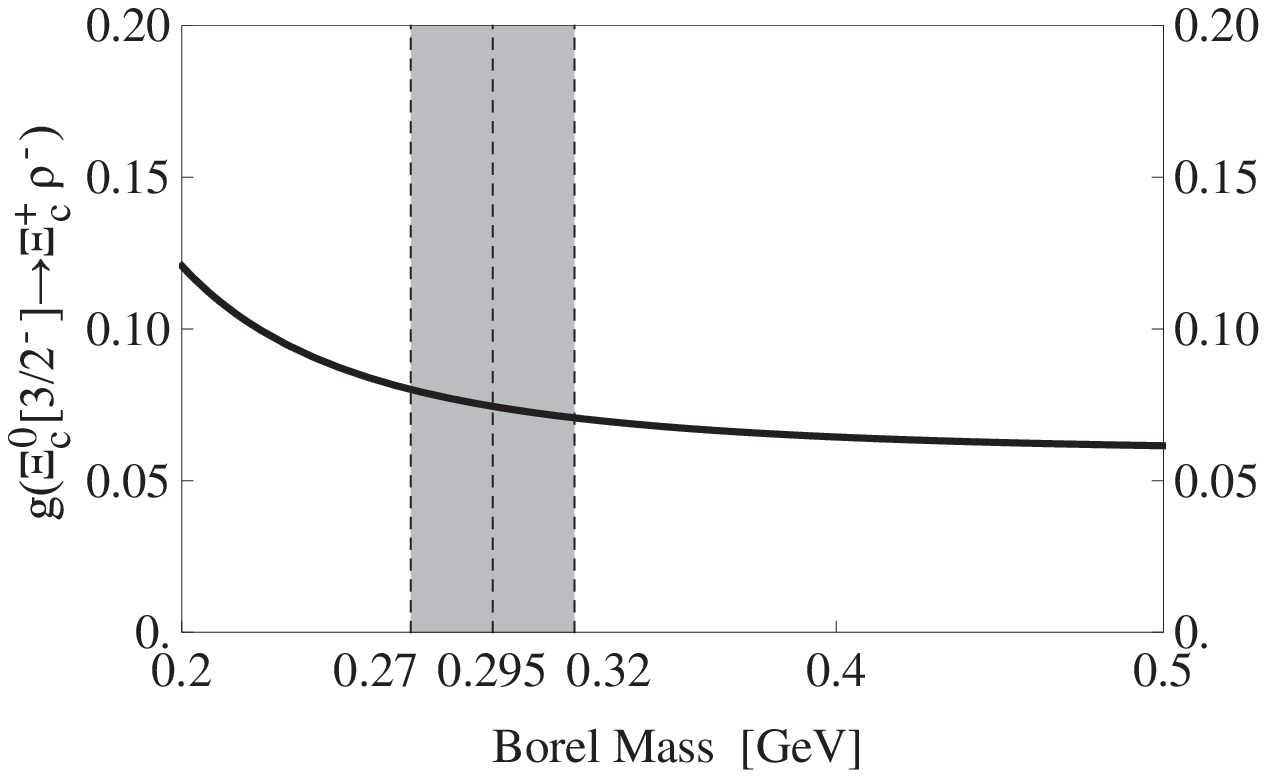}}
\end{center}
\caption{The coupling constants $g_{\Lambda_c^+[{1\over2}^-] \rightarrow \Sigma_c^{++}\pi^-}$ (top-left), $g_{\Xi_c^0[{1\over2}^-] \rightarrow \Xi_c^{\prime+}\pi^-}$ (top-right),
$g_{\Lambda_c^+[{3\over2}^-] \rightarrow \Sigma_c^{*++}\pi^-}$ (middle-left), $g_{\Xi_c^0[{3\over2}^-] \rightarrow \Xi_c^{*+}\pi^-}$ (middle-right),
$g_{\Xi_c^0[{1\over2}^-] \rightarrow \Xi_c^{+}\rho^-}$ (bottom-left) and $g_{\Xi_c^0[{3\over2}^-] \rightarrow \Xi_c^{+}\rho^-}$ (bottom-right) as functions of the Borel mass $T$.
The currents belonging to the baryon doublet $[\mathbf{\bar 3}_F, 1, 1, \rho]$ are used here.
\label{fig:311rho}}
\end{figure}

The $[\mathbf{\bar 3}_F, 1, 1, \rho]$ multiplet contains $\Lambda_c({1\over2}^-/{3\over2}^-)$ and $\Xi_c({1\over2}^-/{3\over2}^-)$. Their sum rules are listed in Appendix~\ref{sec:311rho}, suggesting that their possible decay channels are $(a)$, $(b)$, $(e)$, $(f)$, $(g)$ and $(h)$,
while the other two channels $(c)$ and $(d)$ vanish.
We show the six coupling constants, $g_{\Lambda_c^+[{1\over2}^-] \rightarrow \Sigma_c^{++} \pi^-}$, $g_{\Xi_c^0[{1\over2}^-] \rightarrow \Xi_c^{\prime+}\pi^-}$, $g_{\Lambda_c^+[{3\over2}^-] \rightarrow \Sigma_c^{*++} \pi^-}$, $g_{\Xi_c^0[{3\over2}^-] \rightarrow \Xi_c^{*+} \pi^-}$, $g_{\Xi_c^0[{1\over2}^-] \rightarrow \Xi_c^{+} \rho^-}$ and $g_{\Xi_c^0[{3\over2}^-] \rightarrow \Xi_c^{+} \rho^-}$, as functions of the Borel mass $T$ in Fig.~\ref{fig:311rho}. Using the values of $T$ listed in Table~\ref{tab:pwave1}, we obtain
\begin{eqnarray}
\nonumber &(a)& g_{\Lambda_c^+[{1\over2}^-] \rightarrow \Sigma_c^{++} \pi^-} = 0.25 \, ,
\\
\nonumber &(f)& g_{\Xi_c^0[{1\over2}^-] \rightarrow \Xi_c^{\prime+}\pi^-} = 0.21 \, ,
\\
&(b)& g_{\Lambda_c^+[{3\over2}^-] \rightarrow \Sigma_c^{*++} \pi^-} = 0.033 \, ,
\\
\nonumber &(h)& g_{\Xi_c^0[{3\over2}^-] \rightarrow \Xi_c^{*+} \pi^-} = 0.024 \, ,
\\
\nonumber &(e)& g_{\Xi_c^0[{1\over2}^-] \rightarrow \Xi_c^{+} \rho^-} = 0.11 \, ,
\\
\nonumber &(g)& g_{\Xi_c^0[{3\over2}^-] \rightarrow \Xi_c^{+} \rho^-} = 0.074 \, .
\end{eqnarray}
Using these values and the parameters listed in Sec.~\ref{sec:input}, we further obtain
\begin{eqnarray}
\nonumber &(a,a^\prime)& \Gamma_{\Lambda_c[{1\over2}^-] \rightarrow \Sigma_c \pi (\rightarrow \Lambda \pi \pi)} = 0.39 {\rm~MeV} \, ,
\\
\nonumber &(f)& \Gamma_{\Xi_c[{1\over2}^-] \rightarrow \Xi_c^{\prime}\pi} = 1.6 {\rm~MeV} \, ,
\\
&(b)& \Gamma_{\Lambda_c[{3\over2}^-] \rightarrow \Sigma_c^{*} \pi \to \Lambda_c \pi \pi} = 4 \times 10^{-4} {\rm~MeV} \, ,
\\
\nonumber &(h)& \Gamma_{\Xi_c[{3\over2}^-] \rightarrow \Xi_c^{*} \pi} = 0.01 {\rm~MeV} \, ,
\\
\nonumber &(e)& \Gamma_{\Xi_c[{1\over2}^-] \rightarrow \Xi_c \rho \to  \Xi_c \pi \pi} = 3 \times 10^{-5} {\rm~MeV} \, ,
\\
\nonumber &(g)& \Gamma_{\Xi_c[{3\over2}^-] \rightarrow \Xi_c \rho \to  \Xi_c \pi \pi} = 5 \times 10^{-5} {\rm~MeV} \, .
\end{eqnarray}

\subsection{The baryon doublet $[\mathbf{\bar 3}_F, 2, 1, \rho]$}

\begin{figure}[htb]
\begin{center}
\scalebox{0.6}{\includegraphics{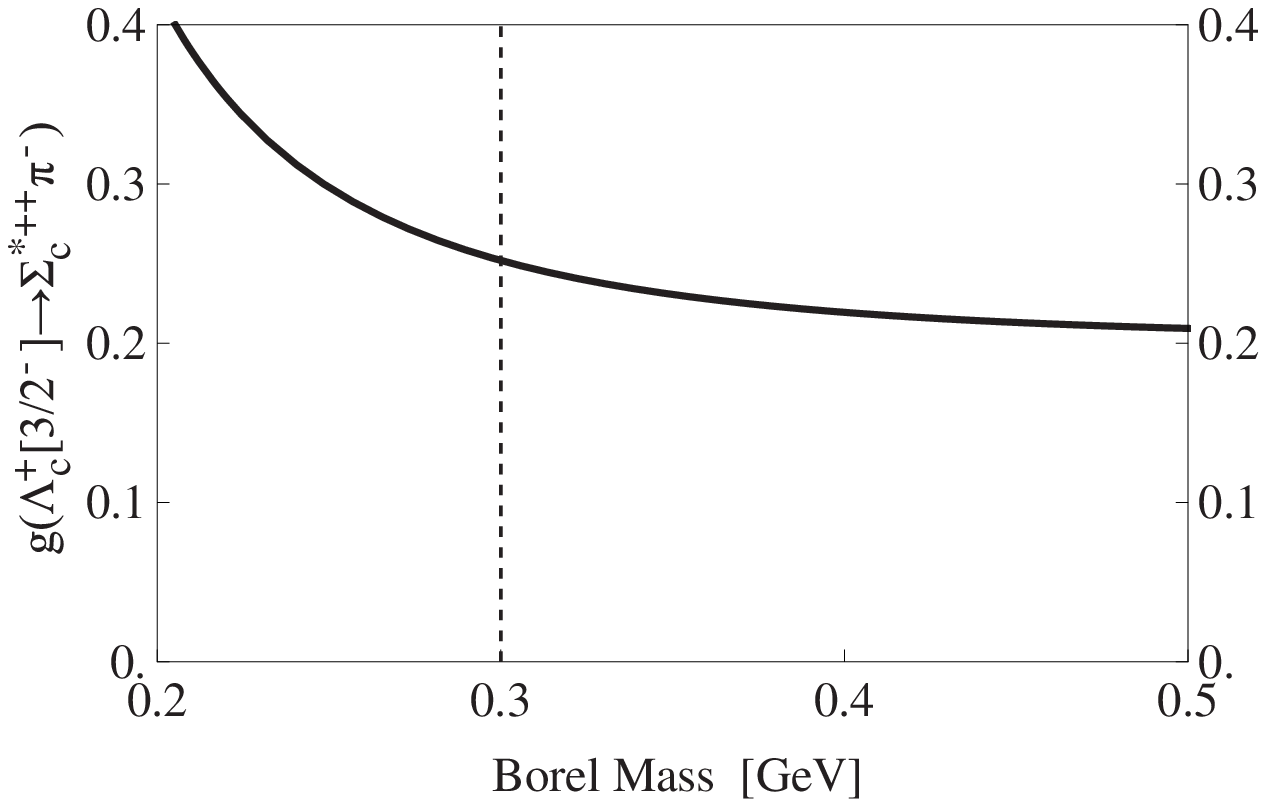}}
\scalebox{0.6}{\includegraphics{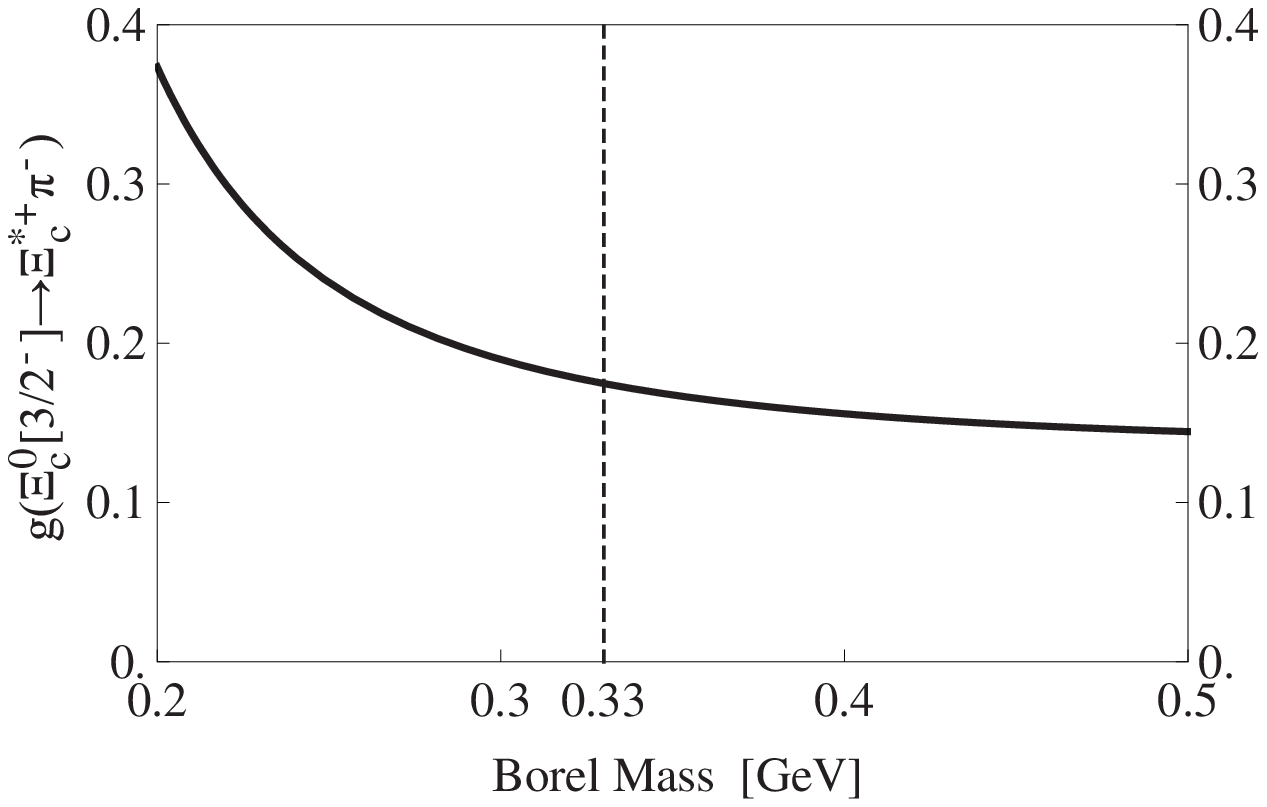}}
\\
\scalebox{0.6}{\includegraphics{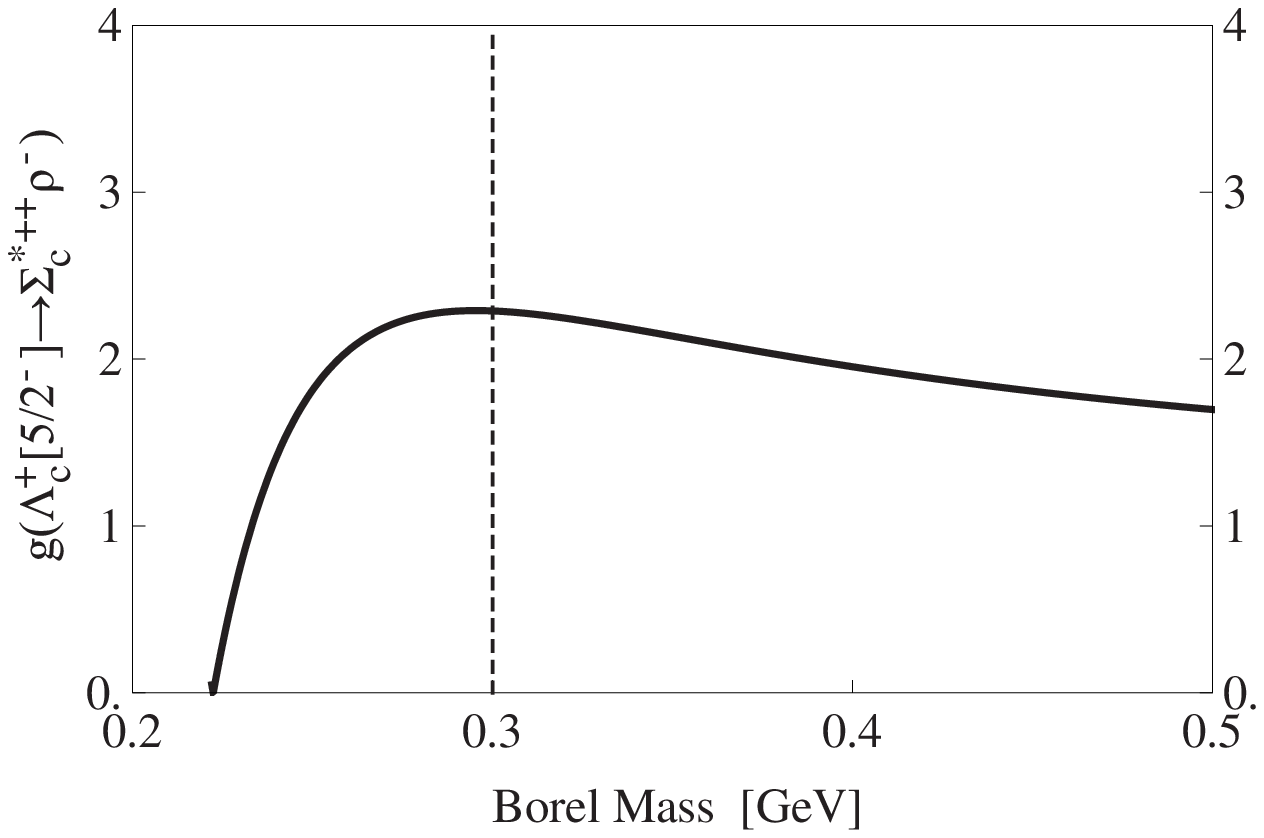}}
\scalebox{0.6}{\includegraphics{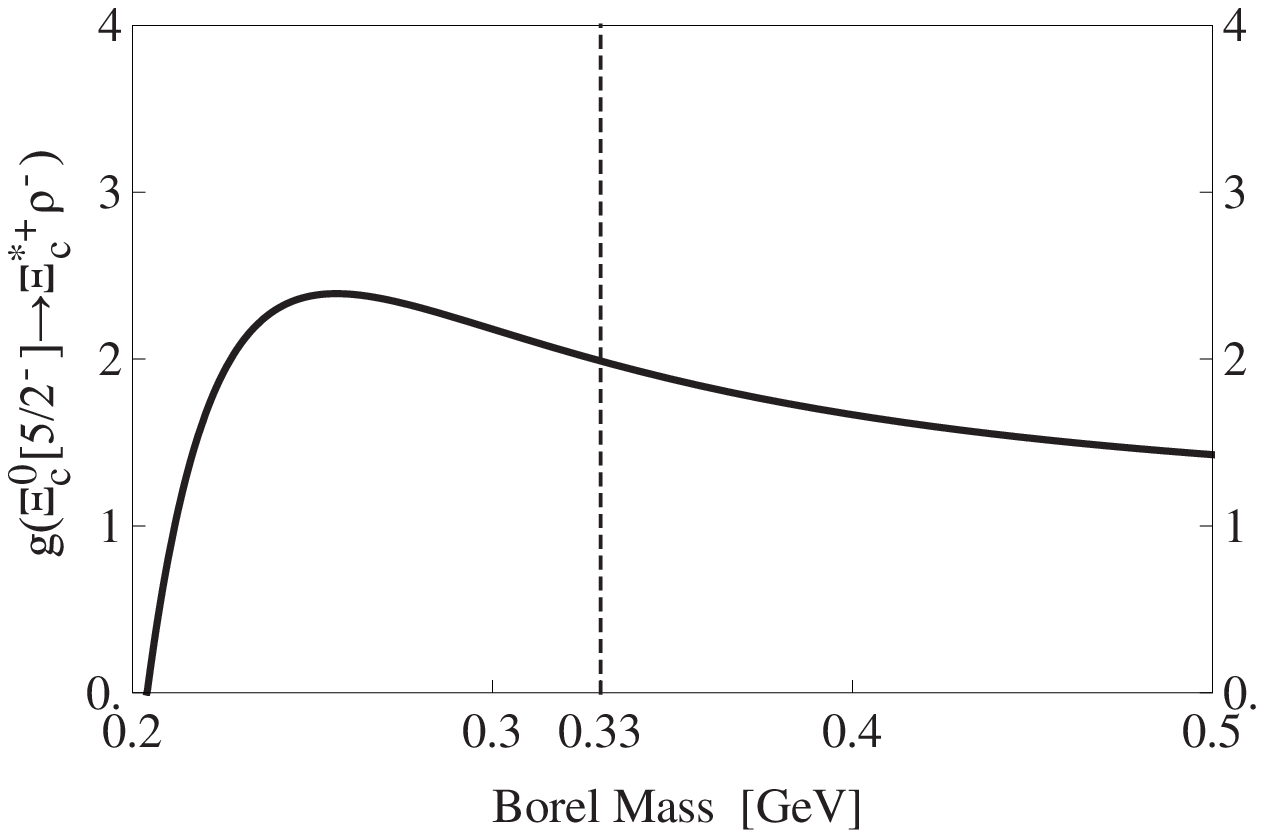}}
\end{center}
\caption{The coupling constants
$g_{\Lambda_c^+[{3\over2}^-] \rightarrow \Sigma_c^{*++}\pi^-}$ (top-left), $g_{\Xi_c^0[{3\over2}^-] \rightarrow \Xi_c^{*+}\pi^-}$ (top-right),
$g_{\Lambda_c^+[{5\over2}^-] \rightarrow \Sigma_c^{*++}\rho^-}$ (bottom-left) and $g_{\Xi_c^0[{5\over2}^-] \rightarrow \Xi_c^{*+}\rho^-}$ (bottom-right) as functions of the Borel mass $T$.
The currents belonging to the baryon doublet $[\mathbf{\bar 3}_F, 2, 1, \rho]$ are used here.
\label{fig:321rho}}
\end{figure}

The $[\mathbf{\bar 3}_F, 2, 1, \rho]$ multiplet contains $\Lambda_c({3\over2}^-/{5\over2}^-)$ and $\Xi_c^+({3\over2}^-/{5\over2}^-)$. Their sum rules are listed in Appendix~\ref{sec:321rho}, suggesting that their possible decay channels are $(b)$, $(h)$, $(i)$ and $(j)$,
while the other channel $(g)$ vanishes.
We show the four coupling constants, $g_{\Lambda_c^+[{3\over2}^-] \rightarrow \Sigma_c^{*++} \pi^-}$, $g_{\Xi_c^0[{3\over2}^-] \rightarrow \Xi_c^{*+} \pi^-}$, $g_{\Lambda_c^+[{5\over2}^-] \rightarrow \Sigma_c^{*++} \rho^-}$ and $g_{\Xi_c^0[{5\over2}^-] \rightarrow \Xi_c^{*+} \rho^-}$, as functions of the Borel mass $T$ in Fig.~\ref{fig:321rho}. Using the values of $T$ listed in Table~\ref{tab:pwave1}, we obtain
\begin{eqnarray}
\nonumber &(b)& g_{\Lambda_c^+[{3\over2}^-] \rightarrow \Sigma_c^{*++} \pi^-} = 0.25 \, ,
\\
&(h)& g_{\Xi_c^0[{3\over2}^-] \rightarrow \Xi_c^{*+} \pi^-} = 0.17 \, ,
\\
\nonumber &(i)& g_{\Lambda_c^+[{5\over2}^-] \rightarrow \Sigma_c^{*++} \rho^-} = 2.3 \, ,
\\
\nonumber &(j)& g_{\Xi_c^0[{5\over2}^-] \rightarrow \Xi_c^{*+} \rho^-} = 2.0 \, .
\end{eqnarray}
Using these values and the parameters listed in Sec.~\ref{sec:input}, we further obtain
\begin{eqnarray}
\nonumber &(b)& \Gamma_{\Lambda_c[{3\over2}^-] \rightarrow \Sigma_c^{*} \pi \to \Lambda_c \pi \pi} = 0.03 {\rm~MeV} \, ,
\\
&(h)& \Gamma_{\Xi_c[{3\over2}^-] \rightarrow \Xi_c^{*} \pi} = 0.69 {\rm~MeV} \, ,
\\
\nonumber &(i)& \Gamma_{\Lambda_c[{5\over2}^-] \rightarrow \Sigma_c^{*} \rho \to \Sigma_c^{*} \pi \pi} = 11 {\rm~MeV} \, ,
\\
\nonumber &(j)& \Gamma_{\Xi_c[{5\over2}^-] \rightarrow \Xi_c^{*} \rho \to \Xi_c^{*} \pi \pi} = 12 {\rm~MeV} \, .
\end{eqnarray}
We note that the two decay widths, $\Gamma_{\Lambda_c[{5\over2}^-] \rightarrow \Sigma_c^{*} \rho}$ and $\Gamma_{\Xi_c[{5\over2}^-] \rightarrow \Xi_c^{*} \rho}$, do
depend significantly on the masses of the $\Lambda_c({5/2}^-)$ and $\Xi_c({5/2}^-)$, which we assumed to be around 2850 MeV and 3000 MeV in Eqs.~(\ref{mass:spin52}). Because their physical masses (if exist) are possibly smaller than these values, these two decay channels might be kinematically forbidden.

\subsection{The baryon doublet $[\mathbf{\bar 3}_F, 1, 0, \lambda]$}

\begin{figure}[htb]
\begin{center}
\scalebox{0.6}{\includegraphics{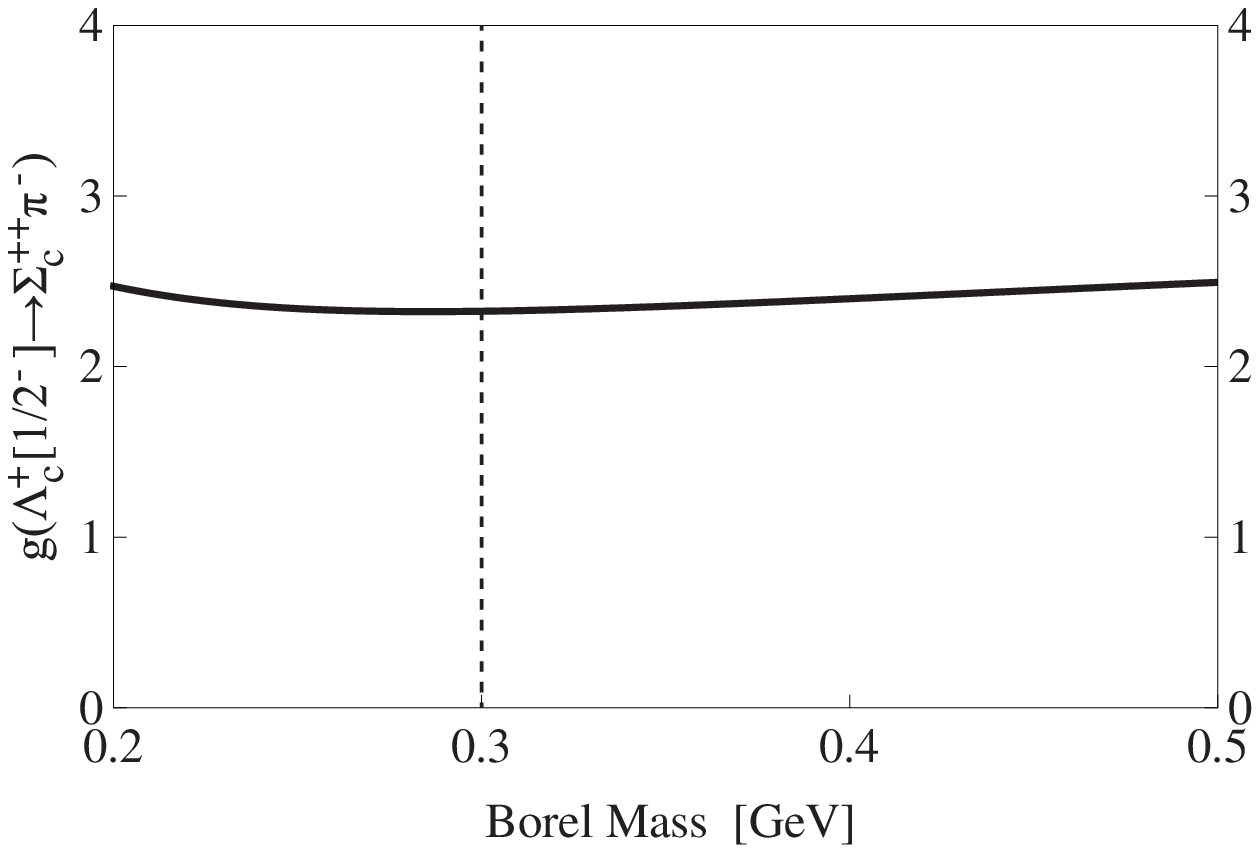}}
\scalebox{0.6}{\includegraphics{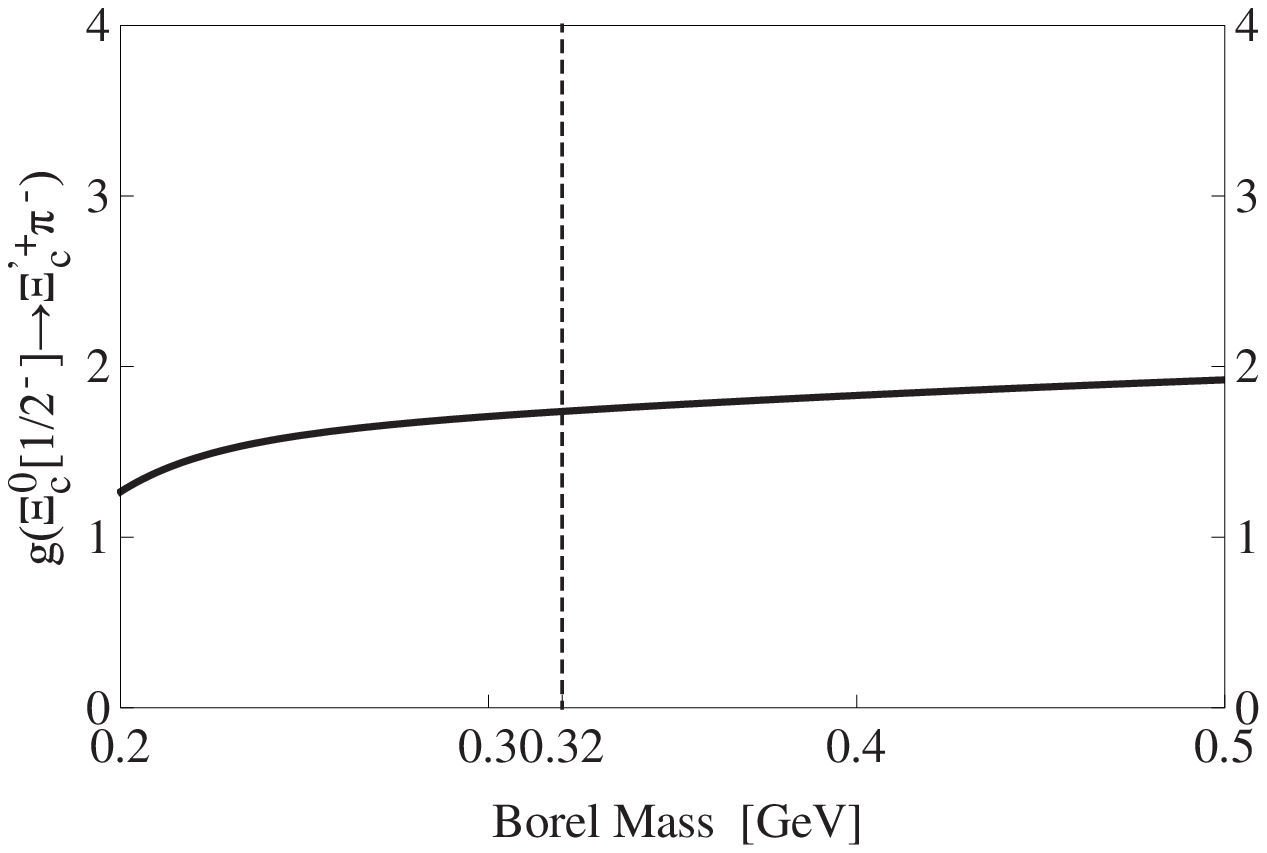}}
\\
\scalebox{0.6}{\includegraphics{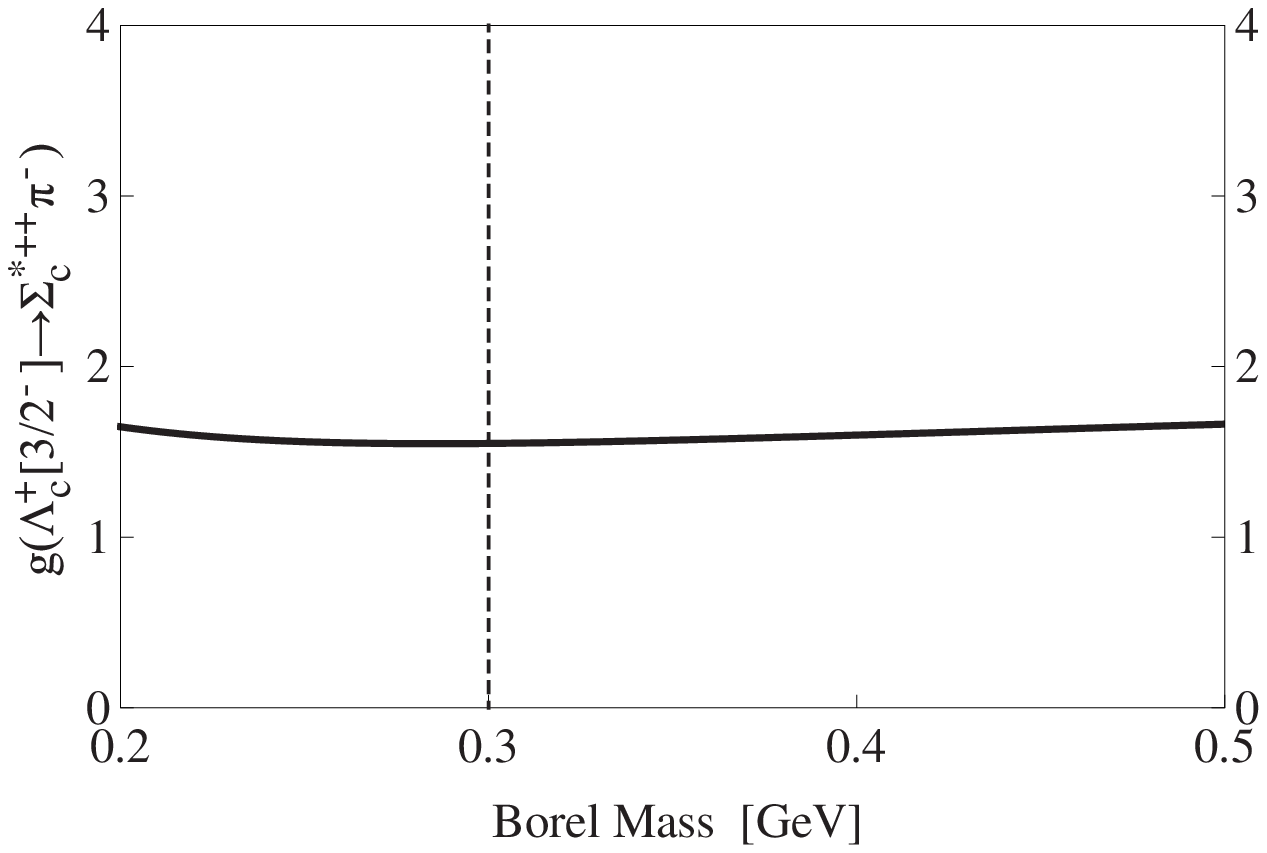}}
\scalebox{0.6}{\includegraphics{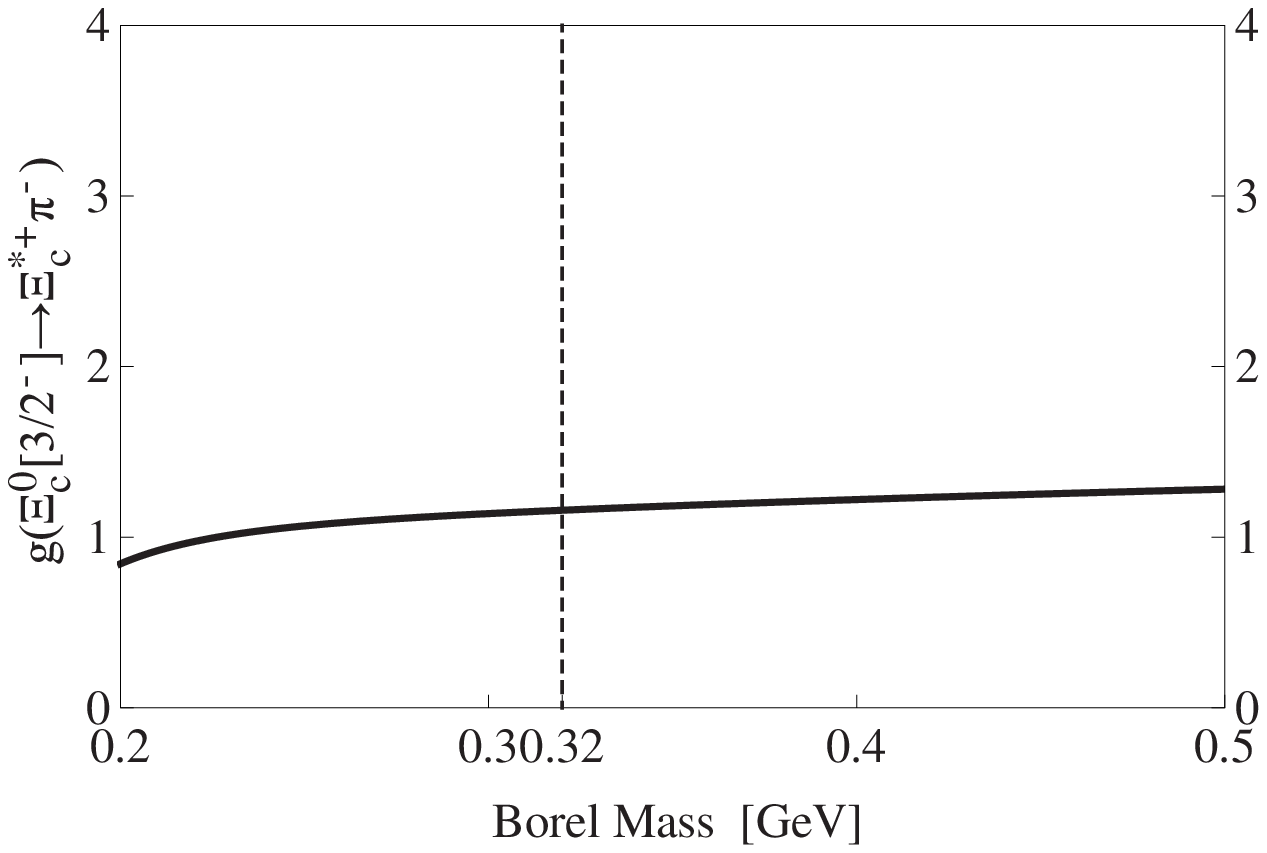}}
\\
\scalebox{0.6}{\includegraphics{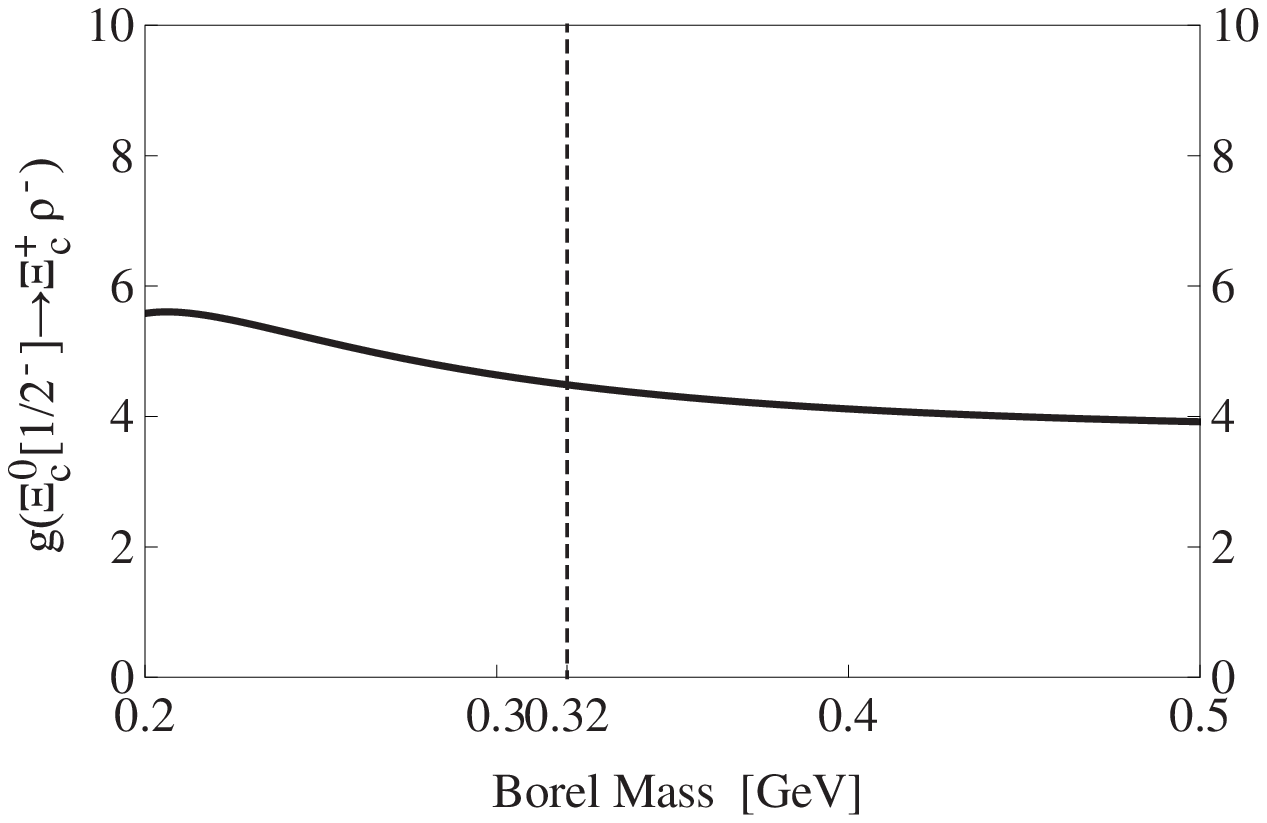}}
\scalebox{0.6}{\includegraphics{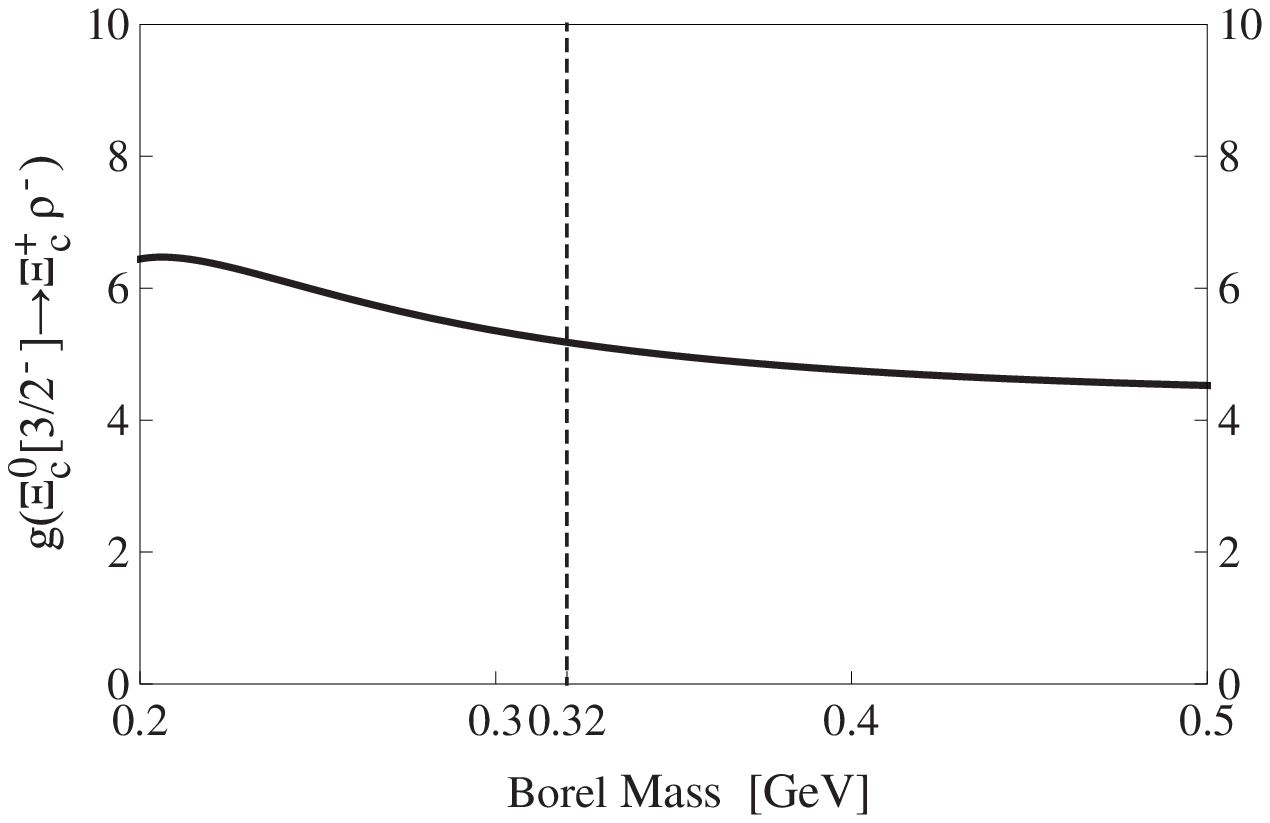}}
\end{center}
\caption{The coupling constants $g_{\Lambda_c^+[{1\over2}^-] \rightarrow \Sigma_c^{++}\pi^-}$ (top-left), $g_{\Xi_c^0[{1\over2}^-] \rightarrow \Xi_c^{\prime+}\pi^-}$ (top-right),
$g_{\Lambda_c^+[{3\over2}^-] \rightarrow \Sigma_c^{*++}\pi^-}$ (middle-left), $g_{\Xi_c^0[{3\over2}^-] \rightarrow \Xi_c^{*+}\pi^-}$ (middle-right),
$g_{\Xi_c^0[{1\over2}^-] \rightarrow \Xi_c^{+}\rho^-}$ (bottom-left) and $g_{\Xi_c^0[{3\over2}^-] \rightarrow \Xi_c^{+}\rho^-}$ (bottom-right) as functions of the Borel mass $T$.
The currents belonging to the baryon doublet $[\mathbf{\bar 3}_F, 1, 0, \lambda]$ are used here.
\label{fig:310lambda}}
\end{figure}

The $[\mathbf{\bar 3}_F, 1, 0, \lambda]$ multiplet contains $\Lambda_c({1\over2}^-/{3\over2}^-)$ and $\Xi_c({1\over2}^-/{3\over2}^-)$. Their sum rules are listed in Appendix~\ref{sec:310lambda}, suggesting that their possible decay channels are $(a)$, $(b)$, $(e)$, $(f)$, $(g)$ and $(h)$,
while the other two channels $(c)$ and $(d)$ vanish.
We show the six coupling constants, $g_{\Lambda_c^+[{1\over2}^-] \rightarrow \Sigma_c^{++} \pi^-}$, $g_{\Xi_c^0[{1\over2}^-] \rightarrow \Xi_c^{\prime+}\pi^-}$, $g_{\Lambda_c^+[{3\over2}^-] \rightarrow \Sigma_c^{*++} \pi^-}$, $g_{\Xi_c^0[{3\over2}^-] \rightarrow \Xi_c^{*+} \pi^-}$, $g_{\Xi_c^0[{1\over2}^-] \rightarrow \Xi_c^{+} \rho^-}$ and $g_{\Xi_c^0[{3\over2}^-] \rightarrow \Xi_c^{+} \rho^-}$, as functions of the Borel mass $T$ in Fig.~\ref{fig:310lambda}. Using the values of $T$ listed in Table~\ref{tab:pwave1}, we obtain
\begin{eqnarray}
\nonumber &(a)& g_{\Lambda_c^+[{1\over2}^-] \rightarrow \Sigma_c^{++} \pi^-} = 2.3 \, ,
\\
\nonumber &(f)& g_{\Xi_c^0[{1\over2}^-] \rightarrow \Xi_c^{\prime+}\pi^-} = 1.7 \, ,
\\
&(b)& g_{\Lambda_c^+[{3\over2}^-] \rightarrow \Sigma_c^{*++} \pi^-} = 1.5 \, ,
\\
\nonumber &(h)& g_{\Xi_c^0[{3\over2}^-] \rightarrow \Xi_c^{*+} \pi^-} = 1.2 \, ,
\\
\nonumber &(e)& g_{\Xi_c^0[{1\over2}^-] \rightarrow \Xi_c^{+} \rho^-} = 4.5 \, ,
\\
\nonumber &(g)& g_{\Xi_c^0[{3\over2}^-] \rightarrow \Xi_c^{+} \rho^-} = 5.2 \, .
\end{eqnarray}
Using these values and the parameters listed in Sec.~\ref{sec:input}, we further obtain
\begin{eqnarray}
\nonumber &(a,a^\prime)& \Gamma_{\Lambda_c[{1\over2}^-] \rightarrow \Sigma_c \pi (\rightarrow \Lambda \pi \pi)} = 32 {\rm~MeV} \, ,
\\
\nonumber &(f)& \Gamma_{\Xi_c[{1\over2}^-] \rightarrow \Xi_c^{\prime}\pi } = 100 {\rm~MeV} \, ,
\\
&(b)& \Gamma_{\Lambda_c[{3\over2}^-] \rightarrow \Sigma_c^{*} \pi \to \Lambda_c \pi \pi} = 0.96 {\rm~MeV} \, ,
\\
\nonumber &(h)& \Gamma_{\Xi_c[{3\over2}^-] \rightarrow \Xi_c^{*} \pi} = 30 {\rm~MeV} \, ,
\\
\nonumber &(e)& \Gamma_{\Xi_c[{1\over2}^-] \rightarrow \Xi_c \rho \to  \Xi_c \pi \pi} = 0.04 {\rm~MeV} \, ,
\\
\nonumber &(g)& \Gamma_{\Xi_c[{3\over2}^-] \rightarrow \Xi_c \rho \to  \Xi_c \pi \pi} = 0.23 {\rm~MeV} \, .
\end{eqnarray}

\section{Decay Properties of flavor $\mathbf{6}_F$ $P$-wave charmed baryons}
\label{sec:sumrule2}

In this section we use the method of light-cone QCD sum rules to study decay properties of the flavor $\mathbf{6}_F$ $P$-wave charmed baryons.
We only study their $S$-wave decays into ground-state charmed baryons accompanied by a pseudoscalar meson ($\pi$ or $K$), including both two-body and three-body decays which are kinematically allowed.
We shall study their $S$-wave decays into ground-state charmed baryons accompanied by a vector meson ($\rho$ or $K^*$) in our future work, but note that the widths of these decays are probably quite small (see the results of the flavor $\mathbf{\bar 3}_F$ $P$-wave charmed baryons).

The possible decay channels are:
\begin{eqnarray}
&(k)& {\bf \Gamma\Big[} \Sigma_c(1/2^-) \rightarrow \Lambda_c(1/2^+) + \pi {\Big ]}
= {\bf \Gamma\Big[} \Sigma_c^{0}(1/2^-) \rightarrow \Lambda_c^{+}(1/2^+) + \pi^- {\Big ]} \, ,
\\ &(l)& {\bf \Gamma\Big[}\Sigma_c(1/2^-) \rightarrow \Sigma_c(1/2^+) + \pi{\Big ]}
= 2 \times {\bf \Gamma\Big[}\Sigma_c^{0}(1/2^-) \rightarrow \Sigma_c^{+}(1/2^+) + \pi^-{\Big ]} \, ,
\\ &(m)& {\bf \Gamma\Big[}\Xi_c^{\prime}(1/2^-) \rightarrow \Xi_c(1/2^+) + \pi{\Big ]}
= {3\over2} \times {\bf \Gamma\Big[}\Xi_c^{\prime 0}(1/2^-) \rightarrow \Xi_c^{+}(1/2^+) + \pi^-{\Big ]} \, ,
\\ &(n)& {\bf \Gamma\Big[} \Xi_c^{\prime}(1/2^-) \rightarrow \Lambda_c(1/2^+) + K {\Big ]}
= {\bf \Gamma\Big[} \Xi_c^{\prime 0}(1/2^-) \rightarrow \Lambda_c^{+}(1/2^+) + K^- {\Big ]} \, ,
\\ &(o)& {\bf \Gamma\Big[}\Xi_c^{\prime}(1/2^-) \rightarrow \Xi_c^{\prime}(1/2^+) + \pi{\Big ]}
= {3\over2} \times  {\bf \Gamma\Big[}\Xi_c^{\prime 0}(1/2^-) \rightarrow \Xi_c^{\prime+}(1/2^+) + \pi^-{\Big ]} \, ,
\\ &(p)& {\bf \Gamma\Big[}\Xi_c^{\prime}(1/2^-) \rightarrow \Sigma_c(1/2^+) + K{\Big ]}
= 3 \times {\bf \Gamma\Big[}\Xi_c^{\prime 0}(1/2^-) \rightarrow \Sigma_c^{+}(1/2^+) + K^-{\Big ]} \, ,
\\ &(q)& {\bf \Gamma\Big[}\Omega_c(1/2^-) \rightarrow \Xi_c(1/2^+) + K{\Big ]}
= 2 \times {\bf \Gamma\Big[}\Omega_c^{0}(1/2^-) \rightarrow \Xi_c^{+}(1/2^+) + K^-{\Big ]} \, ,
\\ &(r)& {\bf \Gamma\Big[}\Omega_c(1/2^-) \rightarrow \Xi_c^{\prime}(1/2^+) + K{\Big ]}
= 2 \times {\bf \Gamma\Big[}\Omega_c^{0}(1/2^-) \rightarrow \Xi_c^{\prime+}(1/2^+) + K^-{\Big ]} \, ,
\\ &(s)& {\bf \Gamma\Big[}\Sigma_c(3/2^-) \rightarrow \Sigma_c^{*}(3/2^+) + \pi {\Big ]}
= 2 \times {\bf \Gamma\Big[}\Sigma_c^{0}(3/2^-) \rightarrow \Sigma_c^{*+}(3/2^+) + \pi^- {\Big ]} \, ,
\\ &(t)& {\bf \Gamma\Big[}\Xi_c^{\prime}(3/2^-) \rightarrow \Xi_c^{*}(3/2^+) + \pi {\Big ]}
= {3\over2} \times {\bf \Gamma\Big[}\Xi_c^{\prime0}(3/2^-) \rightarrow \Xi_c^{*+}(3/2^+) + \pi^- {\Big ]} \, .
\\ &(u)& {\bf \Gamma\Big[}\Xi_c^{\prime}(3/2^-) \rightarrow \Sigma_c^{*}(3/2^+) + K \rightarrow \Lambda_c(1/2^+) + \pi + K {\Big ]}
\\ \nonumber &=& 3 \times {\bf \Gamma\Big[}\Xi_c^{\prime0}(3/2^-) \rightarrow \Sigma_c^{*+}(3/2^+) + K^- \rightarrow \Lambda_c^{+}(3/2^+) + \pi^0 + K^- {\Big ]} \, .
\\ &(v)& {\bf \Gamma\Big[}\Omega_c(3/2^-) \rightarrow \Xi_c^{*}(3/2^+) + K \rightarrow \Xi_c(1/2^+) + \pi + K {\Big ]}
\\ \nonumber &=& 6 \times {\bf \Gamma\Big[}\Omega_c^{0}(3/2^-) \rightarrow \Xi_c^{*+}(3/2^+) + K^- \rightarrow \Xi_c^{+}(1/2^+) + \pi^0 + K^- {\Big ]} \, .
\end{eqnarray}
Their widths can be simply calculated through the two Lagrangians $\mathcal{L}_{X({1/2}^-) \rightarrow Y({1/2}^+) P} = g {\bar X} Y P$ and $\mathcal{L}_{X({3/2}^-) \rightarrow Y({3/2}^+) P} = g {\bar X_\mu} Y_\mu P$.
Especially, we assume the mass of the $\Omega_c(3/2^-)$ state to be 3120 MeV in the case $(v)$ in order to make this decay channel kinematically allowed, but still use 3100 MeV for other cases.

In the following subsections, we shall separately investigate the four $P$-wave charmed baryon multiplets of flavor $\mathbf{6}_F$, $[\mathbf{6}_F, 1, 0, \rho]$, $[\mathbf{6}_F, 0, 1, \lambda]$, $[\mathbf{6}_F, 1, 1, \lambda]$ and $[\mathbf{6}_F, 2, 1, \lambda]$.

\subsection{The baryon doublet $[\mathbf{6}_F, 1, 0, \rho]$}

\begin{figure}[htb]
\begin{center}
\scalebox{0.6}{\includegraphics{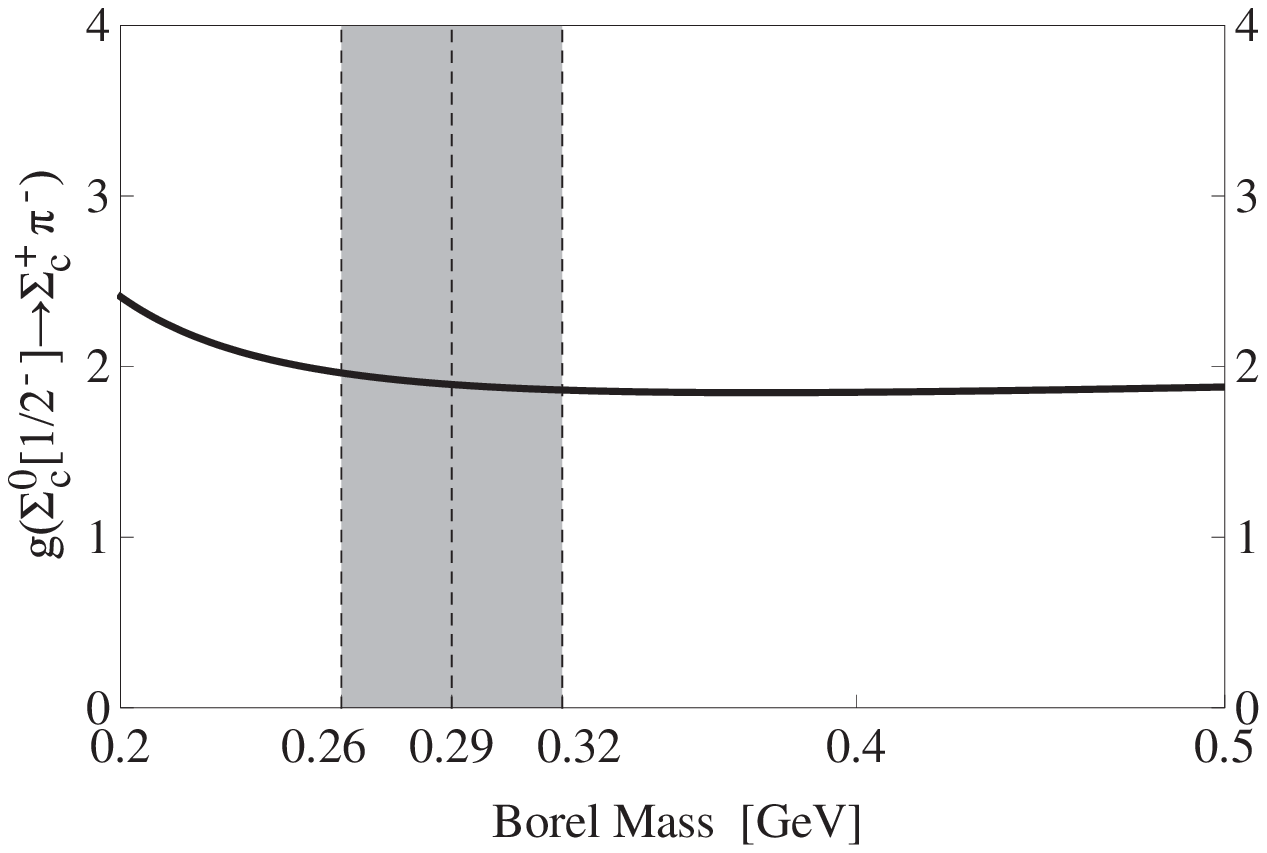}}
\scalebox{0.6}{\includegraphics{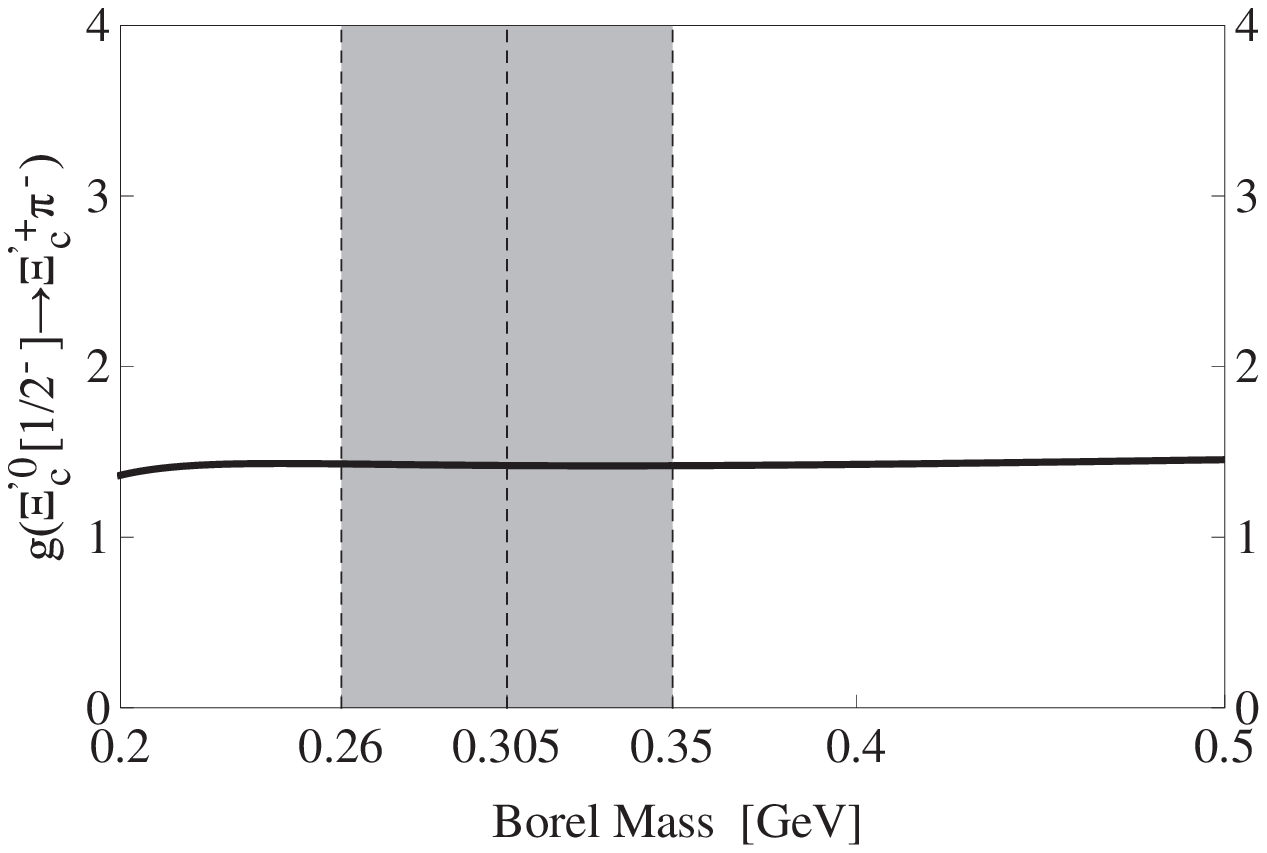}}
\\
\scalebox{0.6}{\includegraphics{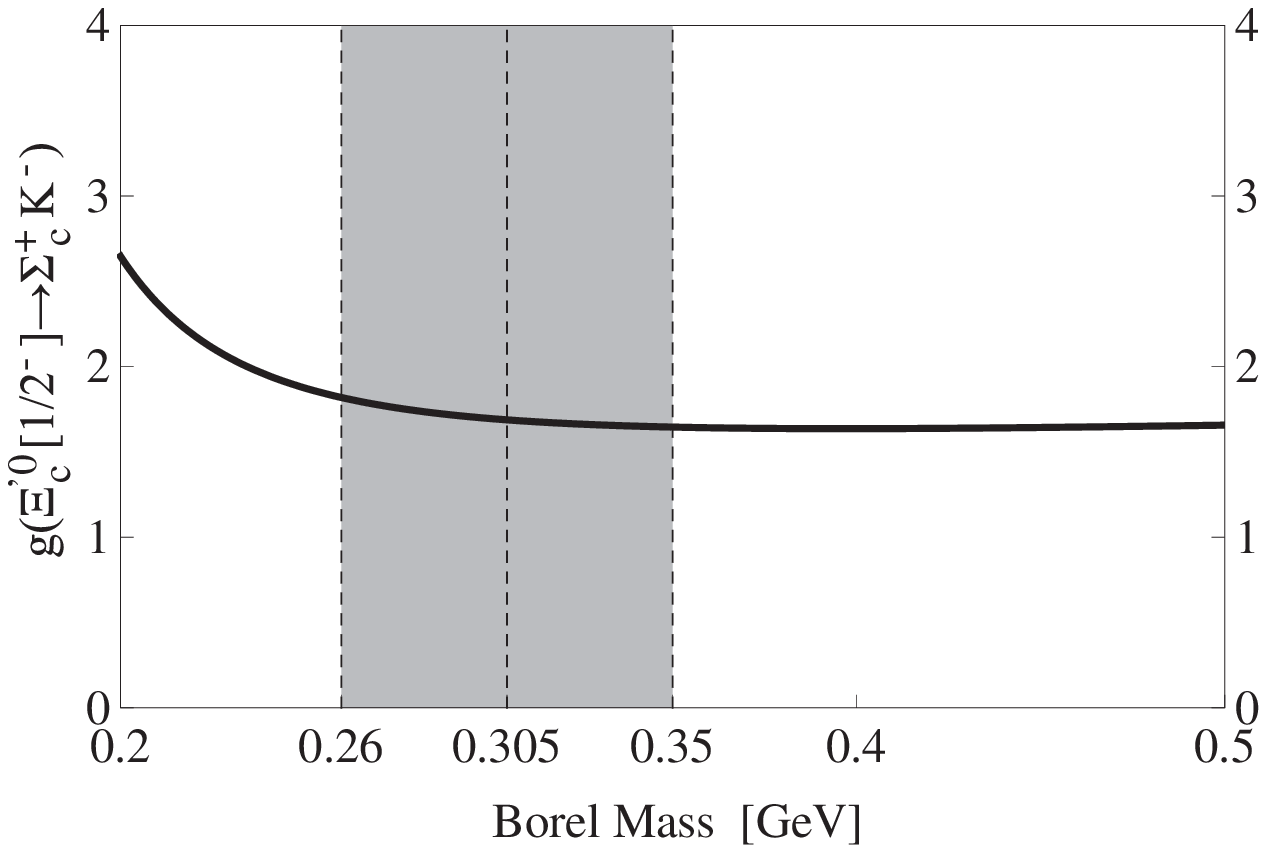}}
\scalebox{0.6}{\includegraphics{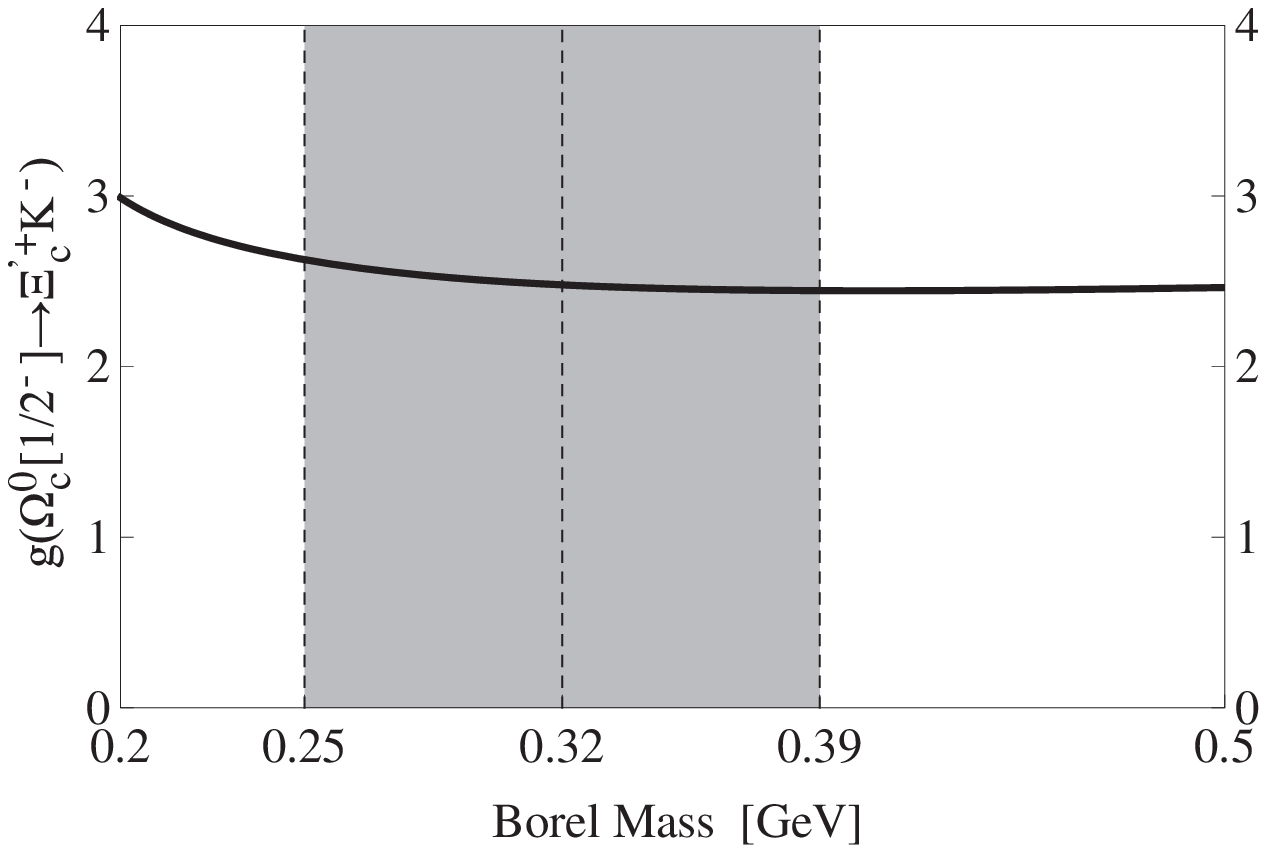}}
\\
\scalebox{0.6}{\includegraphics{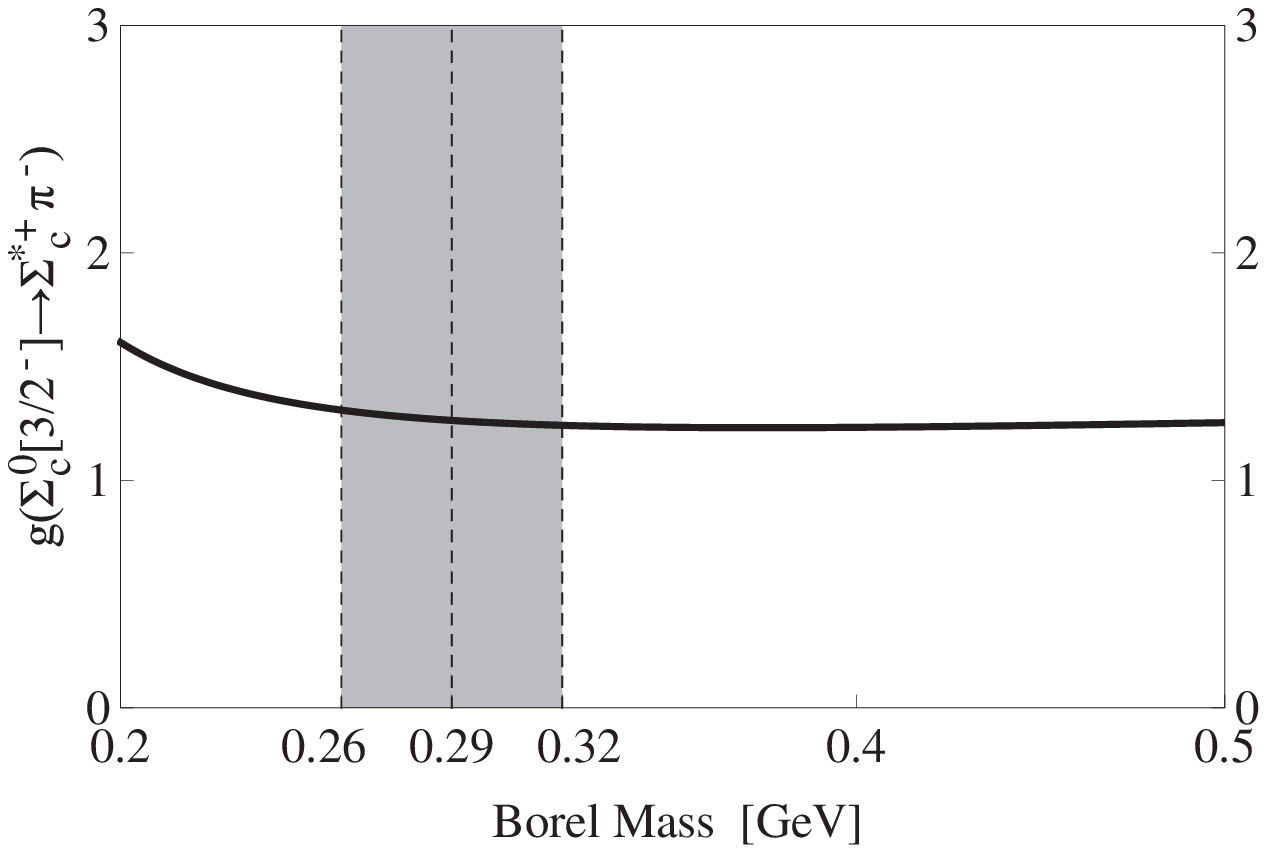}}
\scalebox{0.6}{\includegraphics{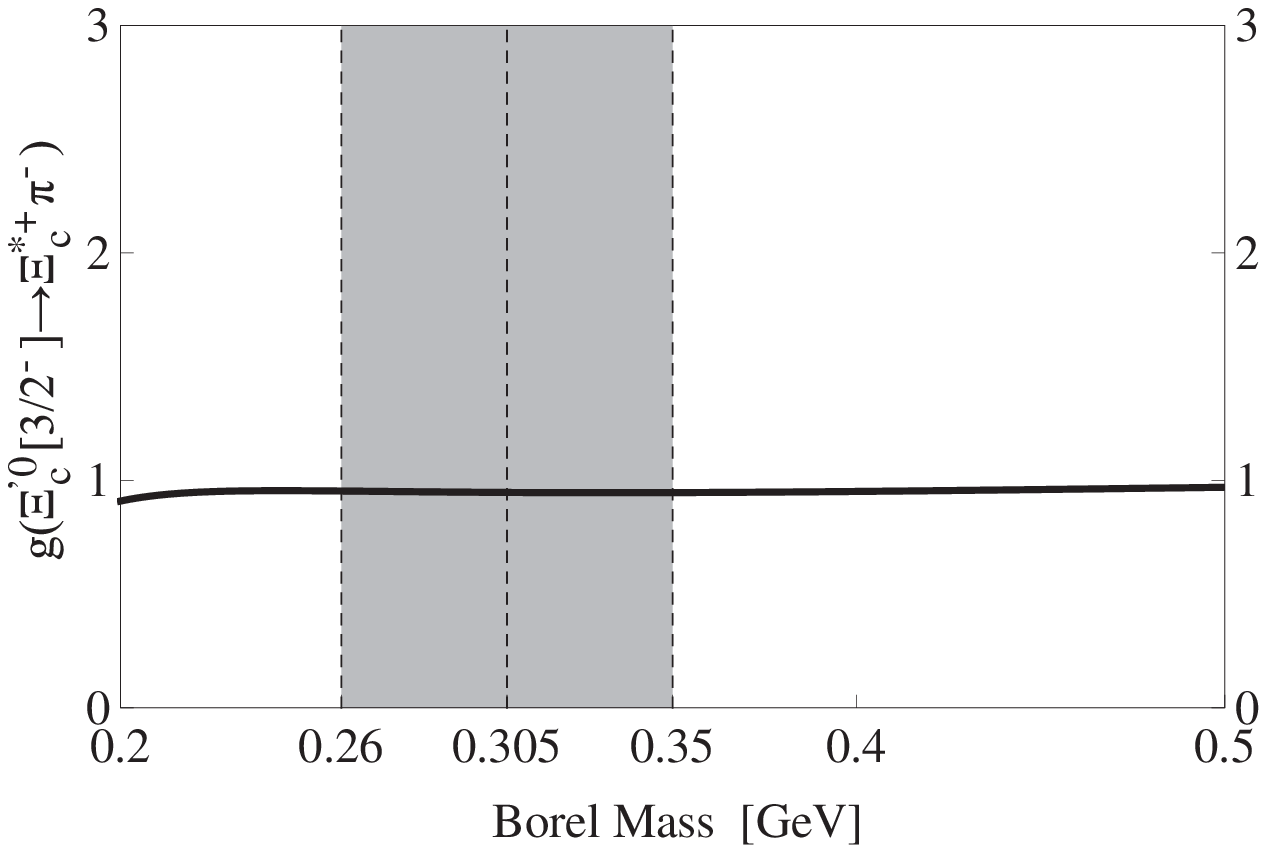}}
\\
\scalebox{0.6}{\includegraphics{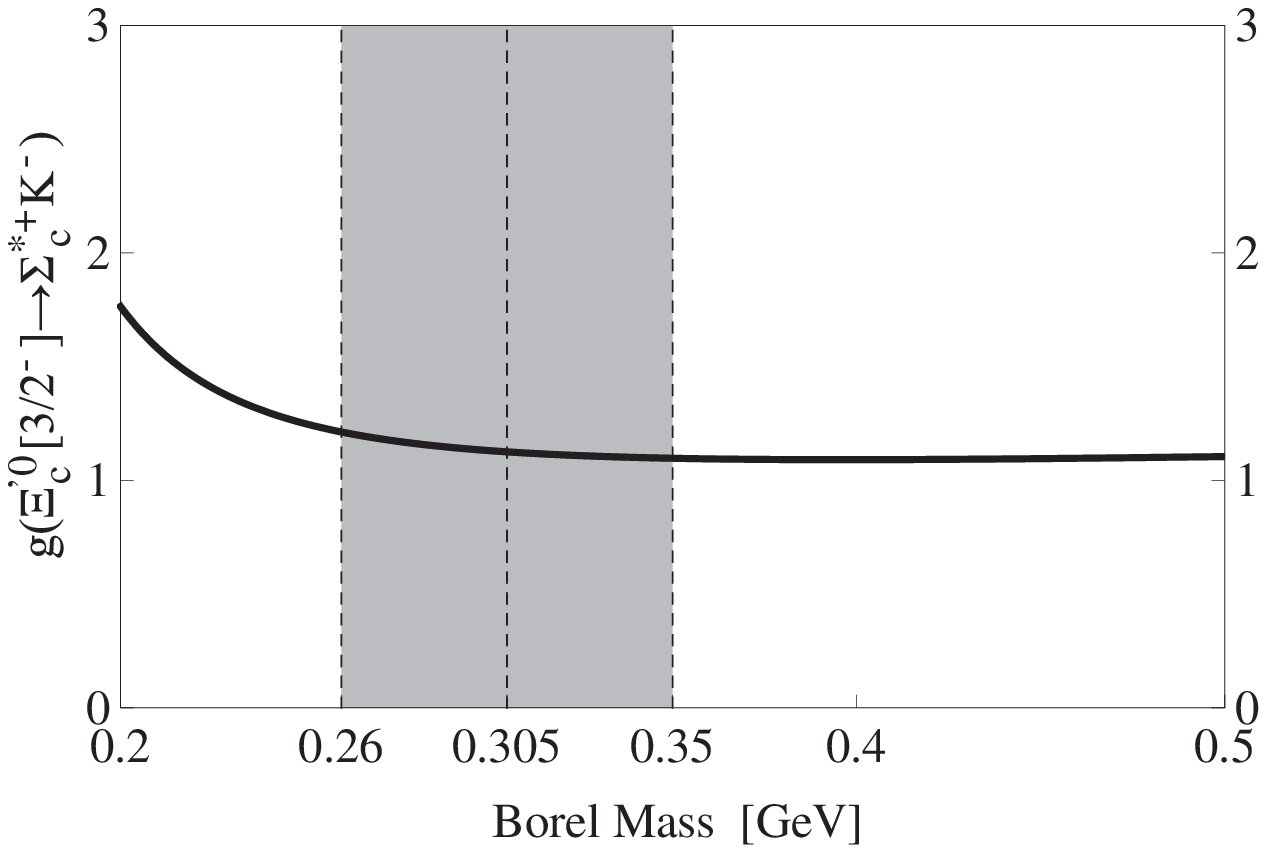}}
\scalebox{0.6}{\includegraphics{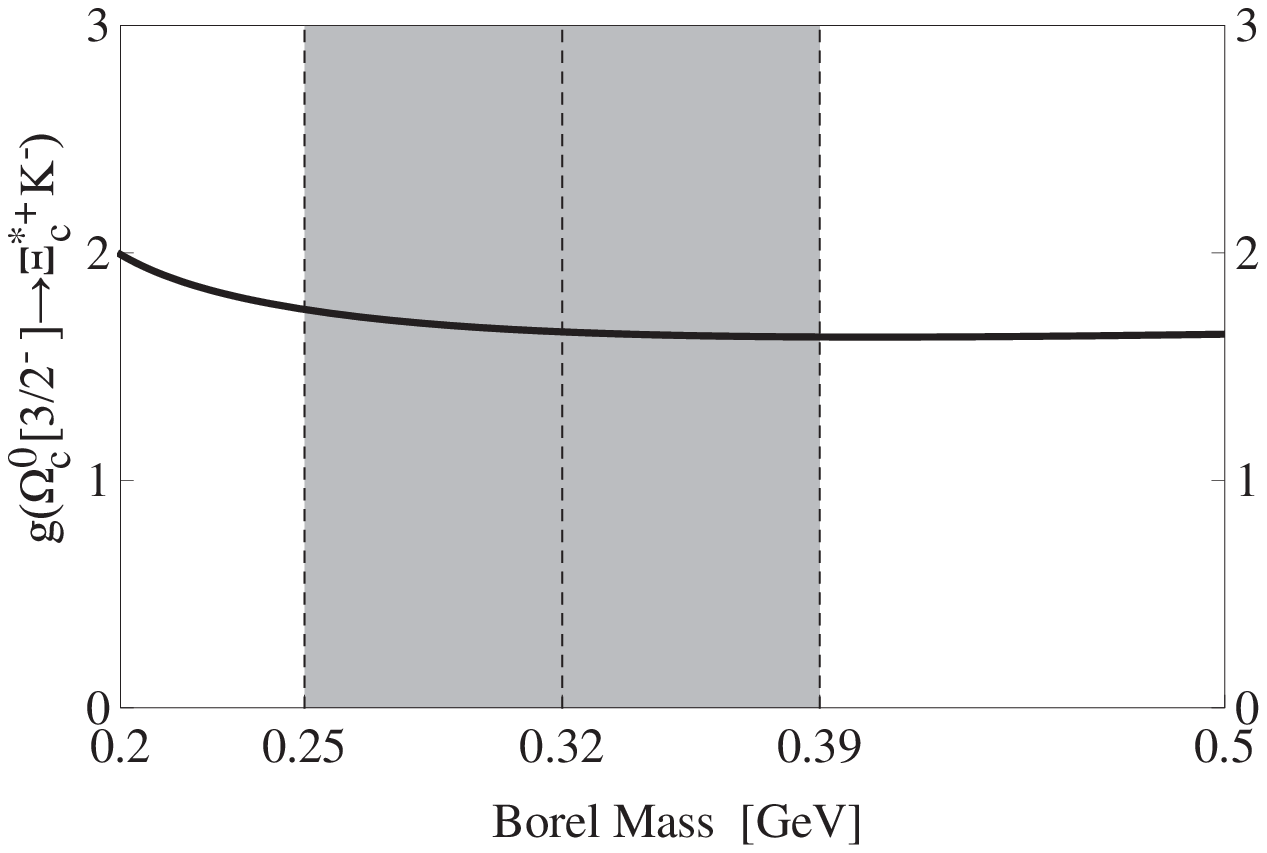}}
\end{center}
\caption{The coupling constants $g_{\Sigma_c^0[{1\over2}^-] \rightarrow \Sigma_c^{+} \pi^-}$ (top-left), $g_{\Xi_c^{\prime0}[{1\over2}^-] \rightarrow \Xi_c^{\prime+}\pi^-}$ (top-right),
$g_{\Xi_c^{\prime0}[{1\over2}^-] \rightarrow \Sigma_c^{+} K^-}$ (middle-left), $g_{\Omega_c^{0}[{1\over2}^-] \rightarrow \Xi_c^{\prime+} K^-}$ (middle-right),
$g_{\Sigma_c^{0}[{3\over2}^-] \rightarrow \Sigma_c^{*+} \pi^-}$ (middle-left), $g_{\Xi_c^{\prime0}[{3\over2}^-] \rightarrow \Xi_c^{*+} \pi^-}$ (middle-right),
$g_{\Xi_c^{\prime0}[{3\over2}^-] \rightarrow \Sigma_c^{*+} K^-}$ (bottom-left) and $g_{\Omega_c^{0}[{3\over2}^-] \rightarrow \Xi_c^{*+} K^-}$ (bottom-right)
as functions of the Borel mass $T$. The currents belonging to the baryon doublet $[\mathbf{6}_F, 1, 0, \rho]$ are used here.
\label{fig:610rho}}
\end{figure}

The $[\mathbf{6}_F, 1, 0, \rho]$ multiplet contains $\Sigma_c({1\over2}^-/{3\over2}^-)$, $\Xi^\prime_c({1\over2}^-/{3\over2}^-)$ and $\Omega_c({1\over2}^-/{3\over2}^-)$. Their sum rules are listed in Appendix~\ref{sec:610rho}, suggesting that their possible decay channels are $(l)$, $(o)$, $(p)$, $(r)$, $(s)$, $(t)$, $(u)$ and $(v)$,
while the other four channels $(k)$, $(m)$, $(n)$ and $(q)$ vanish.
We show the eight coupling constants, $g_{\Sigma_c^0[{1\over2}^-] \rightarrow \Sigma_c^{+} \pi^-}$, $g_{\Xi_c^{\prime0}[{1\over2}^-] \rightarrow \Xi_c^{\prime+}\pi^-}$, $g_{\Xi_c^{\prime0}[{1\over2}^-] \rightarrow \Sigma_c^{+} K^-}$, $g_{\Omega_c^{0}[{1\over2}^-] \rightarrow \Xi_c^{\prime+} K^-}$, $g_{\Sigma_c^{0}[{3\over2}^-] \rightarrow \Sigma_c^{*+} \pi^-}$, $g_{\Xi_c^{\prime0}[{3\over2}^-] \rightarrow \Xi_c^{*+} \pi^-}$, $g_{\Xi_c^{\prime0}[{3\over2}^-] \rightarrow \Sigma_c^{*+} K^-}$ and $g_{\Omega_c^{0}[{3\over2}^-] \rightarrow \Xi_c^{*+} K^-}$, as functions of the Borel mass $T$ in Fig.~\ref{fig:610rho}. Using the values of $T$ listed in Table~\ref{tab:pwave2}, we obtain
\begin{eqnarray}
\nonumber &(l)& g_{\Sigma_c^0[{1\over2}^-] \rightarrow \Sigma_c^{+} \pi^-} = 1.9 \, ,
\\
\nonumber &(o)& g_{\Xi_c^{\prime0}[{1\over2}^-] \rightarrow \Xi_c^{\prime+}\pi^-} = 1.4 \, ,
\\
\nonumber &(p)& g_{\Xi_c^{\prime0}[{1\over2}^-] \rightarrow \Sigma_c^{+} K^-} = 1.7 \, ,
\\
&(r)& g_{\Omega_c^{0}[{1\over2}^-] \rightarrow \Xi_c^{\prime+} K^-} = 2.5 \, ,
\\
\nonumber &(s)& g_{\Sigma_c^{0}[{3\over2}^-] \rightarrow \Sigma_c^{*+} \pi^-} = 1.3 \, ,
\\
\nonumber &(t)& g_{\Xi_c^{\prime0}[{3\over2}^-] \rightarrow \Xi_c^{*+} \pi^-} = 0.95 \, ,
\\
\nonumber &(u)& g_{\Xi_c^{\prime0}[{3\over2}^-] \rightarrow \Sigma_c^{*+} K^-} = 1.1 \, ,
\\
\nonumber &(v)& g_{\Omega_c^{0}[{3\over2}^-] \rightarrow \Xi_c^{*+} K^-} = 1.7 \, .
\end{eqnarray}
Using these values and the parameters listed in Sec.~\ref{sec:input}, we further obtain
\begin{eqnarray}
\nonumber &(l)& \Gamma_{\Sigma_c[{1\over2}^-] \rightarrow \Sigma_c \pi} = 300 {\rm~MeV} \, ,
\\
\nonumber &(o)& \Gamma_{\Xi_c^{\prime}[{1\over2}^-] \rightarrow \Xi_c^{\prime}\pi} = 140 {\rm~MeV} \, ,
\\
\nonumber &(p)& \Gamma_{\Xi_c^{\prime}[{1\over2}^-] \rightarrow \Sigma_c K} = 29 {\rm~MeV} \, ,
\\
&(r)& \Gamma_{\Omega_c[{1\over2}^-] \rightarrow \Xi_c^{\prime} K} = 250 {\rm~MeV} \, ,
\\
\nonumber &(s)& \Gamma_{\Sigma_c[{3\over2}^-] \rightarrow \Sigma_c^{*} \pi} = 110 {\rm~MeV} \, ,
\\
\nonumber &(t)& \Gamma_{\Xi_c^{\prime}[{3\over2}^-] \rightarrow \Xi_c^{*} \pi} = 50 {\rm~MeV} \, ,
\\
\nonumber &(u)& \Gamma_{\Xi_c^{\prime}[{3\over2}^-] \rightarrow \Sigma_c^{*} K \rightarrow \Lambda_c \pi K} = 0.03 {\rm~MeV} \, ,
\\
\nonumber &(v)& \Gamma_{\Omega_c[{3\over2}^-] \rightarrow \Xi_c^{*} K \rightarrow \Xi_c \pi K} = 0.07 {\rm~MeV} \, .
\end{eqnarray}

\subsection{The baryon doublet $[\mathbf{6}_F, 0, 1, \lambda]$}

\begin{figure}[htb]
\begin{center}
\scalebox{0.6}{\includegraphics{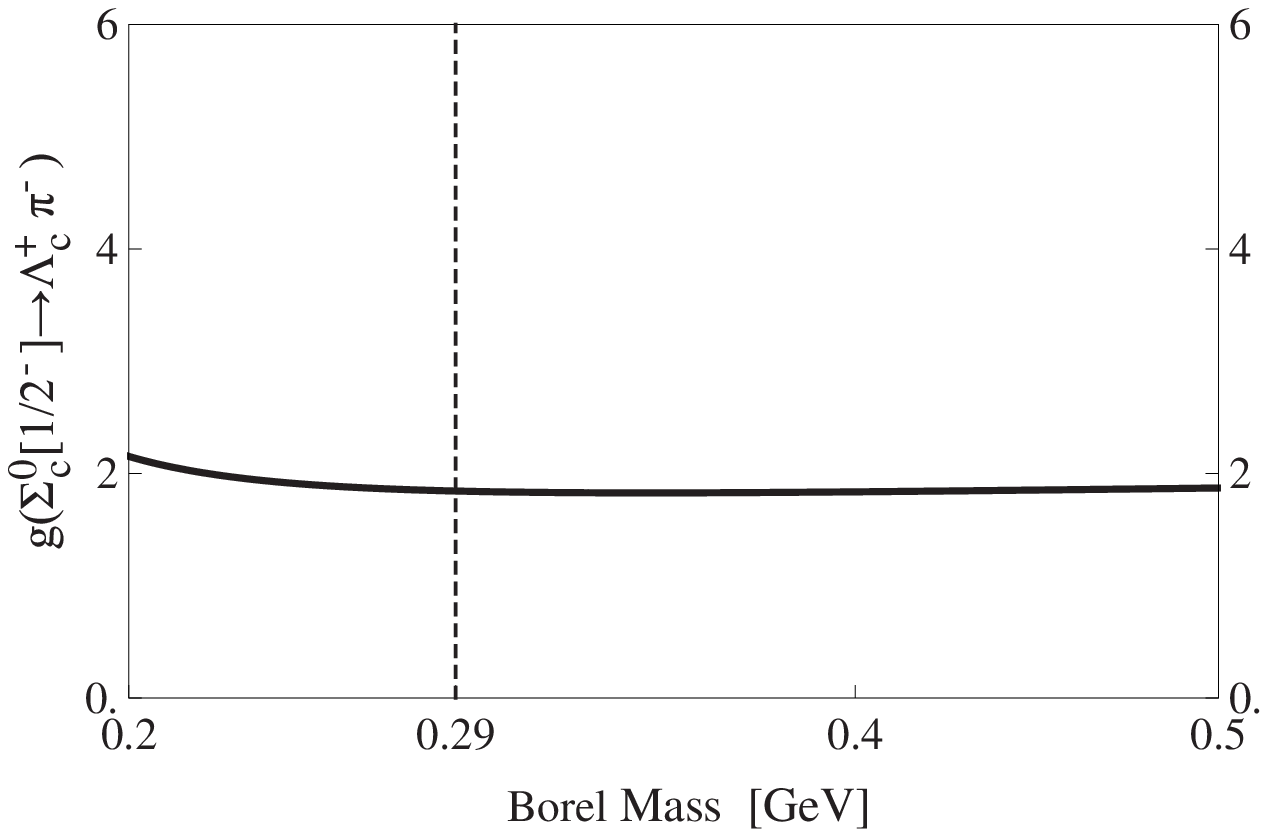}}
\scalebox{0.6}{\includegraphics{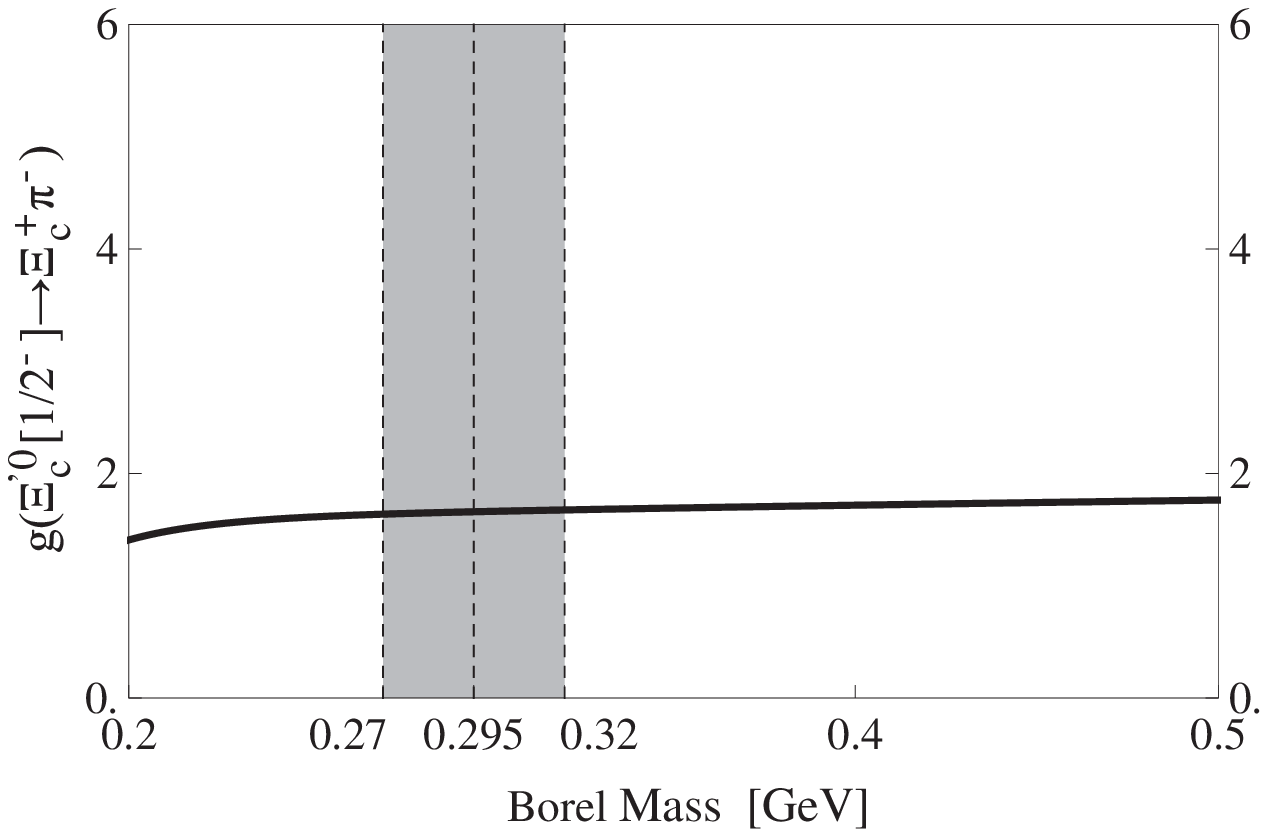}}
\\
\scalebox{0.6}{\includegraphics{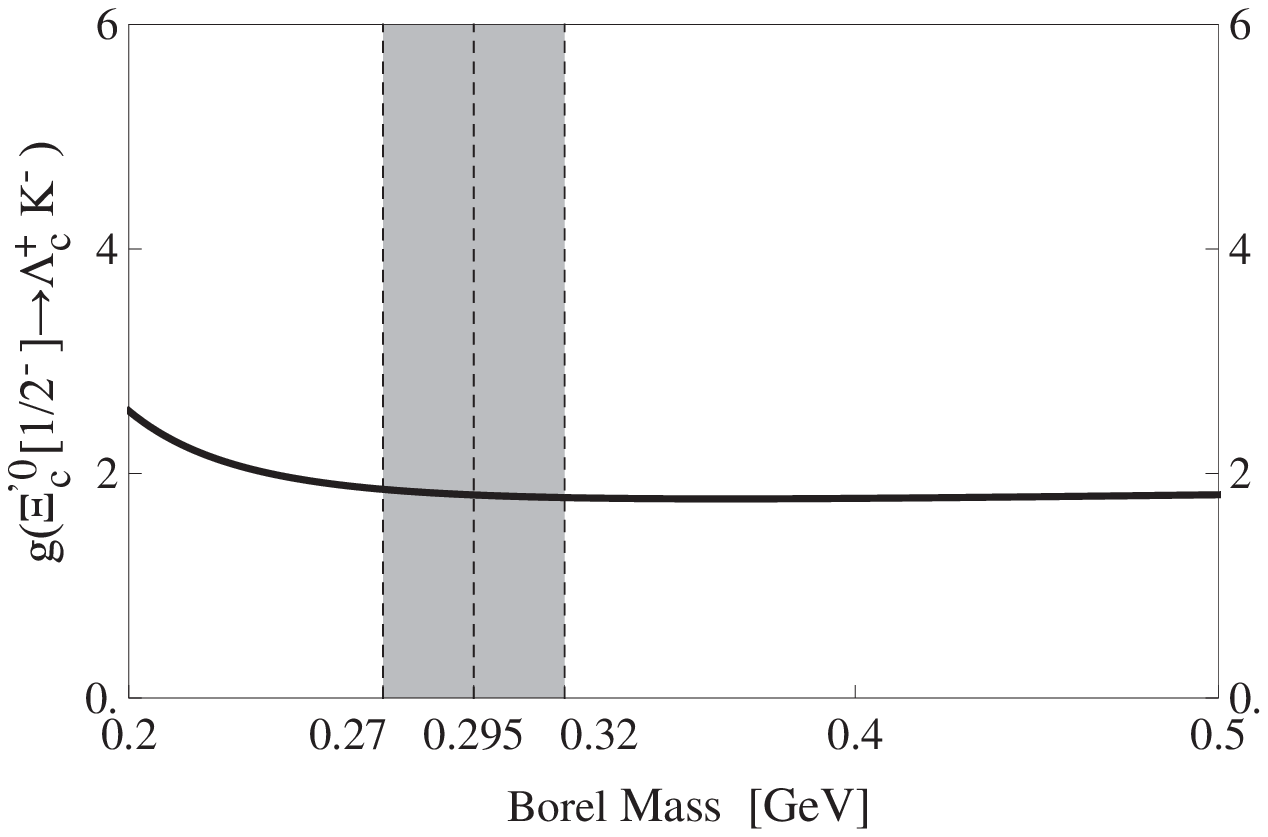}}
\scalebox{0.6}{\includegraphics{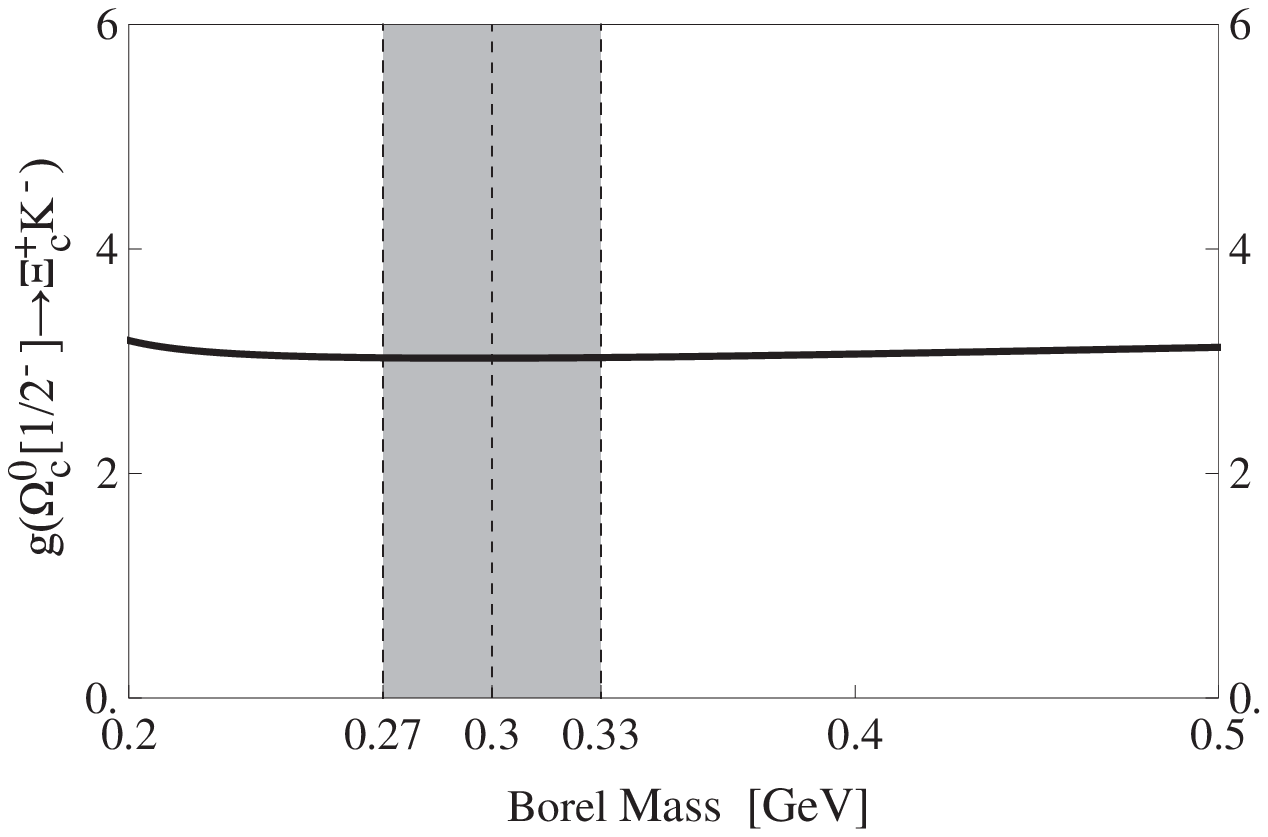}}
\end{center}
\caption{The coupling constants
$g_{\Sigma_c^0[{1\over2}^-] \rightarrow \Lambda_c^{+} \pi^-}$ (top-left), $g_{\Xi_c^{\prime0}[{1\over2}^-] \rightarrow \Xi_c^{+}\pi^-}$ (top-right),
$g_{\Xi_c^{\prime0}[{1\over2}^-] \rightarrow \Lambda_c^{+} K^-}$ (bottom-left) and $g_{\Omega_c^{0}[{1\over2}^-] \rightarrow \Xi_c^{+} K^-}$ (bottom-right)
as functions of the Borel mass $T$. The currents belonging to the baryon doublet $[\mathbf{6}_F, 0, 1, \lambda]$ are used here.
\label{fig:601lambda}}
\end{figure}

The $[\mathbf{6}_F, 0, 1, \lambda]$ multiplet contains $\Sigma_c({1\over2}^-)$, $\Xi^\prime_c({1\over2}^-)$ and $\Omega_c({1\over2}^-)$. Their sum rules are listed in Appendix~\ref{sec:601lambda}, suggesting that their possible decay channels are $(k)$, $(m)$, $(n)$ and $(q)$, while the other four channels $(l)$, $(o)$, $(p)$ and $(r)$ vanish.
We show the four coupling constants, $g_{\Sigma_c^0[{1\over2}^-] \rightarrow \Lambda_c^{+} \pi^-}$, $g_{\Xi_c^{\prime0}[{1\over2}^-] \rightarrow \Xi_c^{+}\pi^-}$, $g_{\Xi_c^{\prime0}[{1\over2}^-] \rightarrow \Lambda_c^{+} K^-}$ and $g_{\Omega_c^{0}[{1\over2}^-] \rightarrow \Xi_c^{+} K^-}$, as functions of the Borel mass $T$ in Fig.~\ref{fig:601lambda}. Using the values of $T$ listed in Table~\ref{tab:pwave2}, we obtain
\begin{eqnarray}
\nonumber &(k)& g_{\Sigma_c^0[{1\over2}^-] \rightarrow \Lambda_c^{+} \pi^-} = 1.8 \, ,
\\
&(m)& g_{\Xi_c^{\prime0}[{1\over2}^-] \rightarrow \Xi_c^{+}\pi^-} = 1.7 \, ,
\\
\nonumber &(n)& g_{\Xi_c^{\prime0}[{1\over2}^-] \rightarrow \Lambda_c^{+} K^-} = 1.8 \, ,
\\
\nonumber &(q)& g_{\Omega_c^{0}[{1\over2}^-] \rightarrow \Xi_c^{+} K^-} = 3.0 \, .
\end{eqnarray}
Using these values and the parameters listed in Sec.~\ref{sec:input}, we further obtain
\begin{eqnarray}
\nonumber &(k)& \Gamma_{\Sigma_c[{1\over2}^-] \rightarrow \Lambda_c \pi} = 200 {\rm~MeV} \, ,
\\
&(m)& \Gamma_{\Xi_c^{\prime}[{1\over2}^-] \rightarrow \Xi_c\pi} = 230 {\rm~MeV} \, ,
\\
\nonumber &(n)& \Gamma_{\Xi_c^{\prime}[{1\over2}^-] \rightarrow \Lambda_c K} = 160 {\rm~MeV} \, ,
\\
\nonumber &(q)& \Gamma_{\Omega_c[{1\over2}^-] \rightarrow \Xi_c K} = 820 {\rm~MeV} \, .
\end{eqnarray}

\subsection{The baryon doublet $[\mathbf{6}_F, 1, 1, \lambda]$}

\begin{figure}[htb]
\begin{center}
\scalebox{0.6}{\includegraphics{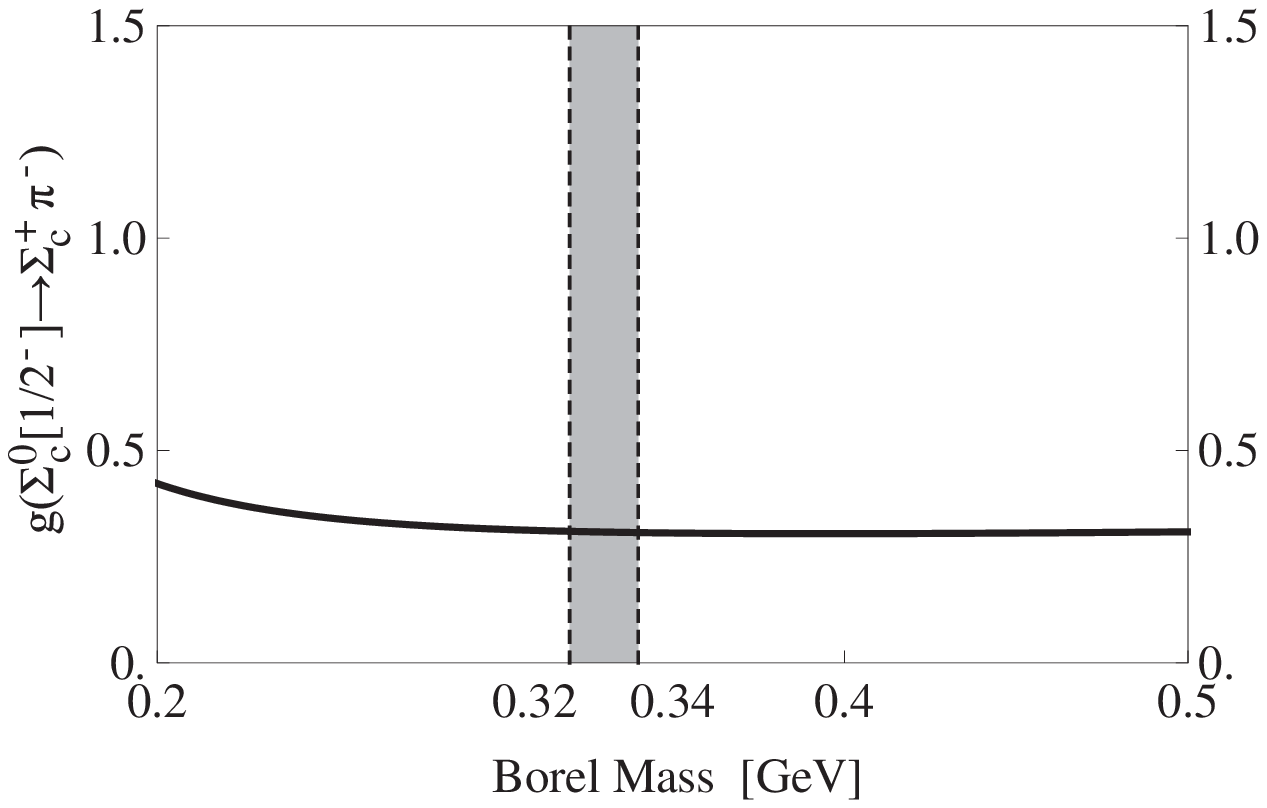}}
\scalebox{0.6}{\includegraphics{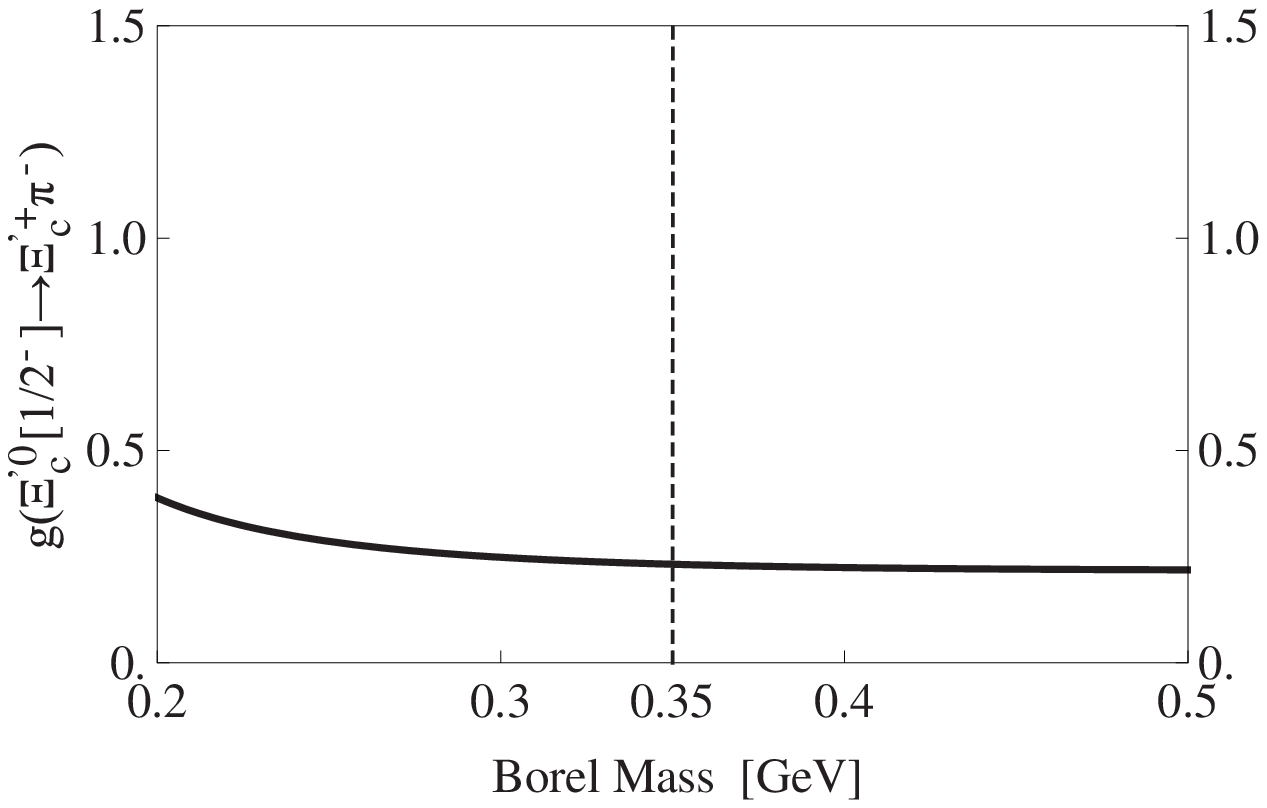}}
\\
\scalebox{0.6}{\includegraphics{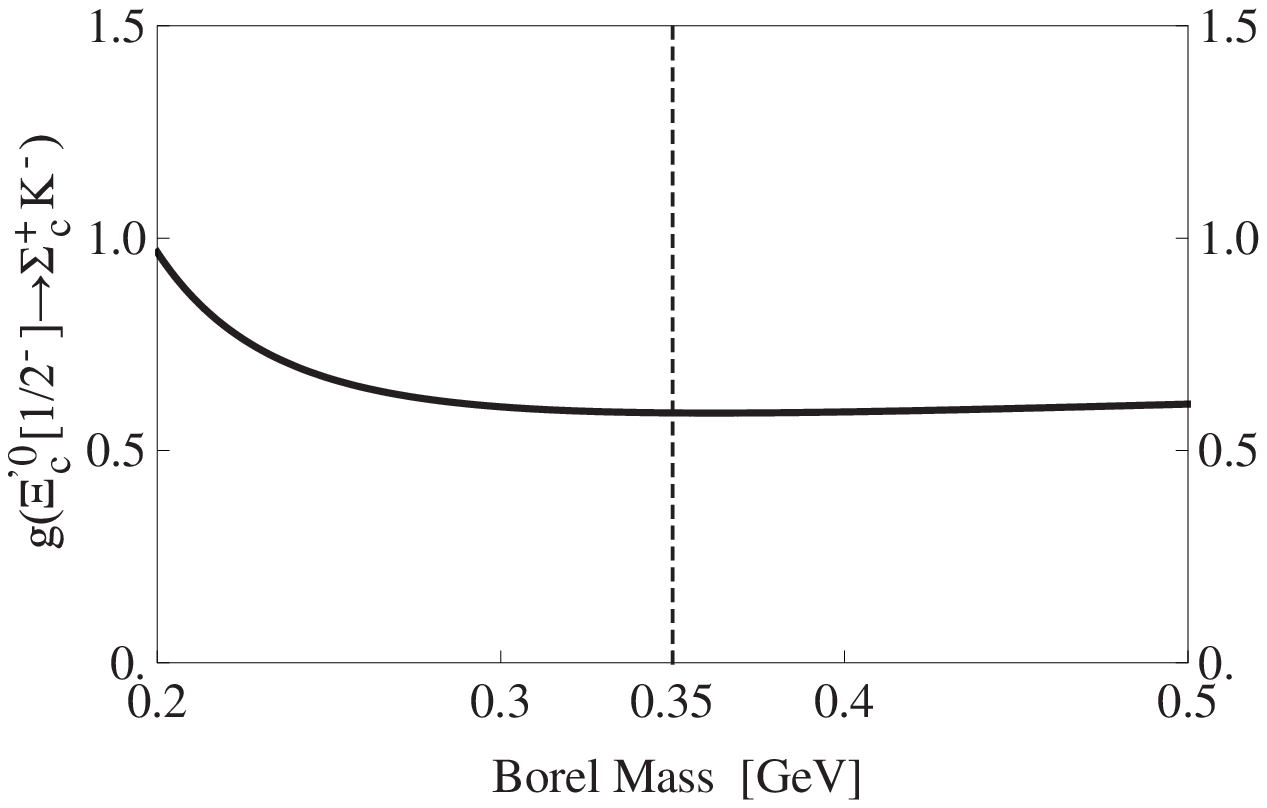}}
\scalebox{0.6}{\includegraphics{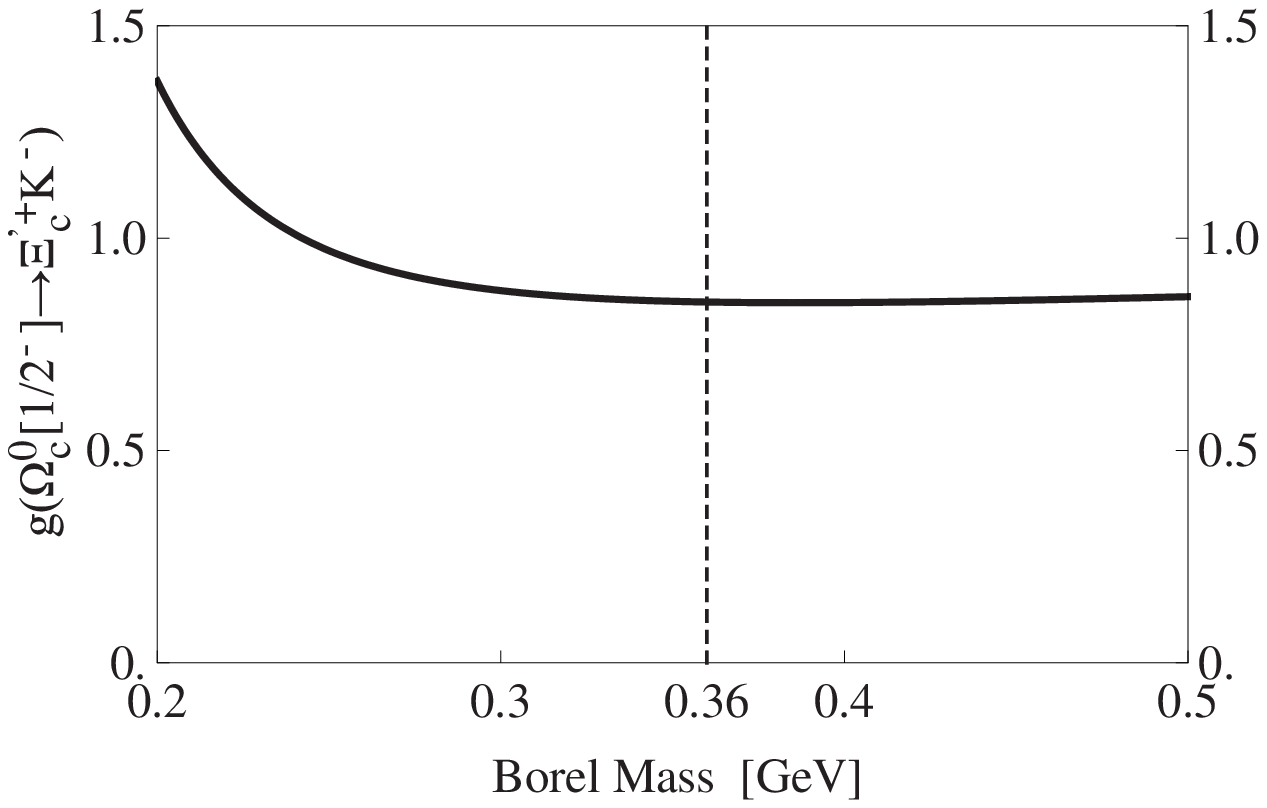}}
\\
\scalebox{0.6}{\includegraphics{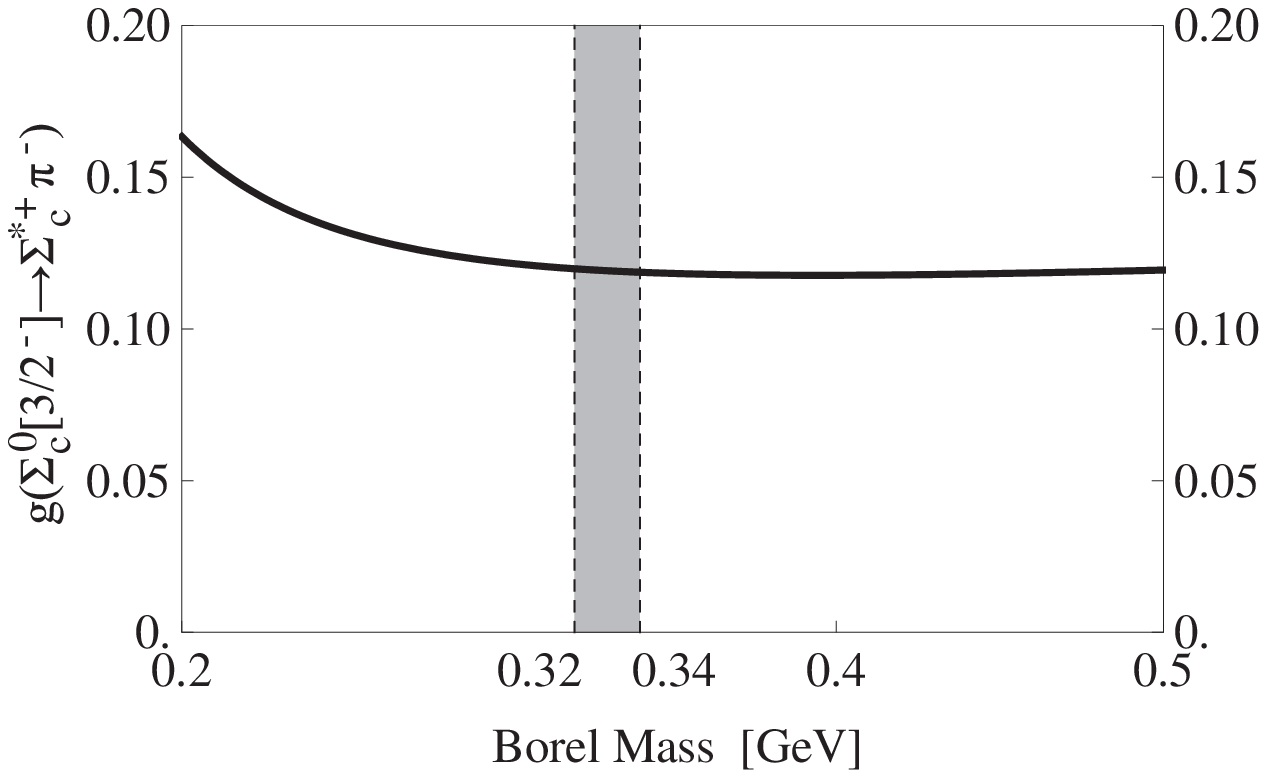}}
\scalebox{0.6}{\includegraphics{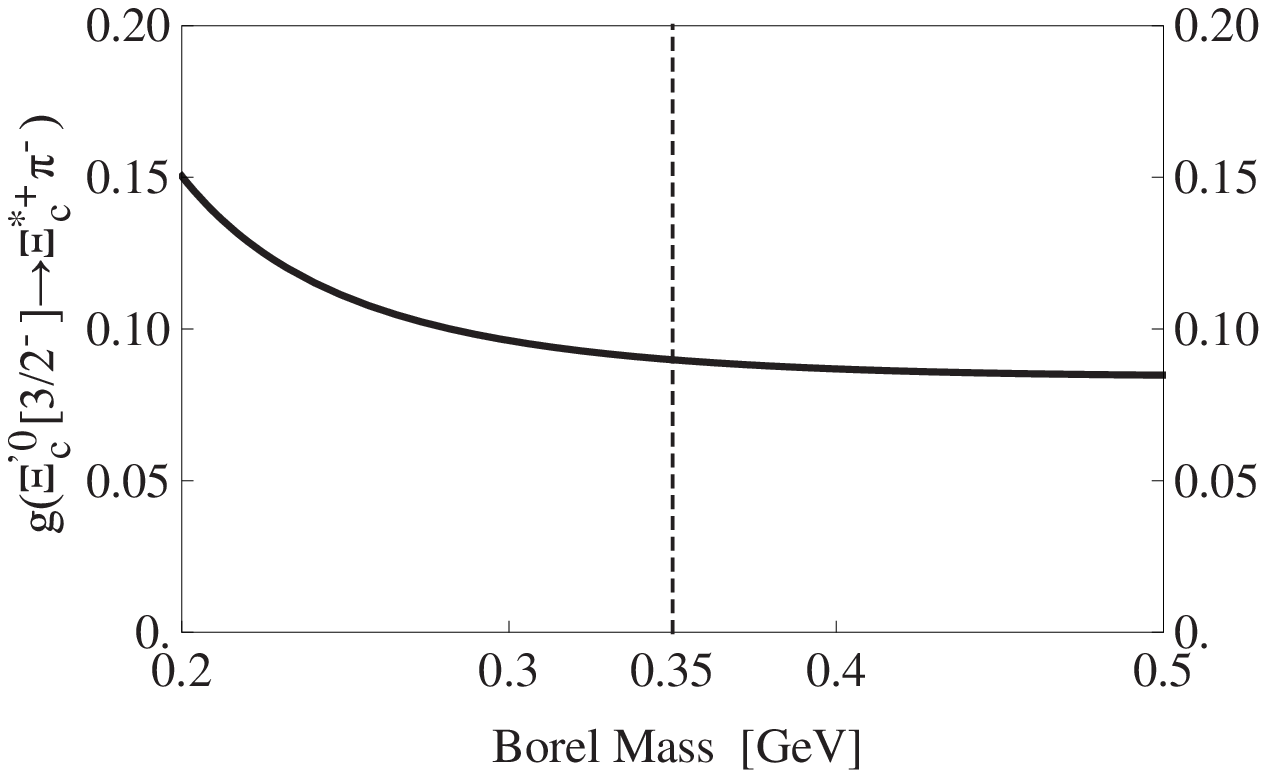}}
\\
\scalebox{0.6}{\includegraphics{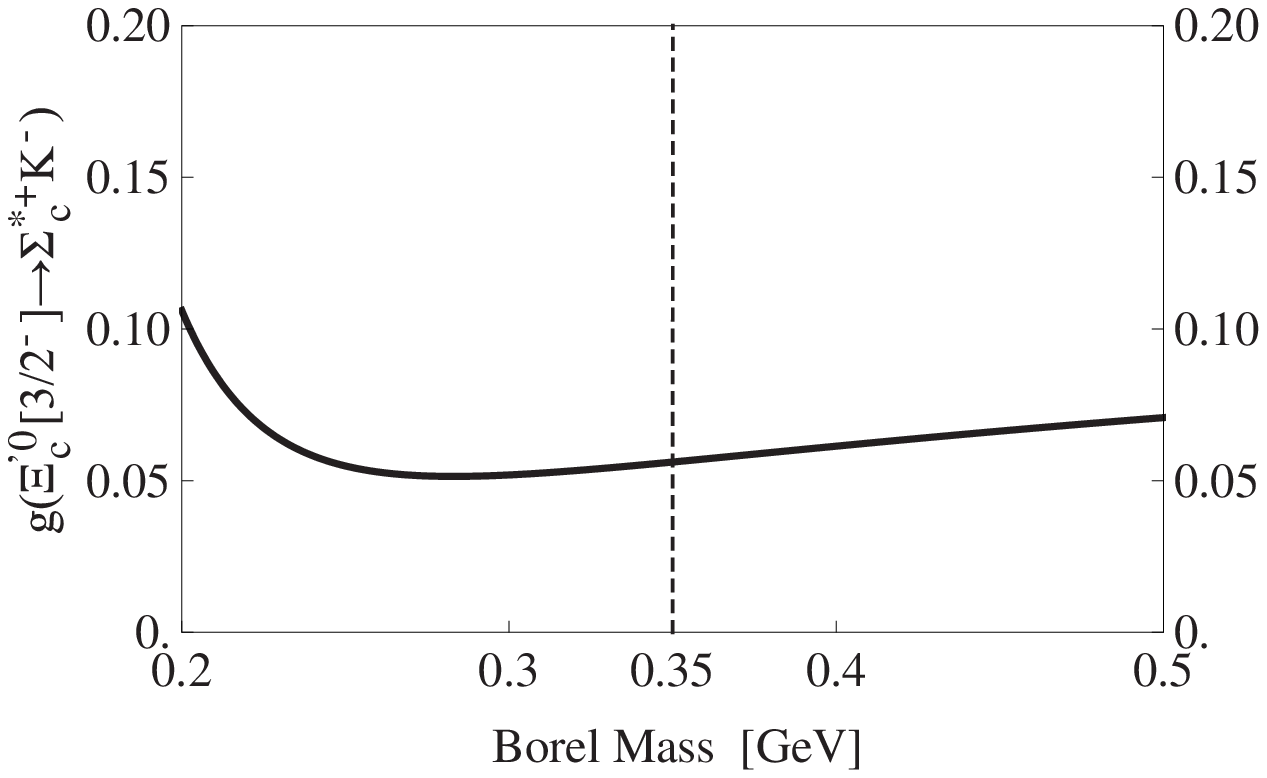}}
\scalebox{0.6}{\includegraphics{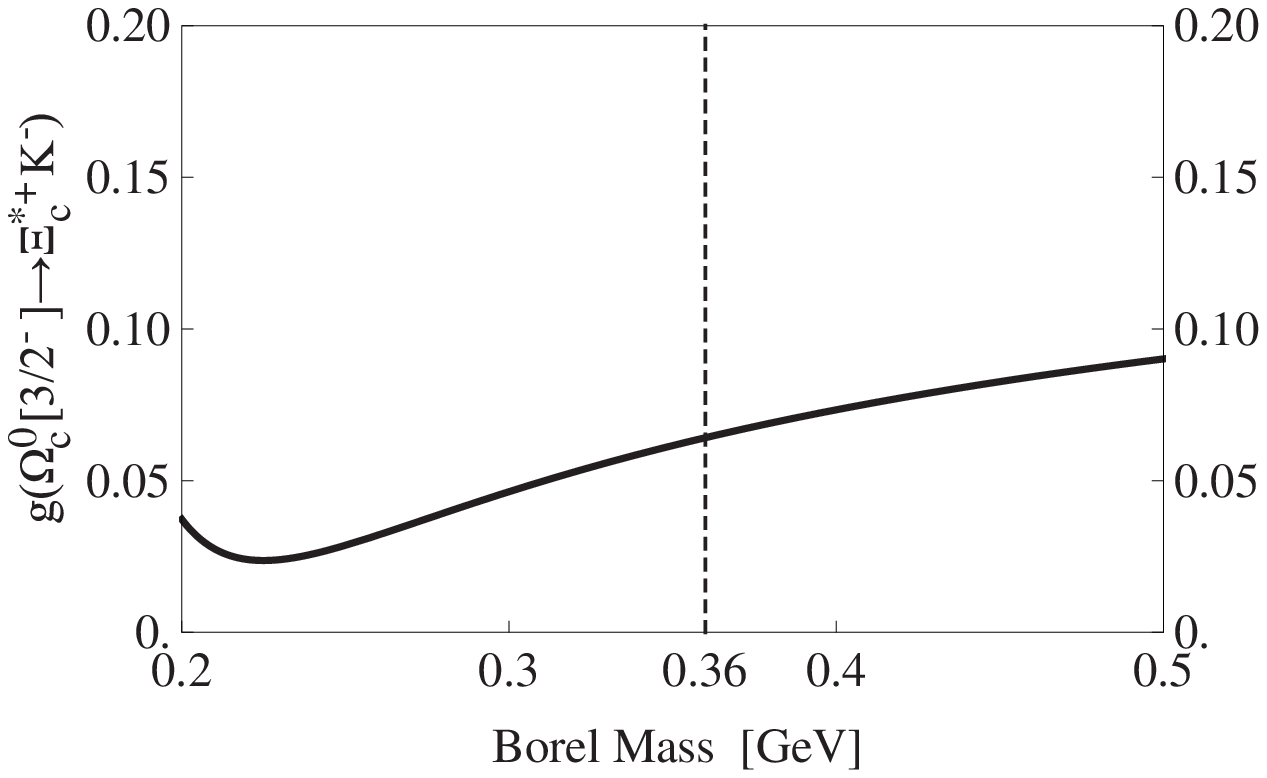}}
\end{center}
\caption{The coupling constants $g_{\Sigma_c^0[{1\over2}^-] \rightarrow \Sigma_c^{+} \pi^-}$ (top-left), $g_{\Xi_c^{\prime0}[{1\over2}^-] \rightarrow \Xi_c^{\prime+}\pi^-}$ (top-right),
$g_{\Xi_c^{\prime0}[{1\over2}^-] \rightarrow \Sigma_c^{+} K^-}$ (middle-left), $g_{\Omega_c^{0}[{1\over2}^-] \rightarrow \Xi_c^{\prime+} K^-}$ (middle-right),
$g_{\Sigma_c^{0}[{3\over2}^-] \rightarrow \Sigma_c^{*+} \pi^-}$ (middle-left), $g_{\Xi_c^{\prime0}[{3\over2}^-] \rightarrow \Xi_c^{*+} \pi^-}$ (middle-right),
$g_{\Xi_c^{\prime0}[{3\over2}^-] \rightarrow \Sigma_c^{*+} K^-}$ (bottom-left) and $g_{\Omega_c^{0}[{3\over2}^-] \rightarrow \Xi_c^{*+} K^-}$ (bottom-right)
as functions of the Borel mass $T$. The currents belonging to the baryon doublet $[\mathbf{6}_F, 1, 1, \lambda]$ are used here.
\label{fig:611lambda}}
\end{figure}

The $[\mathbf{6}_F, 1, 1, \lambda]$ multiplet contains $\Sigma_c({1\over2}^-/{3\over2}^-)$, $\Xi^\prime_c({1\over2}^-/{3\over2}^-)$ and $\Omega_c({1\over2}^-/{3\over2}^-)$. Their sum rules are listed in Appendix~\ref{sec:611lambda}, suggesting that their possible decay channels are $(l)$, $(o)$, $(p)$, $(r)$, $(s)$, $(t)$, $(u)$ and $(v)$,
while the other four channels $(k)$, $(m)$, $(n)$ and $(q)$ vanish.
We show the eight coupling constants, $g_{\Sigma_c^0[{1\over2}^-] \rightarrow \Sigma_c^{+} \pi^-}$, $g_{\Xi_c^{\prime0}[{1\over2}^-] \rightarrow \Xi_c^{\prime+}\pi^-}$, $g_{\Xi_c^{\prime0}[{1\over2}^-] \rightarrow \Sigma_c^{+} K^-}$, $g_{\Omega_c^{0}[{1\over2}^-] \rightarrow \Xi_c^{\prime+} K^-}$, $g_{\Sigma_c^{0}[{3\over2}^-] \rightarrow \Sigma_c^{*+} \pi^-}$, $g_{\Xi_c^{\prime0}[{3\over2}^-] \rightarrow \Xi_c^{*+} \pi^-}$, $g_{\Xi_c^{\prime0}[{3\over2}^-] \rightarrow \Sigma_c^{*+} K^-}$ and $g_{\Omega_c^{0}[{3\over2}^-] \rightarrow \Xi_c^{*+} K^-}$, as functions of the Borel mass $T$ in Fig.~\ref{fig:611lambda}. Using the values of $T$ listed in Table~\ref{tab:pwave2}, we obtain
\begin{eqnarray}
\nonumber &(l)& g_{\Sigma_c^0[{1\over2}^-] \rightarrow \Sigma_c^{+} \pi^-} = 0.31 \, ,
\\
\nonumber &(o)& g_{\Xi_c^{\prime0}[{1\over2}^-] \rightarrow \Xi_c^{\prime+}\pi^-} = 0.23 \, ,
\\
\nonumber &(p)& g_{\Xi_c^{\prime0}[{1\over2}^-] \rightarrow \Sigma_c^{+} K^-} = 0.59 \, ,
\\
&(r)& g_{\Omega_c^{0}[{1\over2}^-] \rightarrow \Xi_c^{\prime+} K^-} = 0.85 \, ,
\\
\nonumber &(s)& g_{\Sigma_c^{0}[{3\over2}^-] \rightarrow \Sigma_c^{*+} \pi^-} = 0.12 \, ,
\\
\nonumber &(t)& g_{\Xi_c^{\prime0}[{3\over2}^-] \rightarrow \Xi_c^{*+} \pi^-} = 0.09 \, ,
\\
\nonumber &(u)& g_{\Xi_c^{\prime0}[{3\over2}^-] \rightarrow \Sigma_c^{*+} K^-} = 0.056 \, ,
\\
\nonumber &(v)& g_{\Omega_c^{0}[{3\over2}^-] \rightarrow \Xi_c^{*+} K^-} = 0.064 \, .
\end{eqnarray}
Using these values and the parameters listed in Sec.~\ref{sec:input}, we further obtain
\begin{eqnarray}
\nonumber &(l)& \Gamma_{\Sigma_c[{1\over2}^-] \rightarrow \Sigma_c \pi} = 7.9 {\rm~MeV} \, ,
\\
\nonumber &(o)& \Gamma_{\Xi_c^{\prime}[{1\over2}^-] \rightarrow \Xi_c^{\prime}\pi} = 3.7 {\rm~MeV} \, ,
\\
\nonumber &(p)& \Gamma_{\Xi_c^{\prime}[{1\over2}^-] \rightarrow \Sigma_c K} = 3.6 {\rm~MeV} \, ,
\\
&(r)& \Gamma_{\Omega_c[{1\over2}^-] \rightarrow \Xi_c^{\prime} K} = 29 {\rm~MeV} \, ,
\\
\nonumber &(s)& \Gamma_{\Sigma_c[{3\over2}^-] \rightarrow \Sigma_c^{*} \pi} = 0.95 {\rm~MeV} \, ,
\\
\nonumber &(t)& \Gamma_{\Xi_c^{\prime}[{3\over2}^-] \rightarrow \Xi_c^{*} \pi} = 0.45 {\rm~MeV} \, ,
\\
\nonumber &(u)& \Gamma_{\Xi_c^{\prime}[{3\over2}^-] \rightarrow \Sigma_c^{*} K \rightarrow \Lambda_c \pi K} = 7 \times 10^{-5} {\rm~MeV} \, ,
\\
\nonumber &(v)& \Gamma_{\Omega_c[{3\over2}^-] \rightarrow \Xi_c^{*} K \rightarrow \Xi_c \pi K} = 1 \times 10^{-4} {\rm~MeV} \, .
\end{eqnarray}

\subsection{The baryon doublet $[\mathbf{6}_F, 2, 1, \lambda]$}

\begin{figure}[htb]
\begin{center}
\scalebox{0.6}{\includegraphics{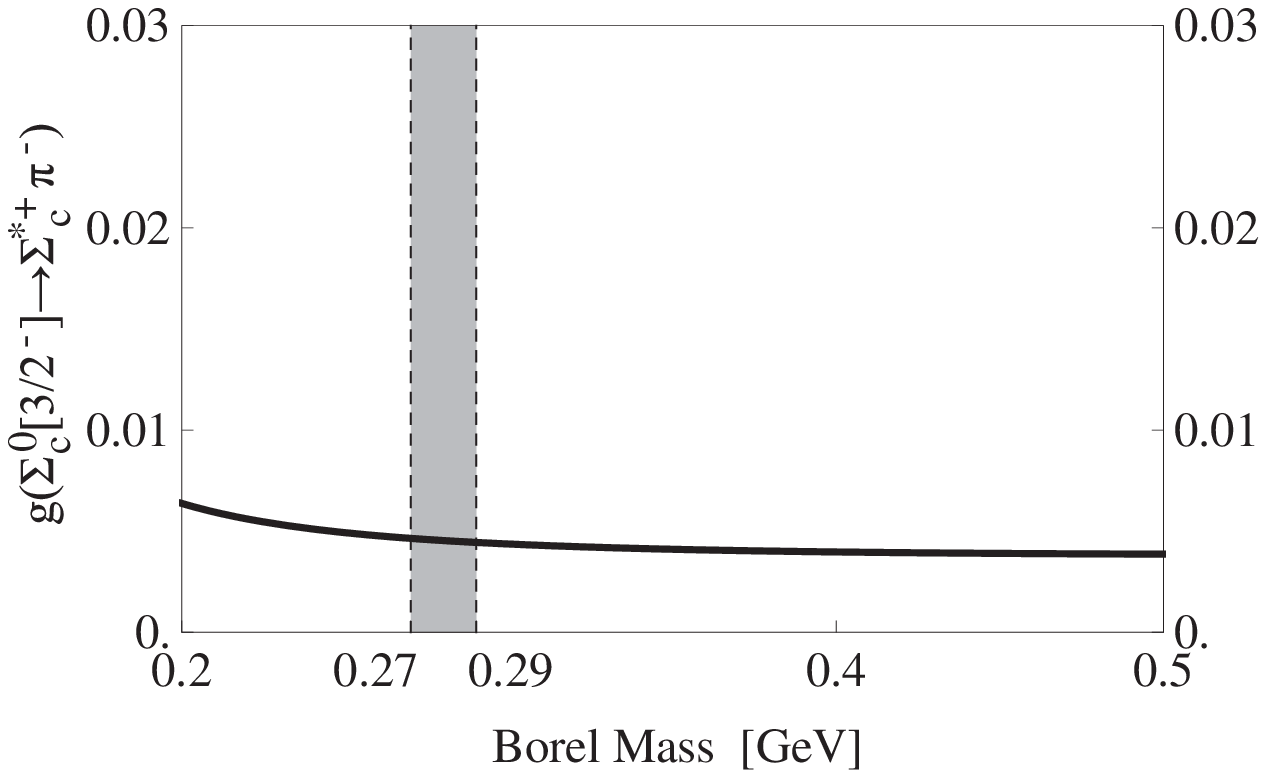}}
\scalebox{0.6}{\includegraphics{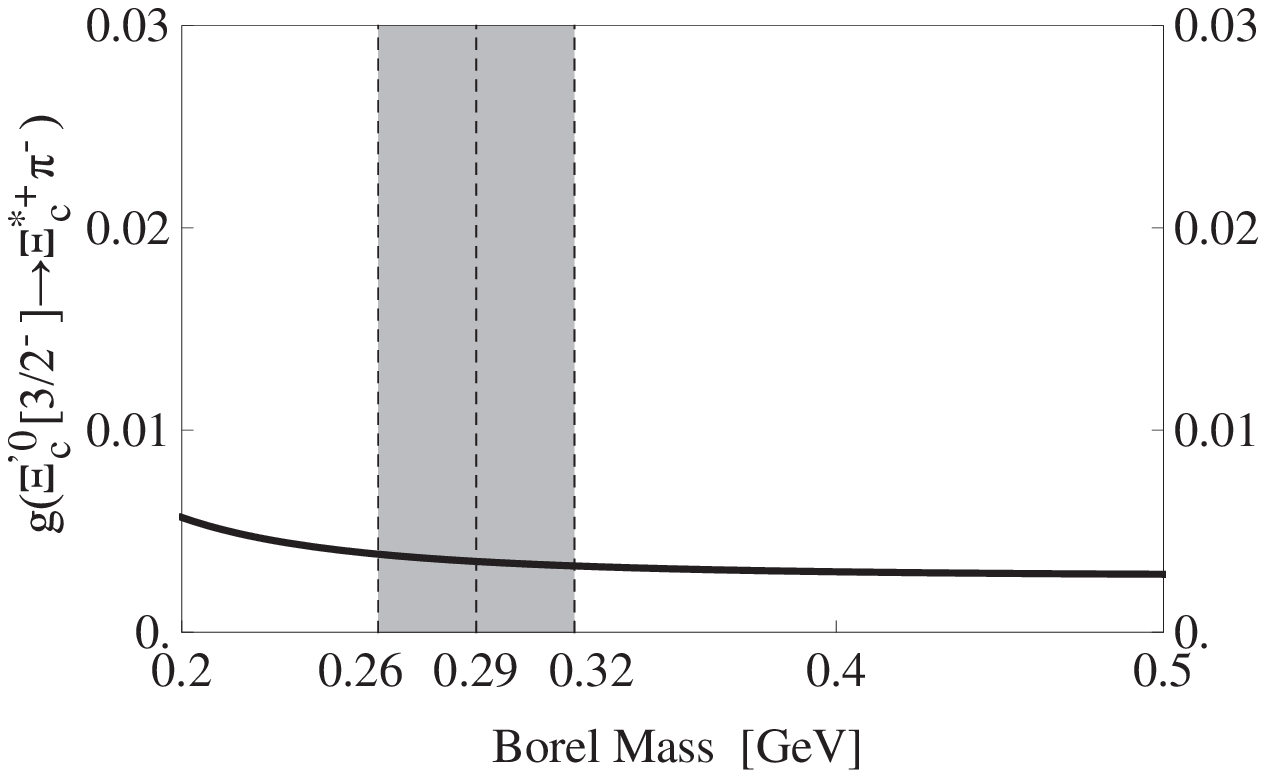}}
\\
\scalebox{0.6}{\includegraphics{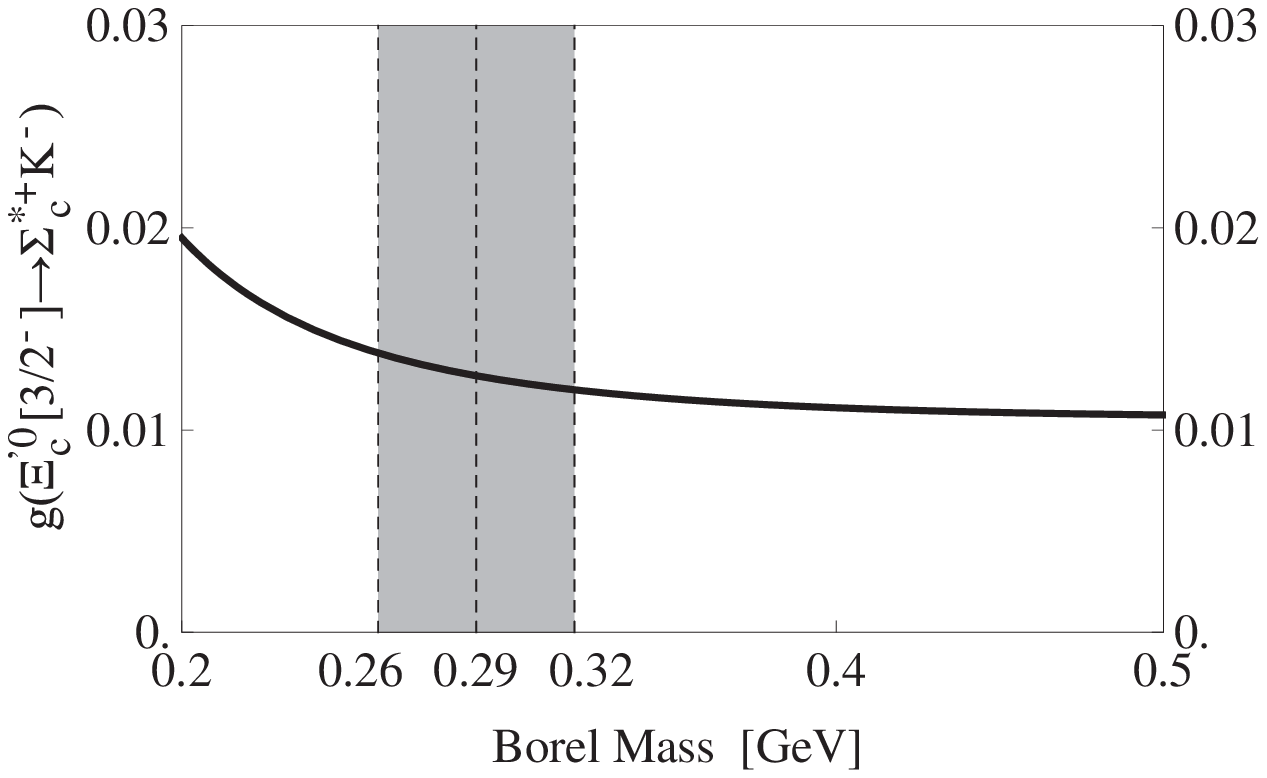}}
\scalebox{0.6}{\includegraphics{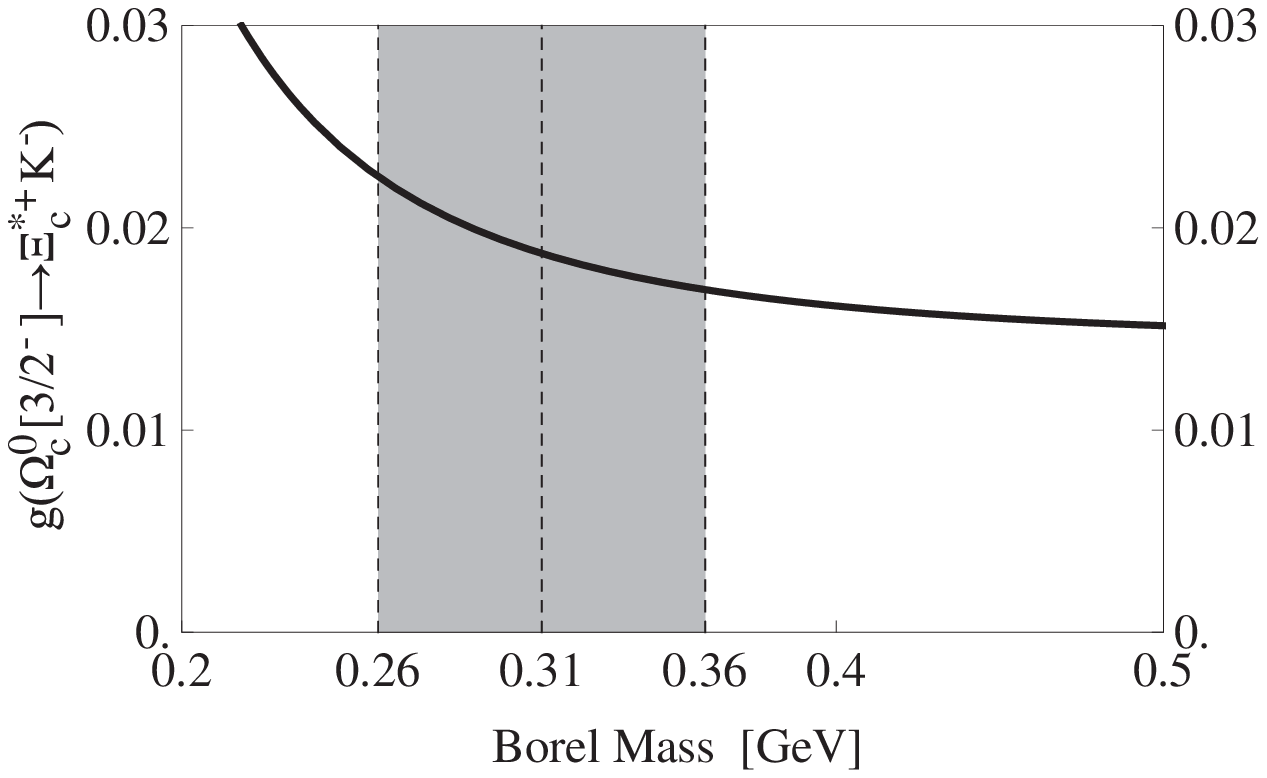}}
\end{center}
\caption{The coupling constants $g_{\Sigma_c^{0}[{3\over2}^-] \rightarrow \Sigma_c^{*+} \pi^-}$ (top-left), $g_{\Xi_c^{\prime0}[{3\over2}^-] \rightarrow \Xi_c^{*+} \pi^-}$ (top-right),
$g_{\Xi_c^{\prime0}[{3\over2}^-] \rightarrow \Sigma_c^{*+} K^-}$ (bottom-left) and $g_{\Omega_c^{0}[{3\over2}^-] \rightarrow \Xi_c^{*+} K^-}$ (bottom-right)
as functions of the Borel mass $T$. The currents belonging to the baryon doublet $[\mathbf{6}_F, 2, 1, \lambda]$ are used here.
\label{fig:621lambda}}
\end{figure}

The $[\mathbf{6}_F, 2, 1, \lambda]$ multiplet contains $\Sigma_c({3\over2}^-/{5\over2}^-)$, $\Xi^\prime_c({3\over2}^-/{5\over2}^-)$ and $\Omega_c({3\over2}^-/{5\over2}^-)$. Their sum rules are listed in Appendix~\ref{sec:621lambda}, suggesting that their possible decay channels are $(s)$, $(t)$, $(u)$ and $(v)$.
We show the four coupling constants, $g_{\Sigma_c^{0}[{3\over2}^-] \rightarrow \Sigma_c^{*+} \pi^-}$, $g_{\Xi_c^{\prime0}[{3\over2}^-] \rightarrow \Xi_c^{*+} \pi^-}$, $g_{\Xi_c^{\prime0}[{3\over2}^-] \rightarrow \Sigma_c^{*+} K^-}$ and $g_{\Omega_c^{0}[{3\over2}^-] \rightarrow \Xi_c^{*+} K^-}$, as functions of the Borel mass $T$ in Fig.~\ref{fig:621lambda}. Using the values of $T$ listed in Table~\ref{tab:pwave2}, we obtain
\begin{eqnarray}
\nonumber &(s)& g_{\Sigma_c^{0}[{3\over2}^-] \rightarrow \Sigma_c^{*+} \pi^-} = 0.005 \, ,
\\
&(t)& g_{\Xi_c^{\prime0}[{3\over2}^-] \rightarrow \Xi_c^{*+} \pi^-} = 0.004 \, ,
\\
\nonumber &(u)& g_{\Xi_c^{\prime0}[{3\over2}^-] \rightarrow \Sigma_c^{*+} K^-} = 0.013 \, ,
\\
\nonumber &(v)& g_{\Omega_c^{0}[{3\over2}^-] \rightarrow \Xi_c^{*+} K^-} = 0.019 \, .
\end{eqnarray}
Using these values and the parameters listed in Sec.~\ref{sec:input}, we further obtain
\begin{eqnarray}
\nonumber &(s)& \Gamma_{\Sigma_c[{3\over2}^-] \rightarrow \Sigma_c^{*} \pi} = 1 \times 10^{-3} {\rm~MeV} \, ,
\\
&(t)& \Gamma_{\Xi_c^{\prime}[{3\over2}^-] \rightarrow \Xi_c^{*} \pi} = 7 \times 10^{-4} {\rm~MeV} \, ,
\\
\nonumber &(u)& \Gamma_{\Xi_c^{\prime}[{3\over2}^-] \rightarrow \Sigma_c^{*} K \rightarrow \Lambda_c \pi K} = 3 \times 10^{-6} {\rm~MeV} \, ,
\\
\nonumber &(v)& \Gamma_{\Omega_c[{3\over2}^-] \rightarrow \Xi_c^{*} K \rightarrow \Xi_c \pi K} = 9 \times 10^{-6} {\rm~MeV} \, .
\end{eqnarray}

\section{Summary and Discussions}
\label{sec:summary}

To summarize this paper, we have used the method of light-cone QCD sum rules to study the decay properties of the $P$-wave charmed baryons.
Firstly we summarize our results on the flavor $\mathbf{\bar 3}_F$ $P$-wave charmed baryons.
We have studied their $S$-wave decays into ground-state charmed baryons accompanied by a pseudoscalar meson ($\pi$ or $K$) or a vector meson ($\rho$ or $K^*$), including both two-body and three-body decays which are kinematically allowed.
The results are listed in Table~\ref{tab:result1}, where the possible decay channels are:
$(a)~\Lambda_c[{1\over2}^-] \rightarrow \Sigma_c \pi (\to \Lambda_c \pi \pi)$,
$(b)~\Lambda_c[{3\over2}^-] \rightarrow \Sigma_c^{*} \pi \to \Lambda_c \pi \pi$,
$(c)~\Xi_c[{1\over2}^-] \rightarrow \Xi_c \pi$,
$(d)~\Xi_c[{1\over2}^-] \rightarrow \Lambda_c K$,
$(e)~\Xi_c[{1\over2}^-] \rightarrow \Xi_c \rho \to  \Xi_c \pi \pi$,
$(f)~\Xi_c[{1\over2}^-] \rightarrow \Xi_c^{\prime}\pi$,
$(g)~\Xi_c[{3\over2}^-] \rightarrow \Xi_c \rho \to  \Xi_c \pi \pi$,
$(h)~\Xi_c[{3\over2}^-] \rightarrow \Xi_c^{*} \pi$,
$(i)~\Lambda_c[{5\over2}^-] \rightarrow \Sigma_c^{*} \rho \to \Sigma_c^{*} \pi \pi$,
and $(j)~\Xi_c[{5\over2}^-] \rightarrow \Xi_c^{*} \rho \to \Xi_c^{*} \pi \pi$.
We note that the uncertainties can be as large as $\Gamma^{+200\%}_{-67\%}$.

\begin{table}[hbt]
\begin{center}
\renewcommand{\arraystretch}{1.5}
\caption{Non-vanishing decay widths of the flavor $\mathbf{\bar 3}_F$ $P$-wave charmed baryons, in units of MeV.
The two mass values with $^*$ are our assumptions, so that the decay channels $(i)$ and $(j)$ are kinematically allowed.
The possible decay channels are:
$(a)~~ \Lambda_c[{1\over2}^-] \rightarrow \Sigma_c \pi (\to \Lambda_c \pi \pi)$,
$(b)~~ \Lambda_c[{3\over2}^-] \rightarrow \Sigma_c^{*} \pi \to \Lambda_c \pi \pi$,
$(c)~~ \Xi_c[{1\over2}^-] \rightarrow \Xi_c \pi$,
$(d)~~ \Xi_c[{1\over2}^-] \rightarrow \Lambda_c K$,
$(e)~~ \Xi_c[{1\over2}^-] \rightarrow \Xi_c \rho \to  \Xi_c \pi \pi$,
$(f)~~ \Xi_c[{1\over2}^-] \rightarrow \Xi_c^{\prime}\pi$,
$(g)~~ \Xi_c[{3\over2}^-] \rightarrow \Xi_c \rho \to  \Xi_c \pi \pi$,
$(h)~~ \Xi_c[{3\over2}^-] \rightarrow \Xi_c^{*} \pi$,
$(i)~~ \Lambda_c[{5\over2}^-] \rightarrow \Sigma_c^{*} \rho \to \Sigma_c^{*} \pi \pi$,
and $(j)~~ \Xi_c[{5\over2}^-] \rightarrow \Xi_c^{*} \rho \to \Xi_c^{*} \pi \pi$.
We use the two-body middle/final states to denote them in the table.
}
\begin{tabular}{c | c | c c c c | c c}
\hline\hline
~~Baryon~~ & ~~Experiments~\cite{pdg}~~ & ~~$[\mathbf{\bar 3}_F, 0, 1, \rho]$~~ & ~~$[\mathbf{\bar 3}_F, 1, 1, \rho]$~~ & ~~$[\mathbf{\bar 3}_F, 2, 1, \rho]$~~ & ~~$[\mathbf{\bar 3}_F, 1, 0, \lambda]$~~ & ~~Mix-$|B_1\rangle$~~ & ~~Mix-$|B_2\rangle$~~
\\ \hline\hline
$\begin{array}{c}
\Lambda_c(2595)
\\
J^P={1\over2}^-
\end{array}$
&
$\begin{array}{c}
\Gamma_{\Lambda_c(2595)} = 2.59
\\
\left[\Sigma_c^{++,0} \pi^\mp\right]: 48\%
\\
\left[\Lambda_c\pi\pi\right]_{{\rm 3body}}: 18\%
\end{array}$
& -- & $\left[\Sigma_c \pi\right] = 0.39$ & -- & $\left[\Sigma_c \pi\right] = 32$
&
\multicolumn{2}{c}{$\begin{array}{c}
\mbox{Input:}~~{\Gamma(\Sigma_c \pi) \over \Gamma(\Lambda_c(2595))} = 0.66
\\
\theta_1 = -20^{\rm o}~~~~~~~~\theta_2 = 7^{\rm o}
\end{array}$}
\\ \hline
$\begin{array}{c}
\Xi_c(2790)
\\
J^P={1\over2}^-
\end{array}$
&
$\begin{array}{c}
\Gamma_{\Xi_c^+(2790)} < 15
\\
\Gamma_{\Xi_c^0(2790)} < 12
\\
\left[\Xi^\prime_c \pi\right]_{{\rm 2body}}: {\rm seen}
\end{array}$
&
$\begin{array}{c}
\left[\Xi_c \pi\right] = 300
\\
\left[\Lambda_c K\right] = 82
\end{array}$
&
$\begin{array}{c}
\left[\Xi_c^\prime \pi\right] = 1.6
\\
\left[\Xi_c \rho\right] = 0.00
\end{array}$
& -- &
$\begin{array}{c}
\left[\Xi_c^\prime \pi\right] = 100
\\
\left[\Xi_c \rho\right] = 0.04
\end{array}$
&
$\begin{array}{c}
\left[\Xi_c^\prime \pi\right] = 4.7
\\
\left[\Xi_c \rho\right] = 0.00
\end{array}$
&
$\begin{array}{c}
\left[\Xi_c^\prime \pi\right] = 6.1
\\
\left[\Xi_c \rho\right] = 0.00
\end{array}$
\\ \hline
$\begin{array}{c}
\Lambda_c(2625)
\\
J^P={3\over2}^-
\end{array}$
&
$\begin{array}{c}
\Gamma_{\Lambda_c(2625)} < 0.97
\\
\left[\Sigma_c \pi\right]_{{\rm 2body}}< 10\%
\\
\left[\Lambda_c\pi\pi\right]_{{\rm 3body}}: {\rm large}
\end{array}$
& -- & $\left[\Sigma_c^* \pi\right] = 0.00$ & $\left[\Sigma_c^* \pi\right] = 0.03$ & $\left[\Sigma_c^* \pi\right] = 0.96$
&
$\left[\Sigma_c^* \pi\right] = 0.11$
&
$\left[\Sigma_c^* \pi\right] = 0.01$
\\ \hline
$\begin{array}{c}
\Xi_c(2815)
\\
J^P={3\over2}^-
\end{array}$
&
$\begin{array}{c}
\Gamma_{\Xi_c^+(2815)} < 3.5
\\
\Gamma_{\Xi_c^0(2815)} < 6.5
\\
\left[\Xi_c \pi \pi\right]_{{\rm 3body}}: {\rm seen}
\end{array}$
& -- &
$\begin{array}{c}
\left[\Xi_c^* \pi\right] = 0.01
\\
\left[\Xi_c \rho\right] = 0.00
\end{array}$
& $\left[\Xi_c^* \pi\right] = 0.69$ &
$\begin{array}{c}
\left[\Xi_c^* \pi\right] = 30
\\
\left[\Xi_c \rho\right] = 0.23
\end{array}$
&
$\begin{array}{c}
\left[\Xi_c^* \pi\right] = 3.0
\\
\left[\Xi_c \rho\right] = 0.03
\end{array}$
&
$\begin{array}{c}
\left[\Xi_c^* \pi\right] = 0.59
\\
\left[\Xi_c \rho\right] = 0.00
\end{array}$
\\ \hline
$\Lambda_c(5/2^-)$ & $M_{\Lambda_c(5/2^-)} \sim 2850^*$ & -- & -- & $\left[\Sigma_c^* \rho\right] = 11$ & -- & -- & --
\\ \hline
$\Xi_c(5/2^-)$ & $M_{\Xi_c(5/2^-)} \sim 3000^*$ & -- & -- & $\left[\Xi_c^* \rho\right] = 12$ & -- & -- & --
\\ \hline \hline
\end{tabular}
\label{tab:result1}
\end{center}
\end{table}

Our calculations are performed based on the heavy quark effective theory (HQET) and separately for the four charmed baryon multiplets of flavor $\mathbf{\bar 3}_F$, $[\mathbf{\bar 3}_F, 0, 1, \rho]$, $[\mathbf{\bar 3}_F, 1, 1, \rho]$, $[\mathbf{\bar 3}_F, 2, 1, \rho]$ and $[\mathbf{\bar 3}_F, 1, 0, \lambda]$.
We find that none of these four multiplets can independently well describe the experimental decay data of the $\Lambda_c(2595)$.
This is somehow in contrast with quark model calculations which describe some of the decay rates, but not all, in a reasonable manner~\cite{Nagahiro:2016nsx}.
It would be a future issue to see further relations of various approaches. See also Refs.~\cite{Liang:2016ydj,Liang:2016exm} for other possible interpretations of the $\Lambda_c(2595)$.
In the present sum rule study, considering the fact that
the heavy quark symmetry is not perfect, the physical states are probably mixed states containing various components with different inner quantum numbers. It is then possible that the $\Lambda_c(2595)$ is an admixture of the above four multiplets.
Thus
we try to use the mixture of $[\mathbf{\bar 3}_F, 1, 1, \rho]$ and $[\mathbf{\bar 3}_F, 1, 0, \lambda]$ as an explanation
and
assume the physical state to be
\begin{eqnarray}
| \Lambda_c(1/2^-) \rangle = \cos \theta \times | 1/2, - , \Lambda_c, 1, 1, \rho \rangle + \sin\theta \times | 1/2, - , \Lambda_c, 1, 0, \lambda \rangle \, ,
\end{eqnarray}
so that we have
\begin{eqnarray}
g_{\Lambda_c[{1\over2}^-] \rightarrow \Sigma_c \pi} = \cos \theta \times g_{| 1/2, - , \Lambda_c, 1, 1, \rho \rangle \to \Sigma_c \pi} + \sin\theta \times g_{| 1/2, - , \Lambda_c, 1, 0, \lambda \rangle \to \Sigma_c \pi} \, ,
\end{eqnarray}
and we can further obtain
\begin{eqnarray}
\sqrt{\Gamma_{\Lambda_c[{1\over2}^-] \rightarrow \Sigma_c \pi (\to \Lambda_c \pi \pi)}} = \cos \theta \times \sqrt{\Gamma_{| 1/2, - , \Lambda_c, 1, 1, \rho \rangle \to \Sigma_c \pi (\to \Lambda_c \pi \pi)}} + \sin\theta \times \sqrt{\Gamma_{| 1/2, - , \Lambda_c, 1, 0, \lambda \rangle \to \Sigma_c \pi (\to \Lambda_c \pi \pi)}} \, .
\end{eqnarray}
Other channels can be similarly evaluated. The mixing angle $\theta$ can be estimated by assuming~\cite{pdg,Albrecht:1997qa}
\begin{eqnarray}
{\Gamma(\Sigma_c \pi) \over \Gamma_{\Lambda_c(2595)}} \approx {\Gamma(\Sigma_c^{++} \pi^- + \Sigma_c^{0} \pi^+) \over \Gamma(\Lambda_c^+ \pi^+ \pi^-)} = 0.66^{+0.13}_{-0.16}\pm0.07 \, .
\end{eqnarray}
There are two possible solutions: $\theta_1 = -20^{\rm o}$ and $\theta_2 = 7^{\rm o}$, which we denote as Mix-$|B_1\rangle$ and Mix-$|B_2\rangle$, respectively. Assuming the mixing angle to be an overall parameter, we evaluate decay widths of the $\Lambda_c(2625)$, $\Xi_c(2790)$ and $\Xi_c(2815)$. The results are listed in Table~\ref{tab:result1}, which are consistent with their experimental decay data, while the Mix-$|B_1\rangle$ seems a bit better. Recall that both $[\mathbf{\bar 3}_F, 1, 1, \rho]$ and $[\mathbf{\bar 3}_F, 1, 0, \lambda]$ can also describe the masses of these states~\cite{Chen:2015kpa}, so these two mixing solutions can well describe both masses and decay properties of the $\Lambda_c(2595)$, $\Lambda_c(2625)$, $\Xi_c(2790)$ and $\Xi_c(2815)$ at the same time. We would like to suggest the Belle/KEK, LHCb, and J-PARC experiments to further examine these values.

Using the same method, we have also studied the decay properties of the flavor $\mathbf{6}_F$ $P$-wave charmed baryons.
We have studied their $S$-wave decays into ground-state charmed baryons accompanied by a pseudoscalar meson ($\pi$ or $K$), including both two-body and three-body decays which are kinematically allowed.
The results are listed in Table~\ref{tab:result2}, where the possible decay channels are:
$(k)~\Sigma_c[{1\over2}^-] \rightarrow \Lambda_c \pi$,
$(l)~\Sigma_c[{1\over2}^-] \rightarrow \Sigma_c \pi$,
$(m)~\Xi_c^{\prime}[{1\over2}^-] \rightarrow \Xi_c \pi$,
$(n)~\Xi_c^{\prime}[{1\over2}^-] \rightarrow \Lambda_c K$,
$(o)~\Xi_c^{\prime}[{1\over2}^-] \rightarrow \Xi_c^{\prime} \pi$,
$(p)~\Xi_c^{\prime}[{1\over2}^-] \rightarrow \Sigma_c K$,
$(q)~\Omega_c[{1\over2}^-] \rightarrow \Xi_c K$,
$(r)~\Omega_c[{1\over2}^-] \rightarrow \Xi_c^{\prime} K$,
$(s)~\Sigma_c[{3\over2}^-] \rightarrow \Sigma_c^{*} \pi$,
$(t)~\Xi_c^{\prime}[{3\over2}^-] \rightarrow \Xi_c^{*} \pi$,
$(u)~\Xi_c^{\prime}[{3\over2}^-] \rightarrow \Sigma_c^{*} K \rightarrow \Lambda_c \pi K$,
and $(v)~\Omega_c[{3\over2}^-] \rightarrow \Xi_c^{*} K \rightarrow \Xi_c \pi K$.
We note again that the uncertainties can be as large as $\Gamma^{+200\%}_{-67\%}$.

\begin{table}[hbt]
\begin{center}
\renewcommand{\arraystretch}{1.5}
\caption{Non-vanishing decay widths of the flavor $\mathbf{6}_F$ $P$-wave charmed baryons, in units of MeV.
The possible decay channels are:
$(k)~~\Sigma_c[{1\over2}^-] \rightarrow \Lambda_c \pi$,
$(l)~~\Sigma_c[{1\over2}^-] \rightarrow \Sigma_c \pi$,
$(m)~~\Xi_c^{\prime}[{1\over2}^-] \rightarrow \Xi_c \pi$,
$(n)~~\Xi_c^{\prime}[{1\over2}^-] \rightarrow \Lambda_c K$,
$(o)~~\Xi_c^{\prime}[{1\over2}^-] \rightarrow \Xi_c^{\prime} \pi$,
$(p)~~\Xi_c^{\prime}[{1\over2}^-] \rightarrow \Sigma_c K$,
$(q)~~\Omega_c[{1\over2}^-] \rightarrow \Xi_c K$,
$(r)~~\Omega_c[{1\over2}^-] \rightarrow \Xi_c^{\prime} K$,
$(s)~~\Sigma_c[{3\over2}^-] \rightarrow \Sigma_c^{*} \pi$,
$(t)~~\Xi_c^{\prime}[{3\over2}^-] \rightarrow \Xi_c^{*} \pi$,
$(u)~~\Xi_c^{\prime}[{3\over2}^-] \rightarrow \Sigma_c^{*} K \rightarrow \Lambda_c \pi K$,
and $(v)~~\Omega_c[{3\over2}^-] \rightarrow \Xi_c^{*} K \rightarrow \Xi_c \pi K$.
We use the two-body middle/final states to denote them in the table.
}
\begin{tabular}{c | c | c c c c | c c}
\hline\hline
~~~Baryon~~~ & ~~~Mass~~~ & ~~~$[\mathbf{6}_F, 1, 0, \rho]$~~~ & ~~~$[\mathbf{6}_F, 0, 1, \lambda]$~~~ & ~~~$[\mathbf{6}_F, 1, 1, \lambda]$~~~ & ~~~$[\mathbf{6}_F, 2, 1, \lambda]$~~~
\\ \hline\hline
$\Sigma_c[{1\over2}^-]$ & $\sim2800$
& $\left[\Sigma_c \pi\right] = 300$ & $\left[\Lambda_c \pi\right] = 200$ & $\left[\Sigma_c \pi\right] = 7.9$ & --
\\ \hline
$\Xi^\prime_c[{1\over2}^-]$ & $\sim2950$
&
$\begin{array}{c}
\left[\Xi_c^\prime \pi\right] = 140
\\
\left[\Sigma_c K\right] = 29
\end{array}$
&
$\begin{array}{c}
\left[\Xi_c \pi\right] = 230
\\
\left[\Lambda_c K\right] = 160
\end{array}$
&
$\begin{array}{c}
\left[\Xi_c^\prime \pi\right] = 3.7
\\
\left[\Sigma_c K\right] = 3.6
\end{array}$
& --
\\ \hline
$\Omega_c[{1\over2}^-]$ & $\sim3100$
& $\left[\Xi_c^\prime K\right] = 250$ & $\left[\Xi_c K\right] = 820$ & $\left[\Xi_c^\prime K\right] = 29$ & --
\\ \hline
$\Sigma_c[{3\over2}^-]$ & $\sim2800$
& $\left[\Sigma_c^* \pi\right] = 110$ & -- & $\left[\Sigma_c^* \pi\right] = 0.95$ & $\left[\Sigma_c^* \pi\right] = 0.00$
\\ \hline
$\Xi^\prime_c[{3\over2}^-]$ & $\sim2950$
&
$\begin{array}{c}
\left[\Xi_c^* \pi\right] = 50
\\
\left[\Sigma_c^* K\right] = 0.03
\end{array}$
& -- &
$\begin{array}{c}
\left[\Xi_c^* \pi\right] = 0.45
\\
\left[\Sigma_c^* K\right] = 0.00
\end{array}$
&
$\begin{array}{c}
\left[\Xi_c^* \pi\right] = 0.00
\\
\left[\Sigma_c^* K\right] = 0.00
\end{array}$
\\ \hline
$\Omega_c[{3\over2}^-]$
& $\sim3120$
& $\left[\Xi_c^* K\right] = 0.07$ & -- & $\left[\Xi_c^* K\right] = 0.00$ & $\left[\Xi_c^* K\right] = 0.00$
\\ \hline
$\Sigma_c[{5\over2}^-]$
& -- & -- & -- & -- & --
\\ \hline
$\Xi^\prime_c[{5\over2}^-]$
& -- & -- & -- & -- & --
\\ \hline
$\Omega_c[{5\over2}^-]$
& -- & -- & -- & -- & --
\\ \hline \hline
\end{tabular}
\label{tab:result2}
\end{center}
\end{table}

Our calculations are done separately for the four charmed baryon multiplets of flavor $\mathbf{6}_F$, $[\mathbf{6}_F, 1, 0, \rho]$, $[\mathbf{6}_F, 0, 1, \lambda]$, $[\mathbf{6}_F, 1, 1, \lambda]$ and $[\mathbf{6}_F, 2, 1, \lambda]$.
The situation in this case is more ambiguous that the previous case of the flavor $\mathbf{\bar 3}_F$ charmed baryons:
\begin{enumerate}

\item The $\Sigma_c(2800)$ is a good $P$-wave charmed baryon candidate of flavor $\mathbf{6}_F$. It has a large width around 70 MeV and was observed in the $\Lambda_c \pi$ decay channel. Our results suggest that it may be interpreted as a $J^P = 1/2^-$ state belonging to the $[\mathbf{6}_F, 0, 1, \lambda]$ multiplet, and it can be better interpreted as a $J^P = 1/2^-$ state containing both $[\mathbf{6}_F, 0, 1, \lambda]$ and $[\mathbf{6}_F, 1, 1, \lambda]$ components.

\item The $\Xi_c(2930)$ has a width around 36 MeV, and it was only observed by the BaBar experiment in the $\Lambda_c K$ decay channel~\cite{Aubert:2007eb}. Our results suggest that it may be interpreted as a $J^P = 1/2^-$ state containing both $[\mathbf{6}_F, 0, 1, \lambda]$ and $[\mathbf{6}_F, 1, 1, \lambda]$ components.

\item The $\Xi_c(2980)$ has a width around 20 MeV. It was observed in the $\Sigma_c(2455)K$ and $\Xi_c(2645)\pi$ decay channels, but was not seen in the $\Lambda_cK$ decay channel. Our results suggest that it may be interpreted as a $J^P = 1/2^-$ state belonging to the $[\mathbf{6}_F, 1, 1, \lambda]$ multiplet but it does not contain $[\mathbf{6}_F, 0, 1, \lambda]$ component.

\end{enumerate}
At present, the $J^P$ quantum number of some states has not been measured. They could also be the candidates of the radial excitations or $D$-wave states. More experiments are also necessary to understand them. Especially, the five excited $\Omega_c$ states recently observed by LHCb~\cite{Aaij:2017nav}, $\Omega_c(3000)$, $\Omega_c(3050)$, $\Omega_c(3066)$, $\Omega_c(3090)$, and $\Omega_c(3119)$, are very helpful to improve our understanding of the excited charmed baryons. Their widths are quite small and were all observed in the $\Xi_c K$ decay channel. We use their masses as inputs and redo the previous calculations. The results are shown in Table~\ref{tab:result3} (note that the $\Xi_c^\prime K$ threshold is 3072 MeV and the $\Xi_c \pi K$ threshold is 3103 MeV):
\begin{enumerate}

\item We may use the $[\mathbf{6}_F, 1, 1, \lambda]$ multiplet together with a tiny $[\mathbf{6}_F, 0, 1, \lambda]$ component to interpret one of these $\Omega_c$ states \big($\Omega_c(3000)$, $\Omega_c(3050)$ or $\Omega_c(3066)$\big) as a $J^P = 1/2^-$ state.

\item The $[\mathbf{6}_F, 2, 1, \lambda]$ multiplet may be used to interpret two of these $\Omega_c$ states as one $J^P = 3/2^-$ state and one $J^P = 5/2^-$ state, but we still need to study their $D$-wave decays into $\Xi_c K$ to check this possibility.

\item Two of these excited $\Omega_c$ states may be interpreted as two $2S$ states of $J^P = 1/2^+$ and $3/2^+$. See the recent reference~\cite{Agaev:2017uky} for more discussions.

\end{enumerate}

\begin{table}[hbt]
\begin{center}
\renewcommand{\arraystretch}{1.5}
\caption{Decay widths of the five excited $\Omega_c$ states recently observed by LHCb~\cite{Aaij:2017nav}, assuming they are $P$-wave charmed baryons. The results are in units of MeV.
The possible decay channels are:
$(q)~~\Omega_c[{1\over2}^-] \rightarrow \Xi_c K$,
$(r)~~\Omega_c[{1\over2}^-] \rightarrow \Xi_c^{\prime} K$,
and $(v)~~\Omega_c[{3\over2}^-] \rightarrow \Xi_c^{*} K \rightarrow \Xi_c \pi K$.
We use the two-body middle/final states to denote them in the table.
}
\begin{tabular}{c | c c | c | c c | c c}
\hline\hline
\multirow{2}{*}{Experiments} & \multicolumn{2}{c|}{$[\mathbf{6}_F, 1, 0, \rho]$} & ~~~$[\mathbf{6}_F, 0, 1, \lambda]$~~~ & \multicolumn{2}{c|}{$[\mathbf{6}_F, 1, 1, \lambda]$} & \multicolumn{2}{c}{$[\mathbf{6}_F, 2, 1, \lambda]$}
\\ \cline{2-8} & ~~~~~~${1/2^-}$~~~~~~ & ~~~~~~${3/2^-}$~~~~~~ & ~~~~~~${1/2^-}$~~~~~~ & ~~~~~~${1/2^-}$~~~~~~ & ~~~~~~${3/2^-}$~~~~~~ & ~~~~~~${3/2^-}$~~~~~~ & ~~~~~~${5/2^-}$~~~~~~
\\ \hline\hline
$\Omega_c(3000)$ & -- & -- & $\left[\Xi_c K\right] = 420$ & -- & -- & -- & --
\\ \hline
$\Omega_c(3050)$ & -- & -- & $\left[\Xi_c K\right] = 650$ & -- & -- & -- & --
\\ \hline
$\Omega_c(3066)$ & -- & -- & $\left[\Xi_c K\right] = 700$ & -- & -- & -- & --
\\ \hline
$\Omega_c(3090)$ & $\left[\Xi_c^\prime K\right] = 200$ & -- & $\left[\Xi_c K\right] = 790$ & $\left[\Xi_c^\prime K\right] = 23$ & -- & -- & --
\\ \hline
$\Omega_c(3119)$ & $\left[\Xi_c^\prime K\right] = 320$ & $\left[\Xi_c^* K\right] = 0.06$ & $\left[\Xi_c K\right] = 870$ & $\left[\Xi_c^\prime K\right] = 38$ & $\left[\Xi_c^* K\right] = 0.00$ & $\left[\Xi_c^* K\right] = 0.00$ & --
\\ \hline \hline
\end{tabular}
\label{tab:result3}
\end{center}
\end{table}

To end this work, we note that we have only investigated the $S$-wave decay properties of these excited charmed baryons in the present study, but their $D$-wave decays can also happen and contribute (although these contributions may be not large). Hence, in our following study we plan to further study their $D$-wave decay properties. We also plan to study the $S$-wave decays of the flavor $\mathbf{6}_F$ $P$-wave charmed baryons into ground-state charmed baryons accompanied by a vector meson ($\rho$ or $K^*$), which have not been done in the present work. We would like to suggest the Belle/KEK, LHCb, and J-PARC experiments to investigate the decays of these excited $\Omega_c$ states into $\Xi_c^\prime K$ to further understand them.

\section*{ACKNOWLEDGMENTS}

This project is supported by
the National Natural Science Foundation of China under Grants No. 11175073, No. 11205011, No. 11222547, No. 11375024, No. 11475015, No. 11575008, No. 11261130311, and No. 11621131001,
the 973 program,
the Ministry of Education of China (SRFDP under Grant No. 20120211110002 and the Fundamental Research Funds for the Central Universities),
the Fok Ying-Tong Education Foundation (Grant No. 131006), and
the National Program for Support of Top-notch Young Professionals.
Q.M. is supported by the Key Natural Science Research Program of Anhui Educational Committee (Grant No. KJ2016A774).
A.H. is supported in part by Grants-in-Aid for Scientific Research of JSPS, No. JP26400273(C).

\appendix

\section{Formulae of decay amplitudes and decay widths}
\label{sec:decaychannels}

The decay widths of $P$-wave charmed baryons can be evaluated based on the Lagrangians $(a$-$j)$ listed in Eqs.~(\ref{lag:swave}), (\ref{lag:ij}) and (\ref{lag:ah}):
\begin{enumerate}

\item The decay amplitude of the two-body decay $(a)~\Lambda_c^+({1/2}^-) \rightarrow \Sigma_c^+ \pi^0$ is
\begin{eqnarray}
&& \mathcal{M} \left( 0 \rightarrow 2 + 1 \right) \equiv \mathcal{M} \left( \Lambda_c^+({1/2}^-) \rightarrow \Sigma_c^{+}(1/2^+) + \pi^0 \right)
= g_{0 \rightarrow 2 + 1} {\bar u_0 u_2} \, ,
\end{eqnarray}
where $0$ denotes the initial state $\Lambda_c^+({1/2}^-)$; $1$ and $2$ denote the finial states $\pi^0$ and $\Sigma_c^{+}(1/2^+)$, respectively. This amplitude can be used to further evaluate its decay width
\begin{eqnarray}
&& \Gamma \left( 0 \rightarrow 2 + 1 \right) \equiv \Gamma \left( \Lambda_c^+({1/2}^-) \rightarrow \Sigma_c^{+}(1/2^+) + \pi^0 \right)
\\ \nonumber &=& { |\vec p_{1}| \over 8 \pi m_{0}^2}
\times g^2_{0 \rightarrow 2 + 1}
\times {1\over2}{\rm Tr}\left[ \left( p\!\!\!\slash_{0} + m_{0} \right) \left( p\!\!\!\slash_{2} + m_{2} \right) \right] \, ,
\end{eqnarray}
where we have used the following formula for the baryon field of spin 1/2
\begin{eqnarray}
\sum_{spin} u(p) \bar u(p) = \left( p\!\!\!\slash + m \right) \, .
\end{eqnarray}
The two-body decays, $(c)$, $(d)$, $(f)$, and $(k)$-$(r)$, can be similarly evaluated.

\item The decay amplitude of the two-body decay $(h)~\Xi_c({3/2}^-) \rightarrow \Xi_c^* \pi$ is
\begin{eqnarray}
\mathcal{M} \left( 0 \rightarrow 2 + 1 \right) \equiv \mathcal{M} \left( \Xi_c^0({3/2}^-) \rightarrow \Xi_c^{*+}(3/2^+) + \pi^- \right) = g_{0 \rightarrow 2 + 1} {\bar u_{0,\mu}} u_{2,\mu} \, ,
\end{eqnarray}
where $0$ denotes the initial state $\Xi_c^0({3/2}^-)$; $1$ and $2$ denote the finial states $\pi^-$ and $\Xi_c^{*+}(3/2^+)$, respectively. This amplitude can be used to further evaluate its decay width
\begin{eqnarray}
&& \Gamma \left( 0 \rightarrow 2 + 1 \right) \equiv \Gamma \left( \Xi_c^0({3/2}^-) \rightarrow \Xi_c^{*+}(3/2^+) + \pi^- \right)
\\ \nonumber &=& { |\vec p_{1}| \over 8 \pi m_{0}^2}
\times g^2_{0 \rightarrow 2 + 1} \times
{1\over4}{\rm Tr}\Big[ \left( g_{\mu^\prime\mu} - {1\over3} \gamma_{\mu^\prime} \gamma_\mu - {p_{2,\mu^\prime}\gamma_\mu - p_{2,\mu}\gamma_{\mu^\prime} \over 3m_{2}} - {2p_{2,\mu^\prime}p_{2,\mu} \over 3m_{2}^2} \right) \left( p\!\!\!\slash_{2} + m_{2} \right) \times
\\ \nonumber && ~~~~~~~~~~~~~~~~~~~~~~~~~~~~ \times \left( g_{\mu\mu^\prime} - {1\over3} \gamma_\mu \gamma_{\mu^\prime} - {p_{0,\mu}\gamma_{\mu^\prime} - p_{0,\mu^\prime}\gamma_\mu \over 3m_{0}} - {2p_{0,\mu}p_{0,\mu^\prime} \over 3m_{0}^2} \right) \left( p\!\!\!\slash_{0} + m_{0} \right) \Big] \, ,
\end{eqnarray}
where we have used the following formula for the baryon field of spin 3/2
\begin{eqnarray}
\sum_{spin} u_{\mu}(p) \bar u_{\mu^\prime}(p) = \left( g_{\mu\mu^\prime} - {1\over3} \gamma_\mu \gamma_{\mu^\prime} - {p_\mu\gamma_{\mu^\prime} - p_{\mu^\prime}\gamma_{\mu} \over 3m} - {2p_{\mu}p_{\mu^\prime} \over 3m^2} \right) \left( p\!\!\!\slash + m \right) \, .
\end{eqnarray}
The two-body decays, $(s)$ and $(t)$, can be similarly evaluated.

\item The decay amplitude of the three-body decay $(a^\prime)~ \Lambda_c^+(1/2^-) \rightarrow \Sigma_c^{++} \pi^- \to \Lambda_c^+ \pi^+ \pi^-$ is
\begin{eqnarray}
&& \mathcal{M} \left( 0 \rightarrow 4 + 1 \rightarrow 3 + 2 + 1 \right)
\equiv \mathcal{M} \left( \Lambda_c^+({1/2}^-) \rightarrow \Sigma_c^{++}(1/2^+) + \pi^- \rightarrow \Lambda_c^{+}(1/2^+) + \pi^+ + \pi^- \right)
\\ \nonumber &=& g_{0 \rightarrow 4 + 1} \times g_{4 \rightarrow 3 + 2}
\times {\bar u_{0}} \times { p\!\!\!\slash_4 + m_4 \over p_4^2 - m_4^2 + i m_4 \Gamma_4 } \times \gamma_\mu \gamma_5 \times u_3 \times p_{2,\mu} \, ,
\end{eqnarray}
where $0$ denotes the initial state $\Lambda_c^+({1/2}^-)$; 4 denotes the middle state $\Sigma_c^{++}(1/2^+)$; $1$, $2$ and $3$ denote the finial states $\pi^-$, $\pi^+$ and $\Lambda_c^{+}(1/2^+)$, respectively. This amplitude can be used to further evaluate its decay width
\begin{eqnarray}
&& \Gamma \left( 0 \rightarrow 4 + 1 \rightarrow 3 + 2 + 1 \right)
\equiv \Gamma \left( \Lambda_c^+({1/2}^-) \rightarrow \Sigma_c^{++}(1/2^+) + \pi^- \rightarrow \Lambda_c^{+}(1/2^+) + \pi^+ + \pi^- \right)
\\ \nonumber &=& {1 \over (2\pi)^3} \times {1 \over 32 m_0^3} \times g^2_{0 \rightarrow 4 + 1} \times g^2_{4 \rightarrow 3 + 2} \times \int d m_{12} d m_{23} \times
\\ \nonumber && ~~~~~ \times {1\over2}{\rm Tr}\Big[ \left( p\!\!\!\slash_3 + m_3 \right) \gamma_{\mu^\prime}\gamma_5 \left( p\!\!\!\slash_4 + m_4 \right) \left( p\!\!\!\slash_0 + m_0 \right) \left( p\!\!\!\slash_4 + m_4 \right) \gamma_{\mu}\gamma_5 \Big]
\times {1 \over |p_4^2 - m_4^2 + i m_4 \Gamma_4|^2 } \times p_{2,\mu} p_{2,\mu^\prime} \, ,
\end{eqnarray}
where we have used the standard Dalitz integration~\cite{pdg}.

\item The decay amplitude of the three-body decay $(b)~\Lambda_c({3/2}^-) \rightarrow \Sigma_c^{*} \pi \rightarrow \Lambda_c \pi \pi$ is
\begin{eqnarray}
&& \mathcal{M} \left( 0 \rightarrow 4 + 1 \rightarrow 3 + 2 + 1 \right)
\equiv \mathcal{M} \left( \Lambda_c^+({3/2}^-) \rightarrow \Sigma_c^{*++}(3/2^+) + \pi^- \rightarrow \Lambda_c^{+}(1/2^+) + \pi^+ + \pi^- \right)
\\ \nonumber &=& g_{0 \rightarrow 4 + 1} \times g_{4 \rightarrow 3 + 2}
\\ \nonumber && ~~~~~ \times {\bar u_{0,\mu}} \times \left( g_{\mu\nu} - {1\over3} \gamma_\mu \gamma_\nu - {p_{4,\mu}\gamma_\nu - p_{4,\nu}\gamma_\mu \over 3m_4} - {2p_{4,\mu}p_{4,\nu} \over 3m_4^2} \right) \times { p\!\!\!\slash_4 + m_4 \over p_4^2 - m_4^2 + i m_4 \Gamma_4 } \times u_3 \times p_{2,\nu} \, ,
\end{eqnarray}
where $0$ denotes the initial state $\Lambda_c^+({3/2}^-)$; 4 denotes the middle state $\Sigma_c^{*++}(3/2^+)$; $1$, $2$ and $3$ denote the finial states $\pi^-$, $\pi^+$ and $\Lambda_c^{+}(1/2^+)$, respectively. This amplitude can be used to further evaluate its decay width
\begin{eqnarray}
&& \Gamma \left( 0 \rightarrow 4 + 1 \rightarrow 3 + 2 + 1 \right)
\equiv \Gamma \left( \Lambda_c^+({3/2}^-) \rightarrow \Sigma_c^{*++}(3/2^+) + \pi^- \rightarrow \Lambda_c^{+}(1/2^+) + \pi^+ + \pi^- \right)
\\ \nonumber &=& {1 \over (2\pi)^3} \times {1 \over 32 m_0^3} \times g^2_{0 \rightarrow 4 + 1} \times g^2_{4 \rightarrow 3 + 2} \times \int d m_{12} d m_{23} \times
\\ \nonumber && ~~~~~ \times {1\over4}{\rm Tr}\Big[ \left( p\!\!\!\slash_3 + m_3 \right) \times
\left( g_{\nu^\prime\mu^\prime} - {1\over3} \gamma_{\nu^\prime} \gamma_{\mu^\prime} - {p_{4,{\nu^\prime}}\gamma_{\mu^\prime} - p_{4,{\mu^\prime}}\gamma_{\nu^\prime} \over 3m_4} - {2p_{4,\nu^\prime}p_{4,\mu^\prime} \over 3m_4^2} \right) \left( p\!\!\!\slash_4 + m_4 \right)
\\ \nonumber && ~~~~~~~~~~~~~~~~~~~~~~~~~~~~ \times
\left( g_{\mu^\prime\mu} - {1\over3} \gamma_{\mu^\prime} \gamma_\mu - {p_{0,\mu^\prime}\gamma_\mu - p_{0,\mu}\gamma_{\mu^\prime} \over 3m_0} - {2p_{0,\mu^\prime}p_{0,\mu} \over 3m_0^2} \right) \left( p\!\!\!\slash_0 + m_0 \right)
\\ \nonumber && ~~~~~~~~~~~~~~~~~~~~~~~~~~~~ \times
\left( g_{\mu\nu} - {1\over3} \gamma_\mu \gamma_\nu - {p_{4,\mu}\gamma_\nu - p_{4,\nu}\gamma_\mu \over 3m_4} - {2p_{4,\mu}p_{4,\nu} \over 3m_4^2} \right) \left( p\!\!\!\slash_4 + m_4 \right) \Big]
\\ \nonumber && ~~~~~ \times {1 \over |p_4^2 - m_4^2 + i m_4 \Gamma_4|^2 } \times p_{2,\nu} p_{2,\nu^\prime} \, .
\end{eqnarray}
The three-body decays, $(u)$ and $(v)$, can be similarly evaluated.

\item The decay amplitude of the three-body decay $(e)~\Xi_c({1/2}^-) \rightarrow \Xi_c \rho \rightarrow \Xi_c \pi \pi$ is
\begin{eqnarray}
&& \mathcal{M} \left( 0 \rightarrow 3 + 4 \rightarrow 3 + 2 + 1 \right)
\equiv \mathcal{M} \left( \Xi_c^0({1/2}^-) \rightarrow \Xi_c^{+}(1/2^+) + \rho^- \rightarrow \Xi_c^{+}(1/2^+) + \pi^0 + \pi^- \right)
\\ \nonumber &=& g_{0 \rightarrow 3 + 4} \times g_{4 \rightarrow 2 + 1}
\times \bar u_{0} \gamma_\mu \gamma_5 u_3 \times \left( g_{\mu\nu} - { p_{4,\mu} p_{4,\nu} \over m_4^2} \right) \times { 1 \over p_4^2 - m_4^2 + i m_4 \Gamma_4 } \times \left( p_{1,\nu} + p_{2,\nu} \right) \, ,
\end{eqnarray}
where $0$ denotes the initial state $\Xi_c^0({1/2}^-)$; 4 denotes the middle state $\rho^{-}$; $1$, $2$ and $3$ denote the finial states $\pi^-$, $\pi^0$ and $\Xi_c^{+}(1/2^+)$, respectively. This amplitude can be used to further evaluate its decay width
\begin{eqnarray}
&& \Gamma \left( 0 \rightarrow 3 + 4 \rightarrow 3 + 2 + 1 \right)
\equiv \Gamma \left( \Xi_c^0({1/2}^-) \rightarrow \Xi_c^{+}(1/2^+) + \rho^- \rightarrow \Xi_c^{+}(1/2^+) + \pi^0 + \pi^- \right)
\\ \nonumber &=& {1 \over (2\pi)^3} \times {1 \over 32 m_0^3} \times g^2_{0 \rightarrow 3 + 4} \times g^2_{4 \rightarrow 2 + 1} \times \int d m_{12} d m_{23} \times
\\ \nonumber && ~~~~~ \times {1\over2}{\rm Tr}\Big[ \left( p\!\!\!\slash_3 + m_3 \right) \gamma_{\mu^\prime} \gamma_5 \left( p\!\!\!\slash_0 + m_0 \right) \gamma_{\mu} \gamma_5 \Big]
\\ \nonumber && ~~~~~ \times \left( g_{\mu\nu} - { p_{4,\mu} p_{4,\nu} \over m_4^2} \right)\left( g_{\mu^\prime\nu^\prime} - { p_{4,\mu^\prime} p_{4,\nu^\prime} \over m_4^2} \right)
\times {1 \over |p_4^2 - m_4^2 + i m_4 \Gamma_4|^2 } \times \left( p_{1,\nu} + p_{2,\nu} \right) \left( p_{1,\nu^\prime} + p_{2,\nu^\prime} \right) \, .
\end{eqnarray}

\item The decay amplitude of the three-body decay $(g)~\Xi_c({3/2}^-) \rightarrow \Xi_c \rho \rightarrow \Xi_c \pi \pi$ is
\begin{eqnarray}
&& \mathcal{M} \left( 0 \rightarrow 3 + 4 \rightarrow 3 + 2 + 1 \right)
\equiv \mathcal{M} \left( \Xi_c^0({3/2}^-) \rightarrow \Xi_c^{+}(1/2^+) + \rho^- \rightarrow \Xi_c^{+}(1/2^+) + \pi^0 + \pi^- \right)
\\ \nonumber &=& g_{0 \rightarrow 3 + 4} \times g_{4 \rightarrow 2 + 1}
\times \bar u_{0,\mu} u_3 \times \left( g_{\mu\nu} - { p_{4,\mu} p_{4,\nu} \over m_4^2} \right) \times { 1 \over p_4^2 - m_4^2 + i m_4 \Gamma_4 } \times \left( p_{1,\nu} + p_{2,\nu} \right) \, ,
\end{eqnarray}
where $0$ denotes the initial state $\Xi_c^0({3/2}^-)$; 4 denotes the middle state $\rho^{-}$; $1$, $2$ and $3$ denote the finial states $\pi^-$, $\pi^0$ and $\Xi_c^{+}(1/2^+)$, respectively. This amplitude can be used to further evaluate its decay width
\begin{eqnarray}
&& \Gamma \left( 0 \rightarrow 3 + 4 \rightarrow 3 + 2 + 1 \right)
\equiv \Gamma \left( \Xi_c^0({3/2}^-) \rightarrow \Xi_c^{+}(1/2^+) + \rho^- \rightarrow \Xi_c^{+}(1/2^+) + \pi^0 + \pi^- \right)
\\ \nonumber &=& {1 \over (2\pi)^3} \times {1 \over 32 m_0^3} \times g^2_{0 \rightarrow 3 + 4} \times g^2_{4 \rightarrow 2 + 1} \times \int d m_{12} d m_{23} \times
\\ \nonumber && ~~~~~ \times {1\over4}{\rm Tr}\Big[ \left( p\!\!\!\slash_3 + m_3 \right) \left( g_{\mu^\prime\mu} - {1\over3} \gamma_{\mu^\prime} \gamma_\mu - {p_{0,\mu^\prime}\gamma_\mu - p_{0,\mu}\gamma_{\mu^\prime} \over 3m_0} - {2p_{0,\mu^\prime}p_{0,\mu} \over 3m_0^2} \right) \left( p\!\!\!\slash_0 + m_0 \right) \Big]
\\ \nonumber && ~~~~~ \times \left( g_{\mu\nu} - { p_{4,\mu} p_{4,\nu} \over m_4^2} \right)\left( g_{\mu^\prime\nu^\prime} - { p_{4,\mu^\prime} p_{4,\nu^\prime} \over m_4^2} \right)
\times {1 \over |p_4^2 - m_4^2 + i m_4 \Gamma_4|^2 } \times \left( p_{1,\nu} + p_{2,\nu} \right) \left( p_{1,\nu^\prime} + p_{2,\nu^\prime} \right) \, .
\end{eqnarray}

\item The decay amplitude of the three-body decay $(i)~\Lambda_c({5/2}^-) \rightarrow \Sigma_c^* \rho \rightarrow \Sigma_c^* \pi \pi$ is
\begin{eqnarray}
&& \mathcal{M} \left( 0 \rightarrow 3 + 4 \rightarrow 3 + 2 + 1 \right)
\equiv \mathcal{M} \left( \Lambda_c^+({5/2}^-) \rightarrow \Sigma_c^{*++}(3/2^+) + \rho^- \rightarrow \Sigma_c^{*++}(3/2^+) + \pi^0 + \pi^- \right)
\\ \nonumber &=& g_{0 \rightarrow 3 + 4} \times g_{4 \rightarrow 2 + 1}
\times \bar u_{0,\mu\rho} u_{3,\rho} \times \left( g_{\mu\nu} - { p_{4,\mu} p_{4,\nu} \over m_4^2} \right) \times { 1 \over p_4^2 - m_4^2 + i m_4 \Gamma_4 } \times \left( p_{1,\nu} + p_{2,\nu} \right) \, ,
\end{eqnarray}
where $0$ denotes the initial state $\Lambda_c^+({5/2}^-)$; 4 denotes the middle state $\rho^{-}$; $1$, $2$ and $3$ denote the finial states $\pi^-$, $\pi^0$ and $\Sigma_c^{*++}(3/2^+)$, respectively. This amplitude can be used to further evaluate its decay width
\begin{eqnarray}
&& \Gamma \left( 0 \rightarrow 3 + 4 \rightarrow 3 + 2 + 1 \right)
\equiv \Gamma \left( \Lambda_c^+({5/2}^-) \rightarrow \Sigma_c^{*++}(3/2^+) + \rho^- \rightarrow \Sigma_c^{*++}(3/2^+) + \pi^0 + \pi^- \right)
\\ \nonumber &=& {1 \over (2\pi)^3} \times {1 \over 32 m_0^3} \times g^2_{0 \rightarrow 3 + 4} \times g^2_{4 \rightarrow 2 + 1} \times \int d m_{12} d m_{23} \times
\\ \nonumber && ~~~~~ \times {1\over6}{\rm Tr}\Big[ \left( g_{\rho^\prime\rho} - {1\over3} \gamma_{\rho^\prime} \gamma_\rho - {p_{3,\rho^\prime}\gamma_\rho - p_{3,\rho}\gamma_{\rho^\prime} \over 3m_3} - {2p_{3,\rho^\prime}p_{3,\rho} \over 3m_3^2} \right) \left( p\!\!\!\slash_3 + m_3 \right)
\\ \nonumber && ~~~~~~~~~~~~~~~ \times \left( {1\over2} g_{\mu\mu^\prime} g_{\rho\rho^\prime} + {1\over2} g_{\mu\rho^\prime} g_{\rho\mu^\prime} -  {1\over3} g_{\mu\rho} g_{\mu^\prime\rho^\prime} \right) \left( p\!\!\!\slash_0 + m_0 \right) \Big]
\\ \nonumber && ~~~~~ \times \left( g_{\mu\nu} - { p_{4,\mu} p_{4,\nu} \over m_4^2} \right)\left( g_{\mu^\prime\nu^\prime} - { p_{4,\mu^\prime} p_{4,\nu^\prime} \over m_4^2} \right)
\times {1 \over |p_4^2 - m_4^2 + i m_4 \Gamma_4|^2 } \times \left( p_{1,\nu} + p_{2,\nu} \right) \left( p_{1,\nu^\prime} + p_{2,\nu^\prime} \right) \, ,
\end{eqnarray}
where we have simply used the following formula for the baryon field of spin 5/2
\begin{eqnarray}
\sum_{spin} u_{\mu\nu}(p) \bar u_{\mu^\prime\nu^\prime}(p) = \left( {1\over2} g_{\mu\mu^\prime} g_{\nu\nu^\prime} + {1\over2} g_{\mu\nu^\prime} g_{\nu\mu^\prime} -  {1\over3} g_{\mu\nu} g_{\mu^\prime\nu^\prime} \right) \left( p\!\!\!\slash + m \right) \, .
\end{eqnarray}
The three-body decay, $(j)~\Xi_c({5/2}^-) \rightarrow \Xi_c^* \rho \rightarrow \Xi_c^* \pi \pi$, can be similarly evaluated.

\end{enumerate}

\section{Light-cone distribution amplitudes of the K meson}
\label{sec:wavefunction}

In this appendix we list the light-cone distribution amplitudes of the $K$ meson as examples. They are taken from Ref.~\cite{Ball:2006wn}, and we refer interested readers to read Refs.~\cite{Ball:1998je,Ball:2006wn,Ball:2004rg,Ball:1998kk,Ball:1998sk,Ball:1998ff,Ball:2007rt,Ball:2007zt} for details. The light-cone distribution amplitudes of the $K$ meson used in the present study are:
\begin{eqnarray}
\langle 0 | \bar q(z)\gamma_\mu\gamma_5 s(-z) | K(q) \rangle
&=& i f_K q_\mu \int_0^1 du \, e^{i (2u-1) q \cdot z} \left(\phi_{2;K}(u) + \frac{1}{4}\,z^2 \phi_{4;K}(u)\right)
\\ \nonumber && ~~~~~~~~~~ + \frac{i}{2}\, f_K\, \frac{1}{q \cdot z}\, z_\mu  \int_0^1 du \, e^{i (2u-1) q \cdot z} \psi_{4;K}(u) \, ,
\\ \langle 0 | \bar q(z) i\gamma_5 s(-z) | K(q) \rangle
&=& \frac{f_K m_K^2}{m_s+m_q}\, \int_0^1 du \, e^{i(2u-1) q \cdot z} \, \phi^{p}_{3;K}(u) \, ,
\\ \langle 0 | \bar q(z) \sigma_{\alpha\beta}\gamma_5 s(-z) | K(q) \rangle &=& -\frac{i}{3}\, \frac{f_K m_K^2}{m_s+m_q}  (q_\alpha z_\beta- q_\beta z_\alpha) \int_0^1 du \, e^{i(2u-1) q \cdot z}\,\phi^{\sigma}_{3;K}(u) \, ,
\\ \langle 0 | \bar q(z)\gamma_\mu\gamma_5 gG_{\alpha\beta}(vz)s(-z)|K(q)\rangle
&=& q_\mu (q_\alpha z_\beta - q_\beta z_\alpha)\, \frac{1}{q \cdot z}\, f_{K} \Phi_{4;K}(v,q \cdot z)
\\ \nonumber && ~~~~~~~~~~ + (q_\beta g_{\alpha\mu}^\perp - q_\alpha g_{\beta\mu}^\perp) f_{K} \Psi_{4;K}(v,q \cdot z) \, ,
\\ \langle 0 | \bar q(z)\gamma_\mu i g\widetilde{G}_{\alpha\beta}(vz)s(-z)| K(q)\rangle\
&=& q_\mu (q_\alpha z_\beta - q_\beta z_\alpha)\, \frac{1}{q \cdot z}\, f_K \widetilde\Phi_{4;K}(v,q \cdot z)
\\ \nonumber && ~~~~~~~~~~ + (q_\beta g_{\alpha\mu}^\perp - q_\alpha g_{\beta\mu}^\perp) f_{K} \widetilde\Psi_{4;K}(v,q \cdot z) \, ,
\\ \langle 0 | \bar q(z) \sigma_{\mu\nu}\gamma_5 gG_{\alpha\beta}(vz) s(-z)| K(q)\rangle
&=& i\,f_{3K} \left(q_\alpha q_\mu g_{\nu\beta}^\perp - q_\alpha q_\nu g_{\mu\beta}^\perp - (\alpha\leftrightarrow\beta) \right) \times
\\ \nonumber && ~~~~~~~~~~ \times \int {\cal D}\underline{\alpha} \, e^{-iq \cdot z(\alpha_2 -\alpha_1 + v\alpha_3)} {\Phi}_{3;K}(\alpha_1,\alpha_2,\alpha_3) \, .
\end{eqnarray}
where $\widetilde{G}_{\mu\nu} = \frac{1}{2}\epsilon_{\mu\nu \rho\sigma} G^{\rho\sigma}$.

\section{Other sum rules}
\label{sec:othersumrule}

In this appendix we show the sum rules for other currents with different quark contents.

\subsection{$[\mathbf{\bar 3}_F, 0, 1, \rho]$}
\label{sec:301rho}

The sum rule for $\Lambda_c^+({1\over2}^-)$ belonging to $[\mathbf{\bar 3}_F, 0, 1, \rho]$ is
\begin{eqnarray}
&& G_{\Lambda_c^+[{1\over2}^-] \rightarrow \Sigma_c^{++}\pi^-} (\omega, \omega^\prime)
= { g_{\Lambda_c^+[{1\over2}^-] \rightarrow \Sigma_c^{++}\pi^-} f_{\Lambda_c^+[{1\over2}^-]} f_{\Sigma_c^{++}} \over (\bar \Lambda_{\Lambda_c^+[{1\over2}^-]} - \omega^\prime) (\bar \Lambda_{\Sigma_c^{++}} - \omega)}
= 0 \, .
\end{eqnarray}

The sum rules for $\Xi_c^0({1\over2}^-)$ belonging to $[\mathbf{\bar 3}_F, 0, 1, \rho]$ are
\begin{eqnarray}
&& G_{\Xi_c^0[{1\over2}^-] \rightarrow \Xi_c^{\prime+}\pi^-} (\omega, \omega^\prime)
= { g_{\Xi_c^0[{1\over2}^-] \rightarrow \Xi_c^{\prime+}\pi^-} f_{\Xi_c^0[{1\over2}^-]} f_{\Xi_c^{\prime+}} \over (\bar \Lambda_{\Xi_c^0[{1\over2}^-]} - \omega^\prime) (\bar \Lambda_{\Xi_c^{\prime+}} - \omega)}
= 0 \, ,
\\
&& G_{\Xi_c^0[{1\over2}^-] \rightarrow \Xi_c^{+}\pi^-} (\omega, \omega^\prime)
= { g_{\Xi_c^0[{1\over2}^-] \rightarrow \Xi_c^{+}\pi^-} f_{\Xi_c^0[{1\over2}^-]} f_{\Xi_c^{+}} \over (\bar \Lambda_{\Xi_c^0[{1\over2}^-]} - \omega^\prime) (\bar \Lambda_{\Xi_c^{+}} - \omega)}
\\ \nonumber &=& \int_0^\infty dt \int_0^1 du e^{i (1-u) \omega^\prime t} e^{i u \omega t} \times 4 \times \Big (
\\ \nonumber && - \frac{3 f_\pi m_\pi^2}{4 \pi^2 t^4 (m_u + m_d)} \phi_{3;\pi}^p(u) + \frac{i f_\pi m_\pi^2 v \cdot q}{8 \pi^2 t^3 ( m_u + m_d )} \phi_{3;\pi}^\sigma(u) - \frac{i f_\pi}{16 t v \cdot q} \langle \bar s s \rangle \psi_{4;\pi}(u) - \frac{i f_\pi t}{256 v\cdot q} \langle g_s \bar s \sigma G s\rangle \psi_{4;\pi}(u)
\\ \nonumber && - \frac{3 i f_\pi}{16 \pi^2 t^3 v\cdot q} m_s \psi_{4;\pi}(u) - \frac{ f_\pi m_\pi^2}{32 (m_u + m_d)} m_s \langle \bar s s \rangle \phi_{3;\pi}^p(u) + \frac{i f_\pi m_\pi^2 t v\cdot q}{192 ( m_u + m_d)} m_s \langle \bar s s \rangle \phi_{3;\pi}^\sigma(u) \Big ) \, ,
\\
&& G_{\Xi_c^0[{1\over2}^-] \rightarrow \Lambda_c^{+} K^-} (\omega, \omega^\prime)
= { g_{\Xi_c^0[{1\over2}^-] \rightarrow \Lambda_c^{+} K^-} f_{\Xi_c^0[{1\over2}^-]} f_{\Lambda_c^{+}} \over (\bar \Lambda_{\Xi_c^0[{1\over2}^-]} - \omega^\prime) (\bar \Lambda_{\Lambda_c^{+}} - \omega)}
\\ \nonumber &=& \int_0^\infty dt \int_0^1 du e^{i (1-u) \omega^\prime t} e^{i u \omega t} \times 4 \times \Big (
\\ \nonumber && - \frac{3 f_K m_K^2}{4 \pi^2 t^4 (m_u + m_s)} \phi_{3;K}^p(u) + \frac{i f_K m_K^2 v \cdot q}{8 \pi^2 t^3 ( m_u + m_s )} \phi_{3;K}^\sigma(u) - \frac{i f_K}{16 t v \cdot q} \langle \bar q q \rangle \psi_{4;K}(u) - \frac{i f_K t}{256 v\cdot q} \langle g_s \bar q \sigma G q\rangle \psi_{4;K}(u) \Big ) \, ,
\\
&& G_{\Xi_c^0[{1\over2}^-] \rightarrow \Xi_c^{+}\rho^-} (\omega, \omega^\prime)
= { g_{\Xi_c^0[{1\over2}^-] \rightarrow \Xi_c^{+}\rho^-} f_{\Xi_c^0[{1\over2}^-]} f_{\Xi_c^{+}} \over (\bar \Lambda_{\Xi_c^0[{1\over2}^-]} - \omega^\prime) (\bar \Lambda_{\Xi_c^{+}} - \omega)}
= 0 \, .
\end{eqnarray}

\subsection{$[\mathbf{\bar 3}_F, 1, 1, \rho]$}
\label{sec:311rho}

The sum rule for $\Lambda_c^+({1\over2}^-)$ belonging to $[\mathbf{\bar 3}_F, 1, 1, \rho]$ is
\begin{eqnarray}
&& G_{\Lambda_c^+[{1\over2}^-] \rightarrow \Sigma_c^{++}\pi^-} (\omega, \omega^\prime)
= { g_{\Lambda_c^+[{1\over2}^-] \rightarrow \Sigma_c^{++}\pi^-} f_{\Lambda_c^+[{1\over2}^-]} f_{\Sigma_c^{++}} \over (\bar \Lambda_{\Lambda_c^+[{1\over2}^-]} - \omega^\prime) (\bar \Lambda_{\Sigma_c^{++}} - \omega)}
\\ \nonumber &=& \int_0^\infty dt \int_0^1 du e^{i (1-u) \omega^\prime t} e^{i u \omega t} \times 8 \times \Big (
\\ \nonumber && \frac{3 i f_\pi v \cdot q }{2 \pi^2 t^4} \phi_{2;\pi}(u) + \frac{3 i f_\pi v \cdot q}{32 \pi^2 t^2} \phi_{4;\pi}(u) + \frac{3 i f_\pi}{2 \pi^2 t^4 v \cdot q} \psi_{4;\pi}(u)
+ \frac{i f_\pi m_\pi^2 v \cdot q}{24 (m_u + m_d)} {\langle \bar q q \rangle} \phi^\sigma_{3;\pi}(u)
\\ \nonumber && + \frac{i f_\pi m_\pi^2 t^2 v \cdot q}{384 (m_u + m_d)} \langle g_s \bar q \sigma G q\rangle \phi_{3;\pi}^\sigma(u) \Big )
\\ \nonumber &+& \int_0^\infty dt \int_0^1 du \int \mathcal{D}\underline{\alpha} e^{i \omega^\prime t (\alpha_2 + u\alpha_3)} e^{i \omega t(1-\alpha_2-u\alpha_3)} \times 8 \times \Big (
\\ \nonumber && \frac{3 i f_\pi v\cdot q}{8 \pi^2 t^2} \Phi_{4;\pi}(\underline{\alpha}) - \frac{i f_\pi v\cdot q}{4 \pi^2 t^2} \Psi_{4;\pi}(\underline{\alpha}) + \frac{i f_\pi v\cdot q}{8 \pi^2 t^2} \widetilde \Phi_{4;\pi}(\underline{\alpha}) + \frac{i f_\pi v\cdot q}{4 \pi^2 t^2} \widetilde \Psi_{4;\pi}(\underline{\alpha})
+ \frac{i f_\pi u v \cdot q }{4 \pi^2 t^2} \Phi_{4;\pi}(\underline{\alpha}) + \frac{i f_\pi u v\cdot q}{2 \pi^2 t^2} \Psi_{4;\pi}(\underline{\alpha}) \Big ) \, .
\end{eqnarray}

One of the sum rules for $\Xi_c^0({1\over2}^-)$ belonging to $[\mathbf{\bar 3}_F, 1, 1, \rho]$ has been given in Eq.~(\ref{eq:sumrule}), and the others are
\begin{eqnarray}
&& G_{\Xi_c^0[{1\over2}^-] \rightarrow \Xi_c^{+}\pi^-} (\omega, \omega^\prime)
= { g_{\Xi_c^0[{1\over2}^-] \rightarrow \Xi_c^{+}\pi^-} f_{\Xi_c^0[{1\over2}^-]} f_{\Xi_c^{+}} \over (\bar \Lambda_{\Xi_c^0[{1\over2}^-]} - \omega^\prime) (\bar \Lambda_{\Xi_c^{+}} - \omega)}
= 0 \, ,
\\
&& G_{\Xi_c^0[{1\over2}^-] \rightarrow \Lambda_c^{+} K^-} (\omega, \omega^\prime)
= { g_{\Xi_c^0[{1\over2}^-] \rightarrow \Lambda_c^{+} K^-} f_{\Xi_c^0[{1\over2}^-]} f_{\Lambda_c^{+}} \over (\bar \Lambda_{\Xi_c^0[{1\over2}^-]} - \omega^\prime) (\bar \Lambda_{\Lambda_c^{+}} - \omega)}
= 0 \, ,
\\
&& G_{\Xi_c^0[{1\over2}^-] \rightarrow \Xi_c^{+}\rho^-} (\omega, \omega^\prime)
= { g_{\Xi_c^0[{1\over2}^-] \rightarrow \Xi_c^{+}\rho^-} f_{\Xi_c^0[{1\over2}^-]} f_{\Xi_c^{+}} \over (\bar \Lambda_{\Xi_c^0[{1\over2}^-]} - \omega^\prime) (\bar \Lambda_{\Xi_c^{+}} - \omega)}
\\ \nonumber &=& \int_0^\infty dt \int_0^1 du e^{i (1-u) \omega^\prime t} e^{i u \omega t} \times 4 \times \Big (
\\ \nonumber && - \frac{ f_{\rho}^\perp v\cdot q}{2 \pi^2 t^4} \phi_{2;\rho}^\perp(u)
+ \frac{ f_{\rho}^\perp m_\rho^2 }{2 \pi^2 t^4 v\cdot q} \phi_{2;\rho}^\perp(u)
- \frac{ f_{\rho}^\perp m_\rho^2 }{2 \pi^2 t^4 v\cdot q} \psi_{4;\rho}^\perp(u)
- \frac{f_{\rho}^\perp m_{\rho}^2 v\cdot q}{32 \pi^2 t^2} \phi_{4;\rho}^\perp(u)
+ \frac{f_{\rho}^\parallel m_{\rho} v\cdot q}{48} \langle \bar s s \rangle \psi_{3;\rho}^\perp(u)
\\ \nonumber &&
+ \frac{f_{\rho}^\parallel m_{\rho} t^2 v\cdot q}{768} \langle g_s \bar s \sigma G s \rangle \psi_{3;\rho}^\perp(u)
+ \frac{ f_{\rho}^\parallel m_{\rho} v\cdot q}{16 \pi^2 t^2} m_s \psi_{3;\rho}^\perp(u)
- \frac{ f_{\rho}^\perp v\cdot q}{48} m_s \langle \bar s s \rangle \phi_{2;\rho}^\perp(u)
+ \frac{ f_{\rho}^\perp m_\rho^2}{48 v\cdot q} m_s \langle \bar s s \rangle \phi_{2;\rho}^\perp(u)
\\ \nonumber &&
- \frac{ f_{\rho}^\perp m_\rho^2}{48 v\cdot q} m_s \langle \bar s s \rangle \psi_{4;\rho}^\perp(u)
- \frac{ f_{\rho}^\perp m_{\rho}^2 t^2 v\cdot q}{768} m_s \langle \bar s s \rangle \phi_{4;\rho}^\perp(u)
 \Big )
\\ \nonumber &+& \int_0^\infty dt \int_0^1 du \int \mathcal{D}\underline{\alpha} e^{i \omega^\prime t (\alpha_2 + u\alpha_3)} e^{i \omega t(1-\alpha_2-u\alpha_3)} \times 4 \times \Big (
\\ \nonumber && - \frac{ f_{\rho}^\perp m_{\rho}^2 v \cdot q}{8 \pi^2 t^2} \Psi_{4;\rho}^\perp(\underline{\alpha}) + \frac{ f_{\rho}^\perp m_{\rho}^2 v \cdot q}{8 \pi^2 t^2} \widetilde \Psi_{4;\rho}^\perp(\underline{\alpha})
- \frac{ f_{\rho}^\perp m_{\rho}^2 u v \cdot q}{8 \pi^2 t^2} \Phi_{4;\rho}^{\perp1}(\underline{\alpha}) + \frac{ f_{\rho}^\perp m_{\rho}^2 u v \cdot q}{8 \pi^2 t^2} \Phi_{4;\rho}^{\perp2}(\underline{\alpha})
- \frac{ f_{\rho}^\perp m_{\rho}^2 u v \cdot q}{4 \pi^2 t^2} \widetilde \Psi_{4;\rho}^{\perp}(\underline{\alpha})
 \Big ) \, .
\end{eqnarray}

The sum rule for $\Lambda_c^+[{3\over2}^-]$ belonging to $[\mathbf{\bar 3}_F, 1, 1, \rho]$ is
\begin{eqnarray}
&& G_{\Lambda_c^+[{3\over2}^-] \rightarrow \Sigma_c^{*++}\pi^-} (\omega, \omega^\prime)
= { g_{\Lambda_c^+[{3\over2}^-] \rightarrow \Sigma_c^{*++}\pi^-} f_{\Lambda_c^+[{3\over2}^-]} f_{\Sigma_c^{*++}} \over (\bar \Lambda_{\Lambda_c^+[{3\over2}^-]} - \omega^\prime) (\bar \Lambda_{\Sigma_c^{*++}} - \omega)}
\\ \nonumber &=& \int_0^\infty dt \int_0^1 du e^{i (1-u) \omega^\prime t} e^{i u \omega t} \times 8 \times \Big (
\\ \nonumber && - \frac{f_\pi v \cdot q }{3 \pi^2 t^4} \phi_{2;\pi}(u) - \frac{f_\pi v \cdot q}{48 \pi^2 t^2} \phi_{4;\pi}(u) - \frac{f_\pi}{3 \pi^2 t^4 v \cdot q} \psi_{4;\pi}(u) - \frac{f_\pi m_\pi^2 v \cdot q}{108 (m_u + m_d)} {\langle \bar q q \rangle} \phi^\sigma_{3;\pi}(u)
\\ \nonumber && - \frac{f_\pi m_\pi^2 t^2 v \cdot q}{1728 (m_u + m_d)} \langle g_s \bar q \sigma G q\rangle \phi_{3;\pi}^\sigma(u) \Big )
\\ \nonumber &+& \int_0^\infty dt \int_0^1 du \int \mathcal{D}\underline{\alpha} e^{i \omega^\prime t (\alpha_2 + u\alpha_3)} e^{i \omega t(1-\alpha_2-u\alpha_3)} \times 8 \times \Big (
\\ \nonumber && - \frac{f_\pi v\cdot q}{24 \pi^2 t^2} \Phi_{4;\pi}(\underline{\alpha}) + \frac{5 f_\pi v\cdot q}{72 \pi^2 t^2} \Psi_{4;\pi}(\underline{\alpha}) - \frac{f_\pi v\cdot q}{72 \pi^2 t^2} \widetilde \Phi_{4;\pi}(\underline{\alpha}) - \frac{5 f_\pi v\cdot q}{72 \pi^2 t^2} \widetilde \Psi_{4;\pi}(\underline{\alpha})
\\ \nonumber && - \frac{f_\pi u v \cdot q }{36 \pi^2 t^2} \Phi_{4;\pi}(\underline{\alpha}) - \frac{7 f_\pi u v\cdot q}{72 \pi^2 t^2} \Psi_{4;\pi}(\underline{\alpha}) + \frac{f_\pi u v \cdot q }{12 \pi^2 t^2} \widetilde \Phi_{4;\pi}(\underline{\alpha}) + \frac{f_\pi u v\cdot q}{24 \pi^2 t^2} \widetilde \Psi_{4;\pi}(\underline{\alpha}) \Big ) \, .
\end{eqnarray}

The sum rules for $\Xi_c^0({3\over2}^-)$ belonging to $[\mathbf{\bar 3}_F, 1, 1, \rho]$ are
\begin{eqnarray}
&& G_{\Xi_c^0[{3\over2}^-] \rightarrow \Xi_c^{*+}\pi^-} (\omega, \omega^\prime)
= { g_{\Xi_c^0[{3\over2}^-] \rightarrow \Xi_c^{*+}\pi^-} f_{\Xi_c^0[{3\over2}^-]} f_{\Xi_c^{*+}} \over (\bar \Lambda_{\Xi_c^0[{3\over2}^-]} - \omega^\prime) (\bar \Lambda_{\Xi_c^{*+}} - \omega)}
\\ \nonumber &=& \int_0^\infty dt \int_0^1 du e^{i (1-u) \omega^\prime t} e^{i u \omega t} \times 4 \times \Big (
\\ \nonumber && - \frac{f_\pi v \cdot q }{3 \pi^2 t^4} \phi_{2;\pi}(u) - \frac{f_\pi v \cdot q}{48 \pi^2 t^2} \phi_{4;\pi}(u) - \frac{f_\pi}{3 \pi^2 t^4 v \cdot q} \psi_{4;\pi}(u) - \frac{f_\pi m_\pi^2 v \cdot q}{108 (m_u + m_d)} {\langle \bar s s \rangle} \phi^\sigma_{3;\pi}(u)
\\ \nonumber &&
- \frac{f_\pi m_\pi^2 t^2 v \cdot q}{1728 (m_u + m_d)} \langle g_s \bar s \sigma G s\rangle \phi_{3;\pi}^\sigma(u)
- \frac{f_\pi m_\pi^2 v \cdot q}{36 \pi^2 t^2 (m_u + m_d)} m_s \phi_{3;\pi}^\sigma(u) - \frac{f_\pi v \cdot q}{72} m_s {\langle \bar s s \rangle} \phi_{2;\pi}(u)
\\ \nonumber &&
- \frac{f_\pi t^2 v\cdot q}{1152} m_s {\langle \bar s s \rangle} \phi_{4;\pi}(u) - \frac{f_\pi }{72 v\cdot q} m_s {\langle \bar s s \rangle} \psi_{4;\pi}(u) \Big )
\\ \nonumber &+& \int_0^\infty dt \int_0^1 du \int \mathcal{D}\underline{\alpha} e^{i \omega^\prime t (\alpha_2 + u\alpha_3)} e^{i \omega t(1-\alpha_2-u\alpha_3)} \times 4 \times \Big (
\\ \nonumber && - \frac{f_\pi v\cdot q}{24 \pi^2 t^2} \Phi_{4;\pi}(\underline{\alpha}) + \frac{5 f_\pi v\cdot q}{72 \pi^2 t^2} \Psi_{4;\pi}(\underline{\alpha}) - \frac{f_\pi v\cdot q}{72 \pi^2 t^2} \widetilde \Phi_{4;\pi}(\underline{\alpha}) - \frac{5 f_\pi v\cdot q}{72 \pi^2 t^2} \widetilde \Psi_{4;\pi}(\underline{\alpha})
\\ \nonumber && - \frac{f_\pi u v \cdot q }{36 \pi^2 t^2} \Phi_{4;\pi}(\underline{\alpha}) - \frac{7 f_\pi u v\cdot q}{72 \pi^2 t^2} \Psi_{4;\pi}(\underline{\alpha}) + \frac{f_\pi u v \cdot q }{12 \pi^2 t^2} \widetilde \Phi_{4;\pi}(\underline{\alpha}) + \frac{f_\pi u v\cdot q}{24 \pi^2 t^2} \widetilde \Psi_{4;\pi}(\underline{\alpha}) \Big ) \, ,
\\
&& G_{\Xi_c^{0}[{3\over2}^-] \rightarrow \Xi_c^{+}\rho^-} (\omega, \omega^\prime)
= { g_{\Xi_c^{0}[{3\over2}^-] \rightarrow \Xi_c^{+}\rho^-} f_{\Xi_c^{0}[{3\over2}^-]} f_{\Xi_c^{+}} \over (\bar \Lambda_{\Xi_c^{0}[{3\over2}^-]} - \omega^\prime) (\bar \Lambda_{\Xi_c^{+}} - \omega)}
\\ \nonumber &=& \int_0^\infty dt \int_0^1 du e^{i (1-u) \omega^\prime t} e^{i u \omega t} \times 4 \times \Big (
\\ \nonumber && \frac{ f_{\rho}^\perp v\cdot q}{3 \pi^2 t^4} \phi_{2;\rho}^\perp(u)
- \frac{ f_{\rho}^\perp m_\rho^2 }{3 \pi^2 t^4 v\cdot q} \phi_{2;\rho}^\perp(u)
+ \frac{ f_{\rho}^\perp m_\rho^2 }{3 \pi^2 t^4 v\cdot q} \psi_{4;\rho}^\perp(u)
+\frac{f_{\rho}^\perp m_{\rho}^2 v\cdot q}{48 \pi^2 t^2} \phi_{4;\rho}^\perp(u)
-\frac{f_{\rho}^\parallel m_{\rho} v\cdot q}{72} \langle \bar s s \rangle \psi_{3;\rho}^\perp(u)
\\ \nonumber &&
-\frac{f_{\rho}^\parallel m_{\rho} t^2 v\cdot q}{1152} \langle g_s \bar s \sigma G s \rangle \psi_{3;\rho}^\perp(u)
- \frac{ f_{\rho}^\parallel m_{\rho} v\cdot q}{24 \pi^2 t^2} m_s \psi_{3;\rho}^\perp(u)
+ \frac{ f_{\rho}^\perp v\cdot q}{72} m_s \langle \bar s s \rangle \phi_{2;\rho}^\perp(u)
- \frac{ f_{\rho}^\perp m_\rho^2}{72 v\cdot q} m_s \langle \bar s s \rangle \phi_{2;\rho}^\perp(u)
\\ \nonumber &&
+ \frac{ f_{\rho}^\perp m_\rho^2}{72 v\cdot q} m_s \langle \bar s s \rangle \psi_{4;\rho}^\perp(u)
+ \frac{ f_{\rho}^\perp m_{\rho}^2 t^2 v\cdot q}{1152} m_s \langle \bar s s \rangle \phi_{4;\rho}^\perp(u)
 \Big )
\\ \nonumber &+& \int_0^\infty dt \int_0^1 du \int \mathcal{D}\underline{\alpha} e^{i \omega^\prime t (\alpha_2 + u\alpha_3)} e^{i \omega t(1-\alpha_2-u\alpha_3)} \times 4 \times \Big (
\\ \nonumber && \frac{ f_{\rho}^\perp m_{\rho}^2 v \cdot q}{12 \pi^2 t^2} \Psi_{4;\rho}^\perp(\underline{\alpha}) - \frac{ f_{\rho}^\perp m_{\rho}^2 v \cdot q}{12 \pi^2 t^2} \widetilde \Psi_{4;\rho}^\perp(\underline{\alpha})
+ \frac{ f_{\rho}^\perp m_{\rho}^2 u v \cdot q}{12 \pi^2 t^2} \Phi_{4;\rho}^{\perp1}(\underline{\alpha}) - \frac{ f_{\rho}^\perp m_{\rho}^2 u v \cdot q}{12 \pi^2 t^2} \Phi_{4;\rho}^{\perp2}(\underline{\alpha})
+ \frac{ f_{\rho}^\perp m_{\rho}^2 u v \cdot q}{6 \pi^2 t^2} \widetilde \Psi_{4;\rho}^{\perp}(\underline{\alpha})
 \Big ) \, .
\end{eqnarray}

\subsection{$[\mathbf{\bar 3}_F, 2, 1, \rho]$}
\label{sec:321rho}

The sum rule for $\Lambda_c^+[{3\over2}^-]$ belonging to $[\mathbf{\bar 3}_F, 2, 1, \rho]$ is
\begin{eqnarray}
&& G_{\Lambda_c^+[{3\over2}^-] \rightarrow \Sigma_c^{*++}\pi^-} (\omega, \omega^\prime)
= { g_{\Lambda_c^+[{3\over2}^-] \rightarrow \Sigma_c^{*++}\pi^-} f_{\Lambda_c^+[{3\over2}^-]} f_{\Sigma_c^{*++}} \over (\bar \Lambda_{\Lambda_c^+[{3\over2}^-]} - \omega^\prime) (\bar \Lambda_{\Sigma_c^{*++}} - \omega)}
\\ \nonumber &=& \int_0^\infty dt \int_0^1 du \int \mathcal{D}\underline{\alpha} e^{i \omega^\prime t (\alpha_2 + u\alpha_3)} e^{i \omega t(1-\alpha_2-u\alpha_3)} \times 8 \times \Big (
\\ \nonumber && \frac{f_\pi v\cdot q}{24 \pi^2 t^2} \Phi_{4;\pi}(\underline{\alpha}) + \frac{f_\pi v\cdot q}{24 \pi^2 t^2} \Psi_{4;\pi}(\underline{\alpha}) - \frac{ f_\pi v\cdot q}{24 \pi^2 t^2} \widetilde \Phi_{4;\pi}(\underline{\alpha}) - \frac{f_\pi v\cdot q}{24 \pi^2 t^2} \widetilde \Psi_{4;\pi}(\underline{\alpha})
\\ \nonumber && - \frac{f_\pi u v \cdot q }{12 \pi^2 t^2} \Phi_{4;\pi}(\underline{\alpha}) - \frac{f_\pi u v\cdot q}{24 \pi^2 t^2} \Psi_{4;\pi}(\underline{\alpha}) + \frac{f_\pi u v \cdot q }{12 \pi^2 t^2} \widetilde \Phi_{4;\pi}(\underline{\alpha}) + \frac{f_\pi u v\cdot q}{24 \pi^2 t^2} \widetilde \Psi_{4;\pi}(\underline{\alpha}) \Big ) \, .
\end{eqnarray}

The sum rules for $\Xi_c^0({3\over2}^-)$ belonging to $[\mathbf{\bar 3}_F, 2, 1, \rho]$ are
\begin{eqnarray}
&& G_{\Xi_c^0[{3\over2}^-] \rightarrow \Xi_c^{*+}\pi^-} (\omega, \omega^\prime)
= { g_{\Xi_c^0[{3\over2}^-] \rightarrow \Xi_c^{*+}\pi^-} f_{\Xi_c^0[{3\over2}^-]} f_{\Xi_c^{*+}} \over (\bar \Lambda_{\Xi_c^0[{3\over2}^-]} - \omega^\prime) (\bar \Lambda_{\Xi_c^{*+}} - \omega)}
\\ \nonumber &=& \int_0^\infty dt \int_0^1 du \int \mathcal{D}\underline{\alpha} e^{i \omega^\prime t (\alpha_2 + u\alpha_3)} e^{i \omega t(1-\alpha_2-u\alpha_3)} \times 4 \times \Big (
\\ \nonumber && \frac{f_\pi v\cdot q}{24 \pi^2 t^2} \Phi_{4;\pi}(\underline{\alpha}) + \frac{f_\pi v\cdot q}{24 \pi^2 t^2} \Psi_{4;\pi}(\underline{\alpha}) - \frac{ f_\pi v\cdot q}{24 \pi^2 t^2} \widetilde \Phi_{4;\pi}(\underline{\alpha}) - \frac{f_\pi v\cdot q}{24 \pi^2 t^2} \widetilde \Psi_{4;\pi}(\underline{\alpha})
\\ \nonumber && - \frac{f_\pi u v \cdot q }{12 \pi^2 t^2} \Phi_{4;\pi}(\underline{\alpha}) - \frac{f_\pi u v\cdot q}{24 \pi^2 t^2} \Psi_{4;\pi}(\underline{\alpha}) + \frac{f_\pi u v \cdot q }{12 \pi^2 t^2} \widetilde \Phi_{4;\pi}(\underline{\alpha}) + \frac{f_\pi u v\cdot q}{24 \pi^2 t^2} \widetilde \Psi_{4;\pi}(\underline{\alpha}) \Big ) \, ,
\\
&& G_{\Xi_c^{0}[{3\over2}^-] \rightarrow \Xi_c^{+}\rho^-} (\omega, \omega^\prime)
= { g_{\Xi_c^{0}[{3\over2}^-] \rightarrow \Xi_c^{+}\rho^-} f_{\Xi_c^{0}[{3\over2}^-]} f_{\Xi_c^{+}} \over (\bar \Lambda_{\Xi_c^{0}[{3\over2}^-]} - \omega^\prime) (\bar \Lambda_{\Xi_c^{+}} - \omega)}
= 0 \, .
\end{eqnarray}

The sum rule for $\Lambda_c^+[{5\over2}^-]$ belonging to $[\mathbf{\bar 3}_F, 2, 1, \rho]$ is
\begin{eqnarray}
&& G_{\Lambda_c^+[{5\over2}^-] \rightarrow \Sigma_c^{*++}\rho^-} (\omega, \omega^\prime)
= { g_{\Lambda_c^+[{5\over2}^-] \rightarrow \Sigma_c^{*++}\rho^-} f_{\Lambda_c^+[{5\over2}^-]} f_{\Sigma_c^{*++}} \over (\bar \Lambda_{\Lambda_c^+[{5\over2}^-]} - \omega^\prime) (\bar \Lambda_{\Sigma_c^{*++}} - \omega)}
\\ \nonumber &=& \int_0^\infty dt \int_0^1 du e^{i (1-u) \omega^\prime t} e^{i u \omega t} \times 8 \times \Big (
\\ \nonumber && -\frac{3 i f_\rho^\parallel m_\rho}{10 \pi^2 t^4} \phi_{2;\rho}^\parallel(u)
+ \frac{3 i f_\rho^\parallel m_\rho^3}{20 \pi^2 t^4 (v\cdot q)^2} \phi_{2;\rho}^\parallel(u)
+ \frac{3 i f_\rho^\parallel m_\rho }{10 \pi^2 t^4} \phi_{3;\rho}^\perp(u)
- \frac{3 i f_\rho^\parallel m_\rho^3}{10 \pi^2 t^4 (v\cdot q)^2} \phi_{3;\rho}^\perp(u)
- \frac{3 f_\rho^\parallel m_\rho v\cdot q}{40 \pi^2 t^3} \psi_{3;\rho}^\perp(u)
\\ \nonumber &&
- \frac{3 i f_\rho^\parallel m_\rho^3}{160 \pi^2 t^2} \phi_{4;\rho}^\parallel(u)
+ \frac{3 i f_\rho^\parallel m_\rho^3}{20 \pi^2 t^4 (v\cdot q)^2} \psi_{4;\rho}^\parallel(u)
+ \frac{f_\rho^\perp m_\rho^2}{40 t v\cdot q} \langle \bar q q \rangle \phi_{2;\rho}^\perp(u)
+ \frac{i f_\rho^\perp m_\rho^2}{40} \langle \bar q q \rangle \psi_{3;\rho}^\parallel(u)
- \frac{f_\rho^\perp m_\rho^2}{40 t v\cdot q} \langle \bar q q \rangle \psi_{4;\rho}^\perp(u)
\\ \nonumber &&
+ \frac{f_\rho^\perp m_\rho^2 t}{640 v\cdot q} \langle g_s \bar q \sigma G q \rangle \phi_{2;\rho}^\perp(u)
+ \frac{ i f_\rho^\perp m_\rho^2 t^2}{640} \langle g_s \bar q \sigma G q \rangle \psi_{3;\rho}^\parallel(u)
- \frac{f_\rho^\perp m_\rho^2 t}{640 v\cdot q} \langle g_s \bar q \sigma G q \rangle \psi_{4;\rho}^\perp(u)
\Big )
\\ \nonumber &+& \int_0^\infty dt \int_0^1 du \int \mathcal{D}\underline{\alpha} e^{i \omega^\prime t (\alpha_2 + u\alpha_3)} e^{i \omega t(1-\alpha_2-u\alpha_3)} \times 8 \times \Big (
\\ \nonumber && \frac{3 i f_\rho^\parallel m_\rho^3}{40 \pi^2 t^2} \Psi_{4;\rho}^\parallel(\underline{\alpha})
- \frac{3 i f_\rho^\parallel m_\rho^3}{40 \pi^2 t^2} \widetilde \Psi_{4;\rho}^\parallel(\underline{\alpha})
+ \frac{3 i f_\rho^\parallel m_\rho^3 u}{20 \pi^2 t^2} \Psi_{4;\rho}^\parallel(\underline{\alpha}) \Big ) \, .
\end{eqnarray}

The sum rule for $\Xi_c^0({5\over2}^-)$ belonging to $[\mathbf{\bar 3}_F, 2, 1, \rho]$ is
\begin{eqnarray}
&& G_{\Xi_c^0[{5\over2}^-] \rightarrow \Xi_c^{*+}{\rho}^-} (\omega, \omega^\prime)
= { g_{\Xi_c^0[{5\over2}^-] \rightarrow \Xi_c^{*+}{\rho}^-} f_{\Xi_c^0[{5\over2}^-]} f_{\Xi_c^{*+}} \over (\bar \Lambda_{\Xi_c^0[{5\over2}^-]} - \omega^\prime) (\bar \Lambda_{\Xi_c^{*+}} - \omega)}
\\ \nonumber &=& \int_0^\infty dt \int_0^1 du e^{i (1-u) \omega^\prime t} e^{i u \omega t} \times 8 \times \Big (
\\ \nonumber && -\frac{3 i f_\rho^\parallel m_\rho}{10 \pi^2 t^4} \phi_{2;\rho}^\parallel(u)
+ \frac{3 i f_\rho^\parallel m_\rho^3}{20 \pi^2 t^4 (v\cdot q)^2} \phi_{2;\rho}^\parallel(u)
+ \frac{3 i f_\rho^\parallel m_\rho }{10 \pi^2 t^4} \phi_{3;\rho}^\perp(u)
- \frac{3 i f_\rho^\parallel m_\rho^3}{10 \pi^2 t^4 (v\cdot q)^2} \phi_{3;\rho}^\perp(u)
- \frac{3 f_\rho^\parallel m_\rho v\cdot q}{40 \pi^2 t^3} \psi_{3;\rho}^\perp(u)
\\ \nonumber &&
- \frac{3 i f_\rho^\parallel m_\rho^3}{160 \pi^2 t^2} \phi_{4;\rho}^\parallel(u)
+ \frac{3 i f_\rho^\parallel m_\rho^3}{20 \pi^2 t^4 (v\cdot q)^2} \psi_{4;\rho}^\parallel(u)
+ \frac{f_\rho^\perp m_\rho^2}{40 t v\cdot q} \langle \bar s s \rangle \phi_{2;\rho}^\perp(u)
+ \frac{i f_\rho^\perp m_\rho^2}{40} \langle \bar s s \rangle \psi_{3;\rho}^\parallel(u)
- \frac{f_\rho^\perp m_\rho^2}{40 t v\cdot q} \langle \bar s s \rangle \psi_{4;\rho}^\perp(u)
\\ \nonumber &&
+ \frac{f_\rho^\perp m_\rho^2 t}{640 v\cdot q} \langle g_s \bar s \sigma G s \rangle \phi_{2;\rho}^\perp(u)
+ \frac{ i f_\rho^\perp m_\rho^2 t^2}{640} \langle g_s \bar s \sigma G s \rangle \psi_{3;\rho}^\parallel(u)
- \frac{f_\rho^\perp m_\rho^2 t}{640 v\cdot q} \langle g_s \bar s \sigma G s \rangle \psi_{4;\rho}^\perp(u)
+ \frac{3 f_{\rho}^\perp m_{\rho}^2}{40 \pi^2 t^3 v\cdot q} m_s \phi_{2;\rho}^\perp(u)
\\ \nonumber &&
+\frac{3 i f_{\rho}^\perp m_{\rho}^2}{40 \pi^2 t^2} m_s \psi_{3;\rho}^\parallel(u)
-\frac{3 f_{\rho}^\perp m_{\rho}^2}{40 \pi^2 t^3 v\cdot q} m_s \psi_{4;\rho}^\perp(u)
-\frac{ i f_{\rho}^\parallel m_{\rho}}{80} m_s \langle \bar s s \rangle \phi_{2;\rho}^\parallel(u)
+\frac{i f_{\rho}^\parallel m_{\rho}^3}{160 (v\cdot q)^2} m_s \langle \bar s s \rangle \phi_{2;\rho}^\parallel(u)
\\ \nonumber &&
+\frac{ i f_{\rho}^\parallel m_{\rho}}{80} m_s \langle \bar s s \rangle \phi_{3;\rho}^\perp(u)
-\frac{i f_{\rho}^\parallel m_{\rho}^3}{80 (v\cdot q)^2} m_s \langle \bar s s \rangle \phi_{3;\rho}^\perp(u)
-\frac{f_{\rho}^\parallel m_{\rho} t v\cdot q}{320} m_s \langle \bar s s \rangle \psi_{3;\rho}^\perp(u)
-\frac{i f_{\rho}^\parallel m_{\rho}^3 t^2 }{1280} m_s \langle \bar s s \rangle \phi_{4;\rho}^\parallel(u)
\\ \nonumber &&
+\frac{i f_{\rho}^\parallel m_{\rho}^3}{160 (v\cdot q)^2} m_s \langle \bar s s \rangle \psi_{4;\rho}^\parallel(u)
\Big )
\\ \nonumber &+& \int_0^\infty dt \int_0^1 du \int \mathcal{D}\underline{\alpha} e^{i \omega^\prime t (\alpha_2 + u\alpha_3)} e^{i \omega t(1-\alpha_2-u\alpha_3)} \times 8 \times \Big (
\\ \nonumber && \frac{3 i f_\rho^\parallel m_\rho^3}{40 \pi^2 t^2} \Psi_{4;\rho}^\parallel(\underline{\alpha})
- \frac{3 i f_\rho^\parallel m_\rho^3}{40 \pi^2 t^2} \widetilde \Psi_{4;\rho}^\parallel(\underline{\alpha})
+ \frac{3 i f_\rho^\parallel m_\rho^3 u}{20 \pi^2 t^2} \Psi_{4;\rho}^\parallel(\underline{\alpha}) \Big ) \, .
\end{eqnarray}

\subsection{$[\mathbf{\bar 3}_F, 1, 0, \lambda]$}
\label{sec:310lambda}

The sum rule for $\Lambda_c^+({1\over2}^-)$ belonging to $[\mathbf{\bar 3}_F, 1, 0, \lambda]$ is
\begin{eqnarray}
&& G_{\Lambda_c^+[{1\over2}^-] \rightarrow \Sigma_c^{++}\pi^-} (\omega, \omega^\prime)
= { g_{\Lambda_c^+[{1\over2}^-] \rightarrow \Sigma_c^{++}\pi^-} f_{\Lambda_c^+[{1\over2}^-]} f_{\Sigma_c^{++}} \over (\bar \Lambda_{\Lambda_c^+[{1\over2}^-]} - \omega^\prime) (\bar \Lambda_{\Sigma_c^{++}} - \omega)}
\\ \nonumber &=& \int_0^\infty dt \int_0^1 du e^{i (1-u) \omega^\prime t} e^{i u \omega t} \times 8 \times \Big (
\\ \nonumber && \frac{3 f_\pi m_\pi^2}{4 \pi^2 t^4 (m_u + m_d)} \phi_{3;\pi}^p(u) - \frac{i f_\pi m_\pi^2 v \cdot q}{8 \pi^2 t^3 ( m_u + m_d )} \phi_{3;\pi}^\sigma(u) + \frac{i f_\pi}{16 t v \cdot q} \langle \bar q q \rangle \psi_{4;\pi}(u) + \frac{i f_\pi t}{256 v\cdot q} \langle g_s \bar q \sigma G q\rangle \psi_{4;\pi}(u) \Big ) \, .
\end{eqnarray}

The sum rules for $\Xi_c^0({1\over2}^-)$ belonging to $[\mathbf{\bar 3}_F, 1, 0, \lambda]$ are
\begin{eqnarray}
&& G_{\Xi_c^0[{1\over2}^-] \rightarrow \Xi_c^{\prime+}\pi^-} (\omega, \omega^\prime)
= { g_{\Xi_c^0[{1\over2}^-] \rightarrow \Xi_c^{\prime+}\pi^-} f_{\Xi_c^0[{1\over2}^-]} f_{\Xi_c^{\prime+}} \over (\bar \Lambda_{\Xi_c^0[{1\over2}^-]} - \omega^\prime) (\bar \Lambda_{\Xi_c^{\prime+}} - \omega)}
\\ \nonumber &=& \int_0^\infty dt \int_0^1 du e^{i (1-u) \omega^\prime t} e^{i u \omega t} \times 4 \times \Big (
\\ \nonumber && \frac{3 f_\pi m_\pi^2}{4 \pi^2 t^4 (m_u + m_d)} \phi_{3;\pi}^p(u) - \frac{i f_\pi m_\pi^2 v \cdot q}{8 \pi^2 t^3 ( m_u + m_d )} \phi_{3;\pi}^\sigma(u) + \frac{i f_\pi}{16 t v \cdot q} \langle \bar s s \rangle \psi_{4;\pi}(u) + \frac{i f_\pi t}{256 v\cdot q} \langle g_s \bar s \sigma G s\rangle \psi_{4;\pi}(u)
\\ \nonumber && + \frac{3 i f_\pi}{16 \pi^2 t^3 v\cdot q} m_s \psi_{4;\pi}(u) + \frac{ f_\pi m_\pi^2}{32 (m_u + m_d)} m_s \langle \bar s s \rangle \phi_{3;\pi}^p(u) - \frac{i f_\pi m_\pi^2 t v\cdot q}{192 ( m_u + m_d)} m_s \langle \bar s s \rangle \phi_{3;\pi}^\sigma(u) \Big ) \, ,
\\
&& G_{\Xi_c^0[{1\over2}^-] \rightarrow \Xi_c^{+}\pi^-} (\omega, \omega^\prime)
= { g_{\Xi_c^0[{1\over2}^-] \rightarrow \Xi_c^{+}\pi^-} f_{\Xi_c^0[{1\over2}^-]} f_{\Xi_c^{+}} \over (\bar \Lambda_{\Xi_c^0[{1\over2}^-]} - \omega^\prime) (\bar \Lambda_{\Xi_c^{+}} - \omega)}
= 0 \, ,
\\
&& G_{\Xi_c^0[{1\over2}^-] \rightarrow \Lambda_c^{+} K^-} (\omega, \omega^\prime)
= { g_{\Xi_c^0[{1\over2}^-] \rightarrow \Lambda_c^{+} K^-} f_{\Xi_c^0[{1\over2}^-]} f_{\Lambda_c^{+}} \over (\bar \Lambda_{\Xi_c^0[{1\over2}^-]} - \omega^\prime) (\bar \Lambda_{\Lambda_c^{+}} - \omega)}
= 0 \, ,
\\ && G_{\Xi_c^{0}[{1\over2}^-] \rightarrow \Xi_c^{+}\rho^-} (\omega, \omega^\prime)
= { g_{\Xi_c^{0}[{1\over2}^-] \rightarrow \Xi_c^{+}\rho^-} f_{\Xi_c^{0}[{1\over2}^-]} f_{\Xi_c^{+}} \over (\bar \Lambda_{\Xi_c^{0}[{1\over2}^-]} - \omega^\prime) (\bar \Lambda_{\Xi_c^{+}} - \omega)}
\\ \nonumber &=& \int_0^\infty dt \int_0^1 du e^{i (1-u) \omega^\prime t} e^{i u \omega t} \times 4 \times \Big (
\\ \nonumber && \frac{i f_{\rho}^\parallel m_{\rho}}{4 \pi^2 t^4} \phi_{2;\rho}^\parallel(u)
- \frac{i f_{\rho}^\parallel m_{\rho}^3}{8 \pi^2 t^4 (v\cdot q)^2} \phi_{2;\rho}^\parallel(u)
+ \frac{i f_{\rho}^\parallel m_{\rho}^3}{4 \pi^2 t^4 (v\cdot q)^2} \phi_{3;\rho}^\perp(u)
- \frac{i f_{\rho}^\parallel m_{\rho}^3}{8 \pi^2 t^4 (v\cdot q)^2} \psi_{4;\rho}^\parallel(u)
+ \frac{i f_{\rho}^\parallel m_{\rho}^3}{64 \pi^2 t^2} \phi_{4;\rho}^\parallel(u)
\\ \nonumber &&
- \frac{i f_{\rho}^\perp m_{\rho}^2}{48} \langle \bar s s \rangle \psi_{3;\rho}^\parallel(u)
- \frac{i f_{\rho}^\perp m_{\rho}^2 t^2}{768} \langle g_s \bar s \sigma G s \rangle \psi_{3;\rho}^\parallel(u)
- \frac{i f_{\rho}^\perp m_{\rho}^2}{16 \pi^2 t^2} m_s \psi_{3;\rho}^\parallel(u)
+ \frac{i f_{\rho}^\parallel m_{\rho}}{96} m_s \langle \bar s s \rangle \phi_{2;\rho}^\parallel(u)
\\ \nonumber &&
- \frac{i f_{\rho}^\parallel m_{\rho}^3}{192 (v\cdot q)^2} m_s \langle \bar s s \rangle \phi_{2;\rho}^\parallel(u)
+ \frac{i f_{\rho}^\parallel m_{\rho}^3}{96 (v\cdot q)^2} m_s \langle \bar s s \rangle \phi_{3;\rho}^\perp(u)
+ \frac{i f_{\rho}^\parallel m_{\rho}^3 t^2}{1536} m_s \langle \bar s s \rangle \phi_{4;\rho}^\parallel(u)
- \frac{i f_{\rho}^\parallel m_{\rho}^3}{192 (v\cdot q)^2} m_s \langle \bar s s \rangle \psi_{4;\rho}^\parallel(u)
 \Big )
\\ \nonumber &+& \int_0^\infty dt \int_0^1 du \int \mathcal{D}\underline{\alpha} e^{i \omega^\prime t (\alpha_2 + u\alpha_3)} e^{i \omega t(1-\alpha_2-u\alpha_3)} \times 4 \times \Big (
\\ \nonumber && - \frac{i f_{\rho}^\parallel m_{\rho}^3}{8 \pi^2 t^2} \Phi_{4;\rho}^\parallel(\underline{\alpha}) - \frac{i f_{\rho}^\parallel m_{\rho}^3}{16 \pi^2 t^2} \Psi_{4;\rho}^\parallel(\underline{\alpha})
- \frac{ i f_{\rho}^\parallel m_{\rho}^3}{8 \pi^2 t^2} \widetilde \Phi_{4;\rho}^\parallel(\underline{\alpha}) - \frac{i f_{\rho}^\parallel m_{\rho}^3}{16 \pi^2 t^2} \widetilde \Psi_{4;\rho}^\parallel(\underline{\alpha})
- \frac{i f_{\rho}^\parallel m_{\rho}^3 u}{4 \pi^2 t^2} \Phi_{4;\rho}^\parallel(\underline{\alpha}) - \frac{i f_{\rho}^\parallel m_{\rho}^3 u}{8 \pi^2 t^2} \Psi_{4;\rho}^\parallel(\underline{\alpha}) \Big ) \, .
\end{eqnarray}

The sum rule for $\Lambda_c^+[{3\over2}^-]$ belonging to $[\mathbf{\bar 3}_F, 1, 0, \lambda]$ is
\begin{eqnarray}
&& G_{\Lambda_c^+[{3\over2}^-] \rightarrow \Sigma_c^{*++}\pi^-} (\omega, \omega^\prime)
= { g_{\Lambda_c^+[{3\over2}^-] \rightarrow \Sigma_c^{*++}\pi^-} f_{\Lambda_c^+[{3\over2}^-]} f_{\Sigma_c^{*++}} \over (\bar \Lambda_{\Lambda_c^+[{3\over2}^-]} - \omega^\prime) (\bar \Lambda_{\Sigma_c^{*++}} - \omega)}
\\ \nonumber &=& \int_0^\infty dt \int_0^1 du e^{i (1-u) \omega^\prime t} e^{i u \omega t} \times 8 \times \Big (
\\ \nonumber && - \frac{f_\pi m_\pi^2}{6 \pi^2 t^4 (m_u + m_d)} \phi_{3;\pi}^p(u) + \frac{i f_\pi m_\pi^2 v \cdot q}{36 \pi^2 t^3 ( m_u + m_d )} \phi_{3;\pi}^\sigma(u) - \frac{i f_\pi}{72 t v \cdot q} \langle \bar q q \rangle \psi_{4;\pi}(u) - \frac{i f_\pi t}{1152 v\cdot q} \langle g_s \bar q \sigma G q\rangle \psi_{4;\pi}(u) \Big ) \, .
\end{eqnarray}

The sum rules for $\Xi_c^0({3\over2}^-)$ belonging to $[\mathbf{\bar 3}_F, 1, 0, \lambda]$ are
\begin{eqnarray}
&& G_{\Xi_c^0[{3\over2}^-] \rightarrow \Xi_c^{*+}\pi^-} (\omega, \omega^\prime)
= { g_{\Xi_c^0[{3\over2}^-] \rightarrow \Xi_c^{*+}\pi^-} f_{\Xi_c^0[{3\over2}^-]} f_{\Xi_c^{*+}} \over (\bar \Lambda_{\Xi_c^0[{3\over2}^-]} - \omega^\prime) (\bar \Lambda_{\Xi_c^{*+}} - \omega)}
\\ \nonumber &=& \int_0^\infty dt \int_0^1 du e^{i (1-u) \omega^\prime t} e^{i u \omega t} \times 4 \times \Big (
\\ \nonumber && - \frac{f_\pi m_\pi^2}{6 \pi^2 t^4 (m_u + m_d)} \phi_{3;\pi}^p(u) + \frac{i f_\pi m_\pi^2 v \cdot q}{36 \pi^2 t^3 ( m_u + m_d )} \phi_{3;\pi}^\sigma(u) - \frac{i f_\pi}{72 t v \cdot q} \langle \bar s s \rangle \psi_{4;\pi}(u) - \frac{i f_\pi t}{1152 v\cdot q} \langle g_s \bar s \sigma G s\rangle \psi_{4;\pi}(u)
\\ \nonumber && - \frac{i f_\pi}{24 \pi^2 t^3 v\cdot q} m_s \psi_{4;\pi}(u) - \frac{ f_\pi m_\pi^2}{144 (m_u + m_d)} m_s \langle \bar s s \rangle \phi_{3;\pi}^p(u) + \frac{i f_\pi m_\pi^2 t v\cdot q}{864 ( m_u + m_d)} m_s \langle \bar s s \rangle \phi_{3;\pi}^\sigma(u) \Big ) \, ,
\\
&& G_{\Xi_c^{0}[{3\over2}^-] \rightarrow \Xi_c^{+}\rho^-} (\omega, \omega^\prime)
= { g_{\Xi_c^{0}[{3\over2}^-] \rightarrow \Xi_c^{+}\rho^-} f_{\Xi_c^{0}[{3\over2}^-]} f_{\Xi_c^{+}} \over (\bar \Lambda_{\Xi_c^{0}[{3\over2}^-]} - \omega^\prime) (\bar \Lambda_{\Xi_c^{+}} - \omega)}
\\ \nonumber &=& \int_0^\infty dt \int_0^1 du e^{i (1-u) \omega^\prime t} e^{i u \omega t} \times 4 \times \Big (
\\ \nonumber && \frac{i f_{\rho}^\parallel m_{\rho}}{6 \pi^2 t^4} \phi_{2;\rho}^\parallel(u)
- \frac{i f_{\rho}^\parallel m_{\rho}^3}{12 \pi^2 t^4 (v\cdot q)^2} \phi_{2;\rho}^\parallel(u)
+ \frac{i f_{\rho}^\parallel m_{\rho}^3}{6 \pi^2 t^4 (v\cdot q)^2} \phi_{3;\rho}^\perp(u)
- \frac{i f_{\rho}^\parallel m_{\rho}^3}{12 \pi^2 t^4 (v\cdot q)^2} \psi_{4;\rho}^\parallel(u)
+ \frac{i f_{\rho}^\parallel m_{\rho}^3}{96 \pi^2 t^2} \phi_{4;\rho}^\parallel(u)
\\ \nonumber &&
- \frac{i f_{\rho}^\perp m_{\rho}^2}{72} \langle \bar s s \rangle \psi_{3;\rho}^\parallel(u)
- \frac{i f_{\rho}^\perp m_{\rho}^2 t^2}{1152} \langle g_s \bar s \sigma G s \rangle \psi_{3;\rho}^\parallel(u)
- \frac{i f_{\rho}^\perp m_{\rho}^2}{24 \pi^2 t^2} m_s \psi_{3;\rho}^\parallel(u)
+ \frac{i f_{\rho}^\parallel m_{\rho}}{144} m_s \langle \bar s s \rangle \phi_{2;\rho}^\parallel(u)
\\ \nonumber &&
- \frac{i f_{\rho}^\parallel m_{\rho}^3}{288 (v\cdot q)^2} m_s \langle \bar s s \rangle \phi_{2;\rho}^\parallel(u)
+ \frac{i f_{\rho}^\parallel m_{\rho}^3}{144 (v\cdot q)^2} m_s \langle \bar s s \rangle \phi_{3;\rho}^\perp(u)
+ \frac{i f_{\rho}^\parallel m_{\rho}^3 t^2}{2304} m_s \langle \bar s s \rangle \phi_{4;\rho}^\parallel(u)
- \frac{i f_{\rho}^\parallel m_{\rho}^3}{288 (v\cdot q)^2} m_s \langle \bar s s \rangle \psi_{4;\rho}^\parallel(u)
 \Big )
\\ \nonumber &+& \int_0^\infty dt \int_0^1 du \int \mathcal{D}\underline{\alpha} e^{i \omega^\prime t (\alpha_2 + u\alpha_3)} e^{i \omega t(1-\alpha_2-u\alpha_3)} \times 4 \times \Big (
\\ \nonumber && - \frac{i f_{\rho}^\parallel m_{\rho}^3}{12 \pi^2 t^2} \Phi_{4;\rho}^\parallel(\underline{\alpha}) - \frac{i f_{\rho}^\parallel m_{\rho}^3}{24 \pi^2 t^2} \Psi_{4;\rho}^\parallel(\underline{\alpha})
- \frac{ i f_{\rho}^\parallel m_{\rho}^3}{12 \pi^2 t^2} \widetilde \Phi_{4;\rho}^\parallel(\underline{\alpha}) - \frac{i f_{\rho}^\parallel m_{\rho}^3}{24 \pi^2 t^2} \widetilde \Psi_{4;\rho}^\parallel(\underline{\alpha})
- \frac{i f_{\rho}^\parallel m_{\rho}^3 u}{6 \pi^2 t^2} \Phi_{4;\rho}^\parallel(\underline{\alpha}) - \frac{i f_{\rho}^\parallel m_{\rho}^3 u}{12 \pi^2 t^2} \Psi_{4;\rho}^\parallel(\underline{\alpha}) \Big ) \, .
\end{eqnarray}

\subsection{$[\mathbf{6}_F, 1, 0, \rho]$}
\label{sec:610rho}

The sum rule for $\Sigma_c^0({1\over2}^-)$ belonging to $[\mathbf{6}_F, 1, 0, \rho]$ is
\begin{eqnarray}
&& G_{\Sigma_c^0[{1\over2}^-] \rightarrow \Lambda_c^{+}\pi^-} (\omega, \omega^\prime)
= { g_{\Sigma_c^0[{1\over2}^-] \rightarrow \Lambda_c^{+}\pi^-} f_{\Sigma_c^0[{1\over2}^-]} f_{\Lambda_c^{+}} \over (\bar \Lambda_{\Sigma_c^0[{1\over2}^-]} - \omega^\prime) (\bar \Lambda_{\Lambda_c^{+}} - \omega)}
= 0 \, ,
\\ && G_{\Sigma_c^0[{1\over2}^-] \rightarrow \Sigma_c^{+}\pi^-} (\omega, \omega^\prime)
= { g_{\Sigma_c^0[{1\over2}^-] \rightarrow \Sigma_c^{+}\pi^-} f_{\Sigma_c^0[{1\over2}^-]} f_{\Sigma_c^{+}} \over (\bar \Lambda_{\Sigma_c^0[{1\over2}^-]} - \omega^\prime) (\bar \Lambda_{\Sigma_c^{+}} - \omega)}
\\ \nonumber &=& \int_0^\infty dt \int_0^1 du e^{i (1-u) \omega^\prime t} e^{i u \omega t} \times 8 \times \Big (
\\ \nonumber &&
\frac{3 f_\pi m_\pi^2}{4 \pi^2 t^4 (m_u + m_d)} \phi_{3;\pi}^p(u)
+ \frac{i f_\pi m_\pi^2 v \cdot q}{8 \pi^2 t^3 ( m_u + m_d )} \phi_{3;\pi}^\sigma(u)
- \frac{i f_\pi}{16 t v \cdot q} \langle \bar q q \rangle \psi_{4;\pi}(u)
- \frac{i f_\pi t}{256 v\cdot q} \langle g_s \bar q \sigma G q\rangle \psi_{4;\pi}(u)
\Big ) \, .
\end{eqnarray}

The sum rule for $\Xi_c^{\prime0}({1\over2}^-)$ belonging to $[\mathbf{6}_F, 1, 0, \rho]$ is
\begin{eqnarray}
&& G_{\Xi_c^{\prime0}[{1\over2}^-] \rightarrow \Xi_c^{+}\pi^-} (\omega, \omega^\prime)
= { g_{\Xi_c^{\prime0}[{1\over2}^-] \rightarrow \Xi_c^{+}\pi^-} f_{\Xi_c^{\prime0}[{1\over2}^-]} f_{\Xi_c^{+}} \over (\bar \Lambda_{\Xi_c^{\prime0}[{1\over2}^-]} - \omega^\prime) (\bar \Lambda_{\Xi_c^{+}} - \omega)}
= 0 \, ,
\\ && G_{\Xi_c^{\prime0}[{1\over2}^-] \rightarrow \Xi_c^{\prime+}\pi^-} (\omega, \omega^\prime)
= { g_{\Xi_c^{\prime0}[{1\over2}^-] \rightarrow \Xi_c^{\prime+}\pi^-} f_{\Xi_c^{\prime0}[{1\over2}^-]} f_{\Xi_c^{\prime+}} \over (\bar \Lambda_{\Xi_c^{\prime0}[{1\over2}^-]} - \omega^\prime) (\bar \Lambda_{\Xi_c^{\prime+}} - \omega)}
\\ \nonumber &=& \int_0^\infty dt \int_0^1 du e^{i (1-u) \omega^\prime t} e^{i u \omega t} \times 4 \times \Big (
\\ \nonumber &&
\frac{3 f_\pi m_\pi^2}{4 \pi^2 t^4 (m_u + m_d)} \phi_{3;\pi}^p(u)
+ \frac{i f_\pi m_\pi^2 v \cdot q}{8 \pi^2 t^3 ( m_u + m_d )} \phi_{3;\pi}^\sigma(u)
- \frac{i f_\pi}{16 t v \cdot q} \langle \bar s s \rangle \psi_{4;\pi}(u)
- \frac{i f_\pi t}{256 v\cdot q} \langle g_s \bar s \sigma G s\rangle \psi_{4;\pi}(u)
\\ \nonumber &&
- \frac{3 i f_\pi}{16 \pi^2 t^3 v\cdot q} m_s \psi_{4;\pi}(u)
+ \frac{ f_\pi m_\pi^2}{32 (m_u + m_d)} m_s \langle \bar s s \rangle \phi_{3;\pi}^p(u)
+ \frac{i f_\pi m_\pi^2 t v\cdot q}{192 ( m_u + m_d)} m_s \langle \bar s s \rangle \phi_{3;\pi}^\sigma(u) \Big ) \, ,
\\ && G_{\Xi_c^{\prime0}[{1\over2}^-] \rightarrow \Lambda_c^{+}K^-} (\omega, \omega^\prime)
= { g_{\Xi_c^{\prime0}[{1\over2}^-] \rightarrow \Lambda_c^{+}K^-} f_{\Xi_c^{\prime0}[{1\over2}^-]} f_{\Lambda_c^{+}} \over (\bar \Lambda_{\Xi_c^{\prime0}[{1\over2}^-]} - \omega^\prime) (\bar \Lambda_{\Lambda_c^{+}} - \omega)}
= 0 \, ,
\\ && G_{\Xi_c^{\prime0}[{1\over2}^-] \rightarrow \Sigma_c^{+}K^-} (\omega, \omega^\prime)
= { g_{\Xi_c^{\prime0}[{1\over2}^-] \rightarrow \Sigma_c^{+}K^-} f_{\Xi_c^{\prime0}[{1\over2}^-]} f_{\Sigma_c^{+}} \over (\bar \Lambda_{\Xi_c^{\prime0}[{1\over2}^-]} - \omega^\prime) (\bar \Lambda_{\Sigma_c^{+}} - \omega)}
\\ \nonumber &=& \int_0^\infty dt \int_0^1 du e^{i (1-u) \omega^\prime t} e^{i u \omega t} \times 4 \times \Big (
\\ \nonumber &&
\frac{3 f_K m_K^2}{4 \pi^2 t^4 (m_u + m_s)} \phi_{3;K}^p(u)
+ \frac{i f_K m_K^2 v \cdot q}{8 \pi^2 t^3 ( m_u + m_s )} \phi_{3;K}^\sigma(u)
- \frac{i f_K}{16 t v \cdot q} \langle \bar q q \rangle \psi_{4;K}(u)
- \frac{i f_K t}{256 v\cdot q} \langle g_s \bar q \sigma G q\rangle \psi_{4;K}(u)
\Big ) \, .
\end{eqnarray}

The sum rule for $\Omega_c^0({1\over2}^-)$ belonging to $[\mathbf{6}_F, 1, 0, \rho]$ is
\begin{eqnarray}
&& G_{\Omega_c^0[{1\over2}^-] \rightarrow \Xi_c^{+}K^{-}} (\omega, \omega^\prime)
= { g_{\Omega_c^0[{1\over2}^-] \rightarrow \Xi_c^{+}K^{-}} f_{\Omega_c^0[{1\over2}^-]} f_{\Xi_c^{+}} \over (\bar \Lambda_{\Omega_c^0[{1\over2}^-]} - \omega^\prime) (\bar \Lambda_{\Xi_c^{+}} - \omega)}
= 0 \, ,
\\ && G_{\Omega_c^0[{1\over2}^-] \rightarrow \Xi_c^{\prime+}K^{-}} (\omega, \omega^\prime)
= { g_{\Omega_c^0[{1\over2}^-] \rightarrow \Xi_c^{\prime+}K^{-}} f_{\Omega_c^0[{1\over2}^-]} f_{\Xi_c^{\prime+}} \over (\bar \Lambda_{\Omega_c^0[{1\over2}^-]} - \omega^\prime) (\bar \Lambda_{\Xi_c^{\prime+}} - \omega)}
\\ \nonumber &=& \int_0^\infty dt \int_0^1 du e^{i (1-u) \omega^\prime t} e^{i u \omega t} \times 8 \times \Big (
\\ \nonumber &&
\frac{3 f_K m_K^2}{4 \pi^2 t^4 (m_u + m_s)} \phi_{3;K}^p(u)
+ \frac{i f_K m_K^2 v \cdot q}{8 \pi^2 t^3 ( m_u + m_s )} \phi_{3;K}^\sigma(u)
- \frac{i f_K}{16 t v \cdot q} \langle \bar s s \rangle \psi_{4;K}(u)
- \frac{i f_K t}{256 v\cdot q} \langle g_s \bar s \sigma G s\rangle \psi_{4;K}(u)
\\ \nonumber &&
- \frac{3 i f_K}{16 \pi^2 t^3 v\cdot q} m_s \psi_{4;K}(u)
+ \frac{ f_K m_K^2}{32 (m_u + m_s)} m_s \langle \bar s s \rangle \phi_{3;K}^p(u)
+ \frac{i f_K m_K^2 t v\cdot q}{192 ( m_u + m_s)} m_s \langle \bar s s \rangle \phi_{3;K}^\sigma(u) \Big ) \, .
\end{eqnarray}

The sum rule for $\Sigma_c^0[{3\over2}^-]$ belonging to $[\mathbf{6}_F, 1, 0, \rho]$ is
\begin{eqnarray}
&& G_{\Sigma_c^0[{3\over2}^-] \rightarrow \Sigma_c^{*+}\pi^-} (\omega, \omega^\prime)
= { g_{\Sigma_c^0[{3\over2}^-] \rightarrow \Sigma_c^{*+}\pi^-} f_{\Sigma_c^0[{3\over2}^-]} f_{\Sigma_c^{*+}} \over (\bar \Lambda_{\Sigma_c^0[{3\over2}^-]} - \omega^\prime) (\bar \Lambda_{\Sigma_c^{*+}} - \omega)}
\\ \nonumber &=& \int_0^\infty dt \int_0^1 du e^{i (1-u) \omega^\prime t} e^{i u \omega t} \times 8 \times \Big (
\\ \nonumber &&
- \frac{f_\pi m_\pi^2}{6 \pi^2 t^4 (m_u + m_d)} \phi_{3;\pi}^p(u)
- \frac{i f_\pi m_\pi^2 v \cdot q}{36 \pi^2 t^3 ( m_u + m_d )} \phi_{3;\pi}^\sigma(u)
+ \frac{i f_\pi}{72 t v \cdot q} \langle \bar q q \rangle \psi_{4;\pi}(u)
+ \frac{i f_\pi t}{1152 v\cdot q} \langle g_s \bar q \sigma G q\rangle \psi_{4;\pi}(u) \Big ) \, .
\end{eqnarray}

The sum rule for $\Xi_c^0({3\over2}^-)$ belonging to $[\mathbf{6}_F, 1, 0, \rho]$ is
\begin{eqnarray}
&& G_{\Xi_c^{\prime0}[{3\over2}^-] \rightarrow \Xi_c^{*+}\pi^-} (\omega, \omega^\prime)
= { g_{\Xi_c^{\prime0}[{3\over2}^-] \rightarrow \Xi_c^{*+}\pi^-} f_{\Xi_c^{\prime0}[{3\over2}^-]} f_{\Xi_c^{*+}} \over (\bar \Lambda_{\Xi_c^{\prime0}[{3\over2}^-]} - \omega^\prime) (\bar \Lambda_{\Xi_c^{*+}} - \omega)}
\\ \nonumber &=& \int_0^\infty dt \int_0^1 du e^{i (1-u) \omega^\prime t} e^{i u \omega t} \times 4 \times \Big (
\\ \nonumber &&
- \frac{f_\pi m_\pi^2}{6 \pi^2 t^4 (m_u + m_d)} \phi_{3;\pi}^p(u)
- \frac{i f_\pi m_\pi^2 v \cdot q}{36 \pi^2 t^3 ( m_u + m_d )} \phi_{3;\pi}^\sigma(u)
+ \frac{i f_\pi}{72 t v \cdot q} \langle \bar s s \rangle \psi_{4;\pi}(u)
+ \frac{i f_\pi t}{1152 v\cdot q} \langle g_s \bar s \sigma G s\rangle \psi_{4;\pi}(u)
\\ \nonumber &&
+ \frac{i f_\pi}{24 \pi^2 t^3 v\cdot q} m_s \psi_{4;\pi}(u)
- \frac{ f_\pi m_\pi^2}{144 (m_u + m_d)} m_s \langle \bar s s \rangle \phi_{3;\pi}^p(u)
- \frac{i f_\pi m_\pi^2 t v\cdot q}{864 ( m_u + m_d)} m_s \langle \bar s s \rangle \phi_{3;\pi}^\sigma(u) \Big ) \, ,
\\ && G_{\Xi_c^{\prime0}[{3\over2}^-] \rightarrow \Sigma_c^{*+}K^-} (\omega, \omega^\prime)
= { g_{\Xi_c^{\prime0}[{3\over2}^-] \rightarrow \Sigma_c^{*+}K^-} f_{\Xi_c^{\prime0}[{3\over2}^-]} f_{\Sigma_c^{*+}} \over (\bar \Lambda_{\Xi_c^{\prime0}[{3\over2}^-]} - \omega^\prime) (\bar \Lambda_{\Sigma_c^{*+}} - \omega)}
\\ \nonumber &=& \int_0^\infty dt \int_0^1 du e^{i (1-u) \omega^\prime t} e^{i u \omega t} \times 4 \times \Big (
\\ \nonumber &&
- \frac{f_K m_K^2}{6 \pi^2 t^4 (m_u + m_s)} \phi_{3;K}^p(u)
- \frac{i f_K m_K^2 v \cdot q}{36 \pi^2 t^3 ( m_u + m_s )} \phi_{3;K}^\sigma(u)
+ \frac{i f_K}{72 t v \cdot q} \langle \bar q q \rangle \psi_{4;K}(u)
+ \frac{i f_K t}{1152 v\cdot q} \langle g_s \bar q \sigma G q\rangle \psi_{4;K}(u) \Big ) \, .
\end{eqnarray}

The sum rule for $\Omega_c^0({3\over2}^-)$ belonging to $[\mathbf{6}_F, 1, 0, \rho]$ is
\begin{eqnarray}
&& G_{\Omega_c^0[{3\over2}^-] \rightarrow \Xi_c^{*+}K^{-}} (\omega, \omega^\prime)
= { g_{\Omega_c^0[{3\over2}^-] \rightarrow \Xi_c^{*+}K^{-}} f_{\Omega_c^0[{3\over2}^-]} f_{\Xi_c^{*+}} \over (\bar \Lambda_{\Omega_c^0[{3\over2}^-]} - \omega^\prime) (\bar \Lambda_{\Xi_c^{*+}} - \omega)}
\\ \nonumber &=& \int_0^\infty dt \int_0^1 du e^{i (1-u) \omega^\prime t} e^{i u \omega t} \times 8 \times \Big (
\\ \nonumber &&
- \frac{f_K m_K^2}{6 \pi^2 t^4 (m_u + m_s)} \phi_{3;K}^p(u)
- \frac{i f_K m_K^2 v \cdot q}{36 \pi^2 t^3 ( m_u + m_s )} \phi_{3;K}^\sigma(u)
+ \frac{i f_K}{72 t v \cdot q} \langle \bar s s \rangle \psi_{4;K}(u)
+ \frac{i f_K t}{1152 v\cdot q} \langle g_s \bar s \sigma G s\rangle \psi_{4;K}(u)
\\ \nonumber &&
+ \frac{i f_K}{24 \pi^2 t^3 v\cdot q} m_s \psi_{4;K}(u)
- \frac{ f_K m_K^2}{144 (m_u + m_s)} m_s \langle \bar s s \rangle \phi_{3;K}^p(u)
- \frac{i f_K m_K^2 t v\cdot q}{864 ( m_u + m_s)} m_s \langle \bar s s \rangle \phi_{3;K}^\sigma(u) \Big ) \, .
\end{eqnarray}

\subsection{$[\mathbf{6}_F, 0, 1, \lambda]$}
\label{sec:601lambda}

The sum rule for $\Sigma_c^0({1\over2}^-)$ belonging to $[\mathbf{6}_F, 0, 1, \lambda]$ is
\begin{eqnarray}
&& G_{\Sigma_c^0[{1\over2}^-] \rightarrow \Lambda_c^{+}\pi^-} (\omega, \omega^\prime)
= { g_{\Sigma_c^0[{1\over2}^-] \rightarrow \Lambda_c^{+}\pi^-} f_{\Sigma_c^0[{1\over2}^-]} f_{\Lambda_c^{+}} \over (\bar \Lambda_{\Sigma_c^0[{1\over2}^-]} - \omega^\prime) (\bar \Lambda_{\Lambda_c^{+}} - \omega)}
\\ \nonumber &=& \int_0^\infty dt \int_0^1 du e^{i (1-u) \omega^\prime t} e^{i u \omega t} \times 8 \times \Big (
\\ \nonumber &&
- \frac{3 f_\pi m_\pi^2}{4 \pi^2 t^4 (m_u + m_d)} \phi_{3;\pi}^p(u)
- \frac{i f_\pi m_\pi^2 v \cdot q}{8 \pi^2 t^3 ( m_u + m_d )} \phi_{3;\pi}^\sigma(u)
+ \frac{i f_\pi}{16 t v \cdot q} \langle \bar q q \rangle \psi_{4;\pi}(u)
+ \frac{i f_\pi t}{256 v\cdot q} \langle g_s \bar q \sigma G q\rangle \psi_{4;\pi}(u) \Big ) \, ,
\\ && G_{\Sigma_c^0[{1\over2}^-] \rightarrow \Sigma_c^{+}\pi^-} (\omega, \omega^\prime)
= { g_{\Sigma_c^0[{1\over2}^-] \rightarrow \Sigma_c^{+}\pi^-} f_{\Sigma_c^0[{1\over2}^-]} f_{\Sigma_c^{+}} \over (\bar \Lambda_{\Sigma_c^0[{1\over2}^-]} - \omega^\prime) (\bar \Lambda_{\Sigma_c^{+}} - \omega)}
= 0 \, .
 \end{eqnarray}

The sum rule for $\Xi_c^{\prime0}({1\over2}^-)$ belonging to $[\mathbf{6}_F, 0, 1, \lambda]$ is
\begin{eqnarray}
&& G_{\Xi_c^{\prime0}[{1\over2}^-] \rightarrow \Xi_c^{+}\pi^-} (\omega, \omega^\prime)
= { g_{\Xi_c^{\prime0}[{1\over2}^-] \rightarrow \Xi_c^{+}\pi^-} f_{\Xi_c^{\prime0}[{1\over2}^-]} f_{\Xi_c^{+}} \over (\bar \Lambda_{\Xi_c^{\prime0}[{1\over2}^-]} - \omega^\prime) (\bar \Lambda_{\Xi_c^{+}} - \omega)}
\\ \nonumber &=& \int_0^\infty dt \int_0^1 du e^{i (1-u) \omega^\prime t} e^{i u \omega t} \times 4 \times \Big (
\\ \nonumber &&
- \frac{3 f_\pi m_\pi^2}{4 \pi^2 t^4 (m_u + m_d)} \phi_{3;\pi}^p(u)
- \frac{i f_\pi m_\pi^2 v \cdot q}{8 \pi^2 t^3 ( m_u + m_d )} \phi_{3;\pi}^\sigma(u)
+ \frac{i f_\pi}{16 t v \cdot q} \langle \bar s s \rangle \psi_{4;\pi}(u)
+ \frac{i f_\pi t}{256 v\cdot q} \langle g_s \bar s \sigma G s\rangle \psi_{4;\pi}(u)
\\ \nonumber &&
+ \frac{3 i f_\pi}{16 \pi^2 t^3 v\cdot q} m_s \psi_{4;\pi}(u)
- \frac{ f_\pi m_\pi^2}{32 (m_u + m_d)} m_s \langle \bar s s \rangle \phi_{3;\pi}^p(u)
- \frac{i f_\pi m_\pi^2 t v\cdot q}{192 ( m_u + m_d)} m_s \langle \bar s s \rangle \phi_{3;\pi}^\sigma(u) \Big ) \, ,
\\ && G_{\Xi_c^{\prime0}[{1\over2}^-] \rightarrow \Xi_c^{\prime+}\pi^-} (\omega, \omega^\prime)
= { g_{\Xi_c^{\prime0}[{1\over2}^-] \rightarrow \Xi_c^{\prime+}\pi^-} f_{\Xi_c^{\prime0}[{1\over2}^-]} f_{\Xi_c^{\prime+}} \over (\bar \Lambda_{\Xi_c^{\prime0}[{1\over2}^-]} - \omega^\prime) (\bar \Lambda_{\Xi_c^{\prime+}} - \omega)}
= 0 \, ,
\\ && G_{\Xi_c^{\prime0}[{1\over2}^-] \rightarrow \Lambda_c^{+}K^-} (\omega, \omega^\prime)
= { g_{\Xi_c^{\prime0}[{1\over2}^-] \rightarrow \Lambda_c^{+}K^-} f_{\Xi_c^{\prime0}[{1\over2}^-]} f_{\Lambda_c^{+}} \over (\bar \Lambda_{\Xi_c^{\prime0}[{1\over2}^-]} - \omega^\prime) (\bar \Lambda_{\Lambda_c^{+}} - \omega)}
\\ \nonumber &=& \int_0^\infty dt \int_0^1 du e^{i (1-u) \omega^\prime t} e^{i u \omega t} \times 4 \times \Big (
\\ \nonumber &&
- \frac{3 f_K m_K^2}{4 \pi^2 t^4 (m_u + m_s)} \phi_{3;K}^p(u)
- \frac{i f_K m_K^2 v \cdot q}{8 \pi^2 t^3 ( m_u + m_s )} \phi_{3;K}^\sigma(u)
+ \frac{i f_K}{16 t v \cdot q} \langle \bar q q \rangle \psi_{4;K}(u)
+ \frac{i f_K t}{256 v\cdot q} \langle g_s \bar q \sigma G q\rangle \psi_{4;K}(u) \Big ) \, ,
\\ && G_{\Xi_c^{\prime0}[{1\over2}^-] \rightarrow \Sigma_c^{+}K^-} (\omega, \omega^\prime)
= { g_{\Xi_c^{\prime0}[{1\over2}^-] \rightarrow \Sigma_c^{+}K^-} f_{\Xi_c^{\prime0}[{1\over2}^-]} f_{\Sigma_c^{+}} \over (\bar \Lambda_{\Xi_c^{\prime0}[{1\over2}^-]} - \omega^\prime) (\bar \Lambda_{\Sigma_c^{+}} - \omega)}
= 0 \, .
 \end{eqnarray}

The sum rule for $\Omega_c^0({1\over2}^-)$ belonging to $[\mathbf{6}_F, 0, 1, \lambda]$ is
\begin{eqnarray}
&& G_{\Omega_c^0[{1\over2}^-] \rightarrow \Xi_c^{+}K^{-}} (\omega, \omega^\prime)
= { g_{\Omega_c^0[{1\over2}^-] \rightarrow \Xi_c^{+}K^{-}} f_{\Omega_c^0[{1\over2}^-]} f_{\Xi_c^{+}} \over (\bar \Lambda_{\Omega_c^0[{1\over2}^-]} - \omega^\prime) (\bar \Lambda_{\Xi_c^{+}} - \omega)}
\\ \nonumber &=& \int_0^\infty dt \int_0^1 du e^{i (1-u) \omega^\prime t} e^{i u \omega t} \times 8 \times \Big (
\\ \nonumber &&
- \frac{3 f_K m_K^2}{4 \pi^2 t^4 (m_u + m_s)} \phi_{3;K}^p(u)
- \frac{i f_K m_K^2 v \cdot q}{8 \pi^2 t^3 ( m_u + m_s )} \phi_{3;K}^\sigma(u)
+ \frac{i f_K}{16 t v \cdot q} \langle \bar s s \rangle \psi_{4;K}(u)
+ \frac{i f_K t}{256 v\cdot q} \langle g_s \bar s \sigma G s\rangle \psi_{4;K}(u)
\\ \nonumber &&
+ \frac{3 i f_K}{16 \pi^2 t^3 v\cdot q} m_s \psi_{4;K}(u)
- \frac{ f_K m_K^2}{32 (m_u + m_s)} m_s \langle \bar s s \rangle \phi_{3;K}^p(u)
- \frac{i f_K m_K^2 t v\cdot q}{192 ( m_u + m_s)} m_s \langle \bar s s \rangle \phi_{3;K}^\sigma(u) \Big ) \, ,
\\ && G_{\Omega_c^0[{1\over2}^-] \rightarrow \Xi_c^{\prime+}K^{-}} (\omega, \omega^\prime)
= { g_{\Omega_c^0[{1\over2}^-] \rightarrow \Xi_c^{\prime+}K^{-}} f_{\Omega_c^0[{1\over2}^-]} f_{\Xi_c^{\prime+}} \over (\bar \Lambda_{\Omega_c^0[{1\over2}^-]} - \omega^\prime) (\bar \Lambda_{\Xi_c^{\prime+}} - \omega)}
= 0 \, .
\end{eqnarray}

\subsection{$[\mathbf{6}_F, 1, 1, \lambda]$}
\label{sec:611lambda}

The sum rule for $\Sigma_c^0({1\over2}^-)$ belonging to $[\mathbf{6}_F, 1, 1, \lambda]$ is
\begin{eqnarray}
&& G_{\Sigma_c^0[{1\over2}^-] \rightarrow \Lambda_c^{+}\pi^-} (\omega, \omega^\prime)
= { g_{\Sigma_c^0[{1\over2}^-] \rightarrow \Lambda_c^{+}\pi^-} f_{\Sigma_c^0[{1\over2}^-]} f_{\Lambda_c^{+}} \over (\bar \Lambda_{\Sigma_c^0[{1\over2}^-]} - \omega^\prime) (\bar \Lambda_{\Lambda_c^{+}} - \omega)}
= 0 \, ,
\\ && G_{\Sigma_c^0[{1\over2}^-] \rightarrow \Sigma_c^{+}\pi^-} (\omega, \omega^\prime)
= { g_{\Sigma_c^0[{1\over2}^-] \rightarrow \Sigma_c^{+}\pi^-} f_{\Sigma_c^0[{1\over2}^-]} f_{\Sigma_c^{+}} \over (\bar \Lambda_{\Sigma_c^0[{1\over2}^-]} - \omega^\prime) (\bar \Lambda_{\Sigma_c^{+}} - \omega)}
\\ \nonumber &=& \int_0^\infty dt \int_0^1 du e^{i (1-u) \omega^\prime t} e^{i u \omega t} \times 8 \times \Big (
\\ \nonumber &&
\frac{3 i f_\pi v \cdot q }{2 \pi^2 t^4} \phi_{2;\pi}(u)
+ \frac{3 i f_\pi v \cdot q}{32 \pi^2 t^2} \phi_{4;\pi}(u)
- \frac{i f_\pi m_\pi^2 v \cdot q}{24 (m_u + m_d)} {\langle \bar q q \rangle} \phi^\sigma_{3;\pi}(u)
- \frac{i f_\pi m_\pi^2 t^2 v \cdot q}{384 (m_u + m_d)} \langle g_s \bar q \sigma G q\rangle \phi_{3;\pi}^\sigma(u) \Big )
\\ \nonumber &+& \int_0^\infty dt \int_0^1 du \int \mathcal{D}\underline{\alpha} e^{i \omega^\prime t (\alpha_2 + u\alpha_3)} e^{i \omega t(1-\alpha_2-u\alpha_3)} \times 8 \times \Big (
\\ \nonumber &&
- \frac{i f_\pi v\cdot q}{8 \pi^2 t^2} \Phi_{4;\pi}(\underline{\alpha})
- \frac{i f_\pi v\cdot q}{4 \pi^2 t^2} \Psi_{4;\pi}(\underline{\alpha})
- \frac{3i f_\pi v\cdot q}{8 \pi^2 t^2} \widetilde \Phi_{4;\pi}(\underline{\alpha})
+ \frac{i f_\pi v\cdot q}{4 \pi^2 t^2} \widetilde \Psi_{4;\pi}(\underline{\alpha})
\\ \nonumber &&
- \frac{3i f_\pi u v \cdot q }{4 \pi^2 t^2} \Phi_{4;\pi}(\underline{\alpha})
+ \frac{i f_\pi u v\cdot q}{2 \pi^2 t^2} \Psi_{4;\pi}(\underline{\alpha}) \Big ) \, .
\end{eqnarray}

The sum rule for $\Xi_c^{\prime0}({1\over2}^-)$ belonging to $[\mathbf{6}_F, 1, 1, \lambda]$ is
\begin{eqnarray}
&& G_{\Xi_c^{\prime0}[{1\over2}^-] \rightarrow \Xi_c^{+}\pi^-} (\omega, \omega^\prime)
= { g_{\Xi_c^{\prime0}[{1\over2}^-] \rightarrow \Xi_c^{+}\pi^-} f_{\Xi_c^{\prime0}[{1\over2}^-]} f_{\Xi_c^{+}} \over (\bar \Lambda_{\Xi_c^{\prime0}[{1\over2}^-]} - \omega^\prime) (\bar \Lambda_{\Xi_c^{+}} - \omega)}
= 0 \, ,
\\ && G_{\Xi_c^{\prime0}[{1\over2}^-] \rightarrow \Xi_c^{\prime+}\pi^-} (\omega, \omega^\prime)
= { g_{\Xi_c^{\prime0}[{1\over2}^-] \rightarrow \Xi_c^{\prime+}\pi^-} f_{\Xi_c^{\prime0}[{1\over2}^-]} f_{\Xi_c^{\prime+}} \over (\bar \Lambda_{\Xi_c^{\prime0}[{1\over2}^-]} - \omega^\prime) (\bar \Lambda_{\Xi_c^{\prime+}} - \omega)}
\\ \nonumber &=&
\int_0^\infty dt \int_0^1 du e^{i (1-u) \omega^\prime t} e^{i u \omega t} \times 4 \times \Big (
\\ \nonumber &&
\frac{3 i f_\pi v \cdot q }{2 \pi^2 t^4} \phi_{2;\pi}(u)
+ \frac{3 i f_\pi v \cdot q}{32 \pi^2 t^2} \phi_{4;\pi}(u)
- \frac{i f_\pi m_\pi^2 v \cdot q}{24 (m_u + m_d)} {\langle \bar s s \rangle} \phi^\sigma_{3;\pi}(u)
- \frac{i f_\pi m_\pi^2 t^2 v \cdot q}{384 (m_u + m_d)} \langle g_s \bar s \sigma G s\rangle \phi_{3;\pi}^\sigma(u)
\\ \nonumber &&
- \frac{i f_\pi m_\pi^2 v \cdot q}{8 \pi^2 t^2 (m_u + m_d)} m_s \phi_{3;\pi}^\sigma(u)
+ \frac{i f_\pi v \cdot q}{16} m_s {\langle \bar s s \rangle} \phi_{2;\pi}(u)
+ \frac{i f_\pi t^2 v\cdot q}{256} m_s {\langle \bar s s \rangle} \phi_{4;\pi}(u) \Big)
\\ \nonumber &+&
\int_0^\infty dt \int_0^1 du \int \mathcal{D}\underline{\alpha} e^{i \omega^\prime t (\alpha_2 + u\alpha_3)} e^{i \omega t(1-\alpha_2-u\alpha_3)} \times 4 \times \Big (
\\ \nonumber &&
- \frac{i f_\pi v\cdot q}{8 \pi^2 t^2} \Phi_{4;\pi}(\underline{\alpha})
- \frac{i f_\pi v\cdot q}{4 \pi^2 t^2} \Psi_{4;\pi}(\underline{\alpha})
- \frac{3i f_\pi v\cdot q}{8 \pi^2 t^2} \widetilde \Phi_{4;\pi}(\underline{\alpha})
+ \frac{i f_\pi v\cdot q}{4 \pi^2 t^2} \widetilde \Psi_{4;\pi}(\underline{\alpha})
\\ \nonumber &&
- \frac{3i f_\pi u v \cdot q }{4 \pi^2 t^2} \Phi_{4;\pi}(\underline{\alpha})
+ \frac{i f_\pi u v\cdot q}{2 \pi^2 t^2} \Psi_{4;\pi}(\underline{\alpha}) \Big ) \, ,
\\ && G_{\Xi_c^{\prime0}[{1\over2}^-] \rightarrow \Lambda_c^{+}K^-} (\omega, \omega^\prime)
= { g_{\Xi_c^{\prime0}[{1\over2}^-] \rightarrow \Lambda_c^{+}K^-} f_{\Xi_c^{\prime0}[{1\over2}^-]} f_{\Lambda_c^{+}} \over (\bar \Lambda_{\Xi_c^{\prime0}[{1\over2}^-]} - \omega^\prime) (\bar \Lambda_{\Lambda_c^{+}} - \omega)}
= 0 \, ,
\\ && G_{\Xi_c^{\prime0}[{1\over2}^-] \rightarrow \Sigma_c^{+}K^-} (\omega, \omega^\prime)
= { g_{\Xi_c^{\prime0}[{1\over2}^-] \rightarrow \Sigma_c^{+}K^-} f_{\Xi_c^{\prime0}[{1\over2}^-]} f_{\Sigma_c^{+}} \over (\bar \Lambda_{\Xi_c^{\prime0}[{1\over2}^-]} - \omega^\prime) (\bar \Lambda_{\Sigma_c^{+}} - \omega)}
\\ \nonumber &=& \int_0^\infty dt \int_0^1 du e^{i (1-u) \omega^\prime t} e^{i u \omega t} \times 4 \times \Big (
\\ \nonumber &&
\frac{3 i f_K v \cdot q }{2 \pi^2 t^4} \phi_{2;K}(u)
+ \frac{3 i f_K v \cdot q}{32 \pi^2 t^2} \phi_{4;K}(u)
- \frac{i f_K m_K^2 v \cdot q}{24 (m_u + m_s)} {\langle \bar q q \rangle} \phi^\sigma_{3;K}(u)
- \frac{i f_K m_K^2 t^2 v \cdot q}{384 (m_u + m_s)} \langle g_s \bar q \sigma G q\rangle \phi_{3;K}^\sigma(u) \Big )
\\ \nonumber &+& \int_0^\infty dt \int_0^1 du \int \mathcal{D}\underline{\alpha} e^{i \omega^\prime t (\alpha_2 + u\alpha_3)} e^{i \omega t(1-\alpha_2-u\alpha_3)} \times 4 \times \Big (
\\ \nonumber &&
- \frac{i f_K v\cdot q}{8 \pi^2 t^2} \Phi_{4;K}(\underline{\alpha})
- \frac{i f_K v\cdot q}{4 \pi^2 t^2} \Psi_{4;K}(\underline{\alpha})
- \frac{3i f_K v\cdot q}{8 \pi^2 t^2} \widetilde \Phi_{4;K}(\underline{\alpha})
+ \frac{i f_K v\cdot q}{4 \pi^2 t^2} \widetilde \Psi_{4;K}(\underline{\alpha})
\\ \nonumber &&
- \frac{3i f_K u v \cdot q }{4 \pi^2 t^2} \Phi_{4;K}(\underline{\alpha})
+ \frac{i f_K u v\cdot q}{2 \pi^2 t^2} \Psi_{4;K}(\underline{\alpha}) \Big ) \, .
 \end{eqnarray}

The sum rule for $\Omega_c^0({1\over2}^-)$ belonging to $[\mathbf{6}_F, 1, 1, \lambda]$ is
\begin{eqnarray}
&& G_{\Omega_c^0[{1\over2}^-] \rightarrow \Xi_c^{+}K^{-}} (\omega, \omega^\prime)
= { g_{\Omega_c^0[{1\over2}^-] \rightarrow \Xi_c^{+}K^{-}} f_{\Omega_c^0[{1\over2}^-]} f_{\Xi_c^{+}} \over (\bar \Lambda_{\Omega_c^0[{1\over2}^-]} - \omega^\prime) (\bar \Lambda_{\Xi_c^{+}} - \omega)}
= 0 \, ,
\\ && G_{\Omega_c^0[{1\over2}^-] \rightarrow \Xi_c^{\prime+}K^{-}} (\omega, \omega^\prime)
= { g_{\Omega_c^0[{1\over2}^-] \rightarrow \Xi_c^{\prime+}K^{-}} f_{\Omega_c^0[{1\over2}^-]} f_{\Xi_c^{\prime+}} \over (\bar \Lambda_{\Omega_c^0[{1\over2}^-]} - \omega^\prime) (\bar \Lambda_{\Xi_c^{\prime+}} - \omega)}
\\ \nonumber &=&
\int_0^\infty dt \int_0^1 du e^{i (1-u) \omega^\prime t} e^{i u \omega t} \times 8 \times \Big (
\\ \nonumber &&
\frac{3 i f_K v \cdot q }{2 \pi^2 t^4} \phi_{2;K}(u)
+ \frac{3 i f_K v \cdot q}{32 \pi^2 t^2} \phi_{4;K}(u)
- \frac{i f_K m_K^2 v \cdot q}{24 (m_u + m_s)} {\langle \bar s s \rangle} \phi^\sigma_{3;K}(u)
- \frac{i f_K m_K^2 t^2 v \cdot q}{384 (m_u + m_s)} \langle g_s \bar s \sigma G s\rangle \phi_{3;K}^\sigma(u)
\\ \nonumber &&
- \frac{i f_K m_K^2 v \cdot q}{8 \pi^2 t^2 (m_u + m_s)} m_s \phi_{3;K}^\sigma(u)
+ \frac{i f_K v \cdot q}{16} m_s {\langle \bar s s \rangle} \phi_{2;K}(u)
+ \frac{i f_K t^2 v\cdot q}{256} m_s {\langle \bar s s \rangle} \phi_{4;K}(u) \Big)
\\ \nonumber &+&
\int_0^\infty dt \int_0^1 du \int \mathcal{D}\underline{\alpha} e^{i \omega^\prime t (\alpha_2 + u\alpha_3)} e^{i \omega t(1-\alpha_2-u\alpha_3)} \times 8 \times \Big (
\\ \nonumber &&
- \frac{i f_K v\cdot q}{8 \pi^2 t^2} \Phi_{4;K}(\underline{\alpha})
- \frac{i f_K v\cdot q}{4 \pi^2 t^2} \Psi_{4;K}(\underline{\alpha})
- \frac{3i f_K v\cdot q}{8 \pi^2 t^2} \widetilde \Phi_{4;K}(\underline{\alpha})
+ \frac{i f_K v\cdot q}{4 \pi^2 t^2} \widetilde \Psi_{4;K}(\underline{\alpha})
\\ \nonumber &&
- \frac{3i f_K u v \cdot q }{4 \pi^2 t^2} \Phi_{4;K}(\underline{\alpha})
+ \frac{i f_K u v\cdot q}{2 \pi^2 t^2} \Psi_{4;K}(\underline{\alpha}) \Big ) \, .
\end{eqnarray}

The sum rule for $\Sigma_c^0[{3\over2}^-]$ belonging to $[\mathbf{6}_F, 1, 1, \lambda]$ is
\begin{eqnarray}
&& G_{\Sigma_c^0[{3\over2}^-] \rightarrow \Sigma_c^{*+}\pi^-} (\omega, \omega^\prime)
= { g_{\Sigma_c^0[{3\over2}^-] \rightarrow \Sigma_c^{*+}\pi^-} f_{\Sigma_c^0[{3\over2}^-]} f_{\Sigma_c^{*+}} \over (\bar \Lambda_{\Sigma_c^0[{3\over2}^-]} - \omega^\prime) (\bar \Lambda_{\Sigma_c^{*+}} - \omega)}
\\ \nonumber &=& \int_0^\infty dt \int_0^1 du e^{i (1-u) \omega^\prime t} e^{i u \omega t} \times 8 \times \Big (
\\ \nonumber &&
- \frac{f_\pi v \cdot q }{3 \pi^2 t^4} \phi_{2;\pi}(u)
- \frac{f_\pi v \cdot q}{48 \pi^2 t^2} \phi_{4;\pi}(u)
+ \frac{f_\pi m_\pi^2 v \cdot q}{108 (m_u + m_d)} {\langle \bar q q \rangle} \phi^\sigma_{3;\pi}(u)
+ \frac{f_\pi m_\pi^2 t^2 v \cdot q}{1728 (m_u + m_d)} \langle g_s \bar q \sigma G q\rangle \phi_{3;\pi}^\sigma(u) \Big )
\\ \nonumber &+& \int_0^\infty dt \int_0^1 du \int \mathcal{D}\underline{\alpha} e^{i \omega^\prime t (\alpha_2 + u\alpha_3)} e^{i \omega t(1-\alpha_2-u\alpha_3)} \times 8 \times \Big (
\\ \nonumber &&
\frac{f_\pi v\cdot q}{72 \pi^2 t^2} \Phi_{4;\pi}(\underline{\alpha})
+ \frac{5 f_\pi v\cdot q}{72 \pi^2 t^2} \Psi_{4;\pi}(\underline{\alpha})
+ \frac{f_\pi v\cdot q}{24 \pi^2 t^2} \widetilde \Phi_{4;\pi}(\underline{\alpha})
- \frac{5 f_\pi v\cdot q}{72 \pi^2 t^2} \widetilde \Psi_{4;\pi}(\underline{\alpha})
\\ \nonumber &&
+ \frac{f_\pi u v \cdot q }{6 \pi^2 t^2} \Phi_{4;\pi}(\underline{\alpha})
- \frac{7 f_\pi u v\cdot q}{72 \pi^2 t^2} \Psi_{4;\pi}(\underline{\alpha})
+ \frac{f_\pi u v\cdot q}{24 \pi^2 t^2} \widetilde \Psi_{4;\pi}(\underline{\alpha}) \Big ) \, .
\end{eqnarray}

The sum rule for $\Xi_c^{\prime0}({3\over2}^-)$ belonging to $[\mathbf{6}_F, 1, 1, \lambda]$ is
\begin{eqnarray}
&& G_{\Xi_c^{\prime0}[{1\over2}^-] \rightarrow \Xi_c^{*+}\pi^-} (\omega, \omega^\prime)
= { g_{\Xi_c^{\prime0}[{1\over2}^-] \rightarrow \Xi_c^{*+}\pi^-} f_{\Xi_c^{\prime0}[{1\over2}^-]} f_{\Xi_c^{*+}} \over (\bar \Lambda_{\Xi_c^{\prime0}[{1\over2}^-]} - \omega^\prime) (\bar \Lambda_{\Xi_c^{*+}} - \omega)}
\\ \nonumber &=& \int_0^\infty dt \int_0^1 du e^{i (1-u) \omega^\prime t} e^{i u \omega t} \times 4 \times \Big (
\\ \nonumber &&
- \frac{f_\pi v \cdot q }{3 \pi^2 t^4} \phi_{2;\pi}(u)
- \frac{f_\pi v \cdot q}{48 \pi^2 t^2} \phi_{4;\pi}(u)
+ \frac{f_\pi m_\pi^2 v \cdot q}{108 (m_u + m_d)} {\langle \bar s s \rangle} \phi^\sigma_{3;\pi}(u)
+ \frac{f_\pi m_\pi^2 t^2 v \cdot q}{1728 (m_u + m_d)} \langle g_s \bar s \sigma G s\rangle \phi_{3;\pi}^\sigma(u)
\\ \nonumber &&
+ \frac{f_\pi m_\pi^2 v \cdot q}{36 \pi^2 t^2 (m_u + m_d)} m_s \phi_{3;\pi}^\sigma(u)
- \frac{f_\pi v \cdot q}{72} m_s {\langle \bar s s \rangle} \phi_{2;\pi}(u)
- \frac{f_\pi t^2 v\cdot q}{1152} m_s {\langle \bar s s \rangle} \phi_{4;\pi}(u) \Big )
\\ \nonumber &+& \int_0^\infty dt \int_0^1 du \int \mathcal{D}\underline{\alpha} e^{i \omega^\prime t (\alpha_2 + u\alpha_3)} e^{i \omega t(1-\alpha_2-u\alpha_3)} \times 4 \times \Big (
\\ \nonumber && \frac{f_\pi v\cdot q}{72 \pi^2 t^2} \Phi_{4;\pi}(\underline{\alpha})
+ \frac{5 f_\pi v\cdot q}{72 \pi^2 t^2} \Psi_{4;\pi}(\underline{\alpha})
+ \frac{f_\pi v\cdot q}{24 \pi^2 t^2} \widetilde \Phi_{4;\pi}(\underline{\alpha})
- \frac{5 f_\pi v\cdot q}{72 \pi^2 t^2} \widetilde \Psi_{4;\pi}(\underline{\alpha})
\\ \nonumber &&
+ \frac{f_\pi u v \cdot q }{6 \pi^2 t^2} \Phi_{4;\pi}(\underline{\alpha})
- \frac{7 f_\pi u v\cdot q}{72 \pi^2 t^2} \Psi_{4;\pi}(\underline{\alpha})
+ \frac{f_\pi u v\cdot q}{24 \pi^2 t^2} \widetilde \Psi_{4;\pi}(\underline{\alpha}) \Big ) \, ,
\\ && G_{\Xi_c^{\prime0}[{1\over2}^-] \rightarrow \Sigma_c^{*+}K^-} (\omega, \omega^\prime)
= { g_{\Xi_c^{\prime0}[{1\over2}^-] \rightarrow \Sigma_c^{*+}K^-} f_{\Xi_c^{\prime0}[{1\over2}^-]} f_{\Sigma_c^{*+}} \over (\bar \Lambda_{\Xi_c^{\prime0}[{1\over2}^-]} - \omega^\prime) (\bar \Lambda_{\Sigma_c^{*+}} - \omega)}
\\ \nonumber &=& \int_0^\infty dt \int_0^1 du e^{i (1-u) \omega^\prime t} e^{i u \omega t} \times 4 \times \Big (
\\ \nonumber &&
- \frac{f_K v \cdot q }{3 \pi^2 t^4} \phi_{2;K}(u)
- \frac{f_K v \cdot q}{48 \pi^2 t^2} \phi_{4;K}(u)
+ \frac{f_K m_K^2 v \cdot q}{108 (m_u + m_s)} {\langle \bar q q \rangle} \phi^\sigma_{3;K}(u)
+ \frac{f_K m_K^2 t^2 v \cdot q}{1728 (m_u + m_s)} \langle g_s \bar q \sigma G q\rangle \phi_{3;K}^\sigma(u) \Big )
\\ \nonumber &+& \int_0^\infty dt \int_0^1 du \int \mathcal{D}\underline{\alpha} e^{i \omega^\prime t (\alpha_2 + u\alpha_3)} e^{i \omega t(1-\alpha_2-u\alpha_3)} \times 4 \times \Big (
\\ \nonumber &&
\frac{f_K v\cdot q}{72 \pi^2 t^2} \Phi_{4;K}(\underline{\alpha})
+ \frac{5 f_K v\cdot q}{72 \pi^2 t^2} \Psi_{4;K}(\underline{\alpha})
+ \frac{f_K v\cdot q}{24 \pi^2 t^2} \widetilde \Phi_{4;K}(\underline{\alpha})
- \frac{5 f_K v\cdot q}{72 \pi^2 t^2} \widetilde \Psi_{4;K}(\underline{\alpha})
\\ \nonumber &&
+ \frac{f_K u v \cdot q }{6 \pi^2 t^2} \Phi_{4;K}(\underline{\alpha})
- \frac{7 f_K u v\cdot q}{72 \pi^2 t^2} \Psi_{4;K}(\underline{\alpha})
+ \frac{f_K u v\cdot q}{24 \pi^2 t^2} \widetilde \Psi_{4;K}(\underline{\alpha}) \Big ) \, .
 \end{eqnarray}

The sum rule for $\Omega_c^0({3\over2}^-)$ belonging to $[\mathbf{6}_F, 1, 1, \lambda]$ is
\begin{eqnarray}
&& G_{\Omega_c^0[{1\over2}^-] \rightarrow \Xi_c^{*+}K^{-}} (\omega, \omega^\prime)
= { g_{\Omega_c^0[{1\over2}^-] \rightarrow \Xi_c^{*+}K^{-}} f_{\Omega_c^0[{1\over2}^-]} f_{\Xi_c^{*+}} \over (\bar \Lambda_{\Omega_c^0[{1\over2}^-]} - \omega^\prime) (\bar \Lambda_{\Xi_c^{*+}} - \omega)}
\\ \nonumber &=&
\int_0^\infty dt \int_0^1 du e^{i (1-u) \omega^\prime t} e^{i u \omega t} \times 8 \times \Big (
\\ \nonumber &&
- \frac{f_K v \cdot q }{3 \pi^2 t^4} \phi_{2;K}(u)
- \frac{f_K v \cdot q}{48 \pi^2 t^2} \phi_{4;K}(u)
+ \frac{f_K m_K^2 v \cdot q}{108 (m_u + m_s)} {\langle \bar s s \rangle} \phi^\sigma_{3;K}(u)
+ \frac{f_K m_K^2 t^2 v \cdot q}{1728 (m_u + m_s)} \langle g_s \bar s \sigma G s\rangle \phi_{3;K}^\sigma(u)
\\ \nonumber &&
+ \frac{f_K m_K^2 v \cdot q}{36 \pi^2 t^2 (m_u + m_s)} m_s \phi_{3;K}^\sigma(u)
- \frac{f_K v \cdot q}{72} m_s {\langle \bar s s \rangle} \phi_{2;K}(u)
- \frac{f_K t^2 v\cdot q}{1152} m_s {\langle \bar s s \rangle} \phi_{4;K}(u) \Big )
\\ \nonumber &+& \int_0^\infty dt \int_0^1 du \int \mathcal{D}\underline{\alpha} e^{i \omega^\prime t (\alpha_2 + u\alpha_3)} e^{i \omega t(1-\alpha_2-u\alpha_3)} \times 8 \times \Big (
\\ \nonumber && \frac{f_K v\cdot q}{72 \pi^2 t^2} \Phi_{4;K}(\underline{\alpha})
+ \frac{5 f_K v\cdot q}{72 \pi^2 t^2} \Psi_{4;K}(\underline{\alpha})
+ \frac{f_K v\cdot q}{24 \pi^2 t^2} \widetilde \Phi_{4;K}(\underline{\alpha})
- \frac{5 f_K v\cdot q}{72 \pi^2 t^2} \widetilde \Psi_{4;K}(\underline{\alpha})
\\ \nonumber &&
+ \frac{f_K u v \cdot q }{6 \pi^2 t^2} \Phi_{4;K}(\underline{\alpha})
- \frac{7 f_K u v\cdot q}{72 \pi^2 t^2} \Psi_{4;K}(\underline{\alpha})
+ \frac{f_K u v\cdot q}{24 \pi^2 t^2} \widetilde \Psi_{4;K}(\underline{\alpha}) \Big ) \, .
\end{eqnarray}

\subsection{$[\mathbf{6}_F, 2, 1, \lambda]$}
\label{sec:621lambda}

The sum rule for $\Sigma_c^0[{3\over2}^-]$ belonging to $[\mathbf{6}_F, 2, 1, \lambda]$ is
\begin{eqnarray}
&& G_{\Sigma_c^0[{3\over2}^-] \rightarrow \Sigma_c^{*+}\pi^-} (\omega, \omega^\prime)
= { g_{\Sigma_c^0[{3\over2}^-] \rightarrow \Sigma_c^{*+}\pi^-} f_{\Sigma_c^0[{3\over2}^-]} f_{\Sigma_c^{*+}} \over (\bar \Lambda_{\Sigma_c^0[{3\over2}^-]} - \omega^\prime) (\bar \Lambda_{\Sigma_c^{*+}} - \omega)}
\\ \nonumber &=& \int_0^\infty dt \int_0^1 du \int \mathcal{D}\underline{\alpha} e^{i \omega^\prime t (\alpha_2 + u\alpha_3)} e^{i \omega t(1-\alpha_2-u\alpha_3)} \times 8 \times \Big (
\\ \nonumber &&
\frac{f_\pi v\cdot q}{24 \pi^2 t^2} \Phi_{4;\pi}(\underline{\alpha})
+ \frac{f_\pi v\cdot q}{24 \pi^2 t^2} \Psi_{4;\pi}(\underline{\alpha})
- \frac{ f_\pi v\cdot q}{24 \pi^2 t^2} \widetilde \Phi_{4;\pi}(\underline{\alpha})
- \frac{f_\pi v\cdot q}{24 \pi^2 t^2} \widetilde \Psi_{4;\pi}(\underline{\alpha})
- \frac{f_\pi u v\cdot q}{24 \pi^2 t^2} \Psi_{4;\pi}(\underline{\alpha})
+ \frac{f_\pi u v\cdot q}{24 \pi^2 t^2} \widetilde \Psi_{4;\pi}(\underline{\alpha}) \Big ) \, .
\end{eqnarray}

The sum rule for $\Xi_c^0({3\over2}^-)$ belonging to $[\mathbf{6}_F, 2, 1, \lambda]$ is
\begin{eqnarray}
&& G_{\Xi_c^{\prime0}[{3\over2}^-] \rightarrow \Xi_c^{*+}\pi^-} (\omega, \omega^\prime)
= { g_{\Xi_c^{\prime0}[{3\over2}^-] \rightarrow \Xi_c^{*+}\pi^-} f_{\Xi_c^{\prime0}[{3\over2}^-]} f_{\Xi_c^{*+}} \over (\bar \Lambda_{\Xi_c^{\prime0}[{3\over2}^-]} - \omega^\prime) (\bar \Lambda_{\Xi_c^{*+}} - \omega)}
\\ \nonumber &=& \int_0^\infty dt \int_0^1 du \int \mathcal{D}\underline{\alpha} e^{i \omega^\prime t (\alpha_2 + u\alpha_3)} e^{i \omega t(1-\alpha_2-u\alpha_3)} \times 4 \times \Big (
\\ \nonumber &&
\frac{f_\pi v\cdot q}{24 \pi^2 t^2} \Phi_{4;\pi}(\underline{\alpha})
+ \frac{f_\pi v\cdot q}{24 \pi^2 t^2} \Psi_{4;\pi}(\underline{\alpha})
- \frac{ f_\pi v\cdot q}{24 \pi^2 t^2} \widetilde \Phi_{4;\pi}(\underline{\alpha})
- \frac{f_\pi v\cdot q}{24 \pi^2 t^2} \widetilde \Psi_{4;\pi}(\underline{\alpha})
- \frac{f_\pi u v\cdot q}{24 \pi^2 t^2} \Psi_{4;\pi}(\underline{\alpha})
+ \frac{f_\pi u v\cdot q}{24 \pi^2 t^2} \widetilde \Psi_{4;\pi}(\underline{\alpha}) \Big ) \, ,
\\ && G_{\Xi_c^{\prime0}[{3\over2}^-] \rightarrow \Sigma_c^{*+}K^-} (\omega, \omega^\prime)
= { g_{\Xi_c^{\prime0}[{3\over2}^-] \rightarrow \Sigma_c^{*+}K^-} f_{\Xi_c^{\prime0}[{3\over2}^-]} f_{\Sigma_c^{*+}} \over (\bar \Lambda_{\Xi_c^{\prime0}[{3\over2}^-]} - \omega^\prime) (\bar \Lambda_{\Sigma_c^{*+}} - \omega)}
\\ \nonumber &=& \int_0^\infty dt \int_0^1 du \int \mathcal{D}\underline{\alpha} e^{i \omega^\prime t (\alpha_2 + u\alpha_3)} e^{i \omega t(1-\alpha_2-u\alpha_3)} \times 4 \times \Big (
\\ \nonumber &&
\frac{f_K v\cdot q}{24 \pi^2 t^2} \Phi_{4;K}(\underline{\alpha})
+ \frac{f_K v\cdot q}{24 \pi^2 t^2} \Psi_{4;K}(\underline{\alpha})
- \frac{ f_K v\cdot q}{24 \pi^2 t^2} \widetilde \Phi_{4;K}(\underline{\alpha})
- \frac{f_K v\cdot q}{24 \pi^2 t^2} \widetilde \Psi_{4;K}(\underline{\alpha})
\\ \nonumber &&
- \frac{f_K u v\cdot q}{24 \pi^2 t^2} \Psi_{4;K}(\underline{\alpha})
+ \frac{f_K u v\cdot q}{24 \pi^2 t^2} \widetilde \Psi_{4;K}(\underline{\alpha}) \Big ) \, .
 \end{eqnarray}

The sum rule for $\Omega_c^0({3\over2}^-)$ belonging to $[\mathbf{6}_F, 2, 1, \lambda]$ is
\begin{eqnarray}
&& G_{\Omega_c^0[{3\over2}^-] \rightarrow \Xi_c^{*+}K^{-}} (\omega, \omega^\prime)
= { g_{\Omega_c^0[{3\over2}^-] \rightarrow \Xi_c^{*+}K^{-}} f_{\Omega_c^0[{3\over2}^-]} f_{\Xi_c^{*+}} \over (\bar \Lambda_{\Omega_c^0[{3\over2}^-]} - \omega^\prime) (\bar \Lambda_{\Xi_c^{*+}} - \omega)}
\\ \nonumber &=&
\int_0^\infty dt \int_0^1 du e^{i (1-u) \omega^\prime t} e^{i u \omega t} \times 8 \times \Big (
\\ \nonumber &&
\frac{f_K v\cdot q}{24 \pi^2 t^2} \Phi_{4;K}(\underline{\alpha})
+ \frac{f_K v\cdot q}{24 \pi^2 t^2} \Psi_{4;K}(\underline{\alpha})
- \frac{ f_K v\cdot q}{24 \pi^2 t^2} \widetilde \Phi_{4;K}(\underline{\alpha})
- \frac{f_K v\cdot q}{24 \pi^2 t^2} \widetilde \Psi_{4;K}(\underline{\alpha})
\\ \nonumber &&
- \frac{f_K u v\cdot q}{24 \pi^2 t^2} \Psi_{4;K}(\underline{\alpha})
+ \frac{f_K u v\cdot q}{24 \pi^2 t^2} \widetilde \Psi_{4;K}(\underline{\alpha}) \Big ) \, .
\end{eqnarray}


\begin{thebibliography}{99}

\bibitem{Aaij:2017nav}
  R.~Aaij {\it et al.} [LHCb Collaboration],
  Observation of five new narrow $\Omega_c^0$ states decaying to $\Xi_c^+ K^-$,
  arXiv:1703.04639 [hep-ex].

\bibitem{pdg}
  C.~Patrignani {\it et al.} [Particle Data Group],
  Review of Particle Physics,
  Chin.\ Phys.\ C {\bf 40}, no. 10, 100001 (2016).

\bibitem{Frabetti:1993hg}
  P.~L.~Frabetti {\it et al.}  [E687 Collaboration],
  An Observation of an excited state of the $\Lambda_c^+$ baryon,
  Phys.\ Rev.\ Lett.\  {\bf 72}, 961 (1994).

\bibitem{Albrecht:1993pt}
  H.~Albrecht {\it et al.}  [ARGUS Collaboration],
  Observation of a new charmed baryon,
  Phys.\ Lett.\ B {\bf 317}, 227 (1993).

\bibitem{Edwards:1994ar}
  K.~W.~Edwards {\it et al.}  [CLEO Collaboration],
  Observation of excited baryon states decaying to $\Lambda_c^+ \pi^+ \pi^-$,
  Phys.\ Rev.\ Lett.\  {\bf 74}, 3331 (1995).

\bibitem{Alexander:1999ud}
  J.~P.~Alexander {\it et al.}  [CLEO Collaboration],
  Evidence of new states decaying into $\Xi_c^* \pi$,
  Phys.\ Rev.\ Lett.\  {\bf 83}, 3390 (1999).

\bibitem{Mizuk:2004yu}
  R.~Mizuk {\it et al.}  [Belle Collaboration],
  Observation of an isotriplet of excited charmed baryons decaying to $\Lambda^+_c \pi$,
  Phys.\ Rev.\ Lett.\  {\bf 94}, 122002 (2005).

\bibitem{Chistov:2006zj}
  R.~Chistov {\it et al.} [Belle Collaboration],
  Observation of new states decaying into $ \Lambda_c^+ K^- \pi^+$ and $\Lambda_c^+ K^0_S \pi^-$,
  Phys.\ Rev.\ Lett.\  {\bf 97}, 162001 (2006).

\bibitem{Aubert:2007dt}
  B.~Aubert {\it et al.} [BaBar Collaboration],
  A Study of Excited Charm-Strange Baryons with Evidence for new Baryons $\Xi_c(3055)^+$ and $\Xi_c(3123)^+$,
  Phys.\ Rev.\ D {\bf 77}, 012002 (2008).

\bibitem{Aubert:2008ax}
  B.~Aubert {\it et al.} [BaBar Collaboration],
  Measurements of $B(\bar B^0 \to \Lambda_c^+ \bar p)$ and $B(B^- \to \Lambda_c^+ \bar p \pi^-)$ and Studies of $\Lambda_c^+ \pi^-$ Resonances,
  Phys.\ Rev.\ D {\bf 78}, 112003 (2008).

\bibitem{Yelton:2016fqw}
  J.~Yelton {\it et al.} [Belle Collaboration],
  Study of Excited $\Xi_c$ States Decaying into $\Xi_c^0$ and $\Xi_c^+$ Baryons,
  Phys.\ Rev.\ D {\bf 94}, 052011 (2016).

\bibitem{Kato:2016hca}
  Y.~Kato {\it et al.} [Belle Collaboration],
  Studies of charmed strange baryons in the $\Lambda$D final state at Belle,
  Phys.\ Rev.\ D {\bf 94}, 032002 (2016).

\bibitem{Aaij:2017vbw}
  R.~Aaij {\it et al.} [LHCb Collaboration],
  Study of the $D^0 p$ amplitude in $\Lambda_b^0\to D^0 p \pi^-$ decays,
  arXiv:1701.07873 [hep-ex].

\bibitem{Chen:2016spr}
  H.~X.~Chen, W.~Chen, X.~Liu, Y.~R.~Liu and S.~L.~Zhu,
  A review of the open charm and open bottom mesons,
  arXiv:1609.08928 [hep-ph].

\bibitem{Capstick:1986bm}
  S.~Capstick and N.~Isgur,
  Baryons in a Relativized Quark Model with Chromodynamics,
  Phys.\ Rev.\ D {\bf 34}, 2809 (1986).

\bibitem{Ebert:2007nw}
  D.~Ebert, R.~N.~Faustov and V.~O.~Galkin,
  Masses of excited charmed baryons in the relativistic quark model,
  Phys.\ Lett.\ B {\bf 659}, 612 (2008).

\bibitem{Garcilazo:2007eh}
  H.~Garcilazo, J.~Vijande and A.~Valcarce,
  Faddeev study of charmed baryon spectroscopy,
  J.\ Phys.\ G {\bf 34}, 961 (2007).

\bibitem{Ebert:2011kk}
  D.~Ebert, R.~N.~Faustov and V.~O.~Galkin,
  Spectroscopy and Regge trajectories of charmed baryons in the relativistic quark-diquark picture,
  Phys.\ Rev.\ D {\bf 84}, 014025 (2011).

\bibitem{Ortega:2012cx}
  P.~G.~Ortega, D.~R.~Entem and F.~Fernandez,
  Quark model description of the $\Lambda_c(2940)^+$ as a molecular $D^*N$ state and the possible existence of the $\Lambda_b(6248)$,
  Phys.\ Lett.\ B {\bf 718}, 1381 (2013).

\bibitem{Shah:2016nxi}
  Z.~Shah, K.~Thakkar, A.~K.~Rai and P.~C.~Vinodkumar,
  Mass spectra and Regge trajectories of $\Lambda_{c}^{+}$, $\Sigma_{c}^{0}$, $\Xi_{c}^{0}$ and $\Omega_{c}^{0}$ baryons,
  Chin.\ Phys.\ C {\bf 40}, 123102 (2016).

\bibitem{GarciaRecio:2012db}
  C.~Garcia-Recio, J.~Nieves, O.~Romanets, L.~L.~Salcedo and L.~Tolos,
  Odd parity bottom-flavored baryon resonances,
  Phys.\ Rev.\ D {\bf 87}, 034032 (2013).

\bibitem{Liang:2014eba}
  W.~H.~Liang, C.~W.~Xiao and E.~Oset,
  Baryon states with open beauty in the extended local hidden gauge approach,
  Phys.\ Rev.\ D {\bf 89}, 054023 (2014).

\bibitem{Liang:2014kra}
  W.~H.~Liang, T.~Uchino, C.~W.~Xiao and E.~Oset,
  Baryon states with open charm in the extended local hidden gauge approach,
  Eur.\ Phys.\ J.\ A {\bf 51}, 16 (2015).

\bibitem{Liang:2016ydj}
  W.~H.~Liang, M.~Bayar and E.~Oset,
  $\Lambda_b \to \pi^- (D_s^-) \Lambda_c(2595),~\pi^- (D_s^-) \Lambda_c(2625)$ decays and $DN,~D^*N$ molecular components,
  Eur.\ Phys.\ J.\ C {\bf 77}, 39 (2017).

\bibitem{Liang:2016exm}
  W.~H.~Liang, E.~Oset and Z.~S.~Xie,
  Semileptonic $\Lambda_b \to \bar \nu_l l \Lambda_c(2595)$ and $\Lambda_b \to \bar \nu_l l \Lambda_c(2625)$ decays in the molecular picture of $\Lambda_c(2595)$ and $\Lambda_c(2625)$,
  Phys.\ Rev.\ D {\bf 95}, 014015 (2017).

\bibitem{Lu:2014ina}
  J.~X.~Lu, Y.~Zhou, H.~X.~Chen, J.~J.~Xie and L.~S.~Geng,
  Dynamically generated $J^P=1/2^-(3/2^-)$ singly charmed and bottom charmed baryons,
  Phys.\ Rev.\ D {\bf 92}, 014036 (2015).

\bibitem{Copley:1979wj}
  L.~A.~Copley, N.~Isgur and G.~Karl,
  Charmed Baryons in a Quark Model with Hyperfine Interactions,
  Phys.\ Rev.\ D {\bf 20}, 768 (1979)
  [Erratum-ibid.\ D {\bf 23}, 817 (1981)].

\bibitem{Karliner:2008sv}
  M.~Karliner, B.~Keren-Zur, H.~J.~Lipkin and J.~L.~Rosner,
  The Quark Model and $b$ Baryons,
  Annals Phys.\  {\bf 324}, 2 (2009).

\bibitem{Bowler:1996ws}
  K.~C.~Bowler {\it et al.}  [UKQCD Collaboration],
  charmed baryon spectroscopy from the lattice,
  Phys.\ Rev.\ D {\bf 54}, 3619 (1996).

\bibitem{Burch:2008qx}
  T.~Burch, C.~Hagen, C.~B.~Lang, M.~Limmer and A.~Sch\"afer,
  Excitations of single-beauty hadrons,
  Phys.\ Rev.\ D {\bf 79}, 014504 (2009).

\bibitem{Brown:2014ena}
  Z.~S.~Brown, W.~Detmold, S.~Meinel and K.~Orginos,
  Charmed bottom baryon spectroscopy from lattice QCD,
  Phys.\ Rev.\ D {\bf 90}, no. 9, 094507 (2014).

\bibitem{Roncaglia:1995az}
  R.~Roncaglia, D.~B.~Lichtenberg and E.~Predazzi,
  Predicting the masses of baryons containing one or two heavy quarks,
  Phys.\ Rev.\ D {\bf 52}, 1722 (1995).

\bibitem{Jenkins:1996de}
  E.~E.~Jenkins,
  Charmed baryon masses in the $1/m_Q$ and $1/N_c$ expansions,
  Phys.\ Rev.\ D {\bf 54}, 4515 (1996).

\bibitem{Roberts:2007ni}
  W.~Roberts and M.~Pervin,
  Charmed baryons in a quark model,
  Int.\ J.\ Mod.\ Phys.\ A {\bf 23}, 2817 (2008).

\bibitem{Chen:2014nyo}
  B.~Chen, K.~W.~Wei and A.~Zhang,
  Assignments of $\Lambda_Q$ and $\Xi_Q$ baryons in the heavy quark-light diquark picture,
  Eur.\ Phys.\ J.\ A {\bf 51}, 82 (2015).

\bibitem{Lu:2016ctt}
  Q.~F.~L\"u, Y.~Dong, X.~Liu and T.~Matsuki,
  Puzzle of the $\Lambda_c$ spectrum,
  arXiv:1610.09605 [hep-ph].

\bibitem{Cheng:2006dk}
  H.~Y.~Cheng and C.~K.~Chua,
  Strong decays of charmed baryons in heavy hadron chiral perturbation theory,
  Phys.\ Rev.\ D {\bf 75}, 014006 (2007).

\bibitem{Chen:2007xf}
  C.~Chen, X.~L.~Chen, X.~Liu, W.~Z.~Deng and S.~L.~Zhu,
  Strong decays of charmed baryons,
  Phys.\ Rev.\ D {\bf 75}, 094017 (2007).

\bibitem{Zhong:2007gp}
  X.~H.~Zhong and Q.~Zhao,
  Charmed baryon strong decays in a chiral quark model,
  Phys.\ Rev.\ D {\bf 77}, 074008 (2008).

\bibitem{Kim:2014qha}
  S.~H.~Kim, A.~Hosaka, H.~C.~Kim, H.~Noumi and K.~Shirotori,
  Pion induced Reactions for Charmed Baryons,
  PTEP {\bf 2014}, 103D01 (2014).

\bibitem{Nagahiro:2016nsx}
  H.~Nagahiro, S.~Yasui, A.~Hosaka, M.~Oka and H.~Noumi,
  Structure of charmed baryons studied by pionic decays,
  Phys.\ Rev.\ D {\bf 95}, 014023 (2017).

\bibitem{Xie:2015zga}
  J.~J.~Xie, Y.~B.~Dong and X.~Cao,
  Role of the $\Lambda^+_c(2940)$ in the $\pi^- p \to D^- D^0 p$ reaction close to threshold,
  Phys.\ Rev.\ D {\bf 92}, 034029 (2015).

\bibitem{Huang:2016ygf}
  Y.~Huang, J.~He, J.~J.~Xie, X.~Chen and H.~F.~Zhang,
  Production of charmed baryon $\Lambda_c(2940)$ by kaon-induced reaction on a proton target,
  arXiv:1610.06994 [hep-ph].
  
\bibitem{Korner:1994nh}
  J.~G.~K\"orner, M.~Kramer and D.~Pirjol,
  Charmed baryons,
  Prog.\ Part.\ Nucl.\ Phys.\  {\bf 33}, 787 (1994).

\bibitem{Bianco:2003vb}
  S.~Bianco, F.~L.~Fabbri, D.~Benson and I.~Bigi,
  A Cicerone for the physics of charm,
  Riv.\ Nuovo Cim.\  {\bf 26N7}, 1 (2003).

\bibitem{Klempt:2009pi}
  E.~Klempt and J.~M.~Richard,
  Baryon spectroscopy,
  Rev.\ Mod.\ Phys.\  {\bf 82}, 1095 (2010).

\bibitem{Crede:2013sze}
  V.~Crede and W.~Roberts,
  Progress towards understanding baryon resonances,
  Rept.\ Prog.\ Phys.\  {\bf 76}, 076301 (2013).

\bibitem{Cheng:2015rra}
  H.~Y.~Cheng,
  Charmed baryons circa 2015,
  Front.\ Phys.\ (Beijing) {\bf 10}, no. 6, 101406 (2015).

\bibitem{Chen:2016qju}
  H.~X.~Chen, W.~Chen, X.~Liu and S.~L.~Zhu,
  The hidden-charm pentaquark and tetraquark states,
  Phys.\ Rept.\  {\bf 639}, 1 (2016).

\bibitem{Liu:2007fg}
  X.~Liu, H.~X.~Chen, Y.~R.~Liu, A.~Hosaka and S.~L.~Zhu,
  Bottom baryons,
  Phys.\ Rev.\ D {\bf 77}, 014031 (2008).

\bibitem{Chen:2015kpa}
  H.~X.~Chen, W.~Chen, Q.~Mao, A.~Hosaka, X.~Liu and S.~L.~Zhu,
  P-wave charmed baryons from QCD sum rules,
  Phys.\ Rev.\ D {\bf 91}, 054034 (2015).

\bibitem{Mao:2015gya}
  Q.~Mao, H.~X.~Chen, W.~Chen, A.~Hosaka, X.~Liu and S.~L.~Zhu,
  QCD sum rule calculation for P-wave bottom baryons,
  Phys.\ Rev.\ D {\bf 92}, 114007 (2015).

\bibitem{Chen:2016phw}
  H.~X.~Chen, Q.~Mao, A.~Hosaka, X.~Liu and S.~L.~Zhu,
  D-wave charmed and bottomed baryons from QCD sum rules,
  Phys.\ Rev.\ D {\bf 94}, 114016 (2016).

\bibitem{Shifman:1978bx}
  M.~A.~Shifman, A.~I.~Vainshtein and V.~I.~Zakharov,
  QCD And Resonance Physics. Sum Rules,
  Nucl.\ Phys.\ B {\bf 147}, 385 (1979).

\bibitem{Reinders:1984sr}
  L.~J.~Reinders, H.~Rubinstein and S.~Yazaki,
  Hadron Properties From QCD Sum Rules,
  Phys.\ Rept.\  {\bf 127}, 1 (1985).

\bibitem{Eichten:1989zv}
  E.~Eichten and B.~R.~Hill,
  An Effective Field Theory for the Calculation of Matrix Elements Involving Heavy Quarks,
  Phys.\ Lett.\ B {\bf 234}, 511 (1990).

\bibitem{Grinstein:1990mj}
  B.~Grinstein,
  The Static Quark Effective Theory,
  Nucl.\ Phys.\ B {\bf 339}, 253 (1990).

\bibitem{Falk:1990yz}
  A.~F.~Falk, H.~Georgi, B.~Grinstein and M.~B.~Wise,
  Heavy Meson Form-factors From {QCD},
  Nucl.\ Phys.\ B {\bf 343}, 1 (1990).

\bibitem{Ivanov:1999bu}
  M.~A.~Ivanov, J.~G.~Korner, V.~E.~Lyubovitskij and A.~G.~Rusetsky,
  $\Lambda_b$ and $\Lambda_c$ baryon decays at finite values of heavy quark masses,
  Phys.\ Lett.\ B {\bf 476}, 58 (2000).

\bibitem{Ivanov:1998ms}
  M.~A.~Ivanov, Y.~L.~Kalinovsky and C.~D.~Roberts,
  Survey of heavy meson observables,
  Phys.\ Rev.\ D {\bf 60}, 034018 (1999).

\bibitem{Shuryak:1981fza}
  E.~V.~Shuryak,
  Hadrons Containing a Heavy Quark and QCD Sum Rules,
  Nucl.\ Phys.\ B {\bf 198}, 83 (1982).

\bibitem{Bagan:1991sg}
  E.~Bagan, P.~Ball, V.~M.~Braun and H.~G.~Dosch,
  QCD sum rules in the effective heavy quark theory,
  Phys.\ Lett.\ B {\bf 278}, 457 (1992).

\bibitem{Neubert:1991sp}
  M.~Neubert,
  Heavy meson form-factors from QCD sum rules,
  Phys.\ Rev.\ D {\bf 45}, 2451 (1992).

\bibitem{Broadhurst:1991fc}
  D.~J.~Broadhurst and A.~G.~Grozin,
  Operator product expansion in static quark effective field theory: Large perturbative correction,
  Phys.\ Lett.\ B {\bf 274}, 421 (1992).

\bibitem{Grozin:1992td}
  A.~G.~Grozin and O.~I.~Yakovlev,
  Baryonic currents and their correlators in the heavy quark effective theory,
  Phys.\ Lett.\ B {\bf 285}, 254 (1992).

\bibitem{Neubert:1993mb}
  M.~Neubert,
  Heavy quark symmetry,
  Phys.\ Rept.\  {\bf 245}, 259 (1994).

\bibitem{Bagan:1993ii}
  E.~Bagan, M.~Chabab, H.~G.~Dosch and S.~Narison,
  Baryon sum rules in the heavy quark effective theory,
  Phys.\ Lett.\ B {\bf 301}, 243 (1993).

\bibitem{Dai:1993kt}
  Y.~B.~Dai, C.~S.~Huang and H.~Y.~Jin,
  Bethe-Salpeter wave functions and transition amplitudes for heavy mesons,
  Z.\ Phys.\ C {\bf 60}, 527 (1993).

\bibitem{Huang:1994zj}
  T.~Huang and C.~W.~Luo,
  Light quark dependence of the Isgur-Wise function from QCD sum rules,
  Phys.\ Rev.\ D {\bf 50}, 5775 (1994).

\bibitem{Dai:1995bc}
  Y.~B.~Dai, C.~S.~Huang, C.~Liu and C.~D.~Lu,
  $1/m$ corrections to charmed baryon masses in the heavy quark effective theory sum rules,
  Phys.\ Lett.\ B {\bf 371}, 99 (1996).

\bibitem{Dai:1996yw}
  Y.~B.~Dai, C.~S.~Huang, M.~Q.~Huang and C.~Liu,
  QCD sum rules for masses of excited heavy mesons,
  Phys.\ Lett.\ B {\bf 390}, 350 (1997).

\bibitem{Dai:1996qx}
  Y.~B.~Dai, C.~S.~Huang and M.~Q.~Huang,
  $O(1/m_Q)$ order corrections to masses of excited heavy mesons from QCD sum rules,
  Phys.\ Rev.\ D {\bf 55}, 5719 (1997).

\bibitem{Groote:1996em}
  S.~Groote, J.~G.~K\"orner and O.~I.~Yakovlev,
  QCD sum rules for charmed baryons at next-to-leading order in $\alpha_s$,
  Phys.\ Rev.\ D {\bf 55}, 3016 (1997).

\bibitem{Colangelo:1998ga}
  P.~Colangelo, F.~De Fazio and N.~Paver,
  Universal $\tau_{1/2}(y)$ Isgur-Wise function at the next-to-leading order in QCD sum rules,
  Phys.\ Rev.\ D {\bf 58}, 116005 (1998).

\bibitem{Lee:2000tb}
  J.~P.~Lee, C.~Liu and H.~S.~Song,
  QCD sum rule analysis of excited $\Lambda_c$ mass parameter,
  Phys.\ Lett.\ B {\bf 476}, 303 (2000).

\bibitem{Huang:2000tn}
  C.~S.~Huang, A.~l.~Zhang and S.~L.~Zhu,
  Excited charmed baryon masses in HQET QCD sum rules,
  Phys.\ Lett.\ B {\bf 492}, 288 (2000).

\bibitem{Wang:2003zp}
  D.~W.~Wang and M.~Q.~Huang,
  Excited charmed baryon masses to order $\Lambda_{\rm QCD}/m_Q$ from QCD sum rules,
  Phys.\ Rev.\ D {\bf 68}, 034019 (2003).

\bibitem{Dai:2003yg}
  Y.~B.~Dai, C.~S.~Huang, C.~Liu and S.~L.~Zhu,
  Understanding the $D^+_{sJ}(2317)$ and $D^+_{sJ}(2460)$ with sum rules in HQET,
  Phys.\ Rev.\ D {\bf 68}, 114011 (2003).

\bibitem{Zhou:2014ytp}
  D.~Zhou, E.~L.~Cui, H.~X.~Chen, L.~S.~Geng, X.~Liu and S.~L.~Zhu,
  The D-wave heavy-light mesons from QCD sum rules,
  Phys.\ Rev.\ D {\bf 90}, 114035 (2014).

\bibitem{Zhou:2015ywa}
  D.~Zhou, H.~X.~Chen, L.~S.~Geng, X.~Liu and S.~L.~Zhu,
  F-wave heavy-light meson spectroscopy in QCD sum rules and heavy quark effective theory,
  Phys.\ Rev.\ D {\bf 92}, 114015 (2015).

\bibitem{Bagan:1991sc}
  E.~Bagan, M.~Chabab, H.~G.~Dosch and S.~Narison,
  The charmed baryons $\Sigma_c$ $\Sigma_b$ from QCD spectral sum rules,
  Phys.\ Lett.\ B {\bf 278}, 367 (1992).

\bibitem{Bagan:1992tp}
  E.~Bagan, M.~Chabab, H.~G.~Dosch and S.~Narison,
  Spectra of charmed baryons from QCD spectral sum rules,
  Phys.\ Lett.\ B {\bf 287}, 176 (1992).

\bibitem{Duraes:2007te}
  F.~O.~Duraes and M.~Nielsen,
  QCD sum rules study of $\Xi_c$ and $\Xi_b$ baryons,
  Phys.\ Lett.\ B {\bf 658}, 40 (2007).

\bibitem{Wang:2007sqa}
  Z.~G.~Wang,
  Analysis of $\Omega_c^*(css)$ and $\Omega_b^*(bss)$ with QCD sum rules,
  Eur.\ Phys.\ J.\ C {\bf 54}, 231 (2008).

\bibitem{Chen:2015moa}
  H.~X.~Chen, W.~Chen, X.~Liu, T.~G.~Steele and S.~L.~Zhu,
  Towards exotic hidden-charm pentaquarks in QCD,
  Phys.\ Rev.\ Lett.\  {\bf 115}, 172001 (2015).

\bibitem{Balitsky:1989ry}
  I.~I.~Balitsky, V.~M.~Braun and A.~V.~Kolesnichenko,
  Radiative Decay $\Sigma^+ \to p \gamma$ in Quantum Chromodynamics,
  Nucl.\ Phys.\ B {\bf 312}, 509 (1989).

\bibitem{Braun:1988qv}
  V.~M.~Braun and I.~E.~Filyanov,
  QCD Sum Rules in Exclusive Kinematics and Pion Wave Function,
  Z.\ Phys.\ C {\bf 44}, 157 (1989)
  [Sov.\ J.\ Nucl.\ Phys.\  {\bf 50}, 511 (1989)]
  [Yad.\ Fiz.\  {\bf 50}, 818 (1989)].

\bibitem{Chernyak:1990ag}
  V.~L.~Chernyak and I.~R.~Zhitnitsky,
  B meson exclusive decays into baryons,
  Nucl.\ Phys.\ B {\bf 345}, 137 (1990).

\bibitem{Ball:1998je}
  P.~Ball,
  Theoretical update of pseudoscalar meson distribution amplitudes of higher twist: The Nonsinglet case,
  JHEP {\bf 9901}, 010 (1999).

\bibitem{Ball:2006wn}
  P.~Ball, V.~M.~Braun and A.~Lenz,
  Higher-twist distribution amplitudes of the K meson in QCD,
  JHEP {\bf 0605}, 004 (2006).

\bibitem{Ball:2004rg}
  P.~Ball and R.~Zwicky,
  $B_{d,s} \to  \rho, \omega, K^*, \phi$ decay form-factors from light-cone sum rules revisited,
  Phys.\ Rev.\ D {\bf 71}, 014029 (2005).

\bibitem{Ball:1998kk}
  P.~Ball and V.~M.~Braun,
  Exclusive semileptonic and rare B meson decays in QCD,
  Phys.\ Rev.\ D {\bf 58}, 094016 (1998).

\bibitem{Ball:1998sk}
  P.~Ball, V.~M.~Braun, Y.~Koike and K.~Tanaka,
  Higher twist distribution amplitudes of vector mesons in QCD: Formalism and twist-three distributions,
  Nucl.\ Phys.\ B {\bf 529}, 323 (1998).

\bibitem{Ball:1998ff}
  P.~Ball and V.~M.~Braun,
  Higher twist distribution amplitudes of vector mesons in QCD: Twist-4 distributions and meson mass corrections,
  Nucl.\ Phys.\ B {\bf 543}, 201 (1999).

\bibitem{Ball:2007rt}
  P.~Ball and G.~W.~Jones,
  Twist-3 distribution amplitudes of $K^*$ and $\phi$ mesons,
  JHEP {\bf 0703}, 069 (2007).

\bibitem{Ball:2007zt}
  P.~Ball, V.~M.~Braun and A.~Lenz,
  Twist-4 distribution amplitudes of the $K^*$ and $\phi$ mesons in QCD,
  JHEP {\bf 0708}, 090 (2007).

\bibitem{Wang:2007mc}
  Z.~G.~Wang,
  Analysis of the vertices $DDV$ and $D^*DV$ with light-cone QCD sum rules,
  Eur.\ Phys.\ J.\ C {\bf 52}, 553 (2007).

\bibitem{Wang:2009hra}
  Y.~M.~Wang, Y.~L.~Shen and C.~D.~Lu,
  $\Lambda_b \to p$, $\Lambda$ transition form factors from QCD light-cone sum rules,
  Phys.\ Rev.\ D {\bf 80}, 074012 (2009).

\bibitem{Aliev:2010uy}
  T.~M.~Aliev, K.~Azizi and M.~Savci,
  Analysis of the $Lambda_{b}\rightarrow \Lambda \ell^+\ell^- $ decay in QCD,
  Phys.\ Rev.\ D {\bf 81}, 056006 (2010).

\bibitem{Sun:2010nv}
  Y.~J.~Sun, Z.~H.~Li and T.~Huang,
  $B_{(s)}\to S$ transitions in the light cone sum rules with the chiral current,
  Phys.\ Rev.\ D {\bf 83}, 025024 (2011).

\bibitem{Khodjamirian:2011jp}
  A.~Khodjamirian, C.~Klein, T.~Mannel and Y.-M.~Wang,
  Form Factors and Strong Couplings of charmed baryons from QCD Light-Cone Sum Rules,
  JHEP {\bf 1109}, 106 (2011).

\bibitem{Han:2013zg}
  H.~Y.~Han, X.~G.~Wu, H.~B.~Fu, Q.~L.~Zhang and T.~Zhong,
  Twist-3 Distribution Amplitudes of Scalar Mesons within the QCD Sum Rules and Its Application to the $B \to S$ Transition Form Factors,
  Eur.\ Phys.\ J.\ A {\bf 49}, 78 (2013).

\bibitem{Offen:2013nma}
  N.~Offen, F.~A.~Porkert and A.~Sch\"afer,
  Light-cone sum rules for the $D_{(s)}\rightarrow \eta^{(\prime)} l \nu_l$ form factor,
  Phys.\ Rev.\ D {\bf 88}, 034023 (2013).

\bibitem{Meissner:2013hya}
  U.~G.~Meissner and W.~Wang,
  Generalized Heavy-to-Light Form Factors in Light-Cone Sum Rules,
  Phys.\ Lett.\ B {\bf 730}, 336 (2014).

\bibitem{Aliev:2016xvq}
  T.~M.~Aliev, T.~Barakat and M.~Savci,
  Analysis of the radiative decays $\Sigma_Q \to \Lambda_Q \gamma$ and $\Xi^\prime_Q \to \Xi_Q \gamma$ in light cone sum rules,
  Phys.\ Rev.\ D {\bf 93},  056007 (2016).

\bibitem{Dai:1996xv}
  Y.~B.~Dai, C.~S.~Huang, M.~Q.~Huang and C.~Liu,
  QCD sum rule analysis for the $\Lambda_b \to \Lambda_c$ semileptonic decay,
  Phys.\ Lett.\ B {\bf 387}, 379 (1996).

\bibitem{Zhu:2000py}
  S.~L.~Zhu,
  Strong and electromagnetic decays of p wave charmed baryons $\Lambda_{c1}$, $\Lambda^*_{c1}$,
  Phys.\ Rev.\ D {\bf 61}, 114019 (2000).

\bibitem{Wei:2005ag}
  W.~Wei, P.~Z.~Huang and S.~L.~Zhu,
  Strong decays of $D_{sJ}(2317)$ and $D_{sJ}(2460)$,
  Phys.\ Rev.\ D {\bf 73}, 034004 (2006).

\bibitem{Huang:2009zy}
  P.~Z.~Huang, L.~Zhang and S.~L.~Zhu,
  The Vector Meson And Heavy Meson Strong Interaction,
  Phys.\ Rev.\ D {\bf 80}, 014023 (2009).

\bibitem{Huang:2009is}
  P.~Z.~Huang, H.~X.~Chen and S.~L.~Zhu,
  Light vector meson and charmed baryon strong interaction,
  Phys.\ Rev.\ D {\bf 80}, 094007 (2009).

\bibitem{Liu:2009sn}
  Y.~L.~Liu, M.~Q.~Huang and D.~W.~Wang,
  Improved analysis on the semi-leptonic decay $\Lambda_c \to \Lambda l^+ \nu$ from QCD light-cone sum rules,
  Phys.\ Rev.\ D {\bf 80}, 074011 (2009).

\bibitem{Huang:2010qa}
  P.~Z.~Huang, L.~Zhang and S.~L.~Zhu,
  Strong and Electromagnetic Decays of The $D$-wave Heavy Mesons,
  Phys.\ Rev.\ D {\bf 81}, 094025 (2010).

\bibitem{Huang:2010dc}
  P.~Z.~Huang, H.~X.~Chen and S.~L.~Zhu,
  The Strong Decay Patterns of the $1^{-+}$ Exotic Hybrid Mesons,
  Phys.\ Rev.\ D {\bf 83}, 014021 (2011).

\bibitem{Hussain:1999sp}
  F.~Hussain, J.~G.~Korner and S.~Tawfiq,
  One pion transitions between heavy baryons in the constituent quark model,
  Phys.\ Rev.\ D {\bf 61}, 114003 (2000).

\bibitem{Mertig:1990an}
  R.~Mertig, M.~B\"ohm and A.~Denner,
  FEYN CALC: Computer algebraic calculation of Feynman amplitudes,
  Comput.\ Phys.\ Commun.\  {\bf 64}, 345 (1991).

\bibitem{Yang:1993bp}
  K.~C.~Yang, W.~Y.~P.~Hwang, E.~M.~Henley and L.~S.~Kisslinger,
  QCD sum rules and neutron proton mass difference,
  Phys.\ Rev.\ D {\bf 47}, 3001 (1993).

\bibitem{Hwang:1994vp}
  W.~Y.~P.~Hwang and K.~C.~Yang,
  QCD sum rules: $\Delta$-$N$ and $\Sigma^0$-$\Lambda$ mass splittings,
  Phys.\ Rev.\ D {\bf 49}, 460 (1994).

\bibitem{Ovchinnikov:1988gk}
  A.~A.~Ovchinnikov and A.~A.~Pivovarov,
  QCD Sum Rule Calculation Of The Quark Gluon Condensate,
  Sov.\ J.\ Nucl.\ Phys.\  {\bf 48}, 721 (1988)
  [Yad.\ Fiz.\  {\bf 48}, 1135 (1988)].

\bibitem{Narison:2002pw}
  S.~Narison,
  QCD as a theory of hadrons (from partons to confinement),
  Camb.\ Monogr.\ Part.\ Phys.\ Nucl.\ Phys.\ Cosmol.\  {\bf 17}, 1 (2002).

\bibitem{Jamin:2002ev}
  M.~Jamin,
  Flavour-symmetry breaking of the quark condensate and chiral corrections to the Gell-Mann-Oakes-Renner relation,
  Phys.\ Lett.\ B {\bf 538}, 71 (2002).

\bibitem{Ioffe:2002be}
  B.~L.~Ioffe and K.~N.~Zyablyuk,
  Gluon condensate in charmonium sum rules with 3-loop corrections,
  Eur.\ Phys.\ J.\ C {\bf 27}, 229 (2003).

\bibitem{Gimenez:2005nt}
  V.~Gimenez, V.~Lubicz, F.~Mescia, V.~Porretti and J.~Reyes,
  Operator product expansion and quark condensate from lattice QCD in coordinate space,
  Eur.\ Phys.\ J.\ C {\bf 41}, 535 (2005).

\bibitem{colangelo}
  P.~Colangelo and A.~Khodjamirian,
  {\it ``At the Frontier of Particle Physics/Handbook of QCD''}
  (World Scientific, Singapore, 2001), Volume 3, 1495.

\bibitem{Albrecht:1997qa}
  H.~Albrecht {\it et al.} [ARGUS Collaboration],
  Evidence for $\Lambda_c^+(2593)$ production,
  Phys.\ Lett.\ B {\bf 402}, 207 (1997).

\bibitem{Aubert:2007eb}
  B.~Aubert {\it et al.} [BaBar Collaboration],
  A Study of $\bar B \to \Xi_c \bar \Lambda_c^-$ and $\bar B \to \Lambda^+_c \bar \Lambda^-_c \bar K$ decays at BABAR,
  Phys.\ Rev.\ D {\bf 77}, 031101 (2008).

\bibitem{Agaev:2017uky}
  S.~S.~Agaev, K.~Azizi and H.~Sundu,
  On the nature of the newly discovered $\Omega^0_c$ states,
  arXiv:1703.07091 [hep-ph].

\end{thebibliography}
\end{document}